%% file: TDE_ms2.tex
\documentclass[iop,numberedappendix,twocolappendix]{emulateapj} 
\slugcomment{Submitted to ApJ} 
\usepackage{natbib,amsmath,booktabs,apjfonts,threeparttable,multirow}
\usepackage{natbib,ifsym}
\usepackage{subfigure}
\usepackage{graphicx}
\usepackage{tabularx}
\usepackage[colorlinks=true,linkcolor=blue,citecolor=blue]{hyperref}
\usepackage{epstopdf}
\usepackage{enumitem}
\usepackage{color}
\usepackage{pdfpages}

\usepackage{longtable} 

\begin{document}
	
\title{New physical insights about Tidal Disruption Events \\ from a comprehensive observational inventory at X-ray wavelengths}

\author{Katie Auchettl\altaffilmark{1,2}, James Guillochon\altaffilmark{3,4}, and Enrico Ramirez-Ruiz\altaffilmark{5}}

\altaffiltext{1}{Center for Cosmology and Astro-Particle Physics, The Ohio State University, 191 West Woodruff Avenue, Columbus, OH 43210, USA}
\altaffiltext{2}{Department of Physics, The Ohio State University, 191 W. Woodruff Avenue, Columbus, OH 43210, USA}
\altaffiltext{3}{Harvard-Smithsonian Center for Astrophysics, 60 Garden St.,Cambridge, MA 02138 USA}
\altaffiltext{4}{Einstein Fellow}
\altaffiltext{5}{Department of Astronomy and Astrophysics, University of California, Santa Cruz, CA 95064, USA}

\begin{abstract}
We perform a comprehensive study of the X-ray emission from 70 transient sources which have been classified as a tidal disruption event (TDE) in the literature. We explore the properties of these candidates using nearly three decades of X-ray observations to quantify the properties and characteristics of X-ray TDEs observationally. We find that the emission from X-ray TDEs increase by two to four orders of magnitude compared to pre-flare constraints, which evolves significantly with time and decays with powerlaw indices that are typically shallower than the canonical $t^{-5/3}$ decay law, implying that X-ray TDEs are viscously delayed.  These events exhibit enhanced column densities relative to Galactic and are quite soft in nature, with no strong correlation between the amount of detected soft and hard emission. At peak, jetted events have an X-ray to optical ratio $\gg$1, while non-jetted events have a ratio $\sim$1, which suggests that these events undergo reprocessing at different rates. X-ray TDEs have long $T_{90}$ values consistent with that expected from a viscously-driven accretion disk formed by the disruption of a main-sequence star by a black hole with a mass $<$$10^{7}M_{\odot}$.  The isotropic luminosities of X-ray TDEs is bimodal such that jetted and non-jetted events are separated by a ``reprocessing valley'' which we suggest is naturally populated by optical/UV TDEs that most likely produce X-rays, but due to reprocessing this emission is ``veiled'' from observations. Our results suggest that non-jetted X-ray TDEs likely originate from partial disruptions and/or disruptions of low mass stars.

\end{abstract}


\section{Introduction}
Black holes (BHs) with masses greater than approximately $10^{5-6} M_{\odot}$ are thought to reside in the central nuclei of all active Galaxies \citep[see e.g.,][and a recent reviews by \citealt{2008ARA&A..46..475H, 2013ARA&A..51..511K, 2016ASSL..418..263G}]{1995ARA&A..33..581K, 1998AJ....115.2285M, 1998bhrs.conf...79R, 2009ApJ...698..198G}. Currently the most direct evidence for the existence of these massive objects comes from the detection of an Active Galactic Nuclei (AGN). AGN activity is characterised generally by recurring luminous X-ray flare emission, and Fe K line variability. The properties of these objects can only be explained by the continual accretion of material onto an object that must have a mass $>10^{6}M_{\odot}$ \citep[e.g.,][]{1969Natur.223..690L, 1963MNRAS.125..169H}. However, based on accretion models of black hole evolution \citep[e.g.,][]{2004MNRAS.351..169M} it is expected that there exists a large number of quiescent, weakly or non-accreting black holes BHs in which gas accretion occurs at a significantly slower rate \citep[e.g.,][]{2001AIPC..586..363K}. Evidence for the presence of these dormant BHs in non-active Galaxies such as the one found in our Milky Way arises from mostly indirect methods \citep[e.g.,][]{2003ApJ...594..812G, 2003ANS...324..527G}, making it difficult to probe the properties of these interesting objects. However, a more direct detection of a dormant massive BH at the center of a non-active Galaxy arises in the form of a tidal disruption event (TDE) \citep{1975Natur.254..295H, 1976MNRAS.176..633F, 1978PThPh..60.1692K, 1979PAZh....5...28L, 1979Natur.280..214G, 1982Natur.296..211C, 1985MNRAS.212...57L, 1988Natur.333..523R}. 

A TDE occurs when a star with a mass $M_{\star}$ that is orbiting around a massive BH of mass $M_{BH}$ approaches the BH at a radius less than the tidal disruption radius $R_{t} = (M_{\rm BH}/M_{\star})^{1/3}$ \citep{1975Natur.254..295H, 1988Natur.333..523R}. At this point the star is subjected to the strong tidal forces of the BH which can exceed the self-gravity of the star, ripping the star apart \citep{1975Natur.254..295H, 1982ApJ...262..120L, 1988Natur.333..523R}\footnote{One should note that for $M_{\rm BH} > 10^{8-9} M_{\odot}$ and a main sequence star, $R_{t}$ resides within the Schwarzschild radius of the black hole. As a consequence no TDE can occur \citep[see][]{2012PhRvD..85b4037K}}. A fraction of the stellar debris of this now destroyed star will be expelled on unbound orbits and escape from the BH, while $\sim0.5M_{\star}$ will be confined to highly eccentric, bound orbits \citep{1988Natur.333..523R, 1989ApJ...346L..13E, 2000ApJ...545..772A}. This material will eventually be accreted onto the BH producing a luminous, short-lived accretion-powered flare \citep{1982ApJ...262..120L, 1988Natur.333..523R, 1989ApJ...346L..13E, 1989Natur.340..595P}, which can emit above the Eddington Luminosity for a BH with $M_{\rm BH} < 10^{7} M_{\odot}$ \citep{2009MNRAS.400.2070S, 2011MNRAS.410..359L, 2012ApJ...760..103D}. This accretion powered flare peaks in the UV and/or soft X-ray bands \citep{1999ApJ...514..180U}\footnote{For BHs with $M_{BH} >10^{7} M_{\odot}$ the peak luminosity will be sub-Eddington \citep{2009MNRAS.400.2070S, 2011MNRAS.410..359L}}. The luminosity of these events is thought to loosely follow a power law decline characterised by $t^{-5/3}$, which is set by the timescale in which the stellar debris eventually returns to pericenter \citep[e.g.,][]{1989ApJ...346L..13E, 1989Natur.340..595P, 1990ApJ...351...38C, 1990Sci...247..817R}. However, exactly how depends heavily on the evolution and physics within the accretion stream \cite[e.g.,][]{1994ApJ...422..508K, 2009ApJ...697L..77R, 2011MNRAS.410..359L, 2013ApJ...767...25G, 2014PhRvD..90f4020C, 2015ApJ...809..166G, 2015ApJ...804...85S}.  In addition to this highly luminous UV/X-ray flare, there are several other sources of radiation that can be produced during this accretion event. This includes the formation of a non-thermal jet \citep[e.g.,][]{2006ApJ...645.1138C, 2011Natur.476..421B, 2011Sci...333..203B, 2016Sci...351...62V}, emission from the collision of tidal streams in bound orbits \citep[e.g.,][]{1999ApJ...519..647K, 2016ApJ...830..125J} or IR/optical/UV emission lines from photoionised ambient medium \citep[e.g.,][]{2009ApJ...701..105K}.

The detection and analysis of the observational properties of TDEs covering a wide range of redshifts can provide a wealth of knowledge about a number of important astrophysical processes \cite[see reviews by][]{2015arXiv150205720K, 2015JHEAp...7..148K}. This includes probing the physics associated with accretion and accretion disc formation under extreme conditions, as well as the formation and evolution of jets. TDEs also provide a way to determine the properties of the dormant BHs (such as mass and spin) in distant Galaxies, as well as aiding in the search for intermediate mass BHs and recoiling BHs. Similarly they can also highlight the properties of the gaseous environment surrounding a BH, allow one to characterise stellar kinematics in different Galaxies and learn about the populations of stars in their centres \citep{2012ApJ...757..134M, 2016MNRAS.461..371K}.

Theoretically, the rate at which TDEs are thought to occur is low. Assuming $M_{BH}\lesssim10^{7.5}M_{\odot}$, the theoretical rate is $\sim10^{-4}$ events per year \citep{1999MNRAS.309..447M, 2004ApJ...600..149W, 2016MNRAS.455..859S} . As a consequence, time domain surveys covering a wide field of view, such as the All-Sky Automated Survey for SuperNovae (ASAS-SN) \citep{2014ApJ...788...48S}, Palomar Transient Factory (PTF) \citep{2009PASP..121.1395L}, Panoramic Survey Telescope and Rapid Response System (PanSTARRS) \citep{2002SPIE.4836..154K} and the Sloan Digital Sky Survey (SDSS) \citep{2011ApJ...741...73V}, have been particularly importantly in detecting, and characterising the optical/UV light curves of these events. However, a significant fraction of the luminosity arising from a TDE accretion powered flare falls within the soft X-ray band, with a maximum luminosity of $10^{45}$ erg s$^{-1}$ for a BH with $M_{\rm BH}<10^{7}M_{\odot}$. As a consequence, the X-ray emission from a TDE will dominate the fainter, extended and more permanent X-ray emission of its host Galaxy (if this emission is present). As such searching for TDEs in the X-ray energy band has proven to be most fruitful in this endeavour, with the number of TDE candidates detected in X-rays now outnumbering those detected in UV/optical alone.

The first soft X-ray TDE candidates were discovered with the X-ray observatory \emph{ROSAT}. Due to its high sensitivity to soft X-rays ($0.1-2.4$ keV), low detector background, all-sky coverage and its eight year mission operation, these properties made it an ideal instrument to detect these transient events. Using the results from the \emph{ROSAT} All-Sky Survey (RASS) that was completed by \emph{ROSAT} during its first year of operation, strong, luminous flares from inactive galaxies\footnote{These Galaxies were classified as inactive based on lack of radio, optical and X-ray emission prior to and after the flare was detected.} were detected. The first soft X-ray TDEs were identified from NGC5905\citep{1996A&A...309L..35B, 1999A&A...343..775K}, IC3599 \citep{1995A&A...299L...5G, 1995MNRAS.273L..47B}, RX J1242-1119 \citep{1999A&A...349L..45K}, RX J1624+7554 \citep{1999A&A...350L..31G} and RX J1420+5334 \citep{2000A&A...362L..25G}. These ``\emph{ROSAT}'' events are characterised by a very soft X-ray flare that peaks with a luminosity of $\sim10^{44}$ erg s$^{-1}$; the X-ray luminosity hardens and declines over a timescale of months to years and appeared to follow the $t^{-5/3}$ law as determined using follow up observations of these sources using current X-ray satellites such as \emph{Chandra} \citep[e.g.,][]{2004ApJ...604..572H, 2004MNRAS.349L...1V, 2004ApJ...603L..17K} and \emph{XMM-Newton} \citep[e.g.,][]{2004ApJ...603L..17K}; are coincident with the center of the host Galaxy (within the error circle of \emph{ROSAT}); and the host galaxies showed no evidence of permanent AGN activity. 

Since \emph{ROSAT}, the \emph{Chandra} X-ray Observatory, \emph{XMM-Newton} Space Observatory and particularly the \emph{Swift} Gamma-ray Burst Mission has dramatically changed our ability to be able to detect, and follow-up potential TDEs in the X-ray energy band. The increased sensitivity, as well as spatial and spectral resolution of these instruments, has allowed us to determine the location, and luminosity and spectral evolution of these events in detail. Using both dedicated observations of TDEs (e.g., \textit{ASASSN-14li}: \citealt{2015Natur.526..542M, 2016MNRAS.455.2918H}; \textit{Swift J1644+57}: \citealt{2011GCN..11847...1B, 2011Sci...333..203B, 2011Natur.476..421B}) and serendipitous discoveries \citep[e.g., \emph{XMM-Newton} slew survey:][]{2007A&A...462L..49E}, this has significantly increased the number of TDE candidates detected in X-rays. In addition, leading to the discovery of rapid variability in the X-ray emission during the first few weeks of detection \citep[e.g.,][]{2012A&A...541A.106S}, the possible discovery of jet formation from these events \citep[e.g.,][]{2011Sci...333..203B, 2011Natur.476..421B}, and the discovery of TDEs occurring in dwarf Galaxies and clusters of Galaxies \citep[e.g.,][]{2009A&A...495L...9C, 2010ApJ...722.1035M, 2013MNRAS.435.1904M, 2014ApJ...792L..29M, 2014ApJ...781...59D}.

Currently, there are $\sim$70 TDE candidates listed in the literature\footnote{see \url{https://tde.space} for a listing of all TDEs so far mentioned in published papers, or Astronomer's Telegram}. Apart from the TDE candidates in which an X-ray observation was triggered through long term monitoring programs such as those run by \emph{Swift}, or via a detailed follow up program such as that completed for the \emph{ROSAT} TDEs, only a handful of these sources have detailed long term X-ray light curves. The vast majority of these TDE candidates have only short time period X-ray data (e.g., only one X-ray data point or an X-ray upperlimit) or no X-ray analysis has been published (this is particularly the case for those TDEs which were originally detected in optical). This makes it difficult to be able to characterise the long term X-ray emission from these events, or from the host Galaxy. One of the main issues associated with current studies of TDEs is that AGN activity can mimic the expected X-ray emission of TDEs. This can make it difficult to be able disentangle the emission from these two different components \citep[][]{2011ApJ...741...73V}. Long term X-ray light curves can help alleviate this issue quickly, as periodic X-ray flares characteristic of AGNs should become obvious in these light curves, thus ruling out particular flaring events as TDEs. 

In addition, due to the differences in focuses, and analysis techniques of current studies of the X-ray emission of TDEs listed in the literature, it is difficult to be able to complete a comparative study of the X-ray emission arising from these events. For the sources in which either short term (e.g., single data point or upper limit) or long term X-ray emission has been published in the literature, the X-ray fluxes and count rates are usually extracted over different energy bands, using different source and background regions, spectral models (i.e., powerlaw or blackbody or something more complicated), and analysis methods. In addition, with the exception for the well-studied X-ray TDEs such as \emph{NGC 5905}, \emph{Swift J1644+57} or \emph{ASASSN-14li}, only a fraction of the available X-ray data for each source has been analysed. 

Thus, to be able to fully characterise the long term X-ray emission from each TDE candidate, such that individual and comparative class studies can be completed, it is imperative to perform a comprehensive and systematic analysis of the X-ray emission from each TDE using \emph{ROSAT}, \emph{Chandra}, \emph{XMM-Newton}, and \emph{Swift}. Here we have undertaken this task in an attempt to characterise the long term X-ray emission from all events which have been classified as a TDE in the literature, allowing us to characterise the properties of X-ray TDEs as a whole. Our analysis method is systematic and takes into account intricacies associated with the X-ray analysis of point sources. This includes taking into account pileup, the encircled energy fraction, binning of spectral data and whether the number of source photons detected is significantly above background (i.e., Poisson fluctuation). Using these data products we study the global and individual properties of these candidates, and classify each candidate based on their derived and literature properties. We also make available all data products (count rates, fluxes, luminosities, light curves) derived in this study publicly available on the \textit{open TDE catalog} which can be found at the following URL: \url{https://tde.space}.

In this paper, we present the details of our systematic analysis of all available X-ray data from either \emph{ROSAT}, \emph{Chandra}, \emph{XMM-Newton}, and/or \emph{Swift} for 70 TDE candidates listed in the literature. In Section \ref{dataanalysis} we describe our data analysis for each of of the four X-ray instruments used in this study. In Section \ref{class} we classify each TDE candidate based on their properties into one of six categories, producing a list of X-ray TDE candidates that best constitute the properties of an X-ray TDE. In Section \ref{props} we analyse the properties of these candidates allowing us to quantify the properties of what defines an X-ray TDE, while in Section \ref{dis} we discuss what this tells us about these type of events. In Section \ref{conclusion} we summarise our results, while in the Appendix we summarise the properties of each individual TDE candidate as derived from this analysis and from the literature, as well as list the products of our analysis.

\section{Data Analysis}\label{dataanalysis}
To perform our analysis, we selected all possible candidates that have been claimed in the literature or inferred by us to potentially be a TDE, regardless of whether the currently favoured interpretation of this source is a TDE or another astrophysical phenomenon. In Table \ref{tdes}, we list the name of the TDE candidates, the host galaxy, the right ascension and declination of the host and the TDE, and redshift of all 70 candidates that we selected for our study. As this list is continually growing as new events are discovered and new observations becoming available all the time, not all potential/confirmed TDEs or their observations, found in Table \ref{tdes}, are included and analysed. In addition, due to the proprietary nature of some observations, not all data has been analysed for all events since at the time of writing, these data were not publicly available (e.g., follow-up \emph{Chandra} observations of \emph{ASASSN-15lh} are not available until the end of 2017: \citealt{2016arXiv161001632M}). 

\input{tdecandidates.tex}

\subsection{Data reduction}
For each TDE candidate listed in Table \ref{tdes}, we searched for and analysed available \emph{ROSAT}, \emph{Chandra}, \emph{XMM-Newton}, and \emph{Swift} observations of these sources. As we are focusing on the 0.2-10.0 keV X-ray emission from these events, we did not use data from \emph{MAXI} or \emph{INTEGRAL} which detect X-ray emission in the 0.7-7.0 keV and 3-35 keV energy bands respectively. Using the positions listed in Table \ref{tdes}, we obtained all available data of each source from these four X-ray missions using the High Energy Astrophysics Science Archive Research Center (\textit{HEASARC}) data archive\footnote{\url{http://heasarc.gsfc.nasa.gov/cgi-bin/W3Browse/w3browse.pl}}, the \emph{ROSAT} X-ray All-Sky Survey (RASS) catalogue\footnote{\url{http://www.xray.mpe.mpg.de/cgi-bin/rosat/rosat-survey}}, and the \emph{XMM-Newton} science archive\footnote{\url{http://xmm.esac.esa.int/xsa/}}. We analysed \emph{ROSAT} PSPB/PSPC pointed, \emph{ROSAT} RASS, \emph{Chandra} ACIS pointed, \emph{XMM} EPIC pointed, \emph{XMM} slew and \emph{Swift} XRT observations for each source, when data was available. 

Due to the low resolution (or large half equivalent width) of the PSF of the Advanced Satellite for Cosmology and Astrophysics (\emph{ASCA}) and \emph{Suzaku} satellites compared to that of \emph{ROSAT}, \emph{XMM}, \emph{Chandra} and \emph{Swift}\footnote{The resolution of the on-axis PSF of \emph{ROSAT} PSPC, \emph{XMM} EPIC, \emph{Chandra} ACIS and \emph{Swift} XRT detectors are $20\arcsec$, $14-15\arcsec$, $0.5\arcsec$ and $18\arcsec$ respectively, while for \emph{ASCA} GIS and \emph{Suzaku} XIS detectors, it is $174\arcsec$ and $\sim 90''$ respectively \citep{2011hxa..book.....A}.}, we did not use observations from these satellites for our analysis. As one of the main aims of this study  is to determine the long term X-ray emission from each TDE candidate, we analysed data from sources that have two or more observations of the source taken at different times. As such out of the 70 candidates that we selected, eight candidates had only one observation overlapping the position of the candidate. This includes NGC 6021, PGC 015259, PGC 1127938, PGC 133344, PGC 170392, UGC 01791, UGC 03317 and TDE1. As such we have excluded these sources from our general analysis below.

\subsubsection{ROSAT}
Nearly all TDE candidates listed in Table \ref{tdes} had either \emph{ROSAT} PSPC/B or \emph{ROSAT} RASS observation overlapping the source of interest. For these sources, we used the screened data from the \textit{HEASARC} or RASS archive that had been quality checked and processed using the \emph{ROSAT} Standard Analysis Software System. To analyse the pointed and RASS observations we used \textit{Xselect} version 2.4c to produce merged event files for observations that occurred around the same Modified Julian Date (MJD) and to extract spectra for observations in which the TDE candidate was bright enough. We considered events in the full 0.1-2.4 keV energy range of \emph{ROSAT}. 

Spectra and count rates were extracted from a circular region with a radius of 100'', as approximately 85-90\% of all source photons at 0.9 keV are enclosed within this extraction region for an on-axis PSPC pointed observation \citep{2000A&AS..141..507B}. For an off-axis PSPC pointed observation roughly 70\% of all source photons at 0.9 keV are enclosed using a region of this size, while for a PSPC Survey such as the RASS, only about 50\% of all photons are enclosed \citep{2000A&AS..141..507B}. We used a circular, source-free background region that has a radius of at least four times that of our source region (i.e., a radius of 400''). We chose this radius so that the background region was sufficiently large such that the uncertainty on the background region is sufficiently small that it can be neglected. This region was placed either immediately surrounding the source of interest (with the source of interest excluded), or nearby the source of interest if there were point sources that contributed to the background.  As the RASS was operated in PSPC scanning mode, data are comprised of a number of individual PSPC fields which results in the exposure time to vary across the field of view. Similarly for pointed PSPC observations the exposure time also varies across the field. For a pointed observation, the exposure time for an on-axis source corresponds to the sum of all the accepted times for each field, while off-axis the exposure time will be less. As a consequence for both RASS and pointed observations, we took this fact into account by positioning the background region such that the exposure time of this region was on average similar to that of the source of interest using the corresponding merged exposure maps of each observation as a guide.

\subsubsection{Chandra}
All \emph{Chandra} data was analysed using version 4.7.0 of the the \emph{Chandra} analysis software, \textit{CIAO}. We reprocessed level one data using \textit{chandra\_repro} to produce new level two event files. All observations were reprocessed using the calibration database CALDB 4.6.9. For observations that occurred around the same MJD, we produced a merged event file using the \textit{CIAO} tool \textit{reproject\_obs} by reprojecting the observations to a common tangent plane based on the World Coordinate System (WCS) information of the earliest \emph{Chandra} observation in our dataset that we wanted to combine. For observations in which we could extract a spectrum we used the \textit{CIAO} task \textit{specextract}. To extract spectra for observations which occurred at the same MJD, we extracted individual spectra from each observation that resulted in the merged event file and we combined these extracted spectra using the \textit{combine} option in \textit{specextract}. Spectra and count rates were extracted from both the ACIS-S and ACIS-I detectors using a circular region of 2''. A region of this size is able to enclose 95\% of all source photons (at 1.496 keV), assuming that the source is on axis\footnote{See section 4.2.3 of \url{http://cxc.harvard.edu/proposer/POG/html/chap4.html}}. We selected a source-free background region with a radius of 20'' which, was positioned either nearby or immediately surrounding the source of interest (which was excluded from the background region if the latter option was chose). Unlike that seen in the \emph{ROSAT} observations, the exposure across the detector does not vary significantly across the detector, allowing us more flexibility for positioning our background region. The only exception to this is when sources are positioned closed to a chip gap or the edge of the detector, however only a handfull of observations fall into this category. Exposure times were derived from the header of the event file from each observation.

\subsubsection{XMM-Newton}
For the \emph{XMM} pointed observations, we started from the observational data files of each observation and used the \textit{XMM-Newton} Science System (SAS) version 14.0.0\footnote{\url{http://xmm.esac.esa.int/sas/current/documentation/}}, and the most up to date calibration files\footnote{\url{http://xmm2.esac.esa.int/external/xmm_sw_cal/calib/index.shtml}} to produce the data products for our study. As \emph{XMM} suffers from periods of high background and/or proton flares, we checked for these time intervals by generating a count rate histogram using events with an energy between 10 - 12 keV for each observation. Before extracting count rates or spectra, we removed the time intervals which are contaminated by a high background or flares producing cleaned event files. As suggested in the current SAS analysis threads\footnote{\url{http://xmm.esac.esa.int/sas/current/documentation/threads/}} and \textit{XMM-Newton} Users Handbook\footnote{\url{http://xmm.esac.esa.int/external/xmm_user_support/documentation/uhb/index.html}}, we reduced the data following the standard screening of events, with single to quadruple pattern events (PATTERN $\le$ 12) chosen for the MOS detectors, while for the PN detectors only single and double patterned events (PATTERN $\le$ 4) were selected.  We also used the standard canned screening set of FLAGS\footnote{A FLAG value provides information about the event condition, such as whether it was detected near a hot pixel or resulted from outside the field.} for both the MOS (\#XMMEA\_EM) and PN (\#XMMEA\_EP) detector respectively. 

For the observations in which the TDE was bright enough to extract spectra, we used the SAS task \textit{evselect} and the cleaned event files from all three EPIC cameras. For each spectrum we extracted, we produced spectral response and effective area files using the tasks \textit{arfgen} and \textit{rmfgen}. For our analysis we consider events between $0.2-10.0$ keV for the MOS detector and $0.2-12.0$ keV for the PN detector. Spectra and count rates were extracted from a circular region with a radius of 30\arcsec. This corresponds to $\sim$85\% of all source photons at 1.9 keV are enclosed by the extraction region for both the MOS and PN detectors (assuming that the source of interest is found on-axis)\footnote{See \url{http://xmm.esac.esa.int/external/xmm_user_support/documentation/uhb_2.5/node17.html} for more information}. We used a circular background region with a radius of 120'', which similar to what was completed with  \emph{Chandra}, was placed immediately surrounding the source of interest (with the source of interest excluded) or placed in source a free region nearby the TDE candidate. To extract the count rates we used only the PN detector, due to its high sensitivity, large effective area and consistent overlap with all source regions (i.e., some sources fell on CCD3 of MOS1 which suffered significant damage after a micro-meteoroid impact\footnote{\url{https://heasarc.gsfc.nasa.gov/docs/xmm/uhb/epic.html}}). For each observation, exposure times were taken from the header of the corresponding event files.

\subsubsection{XMM-Newton slew observations}
In addition to analysing the pointed \emph{XMM} observations of each candidate, we also searched for \emph{XMM}-slew observations that overlap the position of the source. As \emph{XMM} manoeuvres between pointed observations, all three detectors (MOS1, MOS2, PN) are still recording data using the observing mode of the previous pointed observation \citep{2008A&A...480..611S}. In addition the CCD is set with a medium optical blocking filter to prevent contamination from IR, visible or UV photons from points sources with a V-magnitude of m$_{\rm V}$ = 8-10 or less\footnote{\url{http://xmm.esac.esa.int/external/xmm_user_support/documentation/uhb_2.1/node32.html}}. Due to the fast readout time of the PN detector any source detected during the slew observation will not be affected by the motion of the telescope, while the slower readout time of both MOS cameras leads to a highly elongated PSF causing any source detected to appear as long streaks \citep{2001A&A...365L..27T, 2008A&A...480..611S}. As a consequence only data from the PN detector is used to produce a slew observation. 

Starting from the slew data files that are publicly available in the \emph{XMM-Newton} science archive we follow the current SAS analysis thread on how to process EPIC slew data\footnote{\url{http://xmm.esac.esa.int/sas/current/documentation/threads/EPIC_slew_processing_thread.shtml}} and run the command \textit{eslewchain} to produce filtered event files that we use in our analysis. Similar to the analysis for the pointed \emph{XMM} observations we also use SAS version 14.0.0 and the most up to date calibration files, while we use events over the 0.2-12.0 keV range. The number of counts were extracted from a circular source region with a radius of 50\arcsec and a circular background region (with source region excluded if necessary) with a radius of 200\arcsec. Using a source region with a radius of 50\arcsec, $\sim 90$\% of all photons at 1.9 keV are enclosed by our extraction region. Due to the low exposure times of each observation ($\lesssim10-20$ seconds), which was determined using the corresponding exposure files of each observation, we were unable to extract spectra for these objects, even for the brightest of our sources.

\subsubsection{Swift}
Due to \emph{Swift}'s ability to quickly target transient sources, a large number of TDE candidates have \emph{Swift} observations. For each TDE candidate, we analysed all available and overlapping data that was taken in photon-counting model (PC) by \emph{Swift}'s X-Ray Telescope (XRT). Following the \emph{Swift} XRT Data Reduction Guide\footnote{\url{http://swift.gsfc.nasa.gov/analysis/xrt_swguide_v1_2.pdf}}, we reprocessed level one data using the \textit{xrtpipeline} script, producing cleaned event files and exposure maps for each observation. To combine observations which occur around the same MJD, we use \textit{Xselect} version 2.4c. For the brightest sources, we also used \textit{Xselect} to extract spectra from these observations, while exposure times were derived from the header of the event file from each observation.

For each spectrum we produced an ancillary response file (ARF) using the task \textit{xrtmkarf}. This task uses the exposure maps produced during the \textit{xrtpipeline} so that the ARF is corrected for hot columns, bad pixels or loss of counts caused by using an annular extraction region if the source is piled up. To be able to extract spectra from the combined event files, we combined the exposure maps of each observation that went into producing the combined event file before we produced the corresponding ARF file. To combine the exposure maps we used \textit{XIMAGE} version 4.5.0. The response matrix files (RMFs) for each observation were obtained from the CALDB as ready-made files, and was selected such that it matched the suggested RMF file needed in the output of the \textit{xrtmkarf}. To extract spectra and counts, we used a circular source region with a radius of 50\arcsec and a source free, circular background region with a radius of 200\arcsec placed in a similar way as the background sources were position when analysing the \emph{Chandra} and \emph{XMM} observations.

Sources which had a background extracted count rate $>0.5$ counts/s are most likely piled-up, which can lead to issues in the data analysis. For these sources, we followed the \textit{Swift} analysis threads\footnote{\url{http://www.swift.ac.uk/analysis/xrt/pileup.php}} and estimated where pile-up affects the data by fitting the XRT PSF for that particular observation using \textit{XIMAGE}. We then excluded an additional circular region with a radius defined by when the data and model diverge in our source extraction region when extract spectra for these sources.

\subsection{Count rates and is it a background fluctuation or detection? }\label{det}
To determine the number of counts coming from the position of the TDE candidate we used the \textit{funcnts} task that is a part of the FITS library and utility package for astronomical data analysis \textit{FunTools}\footnote{\url{http://hea-www.harvard.edu/RD/funtools/help.html}}. This calculates the background-subtracted source counts, and the number of background counts from an event file and a region file that lists the source and background regions (and any other source that one wants to exclude or include) of interest. For each source we derived the number of source and background counts in the full energy range of all four X-ray satellites for our analysis. In addition, we also derived the number of counts in a soft (0.3-1.0 keV), medium (1.0-2.0 keV), and hard (2.0-10.0 keV or 2.0-2.4 keV for \emph{ROSAT}) energy bands for each observation by filtering the cleaned event file and running \textit{funcnts}. All extracted count rates used for our analysis were corrected for the fact that our regions only enclose a fraction (encircled energy fraction) of the total number of counts arising from the source.

To determine whether the number of X-ray counts we detect arises from a chance background fluctuation or from emission from an X-ray point source, we calculate the probability ($\mathcal{P}$) of having $N$ source (background subtracted) counts given $M$ background counts, using $\mathcal{P}(M,N) = 1 - CDF(M, N-1)$, where CDF is the Cumulative distribution function assuming Poisson statistics. The corresponding detection confidence is calculated by $\mathcal{D}(M,N)=CDF(M, N-1)$. For candidates that have a detection confidence of 0.95 ($\sim2\sigma$) and above, we classified these as a detection of X-ray emission from the source. For the sources that had a detection confidence less than 0.95 ($\sim2\sigma$) we classified this emission as a chance background fluctuation and instead derive the 3$\sigma$ upperlimit. Upperlimits are derived assuming that we would detect a signal if it is $3\sigma$ above background.

In Tables \ref{rosatcounts}--\ref{swiftcounts} we have listed the observation IDs (ObsIDs) of the X-ray data we analysed for each TDE candidate, a label for this ObsID so that we can refer to it in later tables, the time in which the observation was taken in Modified Julian Date (MJD) (or the average MJD of the observations which have been merged) and the total exposure time of the cleaned (and merged) event files.  In addition, we have listed the extracted source and background counts that we obtained over the full energy range of each instrument, the probability of the counts that we detected arise from a chance fluctuation and its detection confidence as calculated in Section \ref{det}. Based on these probabilities, we classified the measured count rate as an upperlimit or a data point, and then derived the corresponding source count rate (i.e., the source counts divided by the total exposure time) in the full instrument energy range of the observation. In these tables, the source and background counts extracted from the source have not been corrected for encircled energy fraction. However, the count rate which we use for our analysis, and is listed in the last column of these tables, has been corrected for encircled energy fraction. For simplicity, we have separated the results we obtained into five tables, with each table corresponding to the count properties derived from the different X-ray instruments we used for this analysis.

In Tables \ref{rosathms}--\ref{swifthms} we have listed the count rates from each source derived in a soft (0.3-1.0 keV), medium (1.0-2.0 keV) and hard (2.0-10.0 keV) energy band for each TDE. For the source counts listed in these tables, we followed the same method as outlined in Section \ref{det} to determine whether the emission we detect in each band is a detection or not. For those which we classified as a detection we derive an one sigma uncertainty, but for those which we classify as a chance fluctuation we derive a three sigma upperlimit. All counts in these tables have been corrected for encircled energy fraction. The ObsID label references the observation that these count rates were taken from, which can be found in Tables \ref{rosatcounts}--\ref{swiftcounts}. We have also separated our soft, medium and hard count results into five tables corresponding to the different instruments we obtained data from.

\subsection{Spectral analysis, and deriving the soft X-ray flux and luminosity of each event.}\label{spec}

For observations in which spectra could be extracted, the spectral fitting was performed using the X-ray analysis software XSPEC version 12.9.0c, over an energy range of 0.3--5.0\, keV. Each spectrum was grouped with a minimum of 20 counts per energy bin using the FTOOLS command \textit{grppha}, and fitted using $\chi^2$ statistics. We fit all spectra using an absorbed powerlaw (\textit{tbabs*powerlaw}) model. This model consists of three parameters: the normalisation, power-law index ($\Gamma$) and the Galactic H\textsc{i} column density ($N_{\rm H}$) assuming \citet{2000ApJ...542..914W} solar abundances. Initially, we let all three parameters be free during our fitting procedure but for a large number of spectra, we find that the fit is unable to constrain the $N_{\rm H}$. As a consequence we freeze the $N_{\rm H}$ to the value derived from the Leiden/Argentine/Bonn (LAB) Survey of Galactic H$\textsc{i}$ \citep{2005A&A...440..775K} in the direction of the source of interest for these fits. For sources in which we were able to extract an X-ray spectrum, we calculate the absorbed flux, with errors, of the best fit absorbed powerlaw model using the XSPEC command \textit{flux} over the energy range of 0.3--2.0\,keV. 

The sources in which we were unable to extract a spectrum due to the low number of source counts, we estimated the X-ray flux using the count rate simulator \textit{WebPimms}\footnote{\url{https://heasarc.gsfc.nasa.gov/cgi-bin/Tools/w3pimms/w3pimms.pl}}. Here one specifies the instrument, the count rate as listed in the last column of Tables \ref{rosatcounts}--\ref{swiftcounts}, the energy range that this count rate was derived, the $N_{\rm H}$ to the source of interest, the redshift and the model parameters assuming a specific model of the source.  As a significant fraction of the sources we analysed did not have enough counts for us to extract a spectrum to characterise the emission from these candidates we had to make an assumption about the emission arising from the source itself. For these candidates we assumed a powerlaw index of 4.5 since such a steep spectrum typically mimics thermal emission over the limited X-ray energy band pass of current X-ray satellites \citep[e.g.,][]{2014ApJ...781...59D}. For sources which had multiple observations from the same instrument, but only a fraction of these had enough counts such that a spectrum could be extracted and modelled, we assumed the average power law index and $N_{\rm H}$ derived from fitting the spectra from these other observation of the source. 

To derive the corresponding X-ray luminosities of each source from our derived fluxes, we assumed a $\Lambda$CDM cosmology with $\Omega_{M} = 0.27$, $\Omega_{\Lambda}=0.73$ and $H_{0} = 71$ km s$^{-1}$ Mpc$^{-1}$. 

In Tables \ref{rosatfits}--\ref{swiftfits} we have listed the $N_{H}$ and powerlaw index $\Gamma$ we derived (or assumed) from the best fit absorbed powerlaw model for each set of data. We have also listed the derived absorbed X-ray flux and corresponding X-ray luminosity in the 0.3-2.0 keV energy band. Again, we have separated the results into five tables, with each table corresponding to a different instrument.  Uncertainties on all parameters listed in these tables are one sigma uncertainties, and the ObsID labels refer to the ObsIDs listed in the second column of Tables \ref{rosatcounts}--\ref{swiftcounts}.

\section{The X-ray properties of the TDE candidates and their classification}\label{class}

\begin{figure}[t]
	\begin{center}
		\includegraphics[width=1.05\columnwidth]{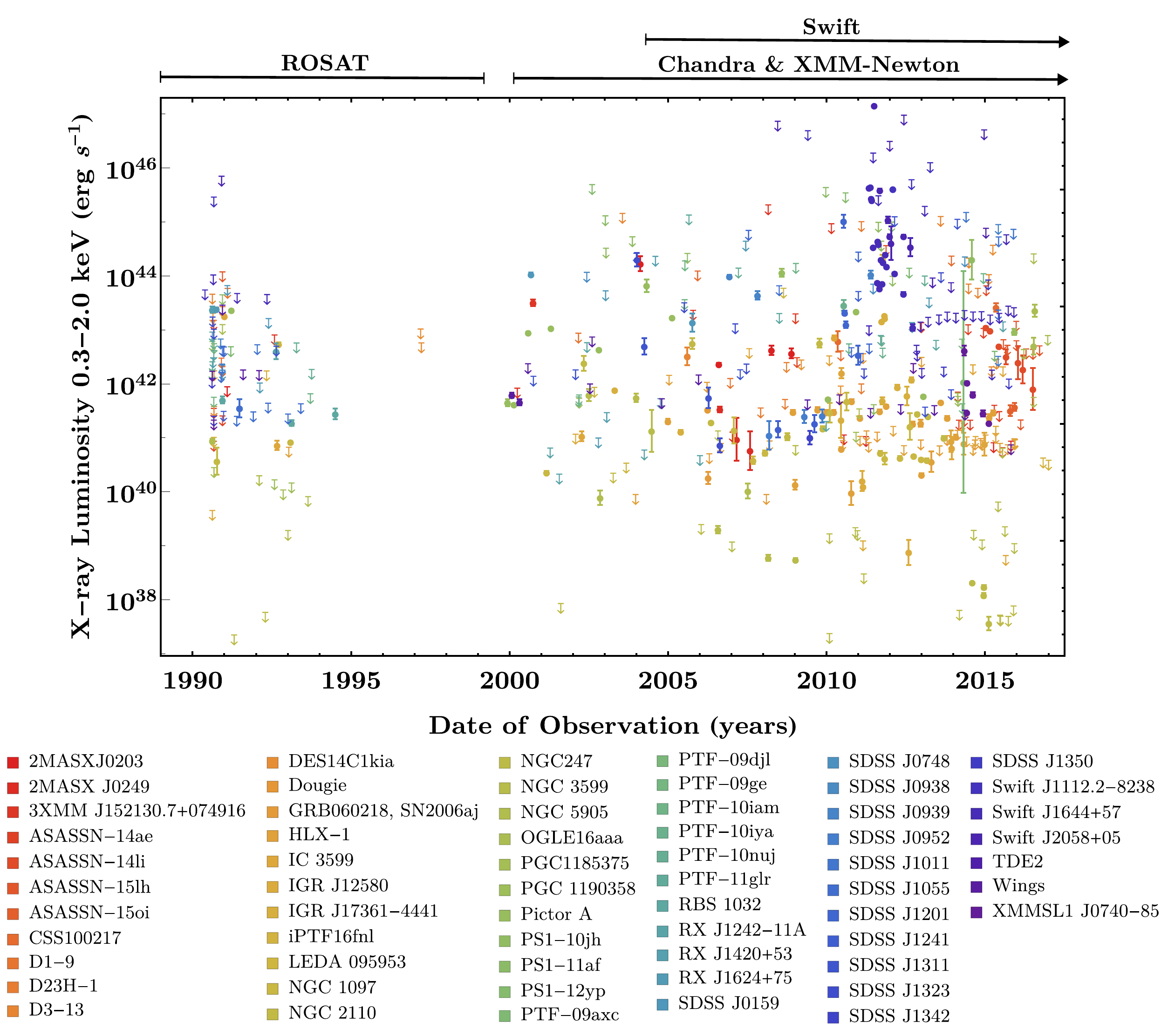}
		\caption{The X-ray light curves of all TDE candidates listed in Table \ref{tdes}. Below the figure is the colour key indicating the different TDE candidates. \label{alltdelightcurves}}
	\end{center}
\end{figure}

A dormant BH at the centre of a quiescent galaxy reveals itself by the detection of emerging flare-like X-ray (or optical/UV) emission that can result from a star being tidally disrupted. However, AGN and numerous other astrophysical processes also result in flare-like emission which can make classifying this emission as a TDE difficult. Long term X-ray light curves can help aid in differentiating recurring flare-like emission from that of TDEs as periodic emission can become apparent over long baselines. Not until recently has X-ray astronomy come into its own, with the advent of a number of high resolution, and high sensitivity X-ray satellites for which we can use to quickly follow up potential candidates.  Since the launch of \emph{ROSAT} in the 1990s until now, we are lucky enough that we have nearly 30 years of available X-ray data that we can use to characterise the X-ray emission from these sources.

Taking advantage of this fact, we used the results derived in Tables \ref{rosatfits}--\ref{swiftfits}, to produce long term 0.3-2.0 keV lightcurves for each of our TDE candidates listed in Table \ref{tdes}. In Figure~\ref{alltdelightcurves} we have overlaid the light curves derived for each event. From Figure~\ref{alltdelightcurves}, one can see that these candidates cover X-ray luminosities over nearly 10 orders of magnitude. 

In the 1990s, \emph{ROSAT} was the only X-ray instrument that was available to search for X-ray emission from TDEs, with the RASS providing the largest number of constraints during this period.  Even though the \emph{ROSAT} mission lasted for over eight years before ending in 1999, the lack of observations seen during 1995 and 1999 arises from the shut down of the PSPC in 1994 to minimise the lost of combustibles. The PSPC was turned back on briefly during 1997 to take a series of pointings to complete the all-sky survey which resulted in the complete use of the remaining detector gas\footnote{\url{https://heasarc.gsfc.nasa.gov/docs/rosat/pspc.html}}. In the early 2000s when \emph{Chandra}, \emph{XMM-Newton} and \emph{Swift} became operational, our ability to localise, detect and constrain emission from these sources dramatically increased. In particular, the ability of  \emph{Swift} to quickly follow up potential TDE candidates has improved our capability to well characterise over shorter timescales the X-ray emission arising from these objects.

In Figure~\ref{lightcurves1}--\ref{lightcurves4} which can be found in the Appendix, we have also plotted the individual X-ray lightcurves of all TDE candidates listed in Table \ref{tdes}. In these figures, we have colour coded each data point/upperlimit based on the instrument in which we derived this measurement. In addition, we have also overlaid on these plots the optical/UV emission that we took from the literature, for sources which were also detected in these wavelengths. From these plots we can: \textit{ (1).} easily rule out different candidates as TDEs due to the presence of recurring X-ray emission, \textit{(2).} highlight sources which do not have enough data to classify it as a TDE,  and \textit{(3).} produce the most comprehensive soft X-ray curves for each of these events. In Appendix Section \ref{individualsources} we have also summarised the properties of each individual TDE candidate listed in Table \ref{tdes}, as well as given an overview of their suggestion/classification as a TDE as presented in the literature.

\subsection{Classifying the TDE candidates}\label{classification}
The large amount of available archival X-ray data, in addition to future data from triggered or serendipitous X-ray observations, opens the door to potentially detecting a large number of X-ray TDEs. However, due to the difficulty in disentangling the X-ray emission arising from a flare from other transient phenomenon such as AGN activity, it is important to determine a set of well-defined properties that allow individuals to classify a potential candidate as an X-ray TDE. 

Apart from taking steps towards removing the degeneracy associated with the nature of these X-ray flares, these well defined characteristics allow us to select a sample of current TDE candidates that encompass the general properties of X-ray TDEs. This is important as this class of events would allow us to (re)define the classification of what constitutes an X-ray TDE observationally. This provides us with the opportunity to perform a global study of the properties of X-ray TDEs and compare these with other transient phenomenon such as AGN (see Auchettl et al. 2017 in preparation).

Based on our analysis, and extending on attempts in the literature to collate the properties of X-ray TDEs (see e.g., \citealt{2015JHEAp...7..148K}), we rank the TDE candidates listed in Table \ref{tdes} based upon the quality and quantity of the available X-ray data into six categories: \textit{TDE X-ray, likely X-ray TDE, possible X-ray TDE, veiled X-ray TDE} and \textit{unknown}. We focus predominantly on their X-ray properties, however we also use other wavelengths (such as optical) to characterise observationally an X-ray TDE from another astrophysical object or process.

\subsubsection{X-ray TDEs}

\begin{figure}[t!]
	\begin{center}
		
		\includegraphics[width=0.49\textwidth]{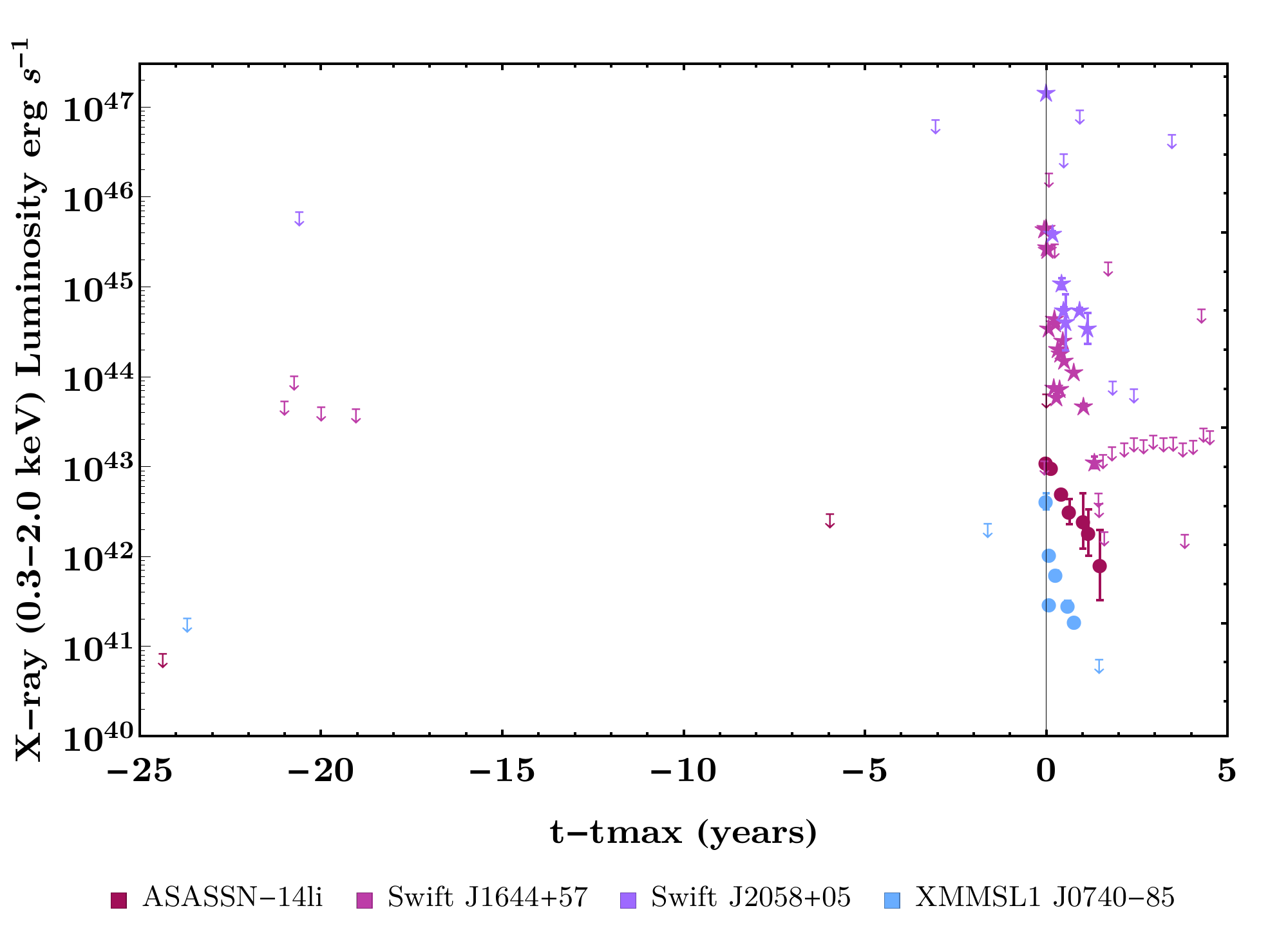}
		\includegraphics[width=0.49\textwidth]{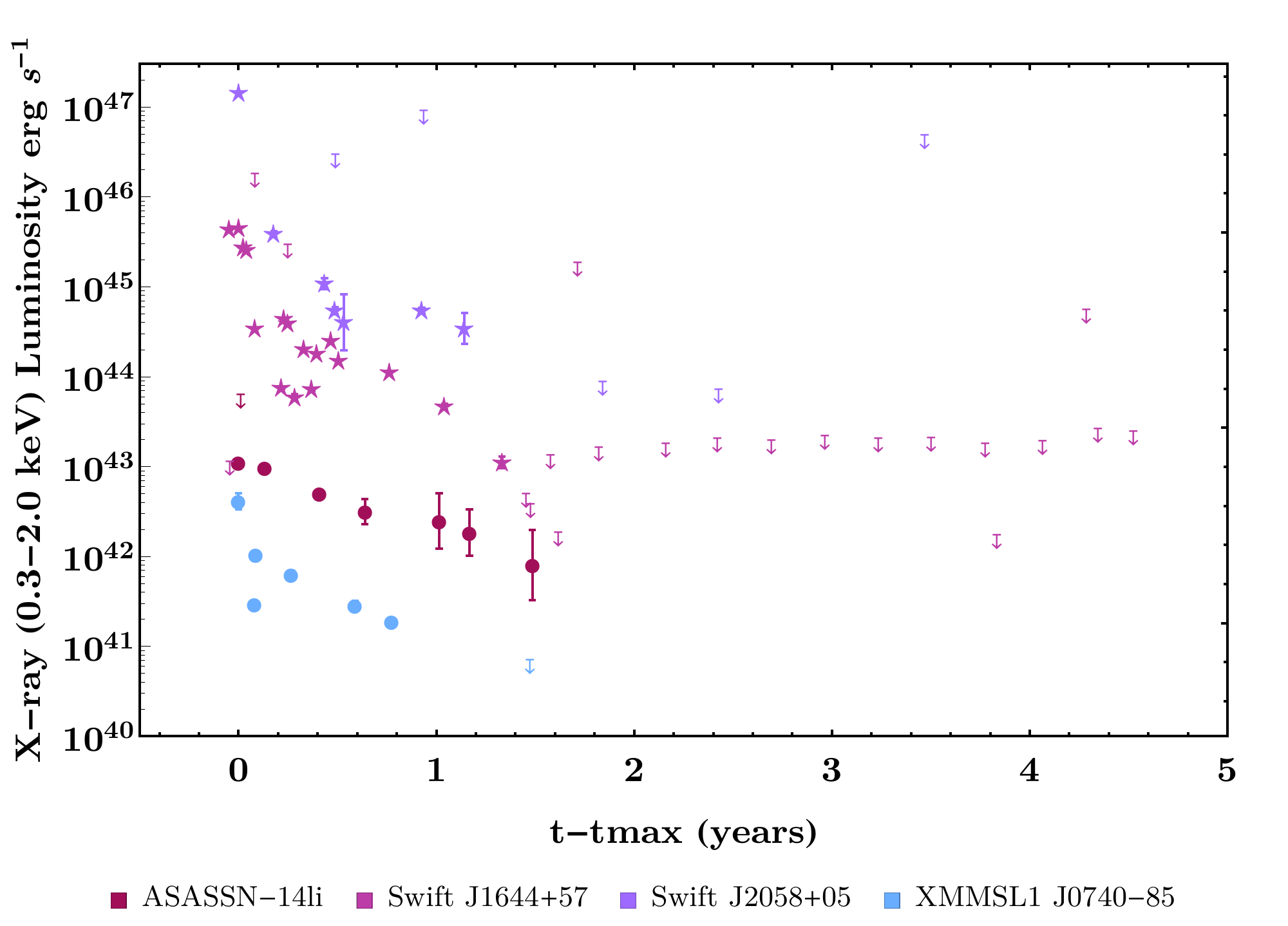}
		\caption{Light curves of the  X-ray TDEs sample scaled such that the zero point occurs at the maximum detected X-ray luminosity ($t_{max}$).  In the \textit{top panel} is the X-ray light curves of the TDE candidates plotted over the full ($\sim$30 year) range of data currently available for each source.  In the \textit{bottom panel} is the X-ray light curve of each X-ray TDE candidates focusing on the detected X-ray emission during and after the original X-ray flare was detected. Sources that are plotted with a star ($\bigstar$) are TDE candidates that have been classified as non-thermal (jetted) TDEs in the literature  (e.g., Swift J1644+57), while those plotted with a filled circle (\textbullet) are either classified as a thermal (non-jetted) TDE in the literature (e.g., ASASSN-14li) or have no classification. Note that this classification using ($\bigstar$) and (\textbullet) will be used throughout the paper. In addition, own arrows are upperlimits, while the colour key for each TDE candidate is shown. \label{goldlight}}
		
	\end{center}
\end{figure}

Our criteria for classifying an X-ray candidate as an \textit{X-ray TDE} are:

\begin{enumerate}[topsep=0pt,itemsep=-1ex,partopsep=1ex,parsep=1ex]
\item The X-ray light curve is well defined, where well defined means that there is sufficient data coverage of the suspected flaring event such that a well defined shape and trend is observable. In addition, there is at least one (but preferably more than one) observation prior to and after the detected flare which can help quantify the pre-flare and post-flare emission of the source.
\item The X-ray light curve shows a rapid increase in X-ray luminosity, which then declines on time-scales of months to years.
\item The general shape of the X-ray light curve decay is monotonically declining, however variability in the X-ray emission on smaller timescales can also be seen but is not necessarily required.
\item The maximum luminosity detected from the event is at least two orders of magnitude larger than the X-ray upperlimit immediately preceeding the discovery of the flare.
\item Over the full time range of X-ray data available for the source of interest, the candidate TDE shows evidence of X-ray emission from only the flare, while no other recurrent X-ray activity is detected.
\item The X-ray flare is coincident with the nucleus of the host galaxy.
\item Based on its optical spectrum or other means, one finds no evidence of AGN activity arising from its host galaxy.
\item The host galaxy shows no evidence of large scale jets in any wavelength.
\item Supernova and Gamma-ray Burst (GRB) origin has been ruled out.
\end{enumerate}

Out of the candidates listed in Table \ref{tdes} there are four candidates which satisfy all the requirements of the \textit{X-ray TDE} category based on their properties summarised in the Appendix Section \ref{individualsources} and presented in Tables \ref{rosatcounts}--\ref{swiftfits}. These are the thermal X-ray TDE \textit{ASASSN-14li} and the non-thermal events  \textit{Swift J1644+57}, \textit{Swift J2058+05} and \textit{XMMSL1 J0740-85}. Due to the rapid follow up of these events after the initial trigger, each of these three sources has a rich data set that allows us to produce a well defined X-ray light curve of its emission. A clear increase in the X-ray luminosity arising from the centre of their host galaxies, followed by a monotonic decay is observed. In addition, variability is also observed on smaller times scales. The maximum luminosity of the flare is also a few orders of magnitude larger than the X-ray upperlimit immediately prior to the flare. Detailed analysis of its host Galaxy and the events have ruled out the presence of AGN, large scale jets or the possibility that it could arise from a supernova or GRB.

In Figure~\ref{goldlight} we have overlaid the X-ray light curves of each of the X-ray TDEs. Here we have plotted the X-ray luminosity as a function of $t-t_{max}$, where $t_{max}$ is the time at which the measured X-ray luminosity was at maximum\footnote{This does not necessarily indicate that this is when the X-ray emission from the source peaked. The source could have peaked a few days to weeks before the first observation in which we detect the maximum measured X-ray luminosity. The difference between the actual and measured X-ray luminosity arises from the limitation of available X-ray satellite resources. As such, we are usually only able to capture emission from these events after the initial peak in X-rays.}. We have plotted these such that one can see how the emission from these TDEs compare over the full $\sim$30 years of available X-ray observations (Figure~
\ref{goldlight} top), as well as focusing directly on the X-ray flare emission (Figure~\ref{goldlight} bottom). One can see in the bottom panel of Figure \ref{goldlight} that these X-ray events are bright for approximately 1-2 years after the measured luminosity was at its maximum. The variability of the \emph{Swift J1644+57}, \emph{Swift J2058+05} and \emph{XMMSL1 J0740-85} (and to a smaller extent for \emph{ASASSN-14li}) is also obvious in these plots, while no X-ray emission before the observed flare is detected. These events also span nearly six orders of magnitude in X-ray luminosity, with the non-thermal X-ray TDEs \emph{Swift J1644+57} and \emph{Swift J2058+05} peaking at higher X-ray luminosities compared to the thermal X-ray TDE \emph{ASASSN-14li} and \emph{XMMSL1 J0740-85} which peak at much lower luminosities.

\subsubsection{Likely X-ray TDEs}

\begin{figure}[t!]
	\begin{center}
		\includegraphics[width=0.49\textwidth]{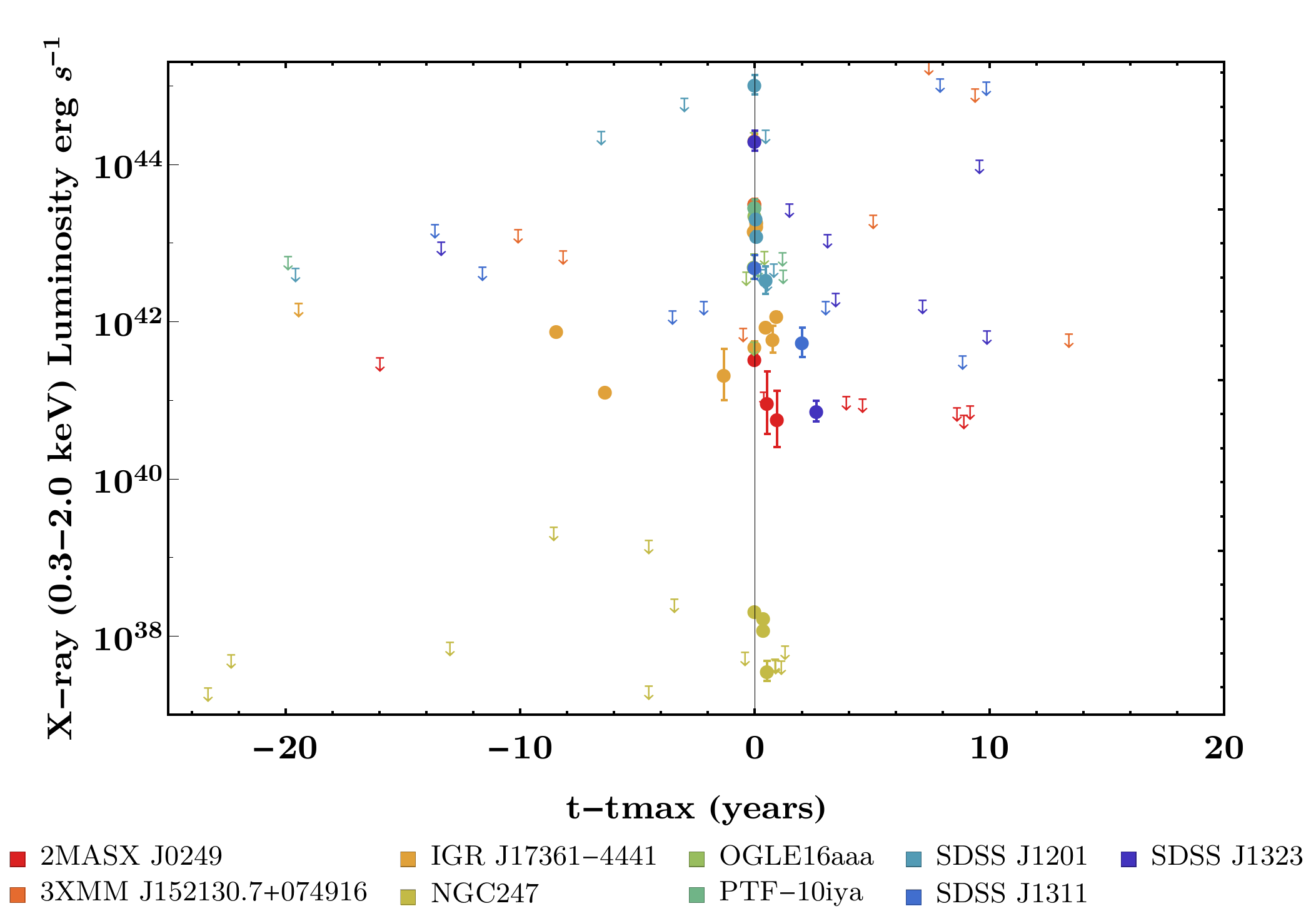}
		\includegraphics[width=0.49\textwidth]{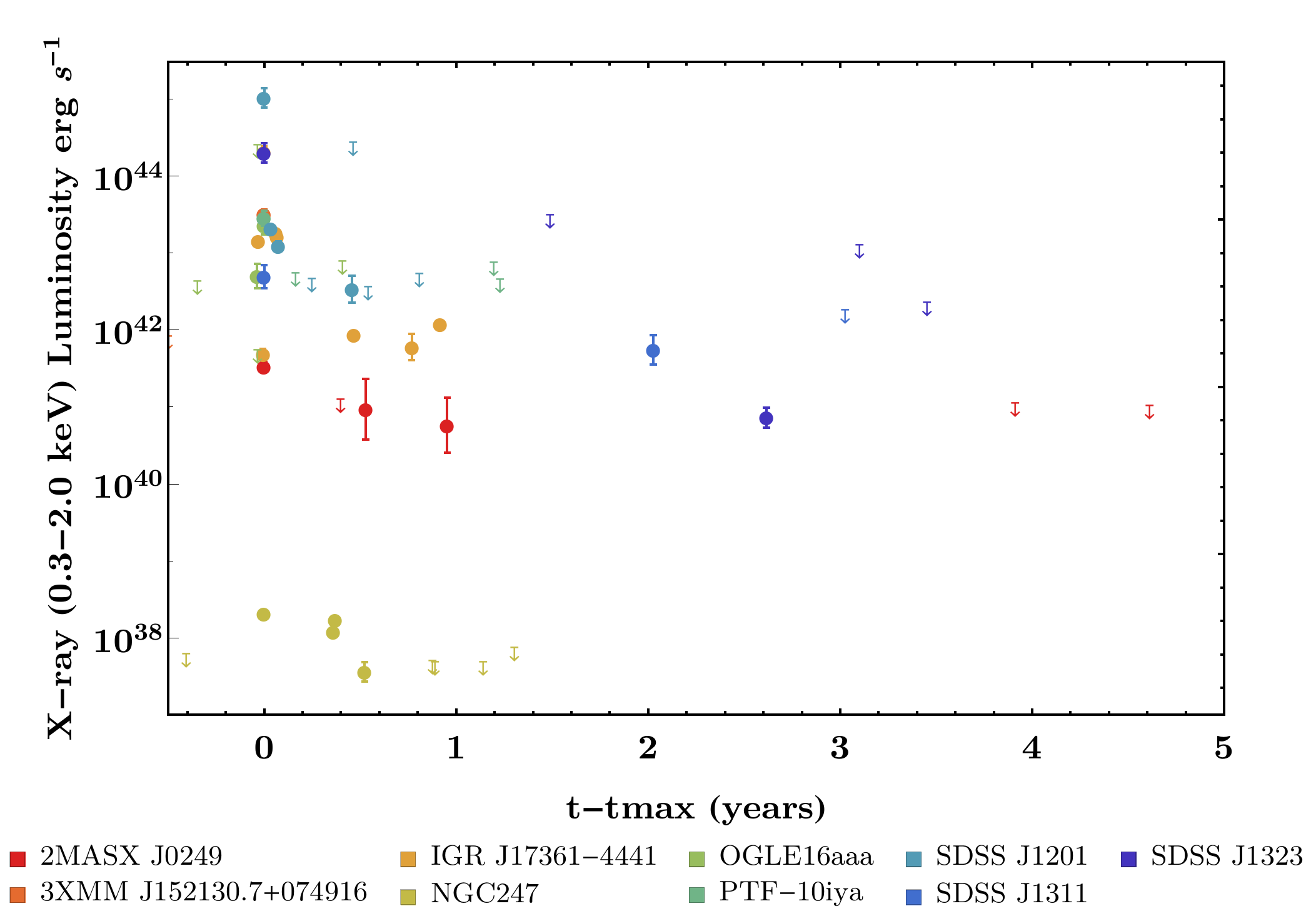}
		\caption{Same as Figure~\ref{goldlight}, however here we have plotted the \textit{likely} X-ray TDEs. \label{silverlight}}
		
	\end{center}
\end{figure}

A candidate falls into the \textit{likely X-ray TDE} category if it has properties nearly identical to those of the \textit{X-ray TDE } category, but with the following differences:

\begin{enumerate}[topsep=0pt,itemsep=-1ex,partopsep=1ex,parsep=1ex]
	\item The X-ray light curve is not very well defined due to the limited number of observations available of the source. Regardless, a general shape and trend is observable even if details of the X-ray emission such as variability, or exactly how the emission evolves is not well defined. There should also still be at least one (but preferably more than one) observation prior to and after the detected flare which can help quantify the pre- and post- flare emission of the source.
	\item The presence of an AGN is ruled out or highly unlikely.
	\item Requirements 2\--6, 8\--9 of the X-ray TDE category.
\end{enumerate}

There are nine TDE candidates listed in Table \ref{tdes} which we classify as likely X-ray TDEs. These are: \textit{2MASX J0249}, \textit{3XMM J152130.7+074916}, \textit{IGR J17361-4441}, \textit{NGC 247}, \textit{OGLE16aaa}, \textit{SDSS J1201}, \textit{SDSS J1311}, \textit{ SDSS J1323}, \textit{PTF-10iya}. All these events have X-ray light curves which show a $\gtrsim1.5\--2$ order of magnitude increase in luminosity compared to previous X-ray upperlimits, which then decays. However, due to the limited amount of available X-ray data from these sources we are able to derive only a general trend for their emission. Observations immediately before and after the flare indicate the presence of no recurrent X-ray emission arising from the source and as of writing, the presence of an AGN is either ruled out or highly unlikely. Other astrophysical sources have also been ruled out as responsible for their emission. 

In Figure~\ref{silverlight} we have overlaid the X-ray light curves of each of the \textit{likely} X-ray TDEs. Similar to Figure~\ref{goldlight}, we have plotted the X-ray luminosity as a function of $t-t_{max}$ over the the full 30 years of available X-ray observations, as well as focusing directly on the X-ray flare emission. One can see that these events cover an even wider range of X-ray luminosities than their X-ray TDE counterparts. These events also tend to decay on much shorter timescales ($\lesssim1$ year).

\subsubsection{Possible X-ray TDEs}

To fall into the \textit{possible X-ray TDE} category, the source would have the following properties:

\begin{enumerate}[topsep=0pt,itemsep=-1ex,partopsep=1ex,parsep=1ex]
	\item Based on the available X-ray data (whether this is limited or not), the emission from the source appears to either peak randomly or unpredictably, or it shows evidence of a periodic pr extend emission signature.
	\item The maximum luminosity detected from the event is one order of magnitude larger than the X-ray upper limit immediately preceding the discovery of the flare.
	\item The event is found to be coincident with the nucleus of the host galaxy.
	\item The host galaxy shows no evidence of large-scale jet like structures in any wavelength.
	\item It appears to be unlikely to arise from an AGN, GRB or a supernova
	\item Although not required, it has been classified in the literature as an optical TDE from a detailed analysis by their authors.
\end{enumerate}

\textit{ ASASSN-15oi}, \textit{ D3-13}, \textit{ LEDA 095953}, \textit{NGC 3599}, \textit{NGC 5905}, \textit{ RBS 1032}, \textit{ RX J1242-11A}, \textit{ RX J1420+53}, \textit{ RX J1624+75}, \textit{ SDSS J0159}, \textit{ Swift J1112-82} and \textit{ Wings} are the candidates which we place into this category. For \textit{ASASSN-15oi} which was classified as an optical/UV TDE by \citet{2016arXiv160201088H}, the limited amount of data available make it is difficult to be able to quantify the evolution of the X-ray emission detected and based on the current results it does not decay significantly. For \textit{D3-13}, which was classified by \citet{2008ApJ...676..944G}  as an optical TDE, the X-ray emission we detect is no more than one order of magnitude larger than the X-ray upperlimit before and after the original flare.  \textit{LEDA 095953} shows evidence of an X-ray flare in the early 90s that is coincident with the centre of the host Galaxy, however the emission appear to be random due to the limited amount of X-ray data around the time of the flare. No follow up X-ray emission has been detected from this source. Even though \textit{NGC 3599} has a significant amount of data that shows a clear increase in the observed X-ray emission by a few orders of magnitude compared to previous X-ray upperlimits which then decays (see Figure~\ref{lightcurves2}), we place \textit{NGC 3599} into the possible X-ray TDE category rather than the likely X-ray TDE category. This is due to the fact that the AGN origin of the detected flare has not been completely ruled out, while unlike other X-ray TDE events which exhibit a fast rise to peak, and then decays within 1-2 years (see Figure~\ref{comp2upper}), the emission from this event exhibits a slow rise and then a long decay over nearly 10 years. Even though slow-rise TDEs are theoretically expected to also be detected, the origin of this flare is not as clear as other events hence leading us to place it in the \textit{possible X-ray TDE} category. The flare arising from \textit{NGC 5905} was detected using \emph{ROSAT}. As a consequence there is no X-ray upperlimit prior to the detected flare making it difficult to quantify the emission immediately before the flare. In addition the detection of late time emission arising from this event is reminiscent of AGN \emph{IC3599}, while \citet{2003ApJ...592...42G, 2004ApJ...601.1159G} found using the Hubble Space Telescope (\emph{HST}) narrow emission lines in the inner nucleus of the host, indicating that there is a low level, non-stellar photoionisation powered by accretion that could be contributing to the observed emission. As the AGN origin of this source is not completely ruled out, and based on current observations a TDE origin is more likely we place this source in the possible X-ray TDE category. \emph{ROSAT} detected an X-ray flare from the centre of inactive Galaxy \textit{RX J1242-11A},  \textit{RX J1420+53} and \textit{RX J1626+75} in the early 90s, however there is limited data which can help us to characterise the evolution of the emission immediately before and after these flares. Follow up observations have ruled out further X-ray emission from these sources at later times, but the data is very sparse making it difficult to determine how this flare-like emission evolved. Even though the emission from \textit{SDSS J0159} shows evidence of a flare which then decays, \citet{2015ApJ...800..144L} showed that the host Galaxy is transitioning from a Type 1 broad-line AGN to a Type 1.9 AGN and the properties of this source could result from the dimming of AGN continuum. However, due to the rarity of this type of dimming, it is possible that this event could arise from a TDE which would be the brightest non-jetted TDE detected. Due to the uncertainty in the actual origin of this event we have placed this event into the possible X-ray TDE category for now. Due to the limited amount of X-ray data and the fact that the soft X-ray emission from each source is of the same order of magnitude as its upperlimit prior to and after the detected X-ray flare, we place \textit{Swift J1112-82} and \emph{Wings} in the possible X-ray TDE category.

\input{goldsilver.tex}

\subsubsection{Veiled X-ray TDE}\label{veiled}

For a candidate to be classified as a \textit{veiled X-ray TDE} these sources must show well defined optical/UV light curves that again show a increase in the optical/UV emission that then decays following a powerlaw or show evidence of coronal lines whose strength decays with time. They must be coincident with the centre of their host Galaxy, their host shows no evidence of AGN activity and they have temperature in the range of $\sim10^{4}$ K as derived from their optical emission. In addition, they either show no X-ray emission at all, or they show evidence of late time X-ray emission well after the original optical flare has disappeared or decayed. To determine whether a source is an optical/UV TDE we rely heavily on the currently accepted view in the literature of the nature of these events as optical/UV TDEs, while using our results to determine their X-ray emission.

The candidates which fall into this category are  \textit{ASASSN-14ae}, \textit{ASASSN-15lh}, \textit{D1-9}, \textit{D23H-1}, \textit{DES14C1kia}, \textit{iPTF16fnl}, \textit{PS1-10jh}, \textit{PS1-11af}, \textit{PS1-12yp}, \textit{PTF-09axc}, \textit{PTF-09djl}, \textit{PTF-09ge}, \textit{SDSS J0748}, \textit{SDSS J0952}, \textit{SDSS J1342}, \textit{SDSS J1350}, and \textit{TDE2}. These events have been classified as an optical/UV TDE in the literature, however our analysis either detects no X-ray emission arising from the position of these source or well after the original flare weak X-ray emission is detected. Even though \emph{PTF-10iya} was first classified as an optical TDE, we did not place this event into this category as its X-ray emission was detected simultaneously with the optical emission from the source, much like that of \emph{ASASSN-14li}, leading it to be placed in the \textit{likely X-ray TDE} category.

\subsubsection{Not a TDE}

The candidates from Table \ref{tdes} which fall into the \emph{not a X-ray TDE} category are either:
\begin{enumerate}[topsep=0pt,itemsep=-1ex,partopsep=1ex,parsep=1ex]
	\item Known AGNs, show evidence of a large scale jet, is a known GRB which shows a clear supernova counterpart or some other astrophysical object such as a low mass X-ray binary. Its classification as one of these astrophysical objects comes from other observations e.g., such as evidence of nuclear emission in the form of optical emission lines or X-ray fluorescence lines indicating an AGN. 
	\item The position of the source is not coincident (within uncertainties) with the centre of the host Galaxy.
	\item In addition, although not necessarily required, the X-ray luminosity of the source does not change significantly across observations. 
	\item Shows evidence of X-ray variability or X-ray emission of the same order of magnitude over long time scales.
\end{enumerate}

There are 15 TDE candidates from Table \ref{tdes} which we categorise as \textit{not a TDE}. This includes \textit{2MASXJ0203}, \textit{CSSS100217}, \textit{GRB060218/SN2006aj}, \textit{HLX1}, \textit{IC3599}, \textit{IGR J12580}, \textit{NGC1097}, \textit{NGC2110}, \textit{Pictor A}, \textit{PTF-10iam}, \textit{SDSS J0938}, \textit{SDSS J0939}, \textit{SDSS J1011}, \textit{SDSS J1055}, and \textit{SDSS J1241}.  Here we summarise why each of these events were placed into this category, while in Appendix Section \ref{individualsources} we go into more detail. 

\textit{2MASXJ0203} was suggested by \citet{2016arXiv160502749S} to be a highly variable AGN and shows evidence of variable X-ray emission which is approximately constant at peak and does not show a powerlaw decay expected for X-ray TDEs. As discussed in Appendix Section \ref{individualsources}, we do not detect any X-ray emission arising from TDE candidate \textit{3XMM}, while the properties derived by \citet{2015ApJ...811...43L} do not match those seen in other TDEs.   Using multi-wavelength observations, \citet{2011ApJ...735..106D} found that \textit{CSS100217} exhibited spectroscopic features representative of Type IIn supernovae. The favoured explanation of \textit{GRB060218/SN2006aj} is an under luminous long GRB with a low ejecta supernova. \textit{HLX-1} shows evidence of variable X-ray emission over 10 years, and radio observations reveal a large scale jet. Using radio observations, \citet{2013ApJ...763...84B} suggested that the properties of \emph{IC3599} is consistent with AGN emission, while \citet{2015ApJ...803L..28G} showed that the periodic X-ray emission observed most likely arises from accretion around the BH not a TDE. \textit{IGR J12580} was classified as a LINER/Seyfert 2 Galaxy based on its optical spectra indicating that this source could be a changing look quasar, while a number of other observations in different wavelengths support the AGN origin of this source. \textit{NGC1097}, and \textit{NGC2110} are thought to be AGN (see e.g., \citealt{1995ApJ...443..617S, 2015MNRAS.447..160M, 2003ApJ...592...42G, 2004ApJ...601.1159G}).  \textit{Pictor A} has a large scale jet \citep{1997A&A...328...12P}, which we believe arises from an underlying AGN.  \citet{2016ApJ...819...35A} suggested that \textit{PTF-10iam} is most likely a peculiar Type II or a hybrid Type Ia-Type II SN.  \citet{2012ApJ...749..115W} originally suggested that \textit{SDSS J0938} was a TDE based on  coronal lines detected from the host, however it is more likely that these lines arise from the presence of an obscured AGN \citep{2013ApJ...774...46Y} . The host of \textit{SDSS J0939} was classified as a Narrow line Seyfert 1 galaxy by \citet{2007A&A...462L..49E}. \citet{2016MNRAS.455.1691R} characterised \textit{SDSS J1011} as a changing look quasar. \textit{SDSS J1055} has narrow line ratios indicating that it is an AGN, while the coronal lines that first lead to \textit{SDSS J1241} being classified as a TDE did not change in strength with time making it more likely to be an AGN \citep{2012ApJ...749..115W}.

\subsubsection{Unknown classification}

Candidates that fall into the \emph{unknown} category show evidence of an X-ray or optical/UV flare in the literature, however due to, e.g., being offset from the center of the host Galaxy or lack of data etc., its classification as a TDE is not certain. There is also no additional information from the literature or archival/additional observations that can either confirm its classification as a optical/UV or X-ray TDE, or rule out the presence of an e.g., AGN etc. The remain five TDE candidates from Table \ref{tdes} that we classify as \textit{unknown} are \textit{Dougie}, \textit{PGC1185375}, \textit{PGC1190358}, \textit{PTF-10nuj}, \textit{PTF-11glr}. 

\textit{Dougie} was suggested to be an optical TDE based on the evolution of its optical light curve, however this source is systematically offset from the center of its host. We also do not detect any X-ray emission from this source and there is no other information in the literature which rules out or confirms its nature.  Even though the \emph{Swift} BAT detected a flare from \textit{PGC1185375} and \textit{PGC1190358}, there is very little soft X-ray data available overlapping the positions of these objects. This makes it difficult to be quantify the nature of these sources. The optical light curve of \textit{PTF-10nuj} and \textit{PTF-11glr} suggests that these source are optical TDEs, however both are found systematically offset from its host Galaxy.  We also do not detect any X-ray emission from the source, and there is very limited amount of data about the host placing these sources into the \textit{unknown} category.

\section{Analysing the X-ray properties of X-ray TDEs}\label{props}
Using the \textit{X-ray TDE} and \textit{likely X-ray TDE} sample summarised in Table \ref{goldsilvers}, we use these events to characterise the properties of X-ray TDEs in a systematic and comprehensive way. 

\subsection{How luminous are X-ray TDEs?}

\begin{figure}[t!]
	\begin{center}
		
		\includegraphics[width=0.5\textwidth]{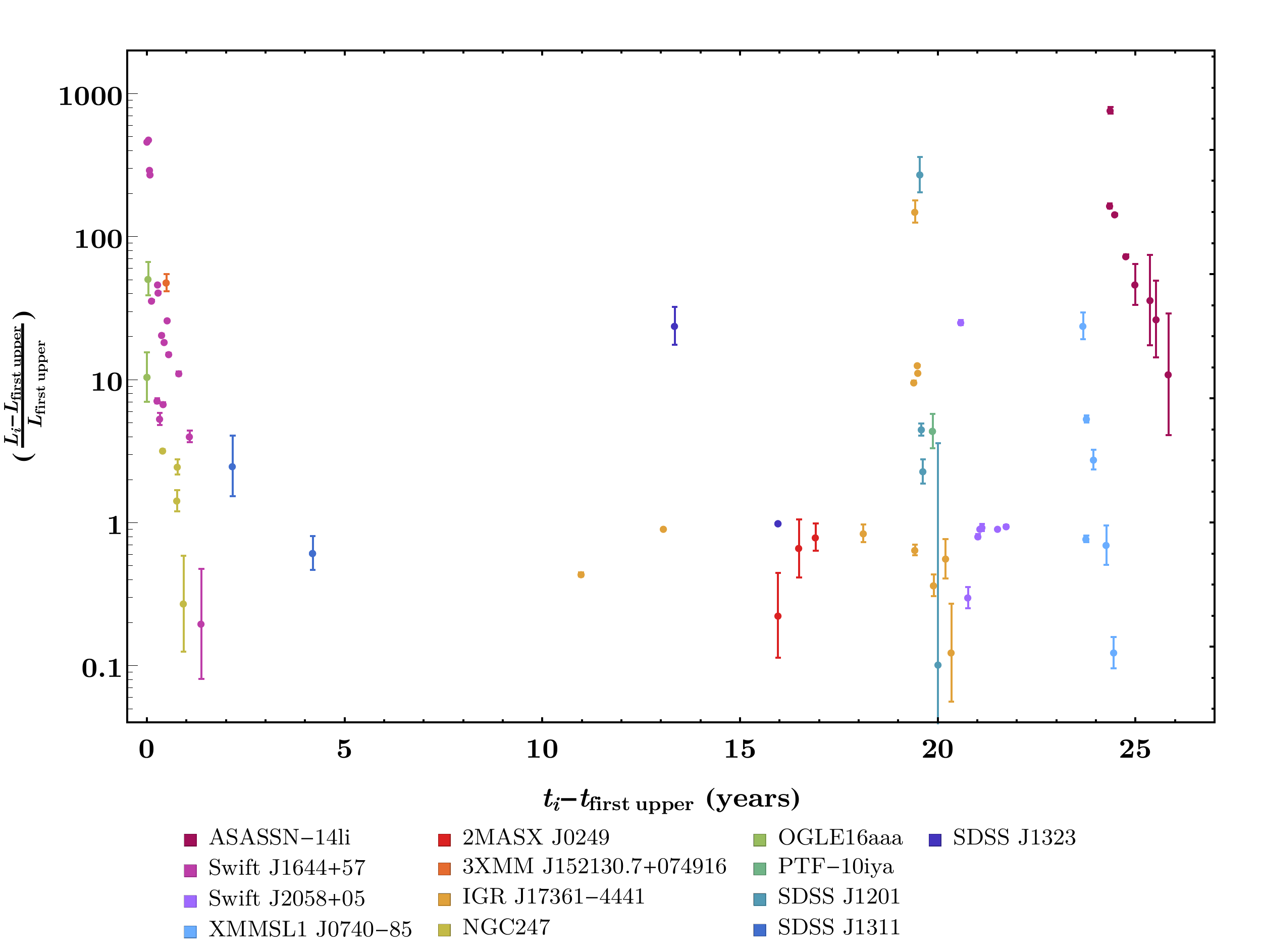}
		\caption{The difference between the X-ray luminosity detected at or after peak compared to the X-ray luminosity derived from the first X-ray upperlimit  $((L_{i}-L_{\rm first upper})/L_{\rm first upper})$. Due to the fact that for a large faction of our events the X-ray upperlimits derived using \emph{XMM-Newton slew} observations are not constraining and are thus larger than the detected X-ray emission from other instruments, we do not use these upperlimits for this plot. This is plotted against the observation date minus the date in which the first upper limit  $(t-t_{first upperlimit})$ was measured and has been normalised to the X-ray luminosity of the first X-ray upperlimit. One can see that the X-ray luminosity of all sources (with the exception of 2MASX J0249) increases by at least one to two orders of magnitude compared to their first X-ray upperlimit. \label{comp2upper}}
		
	\end{center}
\end{figure}

\begin{figure}[t]
	\begin{center}
		
		\includegraphics[width=0.5\textwidth]{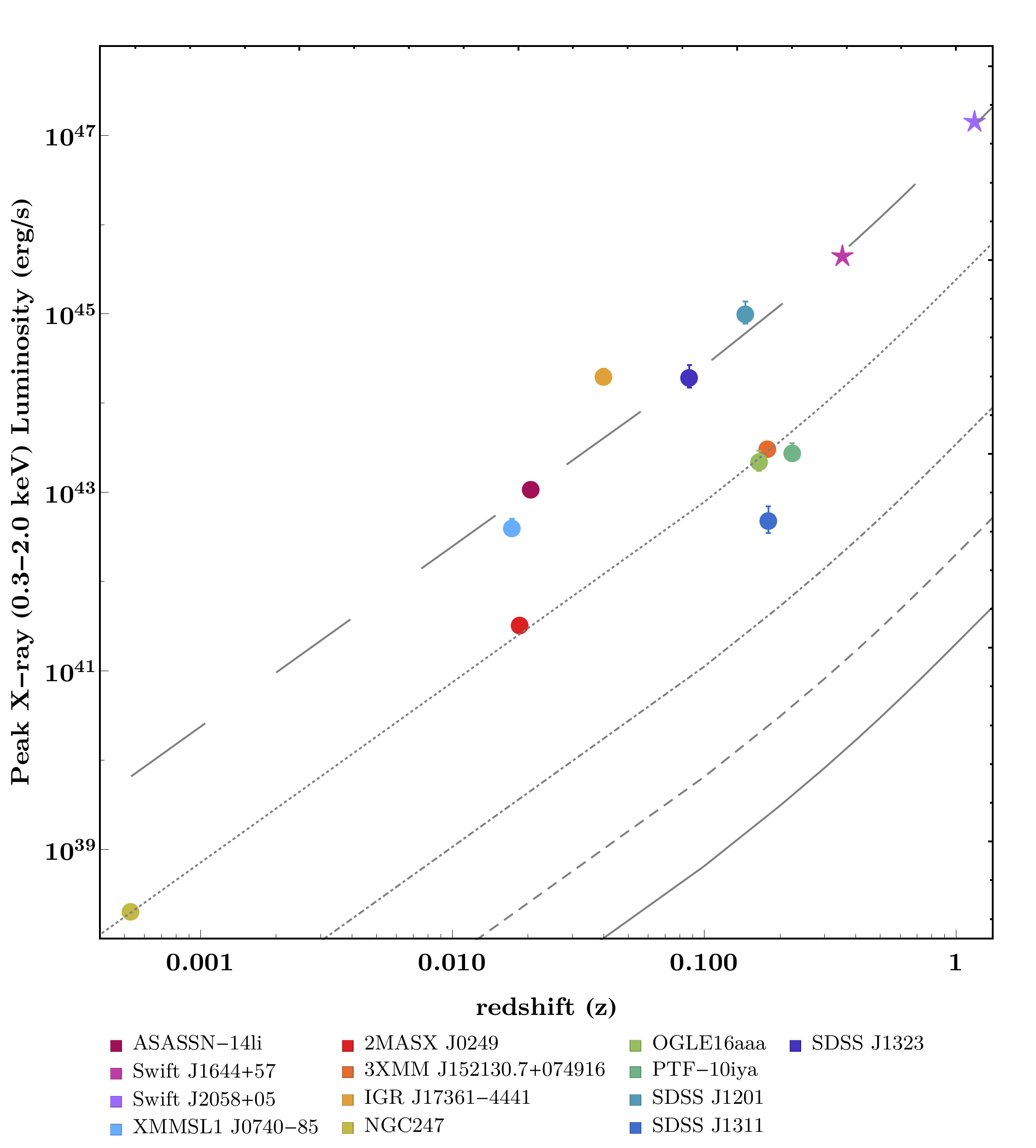}
		\caption{Peak X-ray luminosity as a function of redshift for out X-ray TDE candidates. Overlaid is the sensitivity bands for \textit{ROSAT} all sky survey, \textit{Chandra, XMM} and \textit{Swift}. These bands were taken from the most stringent 0.5-2.0 keV flux limits derived from different extragalactic surveys taken by the different instruments (adapted from Figure~1 of \citealt{2015ApJS..218....8D}). Here the limits derived from the \textit{ROSAT }all sky survey \citep{1999A&A...349..389V}, \textit{Chandra }2Ms Deep field North survey \citep{2003AJ....126..539A},\ \textit{XMM-Newton} Lockmann 0.8 Ms survey \citep{2001A&A...365L..45H}, \textit{ Swift }active galactic nucleus and cluster survey \citep{2015ApJS..218....8D} and the \textit{Swift }BAT 70 month 14\--195 keV all sky survey \citep{2013ApJS..207...19B} are shown as the grey dotted, solid, dashed, dot-dashed and large dashed lines respectively. \label{firstupper}}
		
	\end{center}
\end{figure}

The accretion of stellar material from a star that has been tidally disrupted by its host BH will produce a short-lived, luminous accretion-powered flare. From our systematic analysis we can quantify how luminous these events become relative to their derived pre-flare upperlimits. In Figure~\ref{comp2upper} we have plotted against $t_{i}-t_{\rm first upper}$, the difference between the X-ray luminosity detected at and after peak relative to the luminosity of the upperlimit immediately preceding the first X-ray detection of the flare ($(L_{i}-L_{\rm first upper})/L_{\rm first upper}$). Due to the low exposure times of the \emph{XMM-Newton slew} observations, a large number of the X-ray upperlimits derived from these observations are not very constraining. As such, these upperlimits are either significantly above or equivalent to the peak X-ray emission detected from these TDE candidates and thus provide limited information about the pre- or post-flare emission from these events. As a consequence, we do not use the upperlimits derived using \emph{XMM-Newton slew} to derive Figure~\ref{comp2upper}, and instead rely on the deeper X-ray upperlimits derived using \emph{ROSAT}, \emph{XMM-Newton pointed} observations, \emph{Chandra} or \emph{Swift}. Here $t_{\rm first upper}$ and $L_{\rm first upper}$ is the date and measured luminosity of the first upperlimit before the detected X-ray flare, while $t_{i}$ and $L_{i}$ is the date and luminosity of the $i$-th data point measured during the flare. 

Figure~\ref{comp2upper} highlights the importance of having pre-flare constraints for a TDE candidate that are, at the very least, equivalent to an X-ray upperlimit derived using \emph{ROSAT}. These X-ray upperlimits allow us to characterise how luminous an X-ray TDE becomes during the initial flare, and as it evolves. With the exception of \textit{2MASX J0249} whose first upperlimit before the flare is derived from the shallow \emph{ROSAT} RASS observation, all of the TDE candidates we consider show an increase in their X-ray luminosity between one to three orders of magnitudes at peak. The events that show the most dramatic change in their emission are \textit{ASASSN-14li}, \textit{IGR J17361-4441}, \textit{SDSS J1201} and \textit{Swift J1644+57}, which display an increase in their X-ray luminosity of nearly three orders of magnitude. \textit{Swift J2058+05}, \textit{3XMM}, \textit{OGLE16aaa}, \textit{SDSS J1323} and \textit{XMMSL1 J0740-85} also show a significant increase of nearly two orders of magnitude compared to their pre-flare upperlimit, while all other events differ from their X-ray upperlimits by approximately one order of magnitude. Interestingly, both thermal and non-thermal X-ray TDEs both show this significant increase in their X-ray emission, indicating that X-ray TDEs are intrinsically very luminous events regardless of their nature. For the events that had pre-flare limits nearly immediately before the detected flaring event such as \textit{Swift J1644+57}, we also find that these events dramatically increase by many orders of magnitude over relatively short timescales.  Most of these events are also undetected in X-rays or fall to limits similar to that of their pre-flare limits within a year of the initial flaring event.

In Figure~\ref{firstupper} we have plotted the peak X-ray luminosity of each TDE as a function of redshift. Shown as the solid, dashed, dot-dashed, dotted and large dashed grey lines are the most stringent 0.5-2.0 keV flux limits derived from the \textit{ROSAT} all sky survey \citep{1999A&A...349..389V}, \textit{Chandra} 2Ms deep field north \citep{2003AJ....126..539A}, \emph{XMM-Newton} 0.8Ms Lockman hole survey \citep{2001A&A...365L..45H}, the Swift active galactic nucleus and cluster survey \citep{2015ApJS..218....8D} and the Swift BAT 70 month all sky survey \citep{2013ApJS..207...19B}. One can see that \emph{ASASSN-14li}, \emph{Swift J1644+57}, and \emph{Swift J2058+05}, \emph{IGR J17361-4441}, \emph{SDSS J1201}, and \emph{SDSS J1323} have a peak luminosity that is either above or comparable with the detection limit associated with the \textit{Swift} BAT, indicating that only the most extreme events are going to be detected through trigger of the BAT. The \emph{ROSAT} all sky survey would have detected nearly all of the X-ray TDE candidates we consider, with the exception of \emph{PTF-10iya}, and \emph{SDSS J1311}, which fall below this flux limit. 

Due to the limitations of current X-ray satellites, it is not 100\% surprising that we are currently susceptible in TDE studies to detecting only the most luminous X-ray TDEs found at close redshifts. Most of our current sample of TDEs are detected at redshifts $<0.2$, with very few detected at a redshift greater than $z\sim0.7$. In fact, the events detected at the highest redshift are some of the brightest TDE candidates and are jetted in nature, making these unique events in their own right. X-ray surveys are well designed to search for TDE in the low-redshift universe, while the increase sensitivity of \emph{Chandra}, \emph{XMM-Newton} and \emph{Swift} compared to that of \emph{ROSAT} also allow us to search for fainter TDE candidates at both higher and lower redshifts. 

Due to the observational bias towards detecting the brightest TDEs, this leads to the question of why have we not detected a larger number of lower luminosity TDEs, especially at lower redshifts? The discrepancy between the expected theoretical rate of TDEs and our current rate at which we observe these events is well known \citep[see][ and references therewithin]{2016MNRAS.461..371K}. However, this observational bias might arise from the intrinsic nature of X-ray TDEs themselves. In Section \ref{evolve} and \ref{bhmass}, we suggest that X-ray TDEs are viscously delayed (i.e., the timescale for which material from the disrupted star accretes onto the BH very long). \citet{2015ApJ...809..166G} showed that this process has a dramatic affect on the properties of these events. In particular, if a TDE is drastically viscously delayed most of these events would be sub-Eddington in nature and will peak over timescales of many years. As our current sample of TDEs are found to peak over a few weeks to months (classified as prompt) and exhibit many orders of magnitude increases in their X-ray luminosity, this implies that there is most likely a large population of low luminosity (possibly slow-rise) TDEs that are being missed by current surveys/observations or mistaken for other phenomenon\footnote{For \textit{NGC3599} which \citet{2008A&A...489..543E} and \citet{2015MNRAS.454.2798S} suggest could result from a slow-rise TDE, we find that the emission from this source would exhibit an increase in luminosity similar to those of our brightest, prompt events if we plotted this source on Figure~\ref{comp2upper}. However, rather than increasing and then decreasing in magnitude over a very short time frame, the emission from this candidate would decay over timescales much longer than the other candidates in our sample. If this source is a slow-rise TDE rather than emission arising from thermal instability in the accretion disc of an AGN as suggested by \citet{2015MNRAS.454.2798S}, then this would be a unique event in its own right as it would be the first and the brightest slow-rise TDE detected.}. As a consequence, the viscously slowed nature of X-ray TDEs might explain the current discrepancy between the theoretically expected and observationally detected rate of TDEs \citep{2016MNRAS.455..859S}.

This highlights the need to have a wide range of instruments with quite different capabilities to be able to detect potential X-ray TDEs. Large X-ray surveys provide us with the ability to detect fainter, slow rising and prompt X-ray TDEs over a wide range of redshifts, while monitoring instruments like the \emph{Swift BAT} allow us to detect the most extreme events of these classes at higher redshifts. With upcoming X-ray satellites such as \emph{eROSITA} which have similar capabilities as \emph{ROSAT} but with a larger effective area\footnote{See \url{http://www.mpe.mpg.de/455799/instrument} for more information about \emph{eROSITA}}, will provide us with the ability to detect a wide range of TDEs across the low to high redshift universe. Combined with the continual monitoring of our current X-ray satellites, this will open new doors into studying the formation and evolution of TDEs, as well as the properties and environments of their BHs.

\subsection{How do X-ray TDE decay?}\label{pwldecay}

\begin{figure}[t!]
	\begin{center}
		\includegraphics[width=\columnwidth]{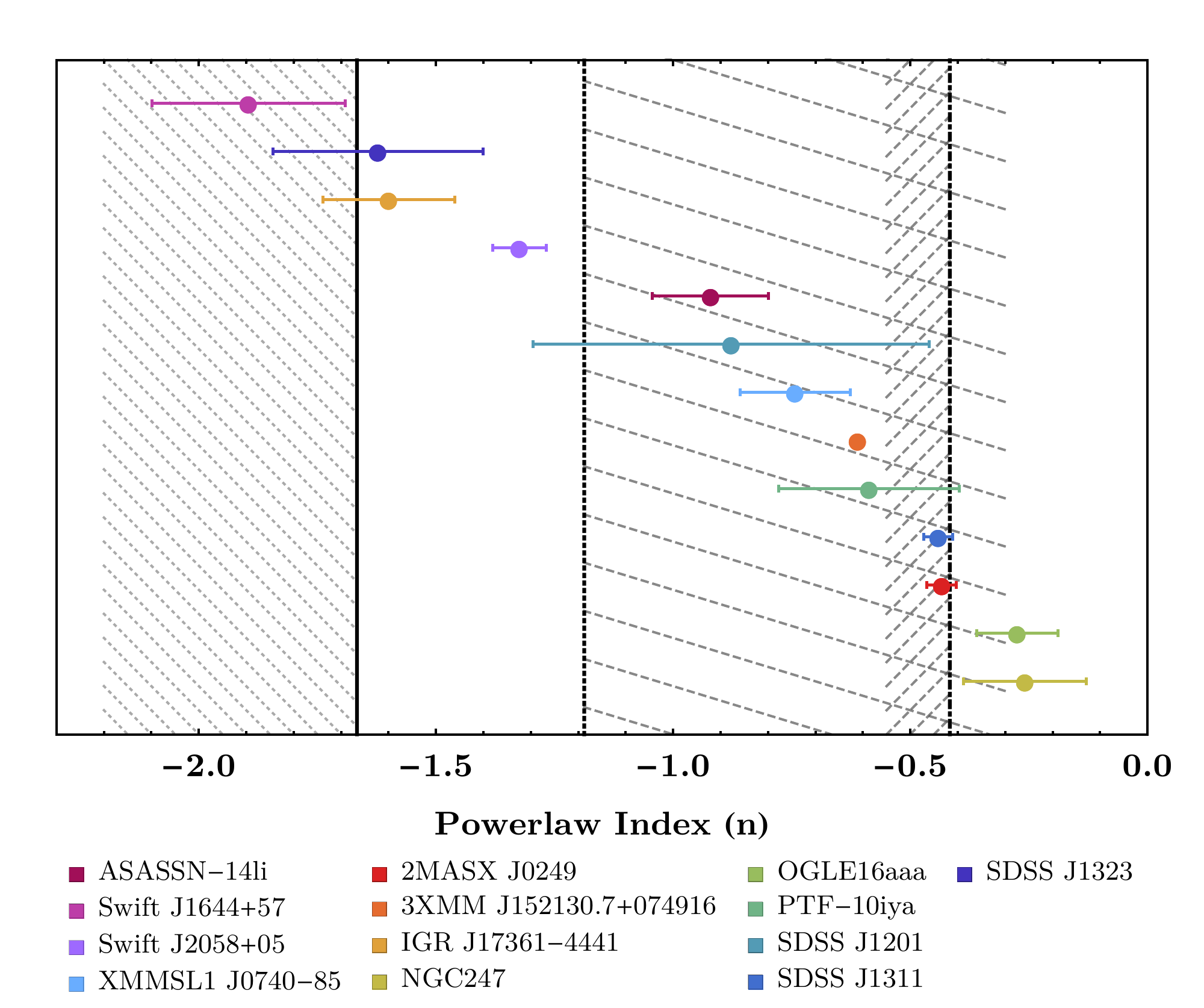}
		\caption{The best fit powerlaw index and its $1\sigma$ uncertainty obtained from fitting the X-ray light curves of our TDE sample as shown in Appendix Section \ref{powerlawmodels}. We have also overlaid the ranges of various powerlaw indexes expected for TDEs. Here the solid black line is the standard $t^{-5/3}$ from fallback (e.g., \citealt{1989ApJ...346L..13E, 1989Natur.340..595P}), while the black dotted line is the powerlaw index $t^{-19/16}$ expected from a viscous disk accretion \citep{1990ApJ...351...38C}. The black dot-dashed line represents the much shallower powerlaw index of  $t^{-5/12}$ representative of disk emission \citep{2011MNRAS.410..359L}.  The shaded regions represent the range of powerlaw indexes expected for TDE as determined by \citet{2013ApJ...767...25G}, assuming either -5/3, -19/16 or -5/12 respectively.\label{pwlindex}}
	\end{center}
\end{figure}

The luminosity of a TDE is assumed to decay following a simple $t^{-5/3}$ powerlaw (e.g., \citealt{1989ApJ...346L..13E, 1989Natur.340..595P}). However, \citet{2009MNRAS.392..332L} showed analytically, while \citet{2013ApJ...767...25G} determined using hydrodynamical simulations, that the powerlaw index for decay depends heavily on the stellar structure (i.e., whether their mass is centrally-concentrated or not) and as such dramatically steeper powerlaw indexes than the assumed $-5/3$ can be obtained. Specifically, \citet{2013ApJ...767...25G} showed that the expected rate of mass return to the black hole for low and high mass stars asymptotes to $\sim-2.2$ (from $-5/3$) for nearly half of all stellar disruptions. In addition, \citet{2011MNRAS.410..359L} showed that at late times the light curves of optical/UV TDEs tend to follow a powerlaw with an index of $-5/12$ , assuming that the observed emission arises from disk emission.

To determine how the X-ray emission from our sample of TDEs decays, we model the full X-ray light curves seen in Figure~\ref{goldlight} and \ref{silverlight} using a simple powerlaw, where we allow the normalisation and the powerlaw index $\Gamma$ to be free parameters. In Figure~\ref{pwlindex} we have plotted the best fit power law index and its $1\sigma$ uncertainty for each TDE candidate we consider. For reference, we have also plotted the individual best fit models and their uncertainties of each of these TDE candidates in Appendix Section \ref{powerlawmodels}. Overlaid on Figure~\ref{pwlindex} is the different powerlaw indexes that one expects to see from TDEs as they decay. Plotted as the solid black line is the standard $t^{-5/3}$, while the black dot-dashed and dotted line represents the much shallower powerlaw index of  $t^{-5/12}$ and $t^{-19/16}$ derived by \citet{2011MNRAS.410..359L} and \citep{1990ApJ...351...38C} respectively. The grey shaded region to the right of the black solid line is the powerlaw indexes derived by \citet{2013ApJ...767...25G}. Plotted as the grey shaded region to the right of the black dot-dashed line is the corresponding band of powerlaw indexes one would expect for disk emission, assuming a similar relationship as that derived by \citet{2013ApJ...767...25G}.

Based on our powerlaw model fits, we find that the X-ray emission of our TDE sample cluster around either the standard powerlaw index expected from accretion ($t^{-5/3}$), or around the index derived assuming disk emission ($t^{-5/12}$). The events which favour the more shallower powerlaw index are thermal in nature. In comparison, \emph{Swift J1644+57}, \emph{SDSS J1323} and \emph{IGR J17361-4441} exhibit a decay that is consistent within uncertainties with the commonly used index of $-5/3$. Uniquely, non-thermal TDE \emph{Swift J2058+05} has a light-curve decays at a rate in between these two characteristic emission properties.

We find that our powerlaw fits, differ somewhat from those listed in the literature for each TDE. For example, \citet{2012ApJ...753...77C} derive a powerlaw decline of $\sim-2.2$ for the early time 0.3\--10.0 keV X-ray emission of \emph{Swift J2058+05}. However we find that taking into account the full X-ray light curve, we derive a shallower index in the 0.3\--2.0 keV energy band. The difference in the results derived in our analysis compared to those in the literature most likely arises from two things. Many papers derive the best fit powerlaw decay index associated with the early time X-ray light curve via the 0.3\--10.0 keV energy band. This is in contrast to our analysis which focuses on the 0.3\--2.0 keV emission of each source and takes advantage of the fact we can now derive and fit the (nearly) complete X-ray light curve of each of these events that include both detected data points and upperlimits. By deriving the X-ray emission in a smaller energy band, we are probing the decay rate of a different component of the TDEs emission which may decay at or contribute at a different rate than that of the harder 2.0-10.0 keV energy band which authors in the literature are also probing. In addition, by combining observations that were taken around the same MJD as discussed in Section \ref{dataanalysis}, we are not as prone to the affect that short term variability and changes in the decay rate has on the derived decay powerlaw index. 

The fact that we find that the powerlaw fits derived from the full X-ray light curve are different from those derived at early times, hints towards the possibility that the decay rate of X-ray TDEs changes with time. Following this idea, in Section \ref{evolve} we investigate the differences between the early and late time decay rates of these events and the question of whether the X-ray emission from these events evolve or not.

\subsection{How quickly do X-ray TDEs release their energy?}\label{t90calc}

\begin{figure}[t!]
	\begin{center}
		\includegraphics[width=\columnwidth]{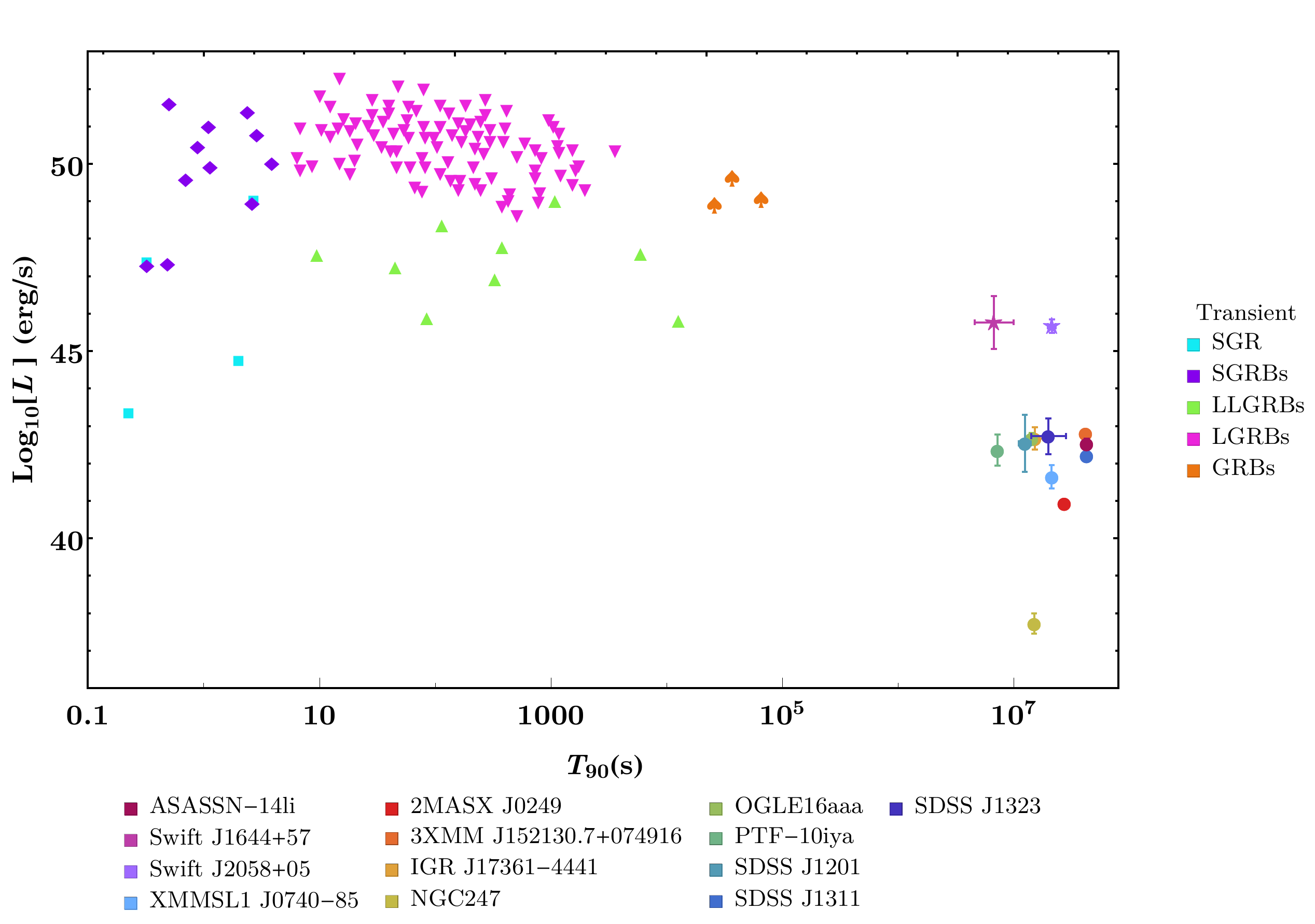}
		\caption{The $T_{90}$ and the corresponding luminosity ($L_{90}$) over this same time period plotted for our TDE candidates. The $T_{90}$ and $L_{90}$ values of the GRB/GRB-like transient events have been adapted from Figure~2 of \citet{2014ApJ...781...13L}. \label{t90s}}
	\end{center}
\end{figure}

\begin{figure}[t]
	\begin{center}
		\includegraphics[width=\columnwidth]{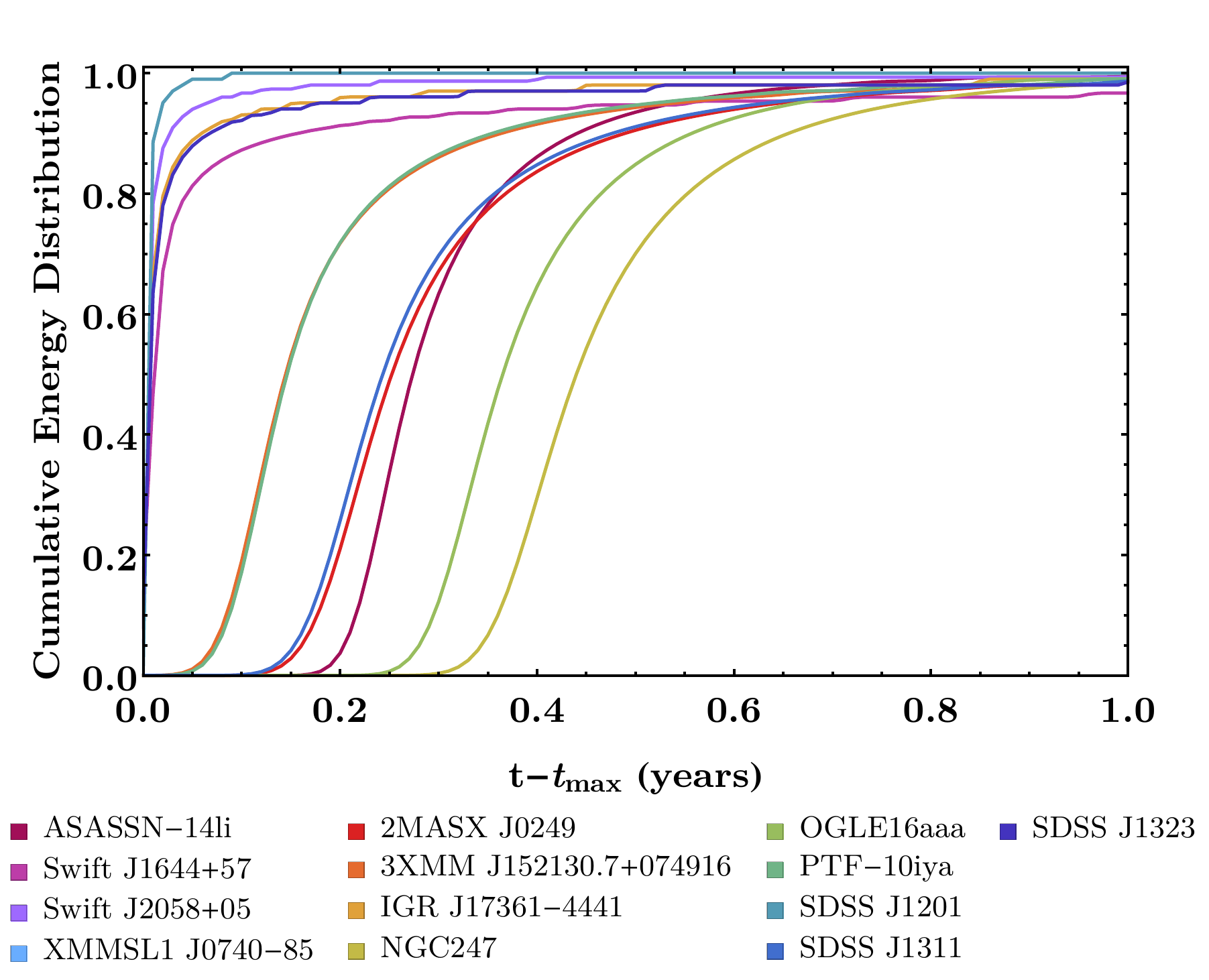}
		\caption{The cumulative energy distribution as a function of time since peak ($t-t_{max}$) for our TDE candidates.  \label{ced}}
	\end{center}
\end{figure}

In studies of GRBs, a commonly derived parameter used to help characterise the properties and type of outburst detected is $T_{90}$. $T_{90}$ represents the time interval in which between 5\% and 95\% of the total fluence from a source is observed (see review by e.g., \citealt{2015JHEAp...7...44L}). Filling in these gaps is not such a problem for short events such as short gamma-ray bursts (SGRBs), since estimating $T_{90}$ is relatively straightforward and a good approximation of the actual $T_{90}$. However for very long events like TDEs, it is difficult for instruments such as \emph{Swift} to capture the full emission structure of an event, leading to gaps in the observed X-ray light curve. However, as the derivation of $T_{90}$ is dominated by the long lived, lower luminosity emission of the source, this means that the $T_{90}$ derived in these cases is more an approximation of the actual $T_{90}$. As a consequence, the longer that a source can be followed, the more energy will be integrated over in the $T_{90}$ calculation leading to larger $T_{90}$ values. Indicating that at early times, for long transient events deriving  $T_{90}$ values provides more of a lower limit to the actual $T_{90}$ of an event. Regardless of this fact, this still gives us insight into the characteristics of these transient events.

For our TDE candidates, we derived $T_{90}$ and the corresponding isotropic luminosity ($L_{90}$) over this same time period. In Figure~\ref{t90s} we compare our TDE $T_{90}$ and $L_{90}$ to other well known transient objects including Galactic soft-gamma repeaters (SGRs), long- and short-duration GRBs (LGRBs and SGRBs), low luminosity GRBs (LLGRBs) and ultra-long GRBs. The values of  $T_{90}$ and  $L_{90}$ for these other transient sources were adapted from Figure~2 of \citet{2014ApJ...781...13L}. Similar to that presented by \citet{2014ApJ...781...13L}, Figure~\ref{t90s} illustrates that TDEs are extremely long lived events compared to other GRB and GRB-like events, significantly differentiated themselves from these other transient events in the $T_{90}$ and $L_{90}$ parameter space. The values of $T_{90}$ are larger than those presented by \citet{2014ApJ...781...13L} as we have the advantage of having the (nearly) full X-ray light curve available to us for each event, while \citet{2014ApJ...781...13L} focus on $T_{90}$ derived from the early time emission of these sources.

Compared to the GRB/GRB-like sample, our sample of TDEs cover a wide range of lower-luminosity values, with the brightest and most extreme TDEs so far detected in our sample (jetted TDEs: \emph{Swift J1644+57} and \emph{Swift J2058+05}) fall towards the lower end of the luminosity distribution of GRB/GRB-like events.  The fact that these events differ significantly from other extreme events such as GRBs, confirms that these flare-like events arise from significantly different progenitors and provides us with a relatively simple diagnostic to separate these events from GRB/GRB-like event based on their $T_{90}$ and $L_{90}$.

Based on our derived $L_{90}$, we find that our sample of events naturally separate into two distinct groups. The non-thermal jetted TDEs have a $L_{90}\sim10^{44}$ erg s$^{-1}$, while the thermal non-jetted events have a $L_{90}\lesssim 10^{42}$ erg s$^{-1}$. This natural separation implies that there is a bimodal distribution in the bolometric luminosities of X-ray TDEs. This raises the question of why are there no X-ray TDEs with a $L_{90}$ intermediate of these two values? One possibility is that there is a population of X-ray TDEs that may bridge the gap in luminosity, however we are missing these TDEs in the X-ray band as their X-rays are being reprocessed into optical or UV wavelengths. This is not unreasonable, as higher luminosities events have significantly more mass surrounding the source. Due to the highly collimated nature of jetted events, these X-ray TDEs are largely unaffected by the large amount of mass surrounding these events (e.g., \emph{Swift J1644+57} was highly extinct, suggesting a significant amount of stellar debris). For the lower luminosity non-jetted X-ray TDEs there is most likely not a large amount of material surrounding these sources and thus these events can ionise the surrounding material quickly making it transparent to X-rays over a short time period. The population of TDEs that could naturally occupy this ``reprocessing valley'' are optical UV/TDEs. A large number of these events exhibit significant reprocessing (i.e., \textit{PS1-10jh}: \citealt{2012Natur.485..217G, 2015ApJ...815L...5G, 2014ApJ...783...23G}, \textit{PS1-11af}: \citet{2014ApJ...780...44C}) and also exhibit bolometric luminosities that fall within this ``reprocessing valley''. These optical/UV TDEs most likely exhibit X-rays, however due to the large amount of material obscuring these events the emission is reprocessed into a lower wavelength.

In Figure~\ref{ced} we have plotted the cumulative energy distribution as a function of time for the TDE candidates we consider. One notices quickly that the non-thermal and thermal TDE candidates release their energy over quite different time scales. Jetted events \emph{Swift J1644+57}, and \emph{Swift J2058+05} release $>$80\% of their energy within their first month. However, the non-jetted events release their energy much more gradually, taking approximately five months to reach the amount released by the jetted events in their first month. Interestingly, \emph{IGR J17361-4441} which was first discovered as a hard X-ray source, and TDE candidates SDSS J1201 and SDSS J1323 also show similar behaviour to the two jetted events in our sample, which could indicate that these sources also exhibit significant non-thermal emission during their flaring event. This highlights that thermal and non-thermal events emit their energy by two quite different mechanisms and that within $\sim$1 month of detecting a non-thermal X-ray TDE, the derived $T_{90}$ value is more or less representative of the actual $T_{90}$ value, while one would need to monitor thermal TDEs over a much long time period to derive $T_{90}$.

\subsection{How absorbed are X-ray TDEs?}\label{absorb}
\begin{figure}[t]
	\begin{center}
		\includegraphics[width=\columnwidth]{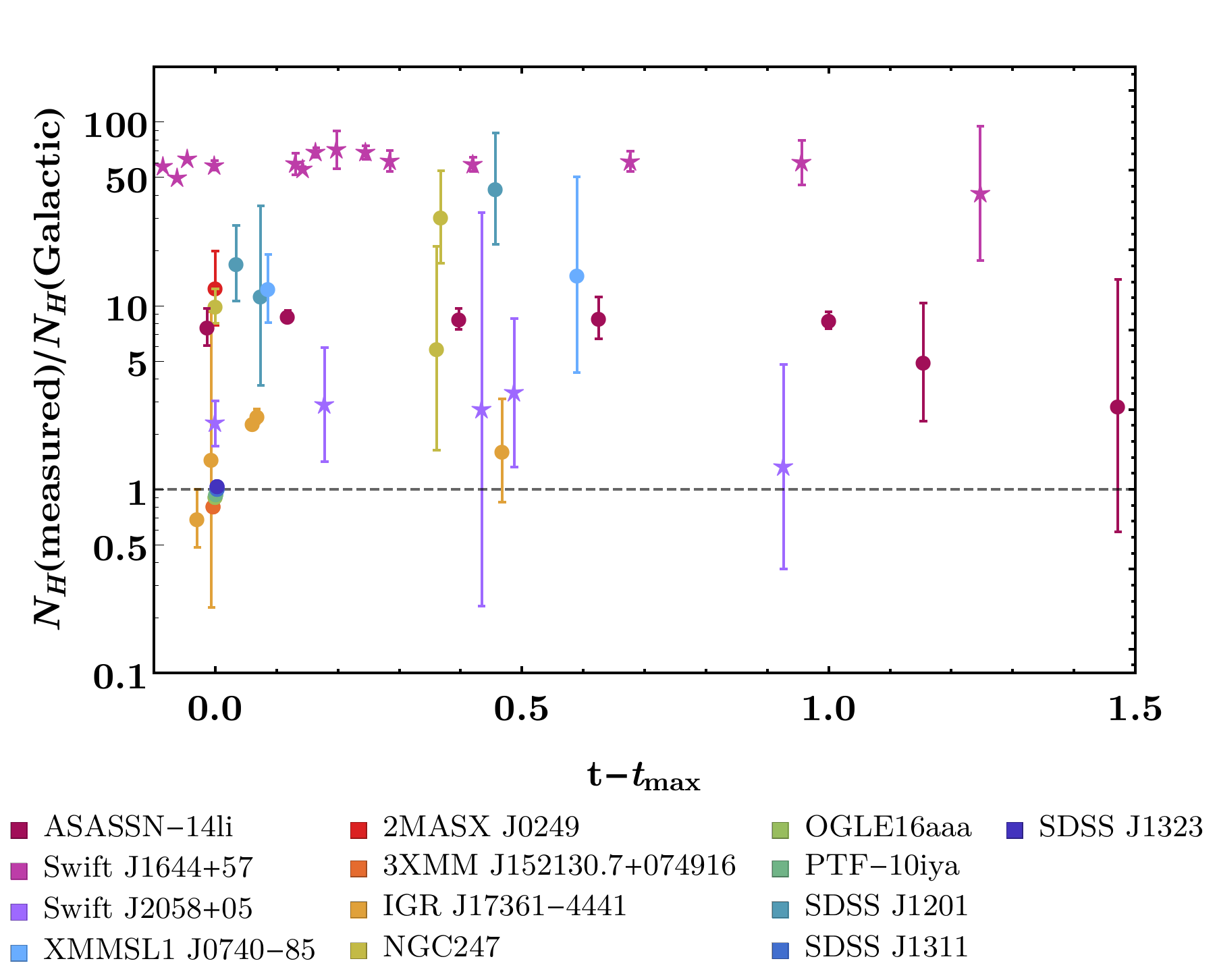}
		
		\caption{The ratio of the measured column density ($N_{H}$) and the Galactic column density  along the line of sight as derived from \citet{2005A&A...440..775K}, as a function of time $t-t_{\rm max}$. The dashed black line corresponds to when the measured $N_{H}$ is equal to the Galactic column density. \label{nhvari} }
	\end{center}
\end{figure}
{2, 5, 6, 8, 9}

From the column densities ($N_{H}$) derived from fitting the X-ray spectra of each TDE (see Tables \ref{rosatfits}--\ref{swiftfits}), we can make a statement about the environment in which these events are found. In Figure~\ref{nhvari} we have plotted as a function of time, the ratio of the measured $N_{H}$ against the Galactic column density along the line of sight as measured by the Leiden/Argentine/Bonn (LAB) Survey of Galactic H$\textsc{i}$ \citep{2005A&A...440..775K}. From this plot one can see that a large fraction of X-ray TDEs have $N_{H}$ values that are at least two times greater than the Galactic column density measured along the line of sight to these events. For \textit{3XMM, OGLE16aaa, PTF-10iya, SDSS J1311} and \textit{SDSS J1323}, which we were unable to constrain $N_{H}$ from the available X-ray observations, and thus assume the $N_{H}$ derived from the LAB Survey. \emph{Swift J1644+57} is the most highly absorbed event out of all our TDE candidates. We find that both jetted and non-jetted events show evidence of this enhanced absorption, with most of the TDE candidates we consider have a $N_{H}$ that is between $\sim2-10$ times that of their Galactic $N_{H}$. 

Based on this analysis, there is no obvious trend in the value of $N_{H}$ that separates thermal (non-jetted) or non-thermal (thermal) TDEs. Both types of events show evidence of strong enhancement in $N_{H}$. We also find that most of the TDE candidates we consider show no significant variation in $N_{H}$ as a function of time. The exception to this is \emph{Swift J1644+57}, which as \citet{2011Natur.476..421B} also highlighted, shows some evidence of variation ($\sim1\sigma$) with $N_{H}$ at early times; however this variation is not significant at later times.

Since we find that nearly all these events are quite absorbed in nature, this indicates that there must be a larger amount of extinction surrounding these sources in their host galaxies. In fact, the values we derive for $N_{H}$ are most likely a lower limit to the actually $N_{H}$ in these hosts. Recently \citet{2014ApJ...793...38A} determined that a significant fraction of optical TDEs are found in post-starburst galaxies. From modelling the optical spectra of these hosts their results imply that optical TDEs occur in galaxies with sub-solar abundances.  However, to derive $N_{H}$ we assume solar abundances, which based on the work by \citet{2014ApJ...793...38A} indicates that we are underestimating the actual $N_{H}$ towards the source. As the $N_{H}$ of each event as a function of time can be well-approximated using a constant, this could also suggest that the material surrounding these events is quite dense and the ionising radiation is unable to change its absorption properties. Even though these events are quite absorbed the amount of material required to produce these column densities implied by our fits is quite small compared to what one would expect from the overall mass/energy budget of the event (i.e., $1-10$\% $M_{\odot}$). The exception to this is \emph{IGR J12580}. However, the low mass estimate implied by these fits most likely arises from the fact we are underestimating the actual $N_{H}$ rather than the amount of material being quite small.

Based on their X-ray light-curves \emph{Swift J1644+57}, \emph{ASASSN-14li}, \emph{XMMSL1 J0740-85} and to a lesser extent \emph{Swift J2058+05} show evidence of variability in their X-ray light curves. This variability is quite pronounced when looking at individual observations of these sources but is also noticeable in our derived light curves (see e.g., Figure~\ref{goldlight}). One possibility is that this variability is driven by absorption. In this case, one would expect to observe dramatic changes in the $N_{H}$ as a function of time which correlated with the observed variability. However, we do not see this in Figure~\ref{nhvari} indicating that the variability seen in the X-ray emission from these events is intrinsic to the source, rather than a result of the environment.

\subsection{How soft are X-ray TDEs and how does this softness evolve?}\label{hrsoft}

\begin{figure}[t!]
	\begin{center}
		\includegraphics[width=0.49\textwidth]{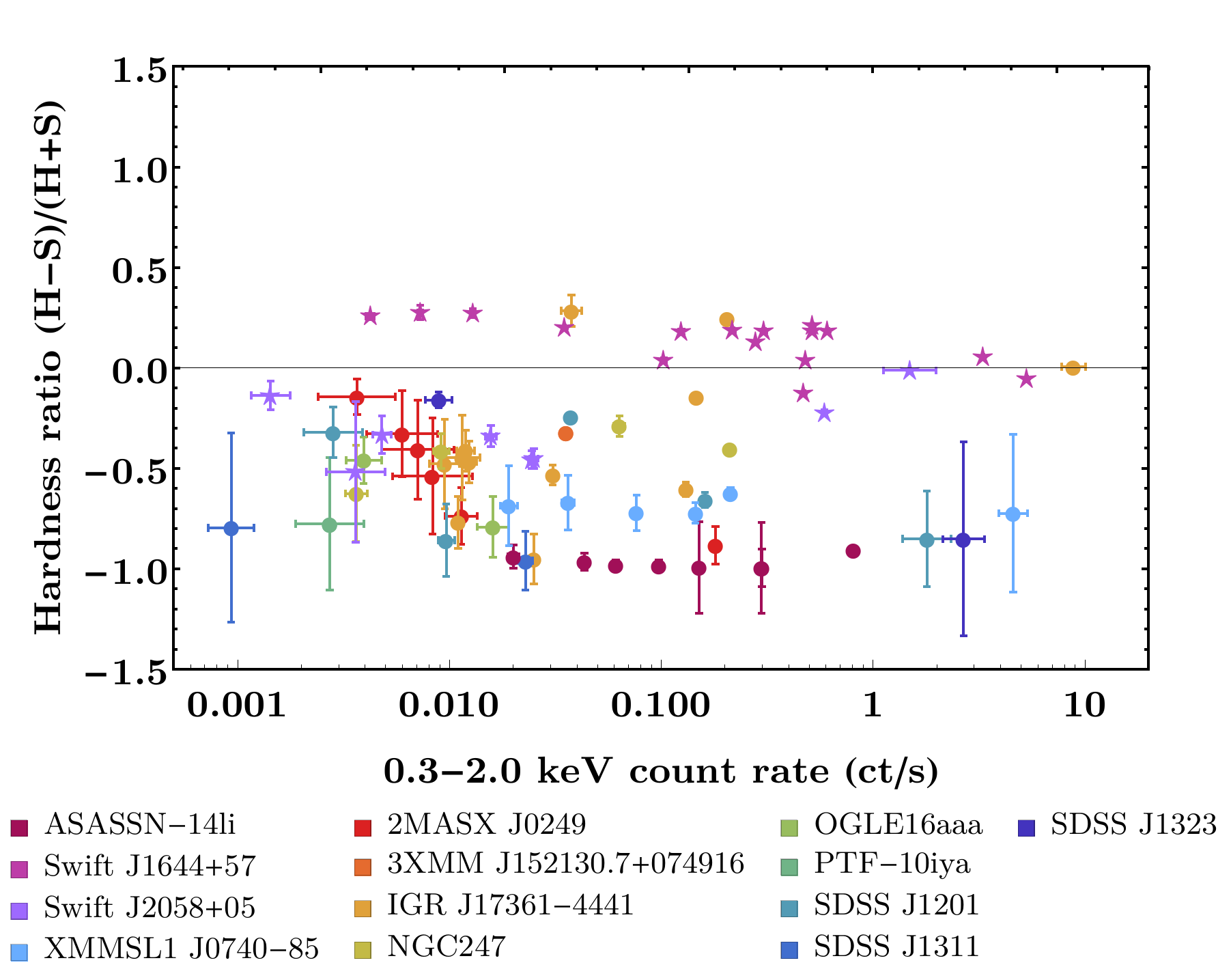}	\includegraphics[width=0.49\textwidth]{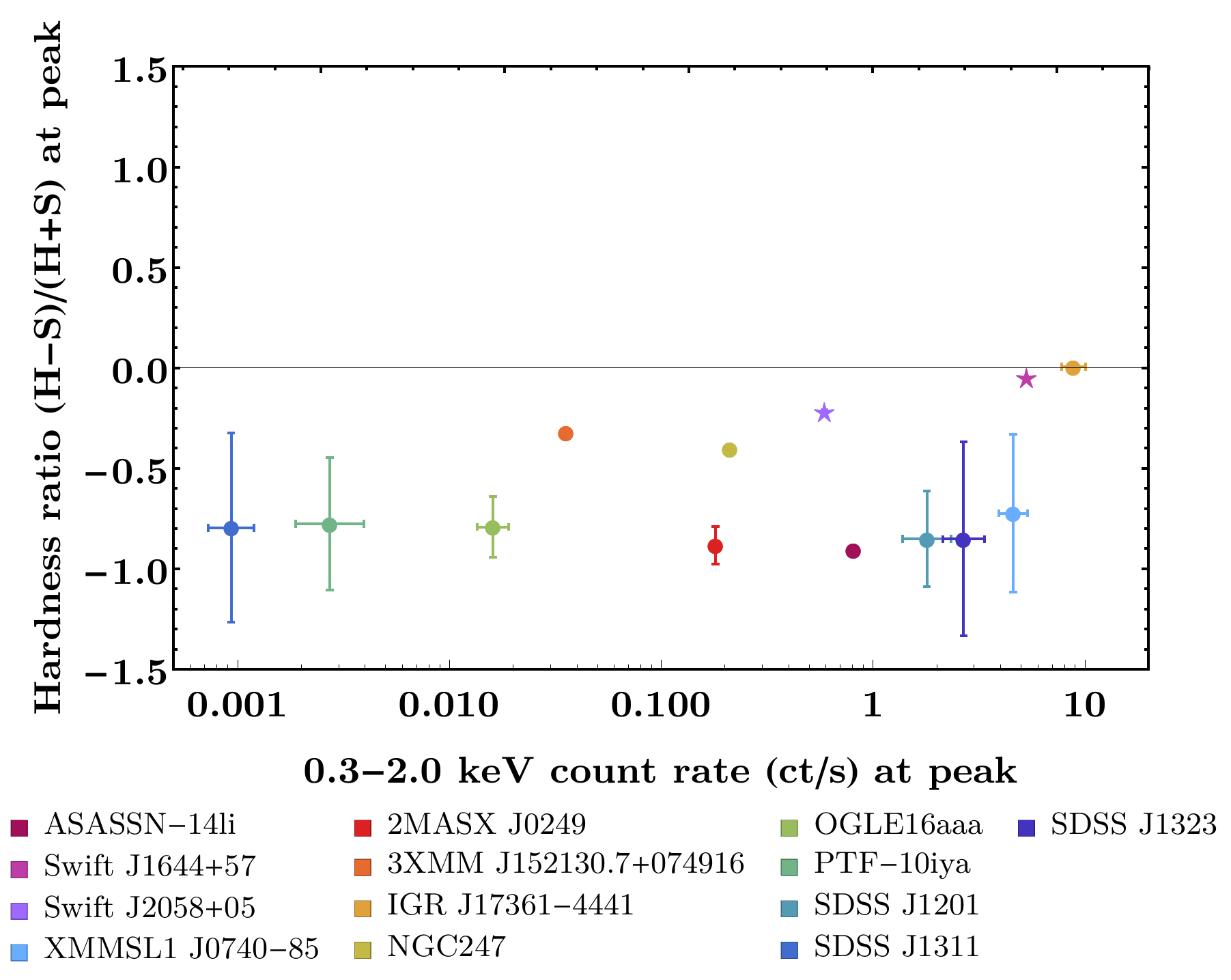}
		\caption{The hardness ratio defined as (H-S)/(H+S), where H is the counts in the equivalent 2.0-10.0 keV energy band and S is the counts in the equivalent 0.3-2.0 keV energy band, plotted against the 0.3-2.0 keV X-ray count rate. The \textit{top panel} highlights the hardness ratio for the full X-ray emission detected for each source, while \textit{bottom panel} shows the hardness ratio when the X-ray emission was measured at its peak. \label{hr}}
	\end{center}
\end{figure}

\begin{figure}[t]
	\begin{center}
		\includegraphics[width=0.49\textwidth]{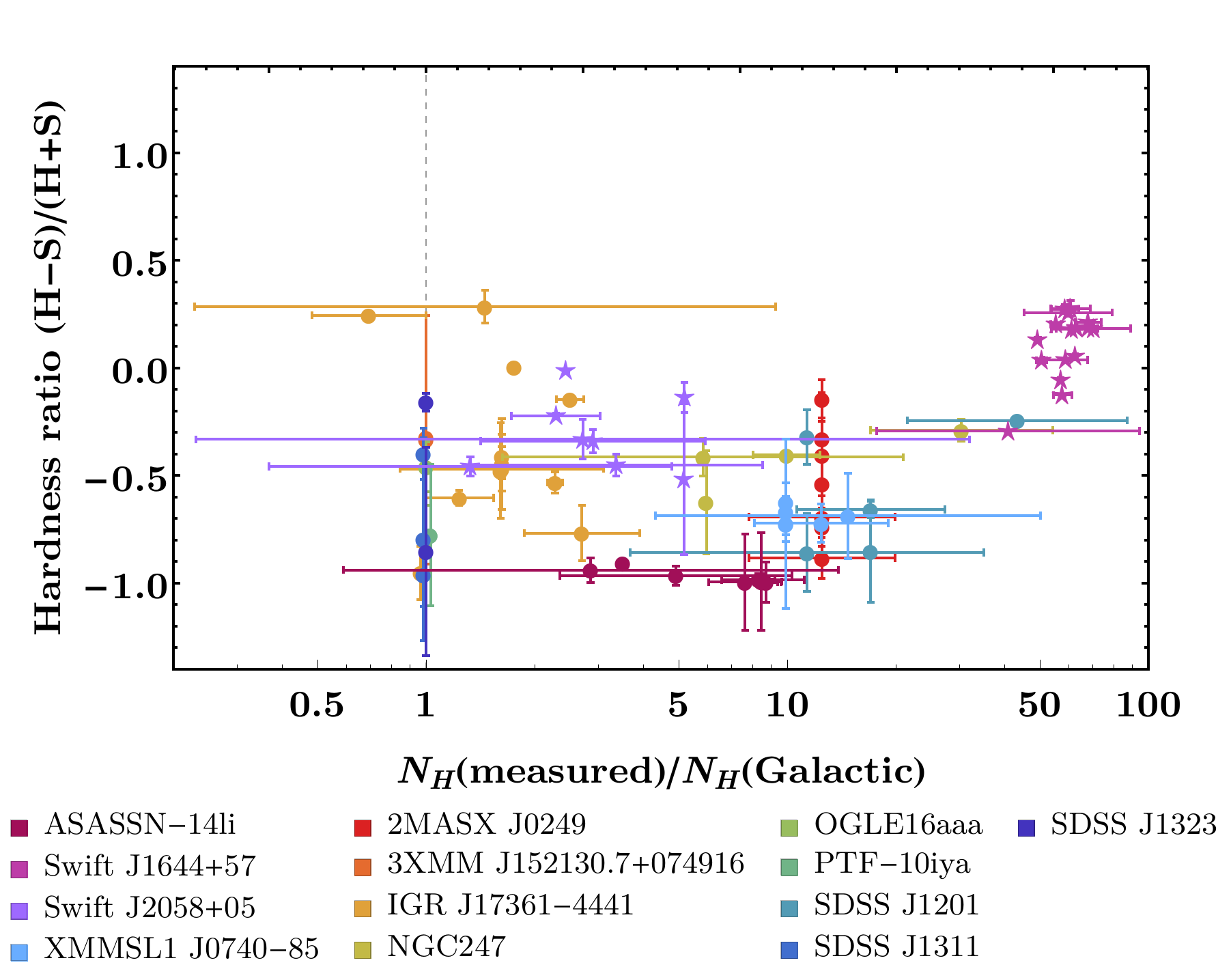}
		\caption{Hardness ratio as a function of the ratio of measured column density, $N_{H} (measured)$ divided by the Galactic column density, $N_{H} (Galactic)$ plotted for all TDE candidates. \label{nhvshr}}
	\end{center}
\end{figure}

Using the soft, medium and hard counts listed in Tables \ref{rosathms}--\ref{swifthms}, we constructed hardness ratios for each of the TDE candidates we consider. The hardness ratio (HR) is defined as (H-S)/(H+S), where H is the counts in the equivalent 2.0-10.0 keV energy band and S is the counts in the equivalent 0.3-2.0 keV energy band. In Figure~\ref{hr} top panel we plotted HR as a function of the soft 0.3-2.0 keV count rate for each TDE candidate, while in Figure~\ref{hr} bottom panel we have plotted the HR ratio as a function of soft count rate when the luminosity was at peak. By plotting the HR against count rate, we are able to determine how the emission from each source evolves.

From Figure~\ref{hr} top panel one can see that the emission from all TDEs is quite soft in nature. These events have a HR ratio that ranges between -1 and +0.3 with most events falls between -1 and 0, while nearly all the emission from \emph{Swift J1644+57} and some of the emission from \emph{IGR J17361-4441} has a HR between 0 and +0.3. Interesting, \emph{ASASSN-14li} and \emph{Swift J1644+57} characterise the most extreme HRs seen from these TDEs. Here \emph{ASASSN-14li} is one of the softest events detected with a HR ratio of $\sim-1$, while \emph{Swift J1644+57} is one of the hardest TDEs with a HR ratio of $\sim0.3$.  At peak, all events have a HR between -1 and 0 (see Figure~\ref{hr} bottom panel). 

Non-thermal jetted TDEs \emph{Swift J1644+57}, and \emph{Swift J2058+05}, along with hard X-ray source \emph{IGR J17361-4441} produce the hardest X-ray emission of our sample, best characterised with a peak HR value $\sim +0.1$, while \emph{ASASSN-14li}, \emph{2MASX J0249}, \emph{SDSS J1201}, and \emph{SDSS J1323} are the softest at peak with a HR of $\sim +1.0$. All other sources fall between these two values. 

\emph{ASASSN-14li} exhibited relatively little hardness evolution as its emission faded, staying extremely soft during its full decay, with \emph{XMMSL1 J0740-85} and \emph{NGC247} also exhibiting similar behaviour even though those sources are not as soft as \emph{ASASSN-14li}. As \emph{Swift J1644+57}, \emph{Swift J2058+05} and \emph{IGR J17361-4441} faded, these sources showed quite a bit of variability in their HRs, especially when the sources were brightest. Even though the early time emission from \emph{Swift J1644+57} varied, a significant fraction of the low count rate emission from this event showed relatively little hardness evolution, consistently staying around a HR $\sim$0.3. This is not the case for \emph{Swift J2058+05} and \emph{IGR J17361-4441}, whose HR varied quite dramatically as it faded. For \emph{2MASX J0249}, \emph{SDSS J1201}  and \emph{SDSS J1323} , these events were soft a peak and became harder as they faded.  However \emph{IGR J17361-4441} which was proposed to be the tidal disruption of a planet, follow the opposite trend, where they are harder at peak and then becoming softer as they decay.

We saw in Section \ref{absorb}, that nearly all X-ray TDEs are quite highly absorbed. As a consequence, the relatively soft HRs that we find for X-ray TDEs could result from the enhanced column densities towards these events. To test this, we plotted HR as function of $N_{H}$ (see Figure~\ref{nhvshr}). If $N_{H}$ was responsible for the soft nature of X-ray TDEs, we would expect to see that the sources which have the largest measured $N_{H}$ value should be the softest sources in our sample (i.e., be predominantly thermal in nature). However, we find that this is not the case with both the hardest and softest sources in our sample exhibiting enhanced $N_{H}$ relative to their Galactic $N_{H}$. As consequence, the relatively soft HRs derived from our analysis is most likely a inherent property of these events, rather than a consequence of the environment as pointed out in Section \ref{absorb}.

\subsection{What is the spectral energy distribution of X-ray TDEs?}

\begin{figure}[t]
	\begin{center}
		\includegraphics[width=\columnwidth]{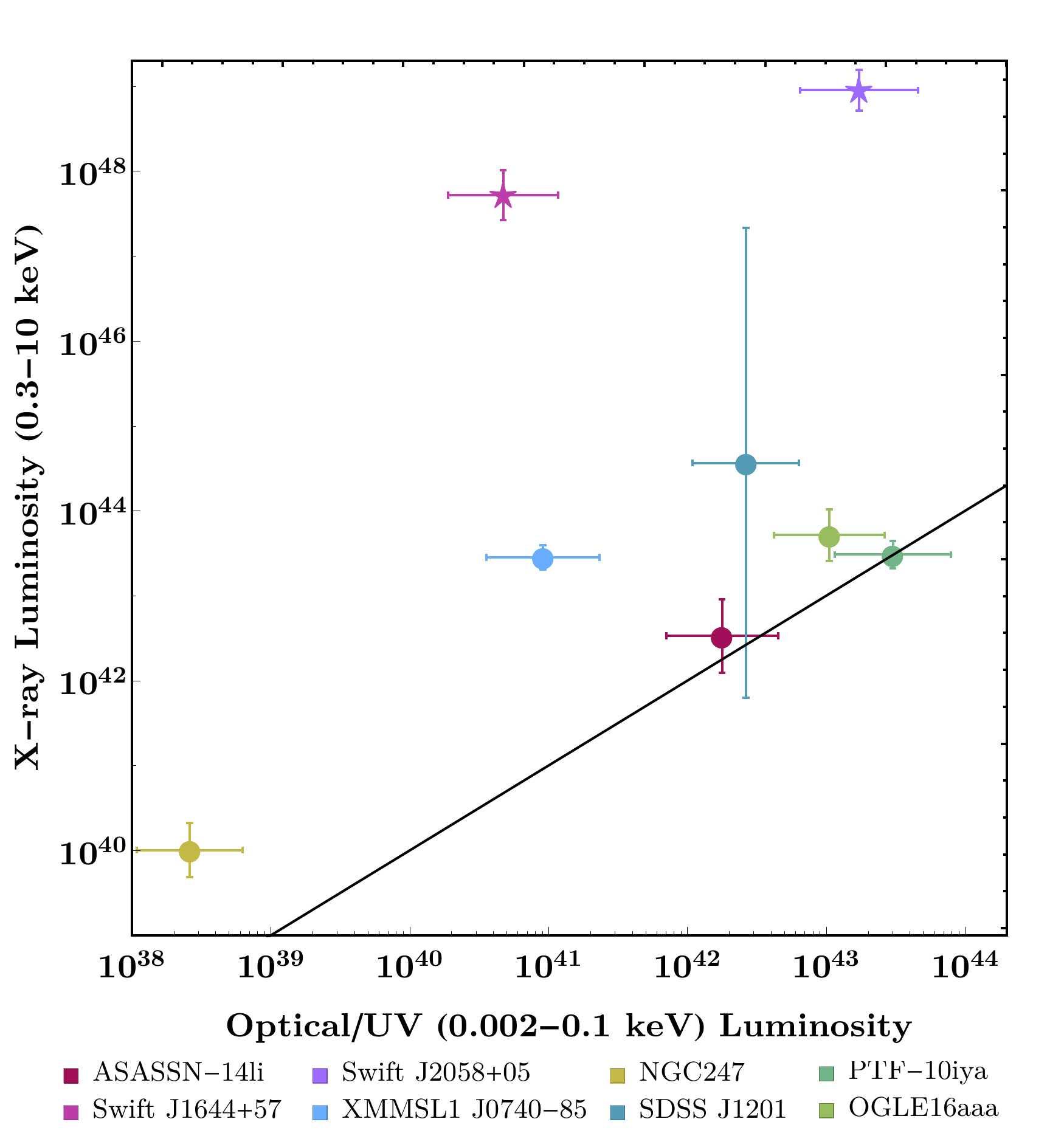}
		\caption{The integrated optical/UV (0.002-0.1 keV) luminosity of each TDE plotted against the integrated X-ray (0.3-10.0 keV) luminosity. Plotted as the black solid line is when the optical/UV and X-ray luminosity are equivalent (i.e., the ratio of these two parameters is unity). Not all of the TDEs listed in Table \ref{goldsilvers} have optical/UV data available around the time the X-ray emission peaked. \label{opticaltoxray}}
	\end{center}
\end{figure}

Using the soft, medium and hard count rates as measured at peak for each TDE candidate, we derive the $\nu F_{\nu}$ spectral energy distribution (SED) for each event. Analysing the SED allows one to determine the amount of energy emitted by each event as a function of wavelength (or energy), as well highlighting in what wavelength (or energy band) each event released most of its energy. As we are interested in the broadband SED of each source, we took from the literature (when it was available), radio, optical, and IR/UV data for each source that was taken simultaneously or close to when the original X-ray flare was detected. In Appendix \ref{individalsed} we have briefly described how we derived the SED and where we obtained the radio, optical/UV data from, while in Figure~\ref{individualnuFnu} which is also found in the Appendix \ref{individalsed}, we have plotted the individual SEDs for each of our TDE sample.

Using Figure~\ref{opticaltoxray}, we can determine whether these events emit most of their energy in the optical/UV or X-ray energy band. To do this we derive the integrated luminosity in both the optical/UV (0.002-0.1 keV) and X-ray (0.3-10.0 keV) energy bands. As not all events have optical/UV emisson around the time the event was discovered, we focus only on the events that have detected optical/UV emission. To derive the integrated luminosity in each energy band, we modelled the emission in the corresponding band using a powerlaw with an exponential cut off which was either left as a free parameter or set to the maximum of our specified energy band (i.e., 0.1 keV in the optical/UV or 10.0 keV in X-rays). We then integrated over the corresponding energy range to obtain the luminosity, while uncertainties are also derived from these model fits. These integrated luminosities are plotted as a function of each other in Figure~\ref{opticaltoxray}. In this figure, the black solid line indicates when the amount of energy released in both the optical/UV and X-ray energy bands is equivalent. Above (Below) this line indicates that the event releases most of this energy in the X-ray (optical/UV) energy band.

The non-thermal jetted X-ray TDEs \emph{Swift J1644+57} and \emph{Swift J2058+05} emit most of their energy in the X-ray energy band. For thermal events like \emph{ASASSN-14li}, \emph{OGLE16aaa} and \emph{PTF-10iya} they emit approximately the same amount of energy in both the optical/UV and X-ray energy bands, while \emph{NGC247}, \emph{SDSS J1201} and \emph{XMMSL1 J0740-85} emit slightly more energy in the X-ray energy band than they do in optical/UV. If veiled X-ray TDEs were to appear on Figure~\ref{opticaltoxray}, it is likely that their X-ray emission would fall below the detection thresholds plotted in Figure~\ref{firstupper}.

\subsection{Do non-thermal jetted and thermal non-jetted TDEs naturally separate?}

Characterising whether an X-ray TDE is thermal or non-thermal in nature provides information about whether the emission one observes arises from the accretion disk/fallback or from the formation of a relativistic jet. In the literature, detailed studies of their properties in multiple wavelengths, has led to the classification of a some of these events as either non-thermal or thermal. However, due to the lack of multi-wavelength data for a number of these events, the classification of some of these events as either thermal or non-thermal is either non-existent or not very clear. Using our analysis, we can specify the common properties non-thermal and thermal X-ray TDEs seem to have, using events such as \emph{Swift J1644+57} and \emph{ASASSN-14li} whose nature is very well accepted in the literature as baselines. Using these common characteristics, we can attempt to classify the TDEs in our sample as either non-thermal or thermal events. 

Based on Tables \ref{rosatfits}--\ref{swiftfits}, we find that non-thermal events tend to have X-ray emission that is best described by a hard powerlaw index when modelling their X-ray spectra. Summarising the results presented in Figures \ref{t90s}, \ref{ced}, \ref{hr}, and \ref{opticaltoxray}, non-thermal TDEs tend to have higher $L_{90}$ values and release a significant amount of their energy over a shorter time period than their thermal counterparts. The emission at peak from a non-thermal TDEs usually has a HR between $\sim -0.5 \-- 0.0$, and a large X-ray to optical luminosity ratio. They also exhibit more variability in their X-ray emission than thermal TDEs. Discussed in more detail in Section \ref{softmedhardcounts}, one can also see that from Figure~\ref{countrates} that non-thermal events are best characterised as having a soft to medium count ratio that is $\lesssim1$.

For thermal X-ray TDEs, we find that their X-ray emission is best described by a very soft powerlaw index when modelling their X-ray spectra. As the very soft powerlaw index mimics thermal emission over the X-ray energy band pass of current X-ray satellites, the very soft index from these events is not unexpected. Most thermal events have an $L_{90} \lesssim 10^{42}$ erg s$^{-1}$ and take much longer to release most of their energy compared to the non-thermal events. These sources are also quite soft, with a HR$\lesssim-0.3$ and have a X-ray to optical luminosity ratio around 1. They also seem to show less amount of variability compared to non-thermal events and they also have a soft to medium count ratio that is $\gtrsim1$.

Using these properties, \emph{Swift J1644+57}, and \emph{Swift J2058+05}, can be classified as non-thermal in nature, while \emph{ASASSN-14li} has properties that seem to characterise the X-ray emission of a thermal TDE. This is consistent with the classification of these events in the literature. However for the \textit{likely X-ray TDEs}, it is not necessarily as clear cut as their properties do not always fall exactly into only one of the categories specified above. However, it seems that \emph{NGC 247}, and \emph{IGR J17361-4441}  have more properties similar to those of the non-thermal TDEs, while \emph{2MASX J0249}, \emph{3XMM}, \emph{SDSS J1201}, \emph{SDSS J1311},  \emph{SDSS J1323},  \emph{OGLE16aaa}, \emph{PTF-10iya}, and \emph{XMM SL1 J0740-85} have more properties common to those of thermal TDEs.

\section{Discussion}\label{dis}

\subsection{The emission from an X-ray TDE peaks in the soft X-ray band.}\label{softmedhardcounts}

\begin{figure*}[h]
	\begin{center}
		\includegraphics[width=0.49\textwidth]{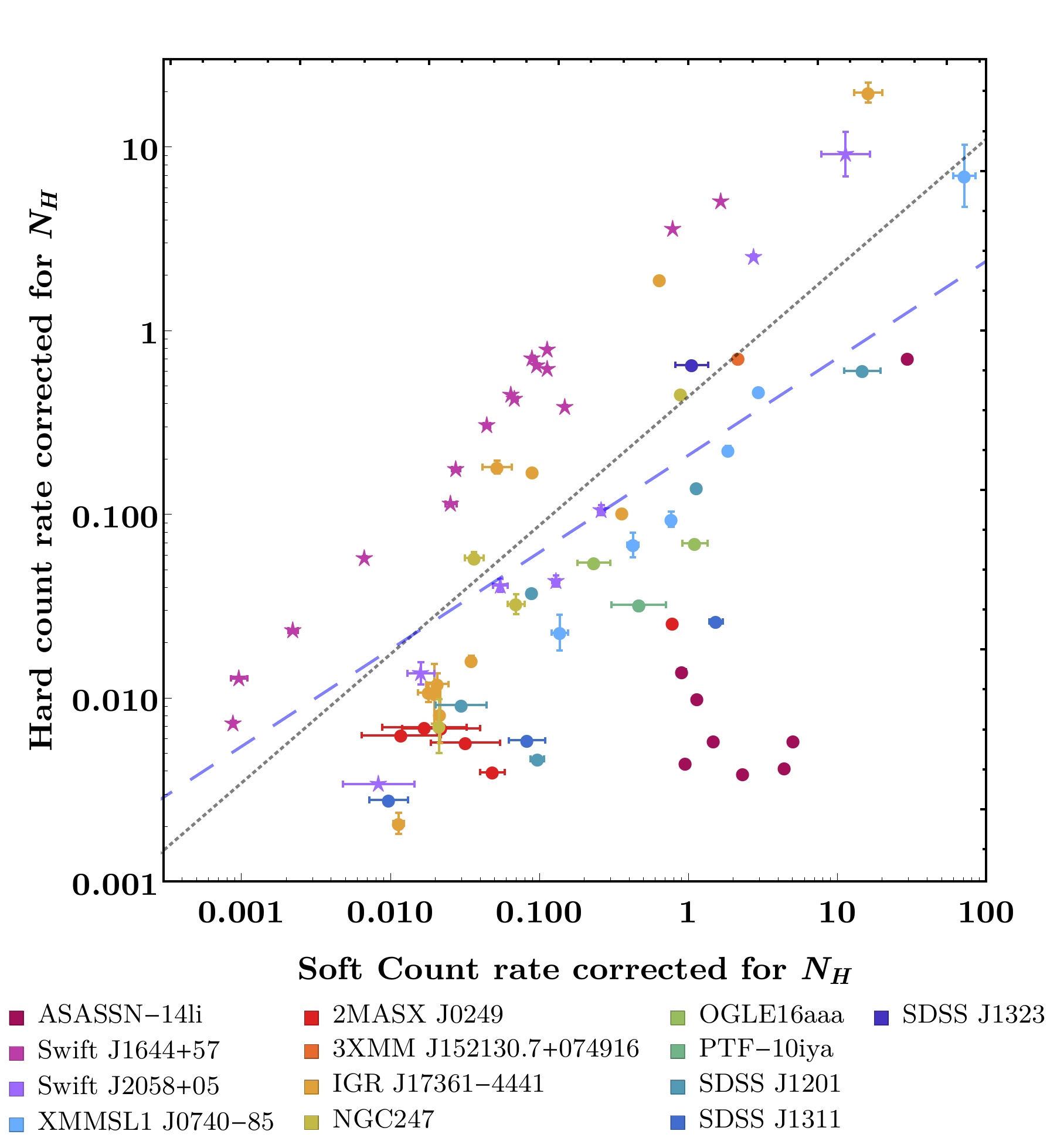}
		\includegraphics[width=0.49\textwidth]{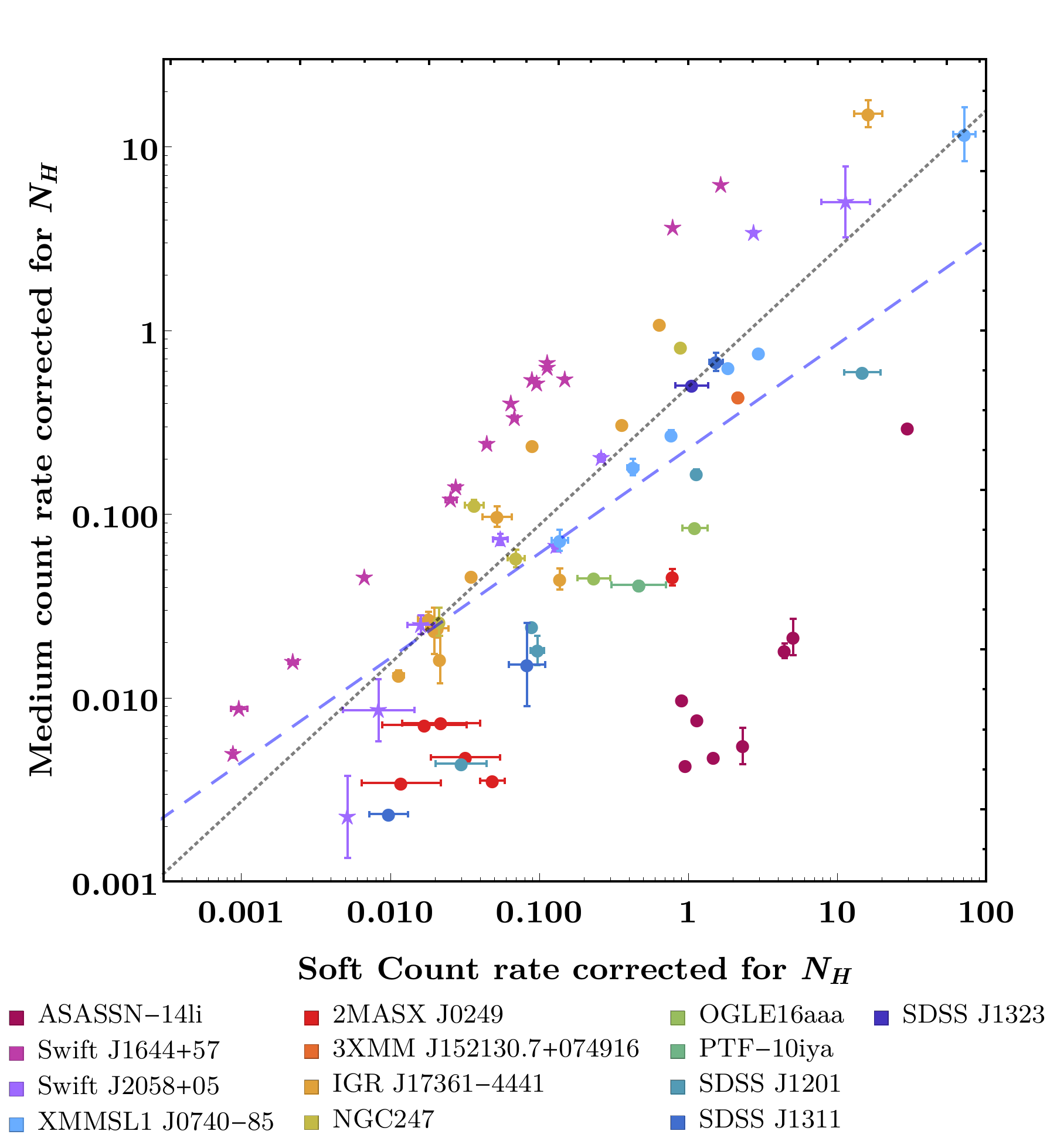}
		\includegraphics[width=0.49\textwidth]{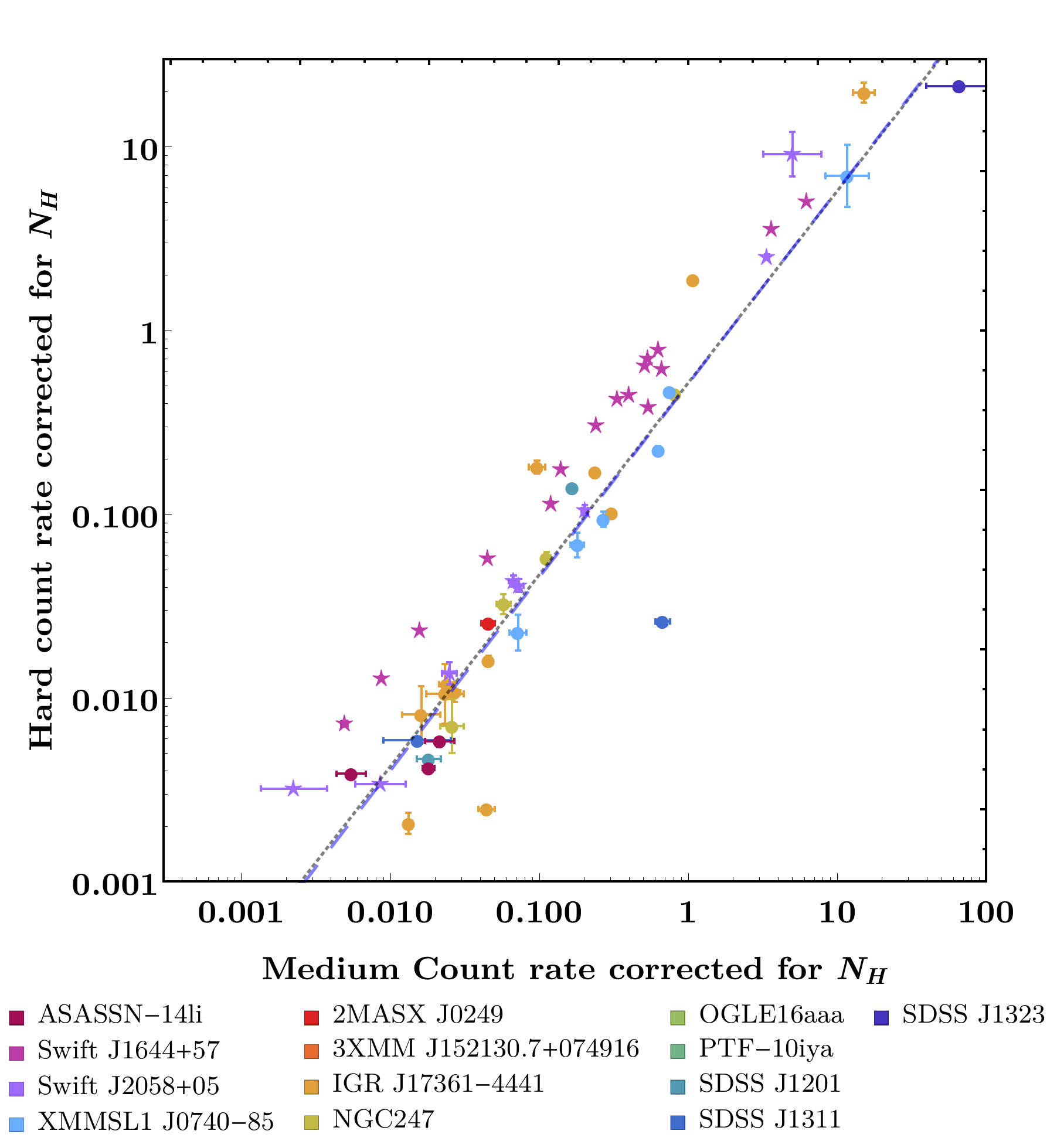}
		\includegraphics[width=0.49\textwidth]{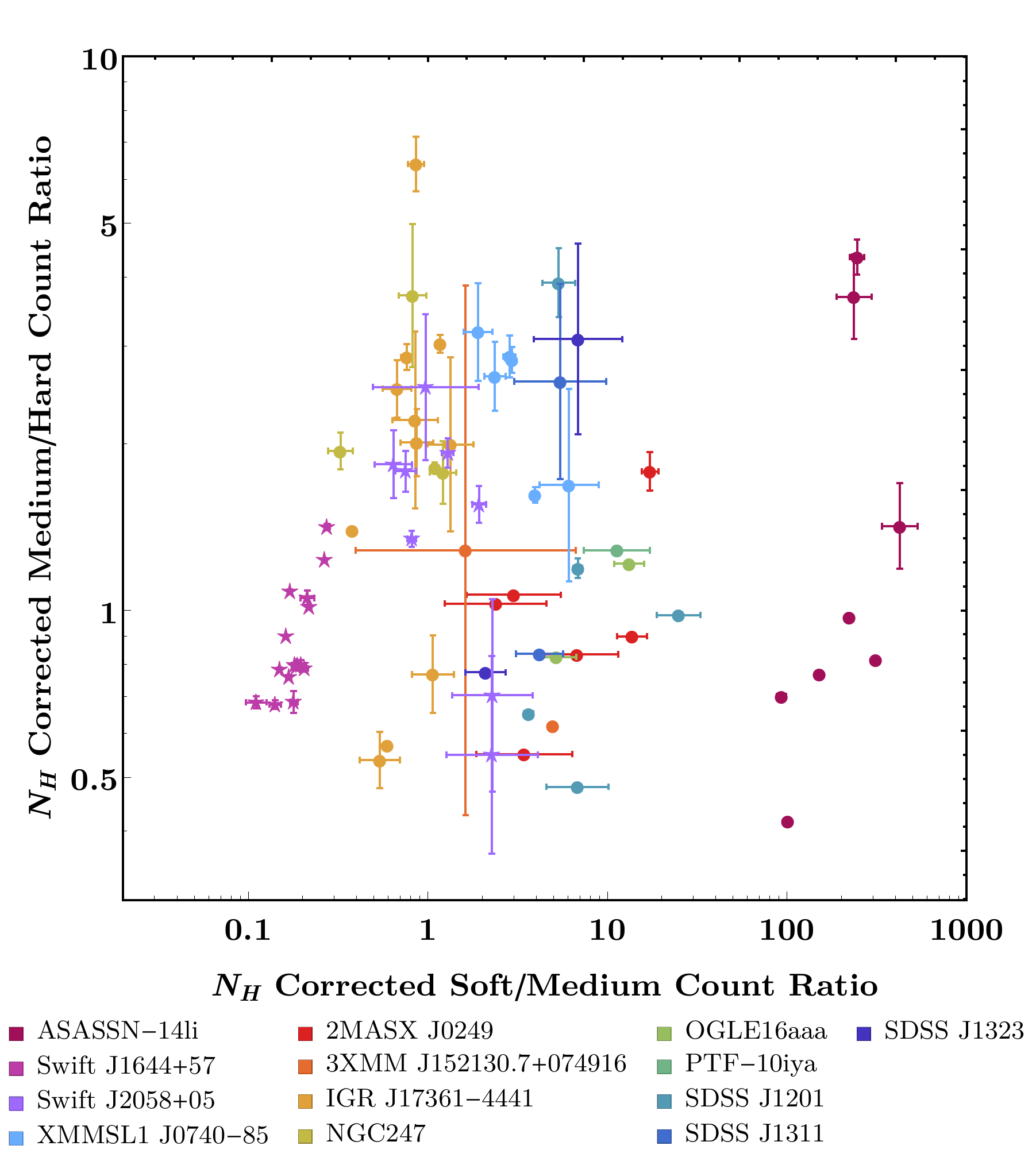}
		\caption{For the TDE candidates we consider, we have plotted the soft ($0.3-1.0$ keV), medium ($1.0-2.0$ keV) and hard ($2.0-10.0$ keV) count rates as a function of one another (top three panels). In the bottom panel we have plotted the ratio of the medium vs. hard counts as a function of soft vs. medium counts.	For all plots, we corrected the number of counts for the effect that column density ($N_{H}$) has on the emission in each energy band by scaling the count rate by (1-$e^{-E_{\text{band}}}$)($N_{H}/(10^{22} \text{cm}^{-2})$). Here (1-$e^{-E_{\text{band}}}$) factor takes into account the fact that the column density affects soft X-ray emission more significantly than other energy bands. By applying this correction, we remove the biasing affect that $N_{H}$ might have on our data. Overlaid on the top three panels is the least squares fit of each plot. Shown as the blue dashed line is the best fit using all data from the TDE candidates we consider, while the dotted black line corresponds to the best fit that is obtained when excluding the data points of ASASSN-14li. The coefficient of determination for each of the blue dashed (black dotted) lines is 0.33, 0.41, 0.89 (0.57, 0.69, 0.89) respectively. 	\label{countrates}}
	\end{center}
\end{figure*}

In Section \ref{hrsoft}, we find that X-ray TDEs are quite soft in nature with HRs $\lesssim+0.3$, while in Section \ref{absorb}, we suggested that variability in column densities measured towards these events was not responsible for their soft HRs. To further test this, we used the soft (0.3-1.0 keV), medium (1.0-2.0 keV) and hard (2.0-10.0 keV) count rates that we derived for each event (for Tables \ref{rosathms}\--\ref{swifthms}) and plotted these as a function of each other. To remove the effect that $N_{H}$ can have on the count rate in the different energy bands, we divided each set of counts by the $N_{H}$ value measured or assumed for that observation (from Tables \ref{rosatfits}--\ref{swiftfits}). As $N_{H}$ has the greatest effect in the lower energy bands, while at higher X-ray energies the effect of absorption is minimal, we also took this into account and scaled $N_{H}$ by 1-$e^{-E_{\text{band}}}$ \citep[e.g.,][]{2000ApJ...542..914W}. In Figure~\ref{countrates}, we have plotted the $N_{H}$ corrected count rate diagrams in which we compare the count rates from energy band, while also plotting the ratio of the soft/medium counts vs. medium/hard counts. We compared these $N_{H}$ corrected count rate diagrams to those we obtain without correcting for $N_{H}$ and we find that apart from different values for the count rates (which is expected), we observe exactly the same trends seen in Figure~\ref{countrates}. This again highlights that the softness of X-ray TDEs is most likely an inherent property of these events. 

In the soft vs. medium and soft vs. hard plots of Figure~\ref{countrates}, we find that jetted and non-jetted events seem to occupy different parts of these diagrams, where \emph{Swift J1644+57} and \emph{ASASSN-14li} naturally provide a boundary in which nearly all X-ray TDEs fall within. One can see that as these events become brighter, the characteristics of the emission from the jetted and non-jetted events diverges from each other. For the same count rate measured in the soft energy band, jetted events increase significantly in the number of medium and hard counts, while non-jetted events tend to have a more flatter evolution in these higher energy bands. 

To determine the correlation between these observables we ran a linear regression and plotted these as the blue dashed and black dotted lines in Figure~\ref{countrates}. To quantify how correlated the count rates are we derive the coefficient of determination (R$^{2}$ value) of each fit, where a high (lower) R$^{2}$ implies that these parameters are (not) well correlated. These values are listed in the caption of Figure~\ref{countrates}. One can see that this divergence of the non-thermal and thermal events seen in both the soft vs. medium and soft. vs hard count rate plots leads to a low R$^{2}$, indicating that there is little correlation between these energy bands. However, when we look at the medium vs. hard count rate plot we find a completely different story. Regardless of the nature of the TDE, there is very little scatter between our X-ray TDE candidates in these higher energy bands, with both thermal and non-thermal TDEs showing a similar evolution of their emission in these bands.  From our linear regression, we find a strong correlation between these count rates, indicating that there is very minimal change in the emission of these events in the 1.0-2.0 keV and 2.0-10.0 keV energies.  This is also seen when one considers the ratio of these count rates in the different energy bands (Figure~\ref{countrates} bottom right). The variation in the medium/hard count ratio is significantly less compared to that seen in the soft/medium energy band, which varies over nearly four orders of magnitude compared to only one. As such, this leads us to conclude that the X-ray emission of TDEs peak predominantly in the soft (0.3-1.0 keV) energy band, producing the large scatter seen in the soft vs. medium and soft vs. hard count rate plots. This is not so surprising since \citet{1999ApJ...514..180U} determined that accretion powered flares from BH with mass $\lesssim10^{7}M_{\odot}$ should radiate in the soft X-ray band.

As we have corrected the count rates of our TDE sample for absorption, the large variation seen in the soft count rate band for predominantly non-jetted TDEs likely suggests that there is significant differences in the reprocessing rates experienced by these events. Due to the lack of variation in these higher energy bands, the enhanced column densities and low X-ray to optical ratios of these events supports the fact that a large fraction of their emission is being reprocessed into either soft X-rays or into optical/UV wavelengths. As a consequence, the enhanced column densities surrounding these sources could lead to significant reprocessing and thus be responsible for the intrinsically soft nature of these X-ray TDEs. This observation also ties in nicely with the fact that there are a number of optical TDEs without X-ray emission (i.e., Veiled TDEs in Section \ref{veiled}), since a large fraction of these events show evidence of significant reprocessing due to a dense surrounding environment (e.g., \textit{PS1-10jh}: \citealt{2012Natur.485..217G}).

In the bottom right panel of Figure~\ref{countrates}, we have plotted the ratio of the $N_{H}$ corrected medium and hard counts ($M/H$) as a function of the $N_{H}$ corrected soft and medium counts ($S/M$). All X-ray TDEs have a $M/H\gtrsim0.5$. However, thermal and non-thermal events naturally separate from each other when looking at their $S/M$. The emission from non-thermal events have a $S/M\lesssim2$, while for thermal X-ray TDEs their emission has a $S/M\gtrsim2$. Things become more complicated around $S/M\sim2$ with emission from both the non-thermal and thermal events begin to overlap as they become fainter. However when most of these sources are relatively bright, their emission is sufficiently different that one could easily classify the type of event based on this diagram. This provides a unique way of observationally categorising X-ray TDEs as either thermal of non-thermal events, especially ones which have limited observational information about the source.

\subsection{The emission from an X-ray TDE evolves with time}\label{evolve}

\begin{figure*}[t]
	\begin{center}
		\includegraphics[width=0.49\textwidth]{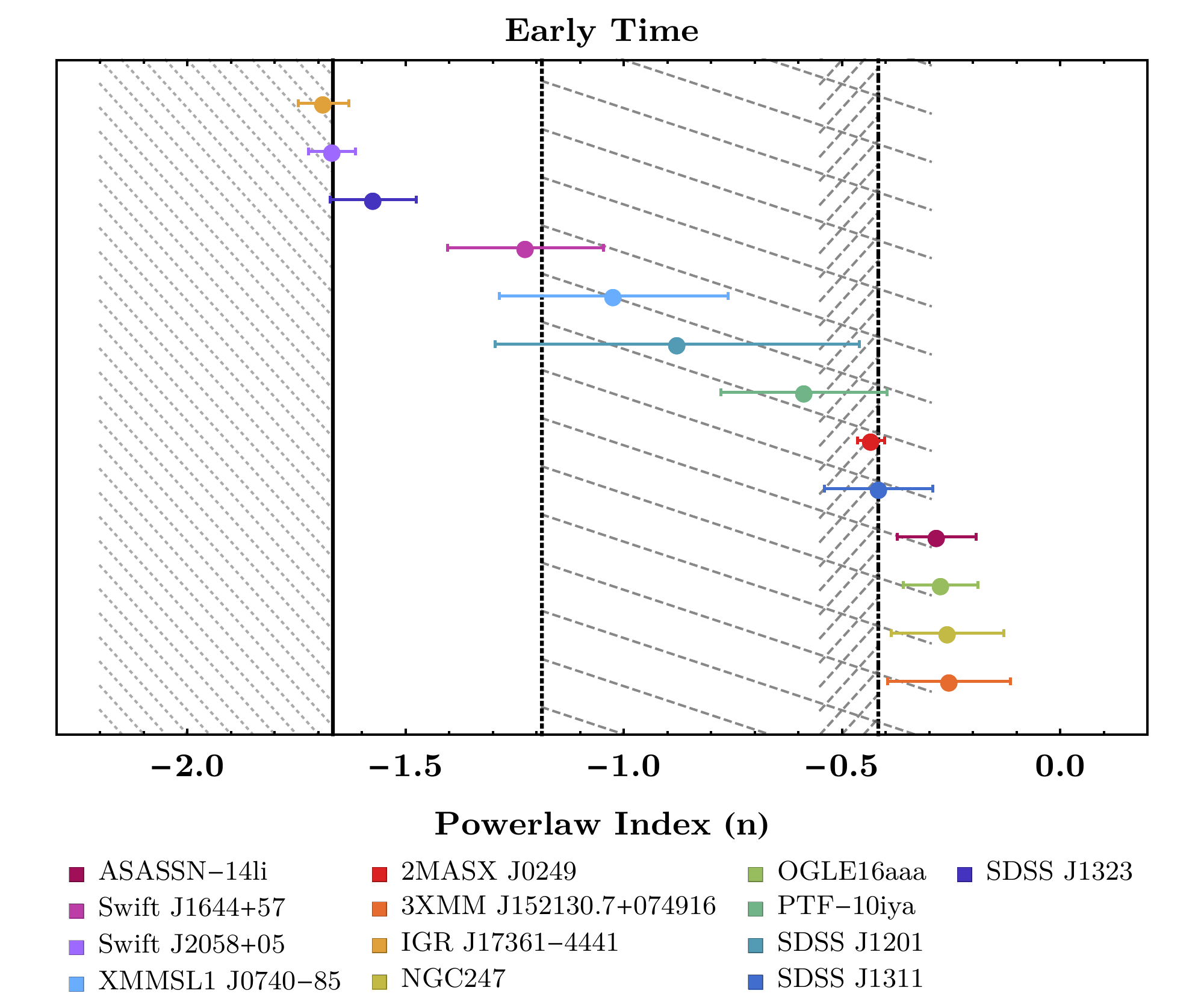}
		\includegraphics[width=0.49\textwidth]{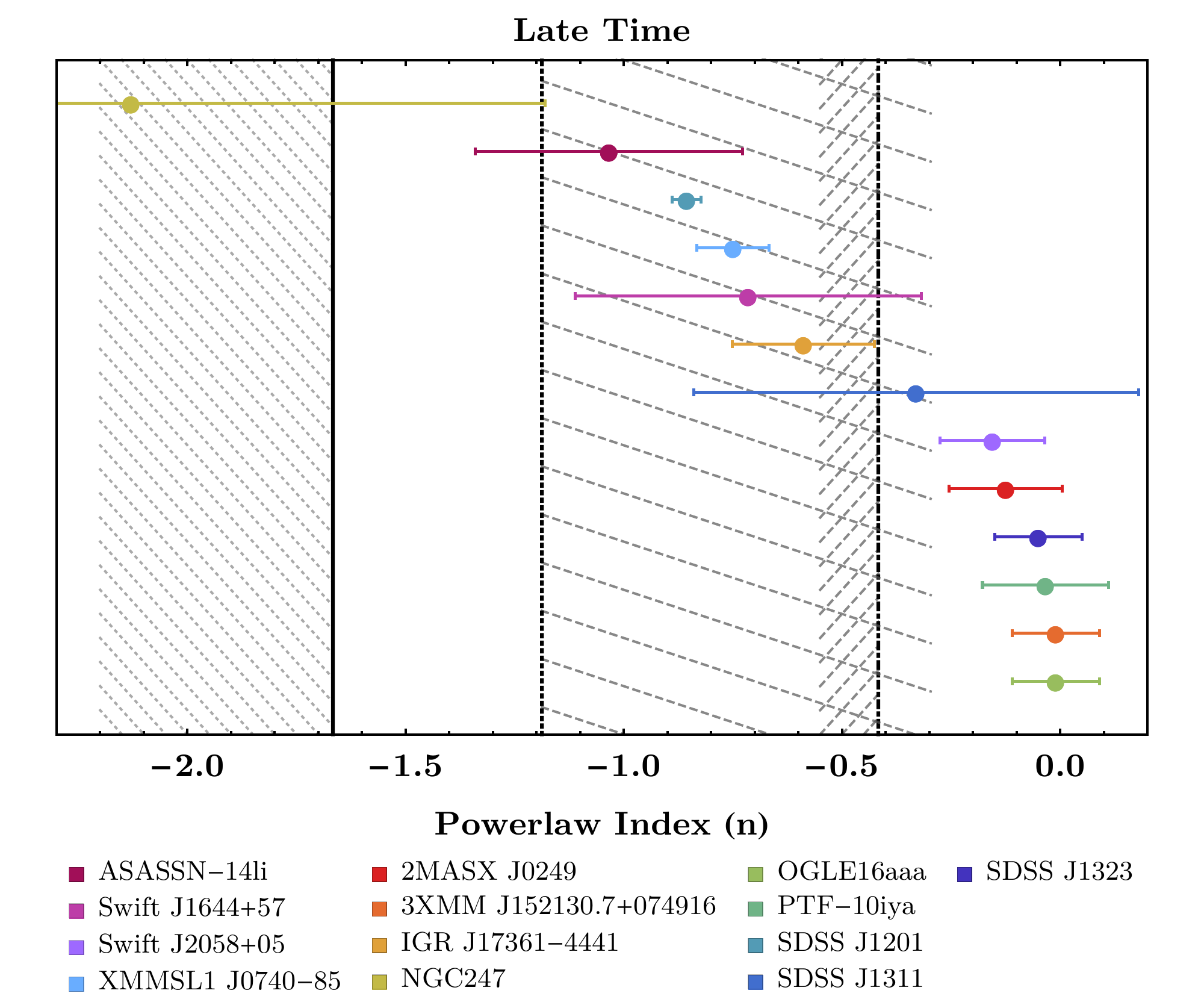}
		\caption{ The best fit powerlaw indexes and their $1\sigma$ uncertainties obtained from fitting the early time emission (\textit{left panel}) and the late time emission (\textit{right panel}) of the TDE candidates we consider. Also overlaid are the different powerlaw indexes expected for TDEs (see the caption of Figure~\ref{pwlindex} for more details). \label{earlylate}	}
	\end{center}
\end{figure*}

\begin{figure}[t]
	\begin{center}
		\includegraphics[width=\columnwidth]{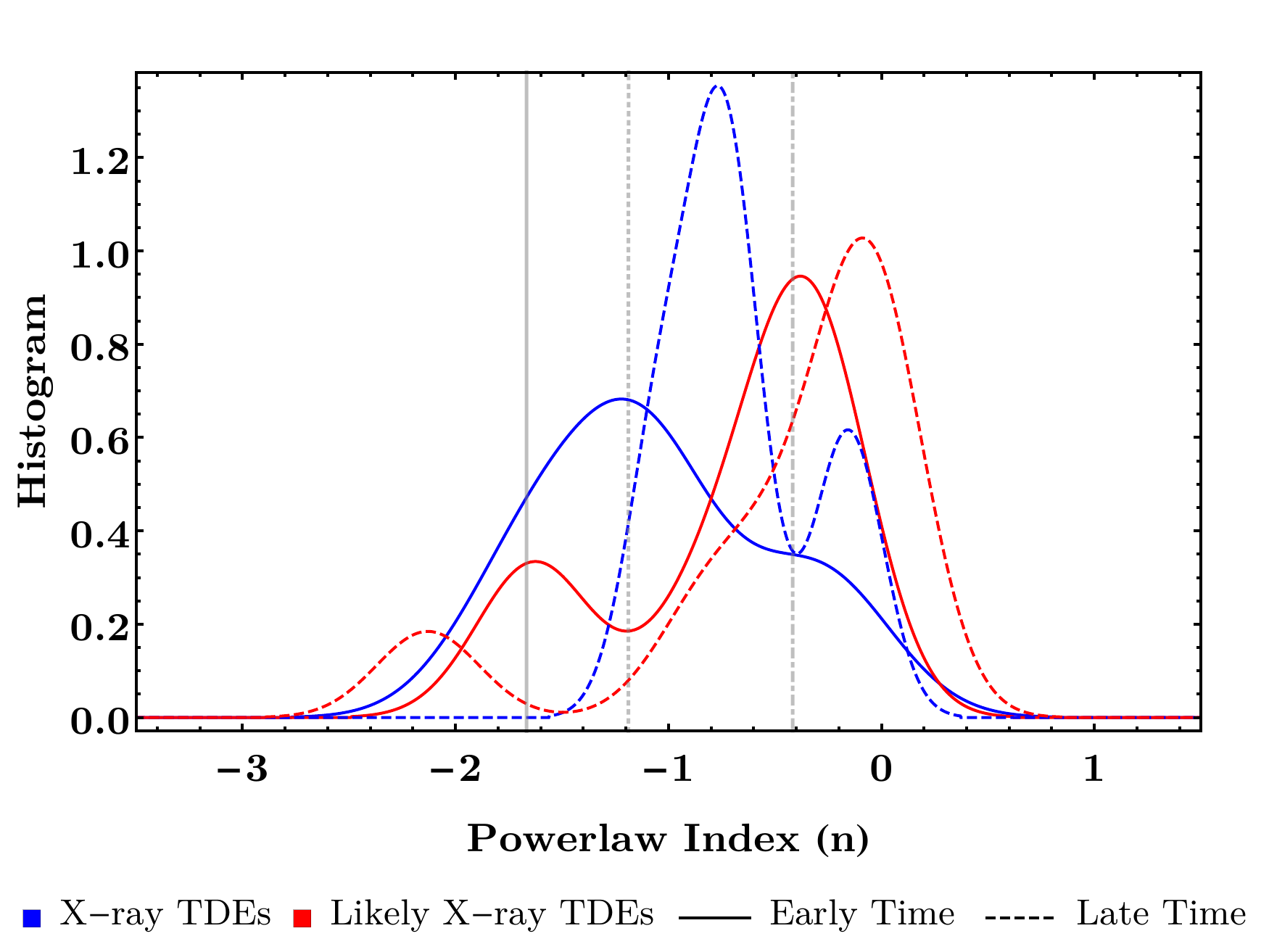}
		\caption{Histogram of the early and late time powerlaw indexes for the TDE candidates we consider seen in red and blue respectively. Here the solid histograms represent the powerlaw index at early times, while the dashed histograms represent the powerlaw index at late times. Plotted as the solid, dotted dash-dotted grey vertical lines are the powerlaw indexes for fallback (-5/3), viscous disk accretion (-19/16) and disk emission (-5/12) respectively. \label{earlylatehist}	}
	\end{center}
\end{figure}

In Section \ref{pwldecay}, we fit the full X-ray light curve of each of the TDE candidates we consider using a simple powerlaw model. We find that the powerlaw index we derive differs from those listed in the literature, which are commonly derived from the early time emission of these sources. As a consequence, this brings into question whether it is reasonable to assume that the emission of each event decays following $t^{-5/3}$ as is done frequently in the literature. To investigate this, we separated the emission from each source into early time and late time intervals and then fit these sets of data using a simple powerlaw with the normalisation and powerlaw index set free, much like that done in Section \ref{pwldecay}.

In Figure~\ref{earlylate} we have plotted, similar to Figure~\ref{pwlindex}, the best fit power law index and its uncertainty for the early and late time emission of each TDE candidate. We have also overlaid the ranges of various powerlaw indexes expected for TDEs.  One can see that at early times the emission from \emph{Swift J2058+05}, and \emph{IGR J17361-4441} is consistent with fallback. For \emph{2MASX J0249, SDSS J1201, SDSS J1311, XMMSL1 J0740-85} and \emph{PTF-10iya}, these events have indexes consistent with disk emission, while \emph{Swift J1644+57} and \emph{SDSS J1323} have indexes that fall between -5/3 and -5/12. In addition, \emph{ASASSN-14li, 3XMM, NGC247} and \emph{OGLE16aaa} have a powerlaw index slightly lower than that expected from disk emission, however within uncertainties they are consistent with -5/12. At late times the emission from each TDE changes quite dramatically. \emph{Swift J2058+05, 2MASX J0249, 3XMM, OGLE16aaa, PTF-10iya, SDSS J1323} have indexes consistent with being flat, while the emission from \emph{SDSS J1311} is still consistent with disk emission. Within uncertainties, \emph{Swift J1644+57},  \emph{ASASSN-14li, XMMSL1 J0740-85, SDSS J1201}, and \emph{IGR J17361-4441} is also consistent with disk emission even though their best fit powerlaw model falls slightly above or below this value. Interestingly, \emph{NGC247} is the only event that has late time emission that is consistent with fall back, however, the large uncertainties associated with this fit also suggests that this event is consistent with viscous disk accretion. The large uncertainties seen for some of these events arises from the light curve being more sparsely sampled, especially at late times.

In Figure~\ref{earlylatehist} we have plotted a histogram of the powerlaw indexes for the TDE candidates we consider at both early and late times. At early times, the emission from the \textit{X-ray TDE} candidates arises mostly from fallback or viscous disk accretion, with both the Swift events dominating the histogram distribution. Here \emph{ASASSN-14li} and \emph{XMMSL1 J0740-85} is the exception, in which at early times it is consistent with disk emission. For the \textit{likely} X-ray TDE sample, the majority of events exhibit emission consistent with disk emission at early times, with the exception of \emph{IGR J17361-4441} which favours fallback like the other Swift events. At late times, we see that the emission from our sample of X-ray TDEs evolves such that nearly all events have emission that converges to a powerlaw index consistent with disk emission or viscous disk accretion. 

One should also notice that, with the exception of \emph{Swift J2058+05} and \emph{IGR J17361-441} at early times, and potentially \emph{NGC 247} at later times, that the emission from these events decays with a powerlaw index much lower than the canonical $t^{-5/3}$ relationship. \citet{2015ApJ...809..166G} showed that if a TDE has a long viscous time scale\footnote{The viscous time scale is defined as the time it takes for material to accrete and is defined by $t_{\rm visc}=\alpha^{-1} (h/r)^{-2} P_{\rm}$, where $\alpha$ is the viscous parameter, $h$ is the scale height on the disk, $P$ is the orbital period and $r$ is the distance from the BH. See \citep{2015ApJ...809..166G} for more details.} that these events will decay at a rate much shallower than the standard $t^{-5/3}$. This implies that TDEs detected in X-rays are most likely viscously slowed and as such the viscous timescale is important in determining the emission of these events, especially at late times.

\begin{figure}[t]
	\includegraphics[width=\columnwidth]{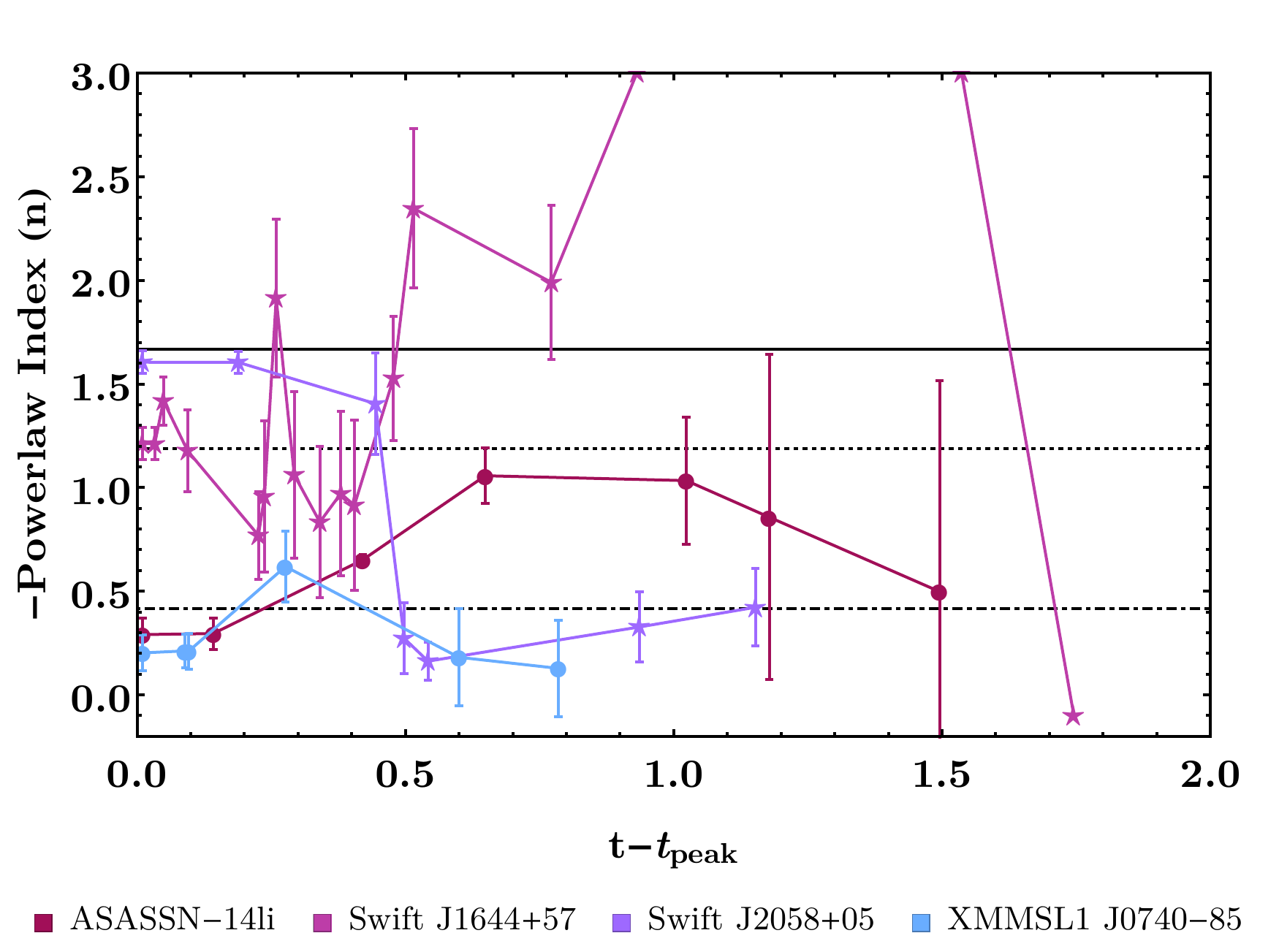}
	\caption{The best fit powerlaw index and its uncertainty for the X-ray TDE sample that was obtained as $t_{0}\rightarrow\infty$. Here we have plotted these indexes relative to $t-t_{peak}$, where $t_{peak}$ is taken as the time when the source was brightest. The lines joining the points are to guide the eye and are not fits. Here the solid, dotted and dot-dashed black horizontal lines correspond to a powerlaw index of -5/3, -19/16, and -5/12 respectively. Due to the sparsity of the X-ray data for the likely TDE candidates we are unable to constrain the powerlaw index as $t_{0}\rightarrow\infty$. \label{tinf} }
\end{figure}

To see whether the transition between early and late times is smooth, we determined the best fit powerlaw index for the \textit{X-ray TDE} candidates as the time of peak goes to infinity. Due to the sparseness of the available X-ray data for the \textit{likely X-ray TDE} candidates, we were unable to determine how the emission from these events evolve in a similar way. In Figure~\ref{tinf} we have plotted the powerlaw index as a function of $t-t_{\rm peak}$, where $t_{ \rm peak}$ is the time in which we detect the its peak luminosity. Straight away, one can see that the non-thermal Swift events shows significant variability in the properties of their emission as a function of time, while the emission from \emph{ASASSN-14li} and \emph{XMMSL1 J0740-85} evolves significantly more smoothly. Even though we find that the powerlaw index for \emph{ASASSN-14} steepens to -1 while it decays, the average emission from \emph{ASASSN-14li} is consistent with disk accretion. Similarly, the emission from \emph{XMMSL1 J0740-85} shows some variation, however it is consistent within uncertainties with viscous disk accretion throughout its full emission. This is not the case for \emph{Swift J1644+57} and \emph{Swift J2058+05}. \emph{Swift J1644+57} initially converges towards an index more consistent with disk emission, but suddenly shows a spike indicating that the emission was more consistent with fallback. After this spike, it again converges towards -5/12, but then steepens dramatically before falling to -5/12.  \emph{Swift J2058+05} was initially consistent with fallback, but approximately half way through its decay, its emission dramatically transitioned such that it was consistent with disk emission. Whether the emission seen for \emph{Swift J2058+05} at both early and late times was smooth or resulted from more extreme variations in the emission of the source as seen in \emph{Swift J1644+57} it is difficult to say due to the sparseness of the data during these periods.

As we observe significant evolution in the powerlaw index of each event on both small time scales (Figure~\ref{tinf}) and larger time scales (Figure~\ref{earlylate}) this indicates that we cannot assume that the X-ray emission from all X-ray TDEs decays follow $t^{-5/3}$. In fact, a large fraction of the emission from X-ray TDEs is consistent with disk accretion at both early and late times, while a majority of the non-thermal jetted events at early times has emission that is more consistent with fallback. Based on Figure~\ref{tinf}, it seems that particularly jetted X-ray TDEs, will transition multiple times between the two emission processes, while thermal TDEs will tend to fluctuate around a powerlaw index that is consistent with one particular emission type.

\subsection{The emission mechanism(s) of X-ray TDEs.}

\begin{figure*}[t]
	\begin{center}
		\includegraphics[width=\textwidth]{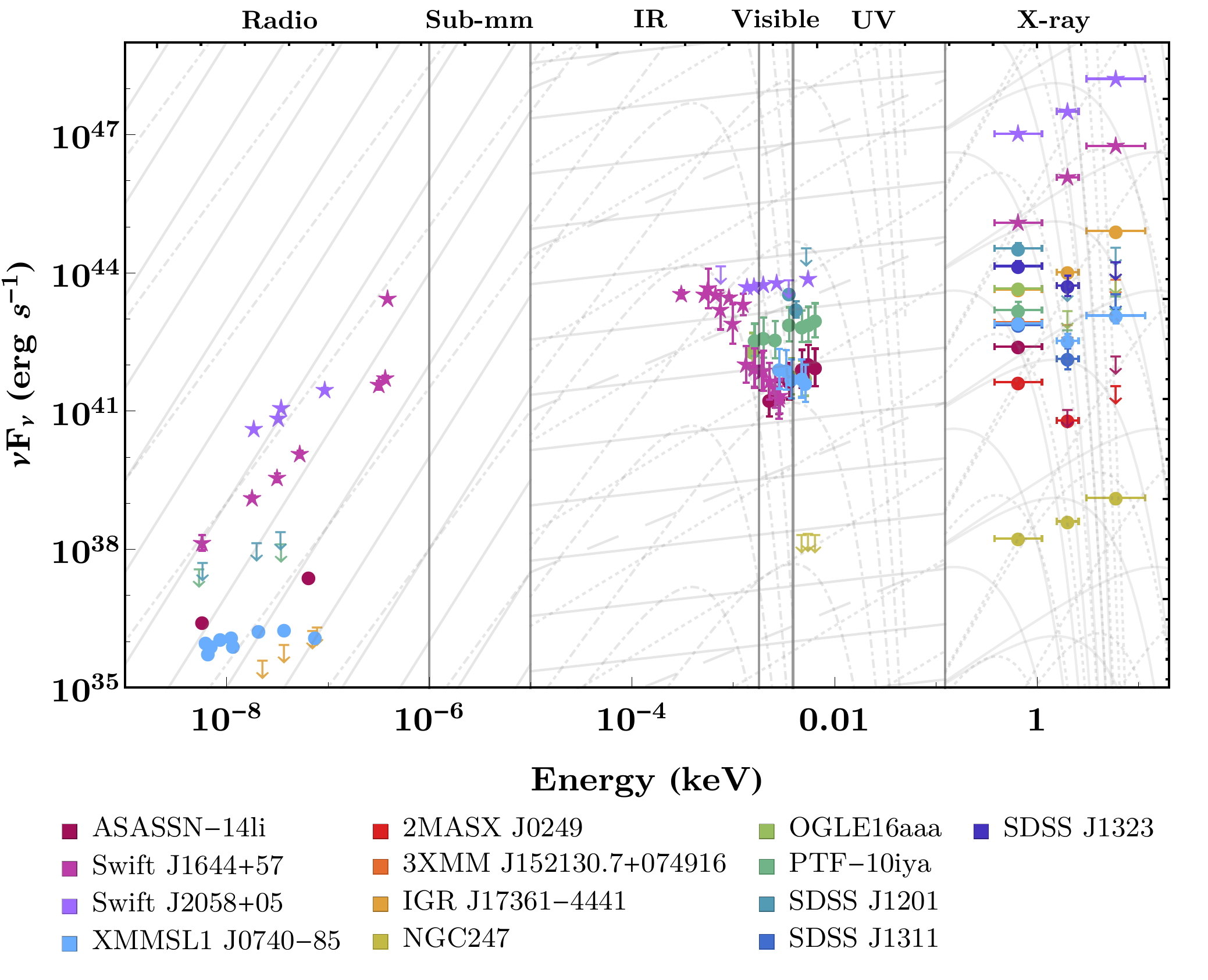}
		\caption{The spectral energy distribution of each TDE candidate plotted together.  In addition, we have also overlaid various radiative astrophysical processes which can reproduce the emission in each energy band. In the radio/sub-mm energy band we have plotted optically thick synchrotron emission ($F_{\nu}\propto\nu^{5/2}$) as the solid line (\----) and Rayleigh-Jeans Law  ($F_{\nu}\propto\nu^{2}$) shown as the dot-dashed line ( $- \cdot - \cdot -$). In the IR/visible/UV energy band we have plotted optically thin synchrotron emission  ($F_{\nu}\propto\nu^{-3/4}$) shown as the solid line (\----), thermal black-body emission represented by Planck's Law ($F_{\nu} \propto \nu^{3}/\exp^{-\nu/(E_{\rm cut}-1)})$ is shown with $E_{\rm cut} = 0.0001, 0.01$ keV as the dot-dashed curve ( $- \cdot - \cdot -$),  optically thin free-free emission  ($F_{\nu}\propto\nu^{-1/10}$) is shown as the dashed line (- - -) and the spectrum of a blackbody accretion disc ($F_{\nu}\propto\nu^{-1/3}$) is shown as the dotted line ($\cdot \cdot \cdot$). In the X-ray band we have plotted saturated inverse Compton scattering which can be approximated by Wein's law ($F_{\nu} \propto \nu^{3}\exp^{-\nu/E_{\rm cut}}$) and is shown with $E_{\rm cut} = 0.1, 1, 10$ keV as the dotted ($\cdot \cdot \cdot$) line. In addition we have also plotted X-ray synchrotron emission which can be approximated using ($F_{\nu}\propto\nu^{1/2} \exp{-\nu/E_{\rm cut}}$). This is plotted  $E_{\rm cut} = 0.1, 1, 10$ keV and is shown as the solid line (\----) in this band. The normalisations of these plots have not been derived by fitting the observed emission, but have been chosen artificially. \label{nuFnu}}
	\end{center}
\end{figure*}

The SED is a powerful tool as it can put strong constraints on the type of mechanism responsible for the emission detected from an astrophysical object. In an attempt to determine the main emission mechanism responsible for the X-ray TDEs we consider, in Figure~\ref{nuFnu} we have plotted the SED of each individual TDE. Overlaid on this plot are various radiative astrophysical processes, such as synchrotron emission, Rayleigh-Jeans, thermal black-body, free-free emission and inverse Compton scattering, which are possible process that can be responsible for the emission observed in each energy band. Details of each radiative process has been described in the caption of Figure~\ref{nuFnu}.

In the X-ray energy band, eight out of the thirteen events do not show significant X-ray emission above 2 keV. These include well known thermal event \emph{ASSASN-14li}, optical TDE \emph{PTF-10iya}, as well as  \emph{2MASX J0249}, \emph{3XMM}, \emph{OGLE16aaa}, \emph{SDSS J1201, SDSS J1311} and \emph{SDSS J1232}. The other five events which do show emission above 2 keV include the non-thermal Swift events, as well as TDEs IGR J17361-4441, NGC 247 and \emph{XMM SL1 J0740-85}. The emission from the first group of events can be well reproduced using either IC scattering or synchrotron emission with a cut-off energy around $\sim$0.1 keV.  However, for the TDEs that fall into the second group, their emission can be more complicated. \emph{Swift J1644+57} can be well reproduced by IC scattering with a cut-off $\sim1$ GeV, while \emph{Swift J2058+05}, \emph{NGC 247}, \emph{XMM SL1 J0740-85} and \emph{IGR J17361-44141} whose emission seems to be better reproduced using synchrotron emission with a cut-off $>10$ GeV.

In the IR to UV energy band the emission from these sources is much more complicated, with each event showing significantly different types of emission. The emission from \emph{Swift J1644+57}, \emph{XMM SL1 J0740-85} and \emph{SDSS J1201} can be well approximated by a thermal blackbody, while \emph{ASASSN-14li}, and \emph{PTF-10iya} are better described by blackbody spectrum from an accretion disc. \emph{Swift J2058+05} has quite different emission in the IR to UV band compared to the other events, as it is better described by optically thin synchrotron emission. These results are consistent with those listed in the literature. 

In the radio/sub-mm energy band, only a few of our X-ray TDE sample have data in this energy band. For a majority of these events, their emission (either constrained directly or through upperlimits) is flatter than what is expected from optically thick synchrotron or Rayleigh-Jeans. The flattening of the radio emission may result from differences in the environment that these events are born into, or in the physics associated with the formation of a jet from these events. \citet{2012ApJ...753...77C} also suggested that a flattening of the radio emission in its SED may also imply a more extended radio source arising from each events. However, even though most of these events seem to diverge from these standard radiative processes, the emission from \emph{Swift J1644+57} can be reasonably well approximated using synchrotron emission which is consistent with the result of \citet{2012MNRAS.420.3528M, 2013ApJ...767..152Z}.

\subsection{What is the BH mass of each X-ray TDE?}

\begin{figure*}[t]
	\begin{center}
		\includegraphics[width=\textwidth]{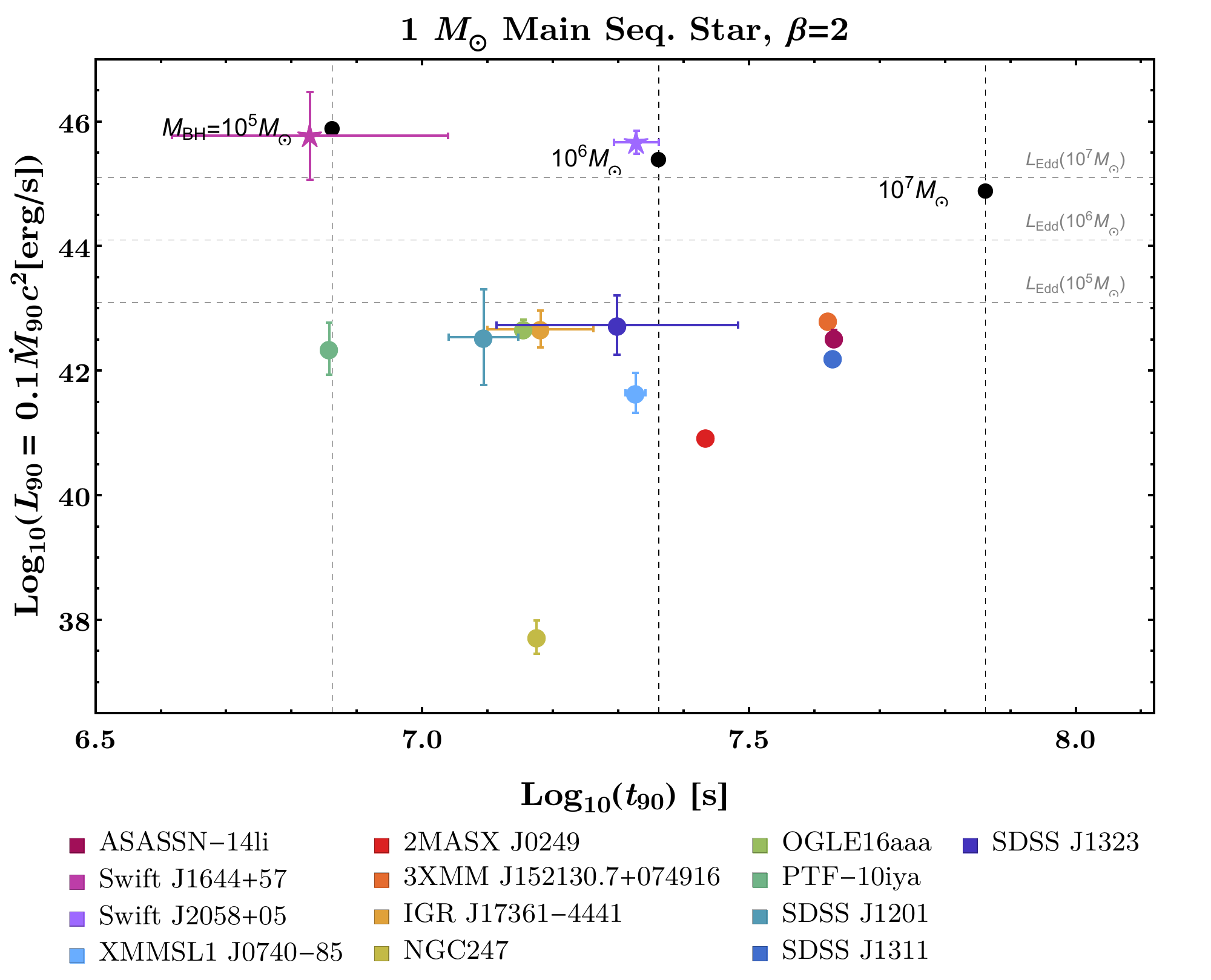}
		\caption{ $L_{90}$ (erg/s) as a function of $t_{90}$ (s) as derived in Section \ref{t90calc}. The solid black symbols represent the equivalent $t_{90}$ and $L_{90}$ values expected for a TDE arising from a low mass (polytropic index $\gamma=5/3$), main sequence star being disrupted by a $M_{BH}=10^{5}M_{\odot}, 10^{6}M_{\odot},10^{7}M_{\odot}$ respectively. We have also assumed that the pericenter distance is two times the tidal disruption radius (i.e., $\beta$=2) implying full disruption.  These results were derived from \citet{2013ApJ...767...25G}, who calculated the fallback accretion rate expected from the disruption of a $1M_{\odot}$ star by a $10^{6}M_{\odot}$ BH. The horizontal dashed lines correspond to the Eddington Luminosity for a $M_{BH}=10^{5}M_{\odot}, 10^{6}M_{\odot},10^{7}M_{\odot}$. \label{bhmass}}
	\end{center}
\end{figure*}

Using the $T_{90}$ and $L_{90}$ values derived from Section \ref{t90calc}, we can attempt to estimate the BH masses of each event. In Figure~\ref{bhmass} we have plotted our derived $L_{90}$ as a function $T_{90}$ for each of the TDE candidates we consider. Assuming the quintessential type of disruption currently used in the literature of a main sequence star being fully disrupted by a black hole with mass $M$, we can derive the expected $T_{90}$ and $L_{90}$ for this type of scenario assuming different BH masses. To estimate these values we use the fallback rate curves produced by \citet{2013ApJ...767...25G}. Here we derive $\dot{M}_{90}$ from these light curves and assume that 10\% of this is accreted on the BH to derive $L_{90}$. To derive $T_{90}$, we follow the same method as listed in \ref{t90calc}. As these curves were derived assuming a BH mass of $10^{6}M_{\odot}$ we scale this to derive the equivalent values for a BH mass of $10^{5}M_{\odot}$ and $10^{7}M_{\odot}$. These expected $T_{90}$ and $L_{90}$ are shown as the solid black data points in Figure~\ref{bhmass}. We also derive the Eddington luminosity for a BH mass of $10^{5}M_{\odot}$, $10^{6}M_{\odot}$ and $10^{7}M_{\odot}$, which are shown as the horizontal lines.

From this plot, one can see that the majority of events have a $T_{90}$ that is consistent with a BH mass between $10^{5}M_{\odot}$ and $10^{7}M_{\odot}$.  With the exception of \emph{Swift J1644+57} and \emph{Swift J2058+05}, all events have luminosities less than the Eddington luminosity of their estimated BH mass. For \emph{Swift J1644+57} and \emph{Swift J2058+05} which are known to be non-thermal jetted TDEs, their $L_{90}$s are super-Eddington, which is consistent with \citet{2012ApJ...760..103D} who showed that super-Eddington accretion rates are required to power the formation of the jet. Since the Eddington luminosity scales linearly with black hole mass, one would think that TDEs arising from low mass BHs would be super-Eddington, while those from higher mass BHs would be sub-Eddington. Under this assumption, one would expect that our $T_{90}$ estimates would suggest that a significant fraction of the TDEs would have a BH $>10^{7}M_{\odot}$, which as seen in Figure~\ref{bhmass} is not the case. However, if the viscous timescales for these events are long, \citet{2015ApJ...809..166G} showed that this assumption does not necessarily hold, with a larger fraction of TDEs arising from lower mass BHs would actually be sub-Eddington. In fact, \citet{2015ApJ...809..166G} determined that a majority of events arising from a BH with a mass below $10^{7}M_{\odot}$ are actually viscously slowed. As all events have a BH mass $<10^{7}M_{\odot}$, and all events, with exception of the jetted \textit{Swift} events have sub-Eddington luminosities, this implies that nearly all X-ray TDEs are viscously slowed.

\section{Conclusion}\label{conclusion}

In this paper we performed a systematic analysis of all (as of writing) publicly available X-ray data from \emph{ROSAT}, \emph{Chandra}, \emph{XMM-Newton} and \emph{Swift} for $\sim$70 TDE candidates currently presented in the literature. Regardless of their literature classification, we characterised the emission from each candidate by extracting the source counts in different energy bands, and when we were able to, their X-ray spectra. Using either the X-ray spectra or the source counts, we derive the 0.3-2.0 keV flux and luminosity of each source, producing multi-decade X-ray light curves for each event. Using the derived X-ray products, a well-defined criteria of the general properties of a TDE, and some guidance from studies of each source found in the literature, we select a set of candidates that allow us to best characterise the properties of the X-ray emission from TDEs. 

We find that the canonical powerlaw index of $t^{-5/3}$ which is commonly assumed as the decay rate of the light curve of a TDE is not necessarily a universal standard in the 0.2-3.0 keV X-ray energy band. Rather, we find that these events have a wide variety of powerlaw indexes, consistent with both fallback and disk emission both over their full X-ray light curve, and during their early and late time emission. We find that the powerlaw index of these TDEs evolves with time, however this evolution is not necessarily smooth and can vary quite dramatically as the TDE decays, particularly for the jetted events. For non-jetted events, we find that at both early times and late times, their emission is consistent with disk emission. However, for jetted events we find that at early times the emission from these events is consistent with fallback, while at later times their emission converges to that seen for non-jetted events. Additionally, we find that for a majority of the events the powerlaw index that we derive is much shallower than $t^{-5/3}$. The shallow nature of their decay is consistent with the emission arising from a TDE that has been viscously slowed, indicating that viscous effects are important for these events. 

Both jetted and non-jetted X-ray TDEs exhibit an increase in their X-ray luminosity at peak of two to three orders of magnitude compared to pre-flare constraints indicating that these events are intrinsically very luminous. In addition these pre-flare limits are crucial for distinguishing these events from AGN activity. We find that both jetted and non-jetted X-ray TDEs are highly absorbed with respect to the Galactic column density along the line of sight of the event. In addition, we find that the absorption for each TDE, within uncertainties, is constant with time. This indicates that the variability in the X-ray emission we observe on smaller timescales, like that seen in e.g., \emph{Swift J1644+57}, is intrinsic to the source, rather than a consequence of variability in absorption.

In addition to quantifying the emission from each candidate in the 0.2-3.0 keV energy range, we derive the count rates of each event in a soft, medium and hard X-ray energy band. By correlating the counts seen in these different energy bands, we find that X-ray TDEs show significant variation in their soft 0.3-1.0 keV X-ray emission. This is in contrast with that seen in the medium and hard energy bands where we find that emission in these bands is highly correlated. Using these count rates we also derive the hardness ratios (HRs) for each event and we find that their emission is quite soft in nature. Throughout their life, these events have a HR between $+0.3$ and $-1$, while at peak their HR is less than 0. The jetted and non-jetted events also naturally separate, with the emission from the non-jetted events being quite soft with a HR between $-1$ and approximately $-0.5$, while the emission from jetted events is well-described using a HR between $-0.5$ and $+0.3$. Due to the significant variation in the soft count rates of each TDE, lack of variation in their medium and hard emission and their enhanced column densities, we suggest that the soft nature of these events implied by these HRs and a number of other properties arises from reprocessing.

For each TDE candidate we consider, we derived the broadband spectral energy distribution (SED) for each event. We find that the X-ray emission from these events is consistent with inverse Compton scattering or synchrotron emission. We find that the non-jetted events tend to have a steeper cutoff in the X-ray band of their SED compared to the jetted events.  For the events which had optical/UV data, we also derived the integrated optical/UV and X-ray energy. We found that the jetted events have a X-ray to optical ratio significantly greater than one, while for non-jetted events this ratio is approximately one. As the non-jetted events had some of the highest measured column densities, and are some of the softest sources in our sample, these low X-ray to optical ratios imply that significant reprocessing must be taking place in these events producing significant optical emission in addition to their detected X-ray emission, as argued by \citet{2014ApJ...783...23G} for PS1-10jh.

The disruption of a star will contribute no more than half of the star's mass to the bound debris surrounding the black hole, with much less being possible in the case of a partial disruption \citep[e.g.,][]{2013ApJ...767...25G, 2014ApJ...783...23G}. \citet{2016ApJ...827....3R} showed that for 0.5 $M_{\odot}$ of bound stellar debris, this material will be optically thick (optical depth $\gg1$) assuming that this envelope of material is dominated by Thomson scattering. Using our derived $T_{90}$ and $L_{90}$ values to estimate the mass of accreted material and Equation [3] from \citet{2016ApJ...827....3R}, we can estimate the optical depth of these events and thus determine the possible nature of the star tidally disrupted. We find that the properties of the sub-Eddington TDEs suggest that the material surrounding these events is optically thin (the optical depth $\ll1$) since the mass of accreted material $\ll$0.5 $M_{\odot}$, implying that sub-Eddington population of TDEs seen in Figure~\ref{bhmass} could potentially arise from either a partial disruption or the disruption of a low mass star.

Compared to other transient events such as GRBs, we find that X-ray TDEs emit the bulk of their emission over significantly longer timescales.  GRBs and GRB-like transients release $5-95$\% of their total fluence on timescales of less than $T_{90}\sim10^{5}$ seconds, while in contrast, X-ray TDEs take $T_{90}\sim10^{7}$ seconds to release the same amount of energy. Since these $T_{90}$s indicate that energy is likely to be injected over much longer timescale, this could be crucial for modelling the dynamics of relativistic jets formed in TDEs \citep{2012ApJ...760..103D}. We also find that there is a ``reprocessing valley'' which separate non-thermal jetted and thermal TDEs based on their derived $L_{90}$. As we argued above, the presence of a solar mass of material about a black hole can easily opaque its inner regions, which naturally produces a gap between the highly-energetic jetted events and events in which the accreted mass is so low that the X-ray emission remains unattenuated. The "veiled" population in between likely represents events for which no jet was produced, but for which the large amount of matter about the black hole absorbs the bulk of the X-ray emission.

We compare our derive $T_{90}$s to the equivalent timescale expected for a main sequence star being disrupted by a BH with a mass between $10^{5}-10^{7}M_{\odot}$ and find that most of these events are consistent with being disrupted by a BH with these masses. We also derive the corresponding Eddington luminosity for the same BH masses and find that the jetted TDEs are super-Eddington, while the non-jetted events are sub-Eddington in nature. 

\citet{2015ApJ...809..166G} recently showed that if the viscous timescales of TDEs are long, a larger fraction of tidal disruptions arising from BHs with a mass $<10^{7}M_{\odot}$ would be sub-Eddington, indicating that these events are most likely viscously slowed. As a large number of our TDE sample are sub-Eddington in nature and arise from a BH with a mass $<10^{7}M_{\odot}$, this is consistent with our conclusion that these events are viscously slowed based on the shallower powerlaw indexes we derived from their X-ray light curves. The super-Eddington luminosities implied for the two non-jetted events \emph{Swift J1644+57} and \emph{Swift J2058+05} are consistent with the picture that super-Eddington accretion rates are required to form a jet. 

In addition, the viscously slowed nature of X-ray TDEs might be able to explain the current discrepancy between the TDE rate derived from theory and observations. Apart from producing sub-Eddington flares, viscous effects will cause the emission from a TDE to peak over long timescales rather than promptly as seen with currently detected events. As we find that our current sample of X-ray TDEs produce very luminous flares, but the properties of their emission implies that they are viscously slowed, there must be a significant population of low luminosity events that have both prompt or long-rise times that current surveys are missing, potentially because they are too dim to be reliably detected (see Figure~\ref{firstupper}). As a consequence, new methods to detect these lower luminosity events might be able to shed light on this problem.

In summary, using over three decades of X-ray data we performed a comprehensive and systematic analysis of the X-ray emission from transient events classified as a TDE in the literature. Using the X-ray products derived from this analysis we were able to produce multi-decade X-ray lightcurves for each event, which allowed us to quantify the decay rates of X-ray TDEs. In particular, we find that the canonical decay rate of $t^{-5/3}$ is not necessarily standard in the soft X-ray energy band, and that there is a reprocessing ``valley'' that separates jetted and non-jetted events that could be naturally populated by optical/UV TDEs.  In addition, we were able to quantify the soft nature of X-ray TDEs from extracting counts in a soft, medium and hard energy band, while determining that viscous effects are important in determining the emission from these events. This provides the community with the first catalogue of X-ray TDE candidates, and all results derived from this analysis are publicly available at the \textit{open TDE catalog} which can be found at \url{https://tde.space}.

\acknowledgements
K. A. is indebted to K. Poppenhaeger for her helpful discussions related to X-ray point-source analysis. The authors would like to thank L. Zoltan Kelley for help with and use of the \textit{AstroCats} python package, and I. Arcavi, P. Maksym, S. van Velzen, D. Lin, O. Graur, and R. Saxton for their helpful comments related some of the text, and data analysis which helped improve the manuscript. This work was supported by Einstein grant PF3-140108 (J. G.), the Packard grant (E. R.), and NASA ATP grant NNX14AH37G (E. R.). We acknowledge the use of NASA-ADS,  HEASARC, the \textit{ROSAT} X-ray All-Sky Survey catalogue, and the \textit{XMM-Newton} Science Archive. The online \url{tde.space} catalog was produced using \textit{AstroCats}.

\input{appendix.tex}

\bibliography{TDE}

\end{document}

%% file: tdecandidates.tex
\begin{table*}[h!]										
	\begin{center}										
		\caption{List name, host, positions and redshifts of TDE candidates and their hosts, irrespective of favoured interpretation. If the right ascension (R.A.) and declination (Decl.) of the TDE is left blank, the host R.A. and Decl. corresponds to the position of the event. \label{tdes}}										
	\begin{tabular}{lllllll}										
		\hline	
		TDE	&	  Host 	&Host	&	 Host	& TDE	& TDE	&	 \\									
	Name 	&	  Name 	&R.A. (J2000)	&	 Decl. (J2000) 	&	R.A. (J2000)	& Decl. (J2000) 	&	 Redshift \\
	\hline \hline										
2MASX J0203&2MASX J02030314$-$0741514&$	02:03:03.14$&$-07:41:51.41	$&	&	&0.0615	\\
2MASX J0249&2MASX J02491731$-$0412521&$	02:49:17.32$&$-04:12:52.20	$&	&	&0.0186	\\
3XMM J152130.7+074916&3XMM J152130.7+074916&$	15:21:30.73$&$+07:49:16.52	$&	$15:21:30.75$& $+07:49:16.70$	&0.17901	\\
ASASSN-14ae&SDSS J110840.11$+$340552.2&$	11:08:40.12$&$+34:05:52.23	$&	$11:08:39.96$&$+34:05:52.70$	&0.0436	\\
ASASSN-14li&SDSS J124815.23$+$174626.4&$	12:48:15.23$&$+17:46:26.44	$&	$12:48:15.23$& $+17:46:26.22$	&0.0206	\\
ASASSN-15oi&2MASX J20390918$-$3045201&$	20:39:09.18$&$-30:45:20.10	$&	$20:39:09.10$& $-30:45:20.71$	&0.0484	\\
ASASSN-15lh	&	APMUKS (BJ) B215839.70-615403.9	&	$22:02:15.39$	&	$-61:39:34.60$	&	$22:02:15.45$& $-61:39:34.64$	&	0.2326 \\
CSS100217 &CSS$100217:102913+404220$& $	10:29:12.56$&$+40:42:20.00	$&	&	&0.148\\
D1-9&GALEX J022517.0$-$043258&$	02:25:17.00$&$-04:32:59.00	$&	&	&0.326	\\
D23H-1&SDSS J233159.53$+$001714.5&$	23:31:59.54$&$+00:17:14.58	$&	&	&0.1855	\\
D3-13&GALEX J141929.8$+$525206&$	14:19:29.81$&$+52:52:06.37	$&	&	&0.3698	\\
DES14C1kia&Uncatalogued, 03:34:47.49 $-$26:19:35.0&$	03:34:47.49$&$-26:19:35.00	$&	&	&0.162	\\
Dougie&SDSS J120847.77$+$430120.1&$	12:08:47.78$&$+43:01:20.27	$&	$12:08:47.87$&$+43:01:20.01$	&0.191	\\
GRB060218, SN2006aj&SDSS J032139.69$+$165201.7&$	03:21:39.69$&$+16:52:01.74	$&	&	&0.0335	\\
HLX-1&ESO 243$-$49&$	01:10:27.75$&$-46:04:27.41	$&	&	&0.0223	\\
IC 3599&IC 3599&$	12:37:41.18$&$+26:42:27.24	$& &		&0.021245	\\
IGR J12580&NGC 4845&$	12:58:01.24$&$+01:34:32.09	$&	$12:58:05.09$&$+01:34:25.70$	&0.00411	\\
IGR J17361-4441&NGC 6388&$	17:36:17.46$&$-44:44:08.34	$&	$17:36:17.42$&$-44:44:05.98$	& 0.04	\\
iPTF16fnl & Mrk950 &$	00:29:57.01$&$32:53:37.24	$&	$00:29:57.04$&$32:53:37.50$	& 0.0163\\
LEDA 095953&LEDA 095953&$	13:47:30.10$&$-32:54:52.00	$&	$13:47:30.33$&$-32:54:50.63$ 	&0.0366	\\
NGC 1097&NGC 1097&$	02:46:19.06$&$-30:16:29.68	$& &		&0.0042	\\
NGC 2110&NGC 2110&$	05:52:11.41$&$-07:27:22.23	$& &		&0.007579	\\
NGC 247&NGC 247&$	00:47:08.55$&$-20:45:37.44	$& &		&0.000531	\\
NGC 3599&NGC 3599&$	11:15:26.95$&$+18:06:37.33	$& &		&0.002699	\\
NGC 5905&NGC 5905&$	15:15:23.32$&$+55:31:01.59	$& &		&0.01124	\\
NGC 6021&NGC 6021&$	15:57:30.68$&$+15:57:22.37	$&	$15:57:30.72$&$+15:57:21.60$	&0.015607	\\
OGLE16aaa &GALEXASC J010720.81-641621.4 &$	01:07:20.88$&$-64:16:20.70	$& &		&0.1655	\\
PGC 015259&2MFGC 3645&$	04:29:21.82$&$-04:45:35.60	$&	$04:29:21.84$&$-04:45:36.00$	&0.014665	\\
PGC 1127938&2SLAQ J011844.35$-$010906.8&$	01:18:44.36$&$-01:09:06.87	$&	$01:18:56.64$&$-01:03:10.80$	&0.02	\\
PGC 1185375&2MASX J15035028$+$0107366&$	15:03:50.29$&$+01:07:36.70	$&	$15:03:50.40$&$+01:07:37.20$	&0.00523	\\
PGC 1190358&N5846$-$162&$	15:05:28.75$&$+01:17:33.17	$&	$15:05:28.56$&$+01:17:31.20$	&0.00766	\\
PGC 133344&6dFGS gJ214256.0$-$300758&$	21:42:55.98$&$-30:07:57.91	$&	$21:42:55.92$&$-30:07:58.80$	&0.02365	\\
PGC 170392&6dFGS gJ222646.4$-$150123&$	22:26:46.35$&$-15:01:23.04	$&	$22:26:46.32$&$-15:01:22.80$	&0.016246	\\
Pictor A&Pictor A&$	05:19:49.72$&$-45:46:43.85	$& &		&0.034	\\
PS1-10jh&SDSS J160928.27$+$534023.9&$	16:09:28.28$&$+53:40:23.99	$&	$16:09:28.29$&$+53:40:23.52$	&0.1696	\\
PS1-11af&SDSS J095726.82$+$031400.9&$	09:57:26.82$&$+03:14:00.94	$&	$09:57:26.82$&$+03:14:01.00$	&0.4046	\\
PS1-12yp&SDSS J133155.90$+$235405.8&$13:31:55.90$&$+23:54:05.8$&$13:31:55.91$&$+23:54:05.70$&	0.581\\
PTF-09axc&SDSS J145313.07$+$221432.2&$	14:53:13.08$&$+22:14:32.27	$&	$14:53:13.06$&$+22:14:32.20$	&0.1146	\\
PTF-09djl&SDSS J163355.97$+$301416.6&$	16:33:55.97$&$+30:14:16.65	$&	$16:33:55.94$&$+30:14:16.30$	&0.184	\\
PTF-09ge&SDSS J145703.17$+$493640.9&$	14:57:03.18$&$+49:36:40.97	$&	$14:57:03.10$&$+49:36:40.80$	&0.064	\\
PTF-10iam&SDSS J154530.83$+$540231.9&$	15:45:30.83$&$+54:02:31.91	$&	$15:45:30.85$&$+54:02:33.00$	&0.109	\\
PTF-10iya&SDSS J143840.98$+$373933.4&$	14:38:40.98$&$+37:39:33.45	$&	$14:38:41.00$&$+37:39:33.60$	&0.22405	\\
PTF-10nuj&SDSS J162624.66$+$544221.4&$	16:26:24.66$&$+54:42:21.44	$&	$16:26:24.70$&$+54:42:21.60$	&0.132	\\
PTF-11glr&SDSS J165406.16$+$412015.4&$	16:54:06.17$&$+41:20:15.45	$&	$16:54:06.13$&$+41:20:14.80$	&0.207	\\
RBS 1032&SDSS J114726.69$+$494257.8&$	11:47:26.80$&$+49:42:59.00	$& &		&0.026	\\
RX J1242-11A&RX J1242.6$-$1119A&$	12:42:36.90$&$-11:19:35.00	$&	$12:42:38.55$&$-11:19:20.80$	&0.05	\\
RX J1420+53&RX J1420.4$+$5334&$	14:20:24.37$&$+53:34:11.72	$&	$14:20:24.20$&$+53:34:11.00$	&0.147	\\
RX J1624+75&RX J1624.9$+$7554&$	16:24:56.66$&$+75:54:56.09	$&	$16:24:56.70$&$+75:54:57.50$	&0.0636	\\
SDSS J0159&SDSS J015957.64$+$003310.4&$	01:59:57.64$&$+00:33:10.49	$& &		&0.31167	\\
SDSS J0748&SDSS J074820.67$+$471214.3&$	07:48:20.67$&$+47:12:14.23	$& &		&0.0615	\\
SDSS J0938&SDSS J093801.64$+$135317.0&$	09:38:01.64$&$+13:53:17.08	$& &		&0.1006	\\
SDSS J0939&SDSS J093922.90$+$370944.0&$	09:39:22.89$&$+37:09:43.90	$& &		&0.18589	\\
SDSS J0952&SDSS J095209.56$+$214313.3&$	09:52:09.56$&$+21:43:13.24	$& &		&0.0789	\\
SDSS J1011&SDSS J101152.98$+$544206.4&$	10:11:52.99$&$+54:42:06.50 	$& &		&0.24608	\\
SDSS J1055&SDSS J105526.41$+$563713.1&$	10:55:26.42$&$+56:37:13.09	$& &		&0.0743	\\
SDSS J1201&SDSS J120136.02$+$300305.5&$	12:01:36.03$&$+30:03:05.52	$& &		&0.146	\\
SDSS J1241&SDSS J124134.25$+$442639.2&$	12:41:34.26$&$+44:26:39.23	$& &		&0.0419	\\
SDSS J1311&SDSS J131122.15$-$012345.6&$13:11:22.15$&$-01:23:45.61$&$13:11:22.18$&$-01:23:45.20$	&0.18	\\
SDSS J1323&SDSS J132341.97$+$482701.3&$	13:23:41.97$&$+48:27:01.26	$& &		&0.08754	\\
SDSS J1342&SDSS J134244.41$+$053056.1&$	13:42:44.42$&$+05:30:56.14	$& &		&0.0366	\\
SDSS J1350&SDSS J135001.49$+$291609.7&$	13:50:01.51$&$+29:16:09.71	$& &		&0.0777	\\
Swift J1112-82&Swift J1112.2$-$8238&$	11:11:47.80$&$-82:38:44.71	$&	$11:11:47.32$&$-82:38:44.20$	&0.89	\\
Swift J1644+57&Swift J164449.3$+$573451&$	16:44:49.30$&$+57:34:51.00	$& &		&0.3543	\\
Swift J2058+05&Swift J205819.7$+$051329&$	20:58:19.85$&$+05:13:33.00	$&	&	&1.1853	\\
UGC 01791&UGC 01791&$	02:19:53.66$&$+28:14:52.60	$&	$02:19:53.52$&$+28:14:52.80$	&0.015881	\\
UGC 03317&UGC 03317&$	05:33:37.54$&$+73:43:26.30	$&	$05:33:37.68$&$+73:43:26.40$	&0.004136	\\
TDE1,VV-1&SDSS J234201.40$+$010629.2&$	23:42:01.41$&$+01:06:29.30	$& &	&0.136	\\
TDE2, VV-2&SDSS J232348.61$-$010810.3&$	23:23:48.62$&$-01:08:10.34	$& &	&0.2515	\\
Wings (A1795)&WINGS J134849.88$+$263557.5&$13:48:49.88$&$+26:35:57.50$&$13:48:49.86$&$+26:35:57.49$	&0.062	\\
XMMSL1 J0740-85&2MASX J2007400785$-$8539307&$07:40:08.09$&$-85:39:31.30$&$07:40:08.43$&$-85:39:31.4$	&0.0173	\\
\hline										
\end{tabular}										
\end{center}										
\end{table*}										

%% file: goldsilver.tex
\begin{table}[t!]
	\begin{center}
		\caption{TDE candidate classified as either \textit{a X-ray TDE}, \textit{likely X-ray TDE}, \textit{possible X-ray TDE}, \textit{veiled TDE}, \textit{not a TDE} or \textit{unknown} based on the requirements listed in Section \ref{classification}. These events have been listed in no particular order. \label{goldsilvers}}
		\resizebox{\columnwidth}{!}{
	\begin{tabular}{llllll}
		\hline
	\textit{X-ray TDE} & \textit{Likely X-ray TDE} & \textit{Possible X-ray TDE} & \textit{Veiled TDE}  & \textit{Not a TDE} & \textit{Unknown}  \\
	\hline \hline
ASASSN-14li	&	2MASX J0249	&	ASASSN-15oi	&	ASASSN-14ae	&	2MASXJ0203	&	Dougie	\\
Swift J1644+57	&	3XMM	&	D3-13	&	ASASSN-15lh	&	CSSS100217	&	PGC1185375	\\
Swift J2058+05	&	IGR J137361	&	LEDA 095953	&	D1-9	&	GRB060218/SN2006aj	&	PGC1190358	\\
XMMSL1 J0740-85	&	NGC247	&	NGC3599	&	D23H-1	&	HLX1	&	PTF-10nuj	\\
&	OGLE16aaa	&	NGC5905	&	DES14C1kia	&	IC3599	&	PTF-11glr	\\
&	PTF-10iya	&	RBS1032	&	iPTF16fnl	&	IGR J12580	&	PTF-11nuj	\\
&	SDSSJ1201	&	RX J1242-11A	&	PS1-10jh	&	NGC1097	&		\\
&	SDSSJ1311	&	RX J1420+53	&	PS1-11af	&	NGC2110	&		\\
&	SDSSJ1323	&	RX J1624+75	&	PS1-12yp	&	Pictor A	&		\\
&		&	SDSSJ0159	&	PTF-09axc	&	PTF-10iam	&		\\
&		&	Swift J1112-82	&	PTF-09djl	&	SDSSJ0938	&		\\
&		&	Wings	&	PTF-09ge	&	SDSSJ0939	&		\\
&		&		&	SDSSJ0748	&	SDSSJ1011	&		\\
&		&		&	SDSSJ0952	&	SDSSJ1055	&		\\
&		&		&	SDSSJ1342	&	SDSSJ1241	&		\\
&		&		&	SDSSJ1350	&		&		\\
&		&		&	TDE2	&		&		\\
	\hline
\end{tabular}
}
\end{center}
\end{table}

%% file: appendix.tex
\begin{appendix}\label{app}

\section{Individual TDE properties and their X-ray light curves}\label{individualsources}

In this section we briefly summarise the properties of each TDE candidate derived from our analysis and as found in the literature. In Figure \ref{lightcurves1}--\ref{lightcurves4} we have also plotted each individual X-ray light curve as derived from our analysis (Tables \ref{rosatfits}--\ref{swiftfits}). Here we have colour coded each data point/upperlimit based on the instrument in which we derived this measurement, with results derived using \emph{ROSAT pointed and RASS} observations shown in red, \emph{Chandra} observations shown in orange, \emph{XMM-Newton slew} observations shown in magenta, \emph{XMM-Newton pointed} observations shown in blue, and \emph{Swift XRT} observations shown in purple. In green we have also plotted the optical/UV emission as taken from the literature for the events which had published data. Here we begin our summary of each event.

\subsection{2MASXJ0203} This candidate was first suggested to be an X-ray TDE by \citet{2007A&A...462L..49E}. These authors compared the count rate derived from the center of its host galaxy 2MASX J02030314$-$0741514 using the \emph{XMM-Newton} Slew Survey Source Catalogue with the count rate derived \emph{ROSAT} PSPC All-Sky Survey. They found that the emission from this host galaxy increased by a factor of 63. Within the error circle of the \emph{XMM-Newton} Slew observation, this detected emission was found to be consistent with the centre of the host galaxy. However recently, \citet{2016arXiv160502749S} suggested that this source could also be a highly variable AGN. We find that the X-ray emission increases by two orders of magnitude compared to the first X-ray detection given by the \emph{XMM-Newton} slew observation. This emission stays approximately constant over a few years, however after 2010 there were no follow-up observations of the source which could help further characterise its emission or to determine whether the emission decays following that expected by a TDE. Our results are similar to that derived by \citet{2016arXiv160502749S}.

\subsection{2MASXJ0249} This candidate was also suggested to be a X-ray TDE by \citet{2007A&A...462L..49E}. By comparing the count rate obtained using their \emph{XMM-Newton} slew observation with the X-ray count rate from the center of galaxy 2MASX J02491731$-$0412521 as derived using \emph{ROSAT}, they found that the X-ray emission increased by a factor of 21. \citet{2016arXiv160502749S} had also suggested that this source could be a highly variable AGN based on the detection of weak [O\textsc{iii}]$\lambda$5007, however the authors highlight that the emission from this source is significantly softer compared to the other candidates in their sample and lacks the standard powerlaw component detected in AGN above $>2$keV making it less likely to an AGN. We find that the X-ray emission from the source peaks and then decreases following a power-law like decay, and shows no recurring periodic emission. However, whether there were flaring emission prior to the detected flare is difficult to say as there is a gap of 15 years in which this host Galaxy was not observed using an X-ray satellite. Similar to \citet{2016arXiv160502749S}, we find that the emission from the source is very soft, as implied by the large powerlaw index derived in our analysis.

\subsection{3XMM} This source was suggested by \citet{2015ApJ...811...43L} to be an X-ray TDE due to its highly transient nature (the authors reported only one data point based on a deep \emph{XMM} observation, however all other observations of the source resulted in upperlimits) and its thermal blackbody (kT$\sim$0.17 keV) X-ray spectrum. However, the black hole mass that is implied from their fits  $M_{BH}=10^{5}-10^{6}M_{\odot}$ is much lower than that expected for a TDE candidate and their derived temperature is significantly higher than that derived for other TDE candidates such as ASASSN-14li or -14ae \citep{2014MNRAS.445.3263H, 2015Natur.526..542M, 2016MNRAS.455.2918H, 2016arXiv160904403B}. In addition, due to the relatively large uncertainty in the error circle of the observation, it is difficult to determine whether the flare is consistent with the centre of its host Galaxy. Much like \citet{2015ApJ...811...43L}, we detect X-ray emission in the 0.3-2.0 keV energy range arising from the position of this source using \emph{XMM}. For all other observations we derive X-ray upperlimits.

\subsection{ASASSN-14ae} This source was classified as a TDE candidate by \citet{2014MNRAS.445.3263H} based on its optical/UV emission using ground-based and follow-up \emph{Swift} observations of the source. This event is the lowest-redshift TDE candidate discovered at optical/UV wavelengths to date (which has been published) and its emission peaks at $10^{43}$ erg/s and decays following an exponential power law. \citet{2014MNRAS.445.3263H} find no X-ray emission arising from the source. We also find no X-ray emission arising from the position of the source and derive upperlimits for all observations.

\subsection{ASASSN-14li} This source was first discovered by ASAS-SN in optical wavelengths and due to its detection is a large number of wavelengths (optical and near UV \citep{2016MNRAS.455.2918H, 2016ApJ...818L..32C, 2016arXiv160904403B}, X-rays \citep{2015Natur.526..542M, 2016MNRAS.455.2918H, 2016arXiv160904403B}, and radio \citep{2016ApJ...819L..25A, 2016Sci...351...62V}) this source is one of the most detailed studied TDEs known. Its classification as an X-ray (and UV/optical) TDE is ubiquitously accepted in the literature. The X-ray emission from the source has been well characterised, and we find that even though we merged X-ray observations rather than consider each individual observation as completed in the literature, we reproduce similar results to those published in the literature. We should also note that radio emission arising from the host Galaxy was detected prior to the detection of ASASSN-14li leading to the possibility of AGN activity, while the detection of narrow [O\textsc{iii}] emission also suggests the presence of a low-luminosity AGN \citep{2016Sci...351...62V}. However, regardless of this fact, the observed properties of ASASSN-14li are inconsistent with that expected of an AGN (see \citealt{2016Sci...351...62V} for more details).

\subsection{ASASSN-15oi} Much like ASASSN-14li and 14ae, this source was first discovered by ASAS-SN and follow up observations using ground based instruments and \emph{Swift} indicate that this source is optical/UV TDE \citep{2016arXiv160201088H}. However, ASASSN-15oi faded significantly more rapidly than other optically discovered TDEs. This source shows evidence of weak X-ray emission, however due to it rapidly fading the X-ray emission from the source decayed much quicker than its UV/optical emission. We find that we also detect weak (compared to the UV/optical emission) X-ray emission from the source, however as we have no late time or early time constraints on its emission, making it difficult to determine how the X-ray emission from this source evolves beyond the two detections we report.

\subsection{ASASSN-15lh} First discovered using ASAS-SN by \citet{2016Sci...351..257D}, this transient event has so far been quite a puzzle. This source had a peak luminosity two times that of any known supernova, and during its early time emission showed features similar to those seen in superluminous supernovae \citep{2016Sci...351..257D}. Follow up observations by \citet{2016arXiv160500645G}, showed that its properties differed significantly from those of known TDEs such as ASASSN-14li and -14ae, putting more weight behind its supernova origin. However, using 10 months of multiwavelength data, \citet{2016arXiv160902927L} showed that the properties of this source is more consistent with a TDE rather than a superluminous supernova based on its temperature evolution, the presence of CNO gas along the line of sight and its location being coincident with the centre of a passive Galaxy. However, the mass implied by their analysis is $>10^{8}M_{\odot}$ making it one of the largest BHs in which a TDE has been detected. Recently \citet{2016arXiv161001632M} presented deep \emph{Chandra} and \emph{Swift} observations of this source and detected persistent soft X-ray emission consistent with the position of the optical transient. In conjunction with their multi-wavelength campaign in which they also study the optical and UV emission arising from the host, they conclude that if this X-ray source is coincident with the optical transient originally detected by ASAS-SN, then this event is consistent with a TDE of a main sequence star by a massive spinning black hole.  Using the \emph{Chandra} data that was available at the time of writing, we do not detect significant X-ray emission arising from the source over the full instrument energy band. Using publicly available \emph{Swift} observations of ASASSN-15lh, we find no significant X-ray emission arising from the position of the source compared to background fluctuation.  However, when considering the emission in the soft, medium and hard energy bands we use for our analysis, we do detect faint X-ray emission in the 0.3-1.0 keV energy band arising from the source using one of the later \emph{Swift} observations, consistent with \citet{2016arXiv161001632M}. The differences in our analysis most likely arises from using different energy ranges, background regions and our different requirements of what constitutes a detection.  

\subsection{CSS100217} This source was originally discovered by the Catalina Real-time Transient Survey as an extremely luminous optical transient arising from the centre of a narrow line Seyfert 1 galaxy \citep{2010ATel.2544....1D, 2011ApJ...735..106D}. \citet{2011ApJ...735..106D} performed extensive multi-wavelength follow up observations of this source and found that it is coincident with the centre of its host Galaxy, and spectroscopically exhibited strong narrow Balmer features representative of other Type IIn supernovae such as SN2008iy, SN2007rt and SN1997ab. The detected X-ray luminosity of the event, and its derived temperature is similar to that of luminous Type IIn supernovae such as SN2006gy, while the lack of gamma-ray emission detected from the source rule out the nature of this source as a Type Ib/c or GRB. Due to its coincidence with the centre of its host Galaxy,  \citet{2011ApJ...735..106D} also suggested that this event could be consistent with a TDE, however its optical lightcurve, peak optical brightness, and temperature varies greatly from that theoretically expected for TDEs. Similar to \citet{2011ApJ...735..106D}, we detect X-ray emission arising from the source coincident with the optical flare detected from this event. Using a follow up \emph{Swift} observation that was taken approximately five years later, we again detect X-ray emission arising from the source. This emission is of the same order of magnitude as the first X-ray emission detection detected, while shallow \emph{XMM slew} observations of this source did not detect any X-ray emission from this event. This makes it unlikely that this X-ray emission arises from a TDE.

\subsection{D1-9} This source was suggested to be a optical/UV TDE based on the detection of a UV/optical flare using \emph{GALEX} from the center of a quiescient early-type galaxy, which then decayed following a power law \citep{2008ApJ...676..944G}. The authors triggered a \emph{Chandra} TOO and detected 4 X-ray photons between 0.2-0.4 keV with a detection confidence of 0.93. Due to our more stringent classification of requiring a detection and measuring the emission over a larger energy range we derive only upperlimits to the X-ray emission of the source and find that the 4 photons that these authors detected using \emph{Chandra}, are more likely arose from Poisson fluctuations. 

\subsection{D23H-1} \citet{2009ApJ...698.1367G} discovered a large magnitude optical/UV flare coincident with the centre of a star forming galaxy using \emph{GALEX}. The spectral energy distribution (SED) from this flare can be well described by a powerlaw decline and is best described by a soft blackbody. Even though a low-luminosity AGN can't be ruled out, the broadband properties of the flare deviate from the average properties observed for AGN and are consistent with that of an optical/UV TDE. They detect no X-ray emission arising from the source and derive 3$\sigma$ upperlimits. We also find that even at later times, no X-ray emission from the position of this source is detected and derive only 3$\sigma$ upperlimits.

\subsection{D3-13} Similar to D1-9 and D23H-1, this flare like emission representative of an optical/UV TDE was first discovered by \citet{2008ApJ...676..944G} using \emph{GALEX}. The emission rose sharply and decayed monotonically as expected from an optical TDE, however the light curve is incomplete making it difficult to get a full picture of the emission from the source. In addition the error circle associated with the position of the flare seems to be slightly off-center from the host galaxy.  Similar to our analysis, the authors detect X-ray emission from a \emph{Chandra} observation taken about a year after the optical/UV flare was detected. Later observations presented by \citet{2008ApJ...676..944G} and analysed in this work only produce upperlimits to the X-ray emission from the source, similar to our analysis.

\subsection{DES14C1kia} A possible optical/UV TDE candidate detected in the Dark Energy Survey \citep{2015ATel.6877....1F}. The emission from this source rose for seven weeks prior to peak brightness and then decayed, while not undergoing rapid colour evolution. Based on optical spectroscopy of the host Galaxy, it is thought that the flare arose from a passive Galaxy. Follow up observations in X-rays \citep{2015ATel.6887....1Y} and Radio \citep{2015ATel.6904....1R} detect no emission in these energy bands. We find no X-ray emission from this source.

\subsection{Dougie} An optical transient that was first discovered using \emph{ROTSE} and followed up in the optical using \emph{ROTSE-IIIb} and in the UV using \emph{Swift} \citep{2015ApJ...798...12V}. Its optical light curve has a quick rise, followed by a reasonably quick decline of approximately a month. However, the source is systematically offset from the centre of its host Galaxy. We analyse all available X-ray data that overlap this source and we find no X-ray emission arising from the position of the transient. Unlike nearly all other TDE candidates in our sample, this source was not covered by \emph{ROSAT}. As a consequence we only have a few year constraint on the X-ray emission associated with the source.

\subsection{GRB060218, SN2006aj} This TDE candidate was first discovered by \emph{Swift} and was first classified as an under-luminous very long GRB, which is thought to be accompanied by a fast, low ejecta mass supernova SN2006aj \citep{2006Natur.442.1008C, 2006Natur.442.1014S, 2006Natur.442.1018M}. X-ray emission from this source was also detected which \citet{2006Natur.442.1008C} interpreted as arising from shock break-out. However even though the GRB/SN is scenario is the most favoured case to explain the properties of this transient source, \citet{2013ApJ...769...85S} was also able to show that the unique properties of this source could be equally well described by a TDE from an intermediate black hole at the centre of a dwarf galaxy. From our analysis we detect the increase in the X-ray emission as seen by \citet{2006Natur.442.1008C}. Using a follow up \emph{Chandra} observation that was taken immediately after the \emph{Swift} detection, we detect X-ray emission that is two orders of magnitude less than that seen using \emph{Swift}. All later observations of the source produced upper limits and no recurrent emission has been detected.

\subsection{HLX-1} This source is an ultra-luminous intermediate-mass black-hole  (IMBH) system which exhibits variability with a possible recurrence time of a few hundred days \citep{2011ApJ...735...89L}. The high luminosity, light curve and
X-ray spectrum evolution of HLX-1 can be explained by the recurring mass-transfer that results from the tidal stripping of a star in an eccentric orbit around the IMBH (i.e., a recurring TDE). However, this source shows hard-to-soft X-ray transitions \citep{2011ApJ...743....6S}, while also showing evidence of a radio jet emission \citep{2012Sci...337..554W}, which are typically observed for galactic black hole binaries. This makes it difficult to explain the observed variablity using a recurring TDE \citep{2013ATel.5439....1G}. From our analysis, we also detect significant periodic variability over a large number of years.

\subsection{IC3599} This source was first characterised as an X-ray TDE by  \citet{1995MNRAS.273L..47B,1995A&A...299L...5G} using \emph{ROSAT}, in which they discovered a rapid decrease in the X-ray flux by two orders of magnitude over a year. \citet{2015A&A...581A..17C} later discovered using follow up \emph{Swift} observations, a recurring flare like event of similar magnitude as that seen using \emph{ROSAT}, which these authors suggest is a periodic, partial TDE. However, \citet{2015ApJ...803L..28G} showed that this periodic emission is most likely consistent with an accretion disk instability around a BH. In addition, its mid-IR  \citep{2010MNRAS.403.1246S} and radio \citep{2013ApJ...763...84B} emission is consistent with that of an AGN. From our analysis we also detect the periodic X-ray emission from this source consistent with that in the literature.

\subsection{IGR J12580} First detected as a strong, hard X-ray flare using \emph{INTEGRAL},  \citet{2013A&A...552A..75N} classified this event as a X-ray TDE of a super Jupiter by a central supermassive BH. \citet{2015ApJ...809..172I} discovered transient radio emission arising from the position of the TDE candidate, while also highlighting that this source has been classified as a LINER/Seyfert 2 Galaxy based on its optical spectra indicating that this source could be a changing look quasar.  The original radio emission detected by \citet{2015ApJ...809..172I}, was detected before the hard X-ray flare and shows evidence of variability, indicative of an AGN jet. In addition, the WISE colours for this event is consistent with those of a luminous AGN \citep{2012ApJ...753...30S}, while based on pre-flare data this source was classified as an AGN in the VCV catalogue \citep{2010A&A...518A..10V}. Irrespectively of the source being more likely to be an AGN, \citet{2016ApJ...816...20L} reinforced the planet TDE interpretation by showing that the observed emission most likely arises from an off-beam relativistic radio jet that formed during the original TDE. Using 37 months of data from the Monitor of All-sky X-ray Image (\emph{MAXI}), \citet{2016PASJ...68...58K} detected significant X-ray flare emission in the 4\-10 keV energy band arising from the position of this source. Similar to \citet{2013A&A...552A..75N} we detect an rapid rise in the X-ray emission by several orders of magnitude and then a gradual decay consistent with an X-ray flare. However, we believe the detected X-ray emission most likely arises from AGN activity rather than a TDE.

\subsection{IGR J17361-4441} This source was first discovered as a hard X-ray source using \emph{INTEGRAL} near the centre of a globular cluster. Follow up observations showed that the X-ray light curve decays following a powerlaw consistent with that of a X-ray TDE, while its thermal component does not evolve significantly with time. \citet{2014MNRAS.444...93D} classified this transient event as the tidal disruption of a free-floating terrestrial icy planet by a white dwarf due to the fact that it was located slightly off-center from the host Galaxy. From our analysis we find that the X-ray emission from the position of the source rapidly rises over a short time period, then decays over less than a year. We also find that there is some periodic X-ray emission from the position of the source, which is consistent within uncertainties with the low state X-ray emission of the flare-like event. We detect no other periodic emission, while no AGN or other transient phenomenon has been suggested as the origin of this event so far.

\subsection{iPTF16fnl} Discovered as a nuclear transient by PTF from the nearby Galaxy Mrk 950. Follow up spectra detected a blue continuum, and strong broad He[\textsc{II}] 4686 emission consistent with that of a TDE \citep{2016ATel.9433....1G}. We detect no X-ray emission arising from the position of iPFT16fnl using \emph{Swift} observations taken after the initial Astronomers Telegram.

\subsection{LEDA 095953} \citet{2009A&A...495L...9C} serendipitously discovered flare-like X-ray emission arising from a Galaxy found within a Galaxy cluster. This flare is found significantly off center from the center of the galaxy cluster, however within uncertainties it is consistent with the center of its host Galaxy. Using follow-up \emph{Chandra} and \emph{XMM} observations of the source, they derive a data point and an upperlimit respectively. They also use \emph{ROSAT} HRI observations of the source which we do not analyse and detect emission from the source. Based on these observations, they interpret that the X-ray emission that they detect is the result of an X-ray TDE. From our analysis we only detect X-ray emission from the \emph{ROSAT} observation of the source, while for the late time \emph{Chandra} and \emph{XMM} observation, we derive only a 3$\sigma$ upperlimit. The difference between our results and the results of \citet{2009A&A...495L...9C} most likely arises from defining the different background regions in our analysis and our more stringent requirements for what constitutes a detection.

\subsection{NGC1097} Monitoring observations by \citet{1995ApJ...443..617S} of this Galaxy showed strong variation in the optical flux and broad double peaked H$\alpha$ emission lines of this host Galaxy. To explain both the variation in the flux and shape of the optical emission, \citet{1995ApJ...443..617S} suggested that it arose from an elliptical ring of material that arose from the tidal disruption of a star by the central BH. However, it is widely accepted that NGC1097 is an AGN  (see e.g., \citealt{1995ApJ...443..617S}), which is made quite evident in our analysis of the X-ray emission from this source. One can see that there is strong variability in the observed X-ray emission at late times, with the emission in the high and low state being consistent with each other over a number of years.

\subsection{NGC2110} \citet{2007ApJ...668L..31M} discovered using polarisation measurements of its optical spectrum, a transient broad, double peaked H$\alpha$ feature arising from NGC2110, which is commonly seen from optical TDE. The discovery of Fe K$\alpha$ lines that vary over timescales of years \citep[e.g.,][]{2015MNRAS.447..160M} and several of its other properties imply that this source is most likely a prototypical double peaked emission-line AGN. From our analysis, we find that the X-ray emission from this source varies over years similar to other AGNs, with the high and low states having similar orders of magnitude.

\subsection{NGC247} Using \emph{XMM-Newton}, \citet{2015ApJ...807..185F} serendipitously discovered a strong X-ray flare from the centre of the inactive Galaxy NGC247. UV spectroscopy shows no evidence for an AGN. Follow up observations of the source with \emph{Swift} detected an increase in the X-ray luminosity of the source which then peaked and decayed exponentially.  \citet{2015ApJ...807..185F} concluded that the properties of this source could result from either an outburst from a low mass X-ray binary with a stellar-mass black hole emitting near its Eddington luminosity, or from a X-ray TDE being accreted onto a $10^{5}M_{\odot}$ nuclear BH. We reproduce the X-ray emission from NGC247 as presented by \citet{2015ApJ...807..185F}, while also placing strong constraints on prior X-ray emission from this source, for which we find no prior X-ray emission before the detected flare.

\subsection{NGC3599} This candidate was also suggested to be an X-ray TDE based on its detection in the \emph{XMM-Newton} slew survey by \citet{2007A&A...462L..49E}. Within the error circle of the \emph{XMM} observation this source was consistent with the centre of its host Galaxy and increased by a factor of 88 compared to its \emph{ROSAT} detection. \citet{2008A&A...489..543E} obtained follow up \emph{XMM-Newton} and \emph{Swift} observations of the source and detected emission that was consistent with the canonical $t^{-5/3}$. \citet{2015MNRAS.454.2798S} showed using archival X-ray data that this source flared 18 months before the detection by \citet{2007A&A...462L..49E}, but was still consistent with a slow rising TDE or possibly from a thermal instability in the accretion disc of an AGN. Optical observations of the host indicated that the source is a low-luminosity Seyfert/low-ionization nuclear emission-line region (LINER) \citet{2008A&A...489..543E}. We similarly find that the source was X-ray bright before the original flaring event was discovered by \citet{2007A&A...462L..49E} and \citet{2015MNRAS.454.2798S}. Unlike \citet{2008A&A...489..543E} we only derive an upperlimit for the \emph{Swift} observations. This difference most likely arises from using different regions to define the source and background, while we required a source to be detected with a significance of 2$\sigma$ or more before it was classified as a detection.

\subsection{NGC5905} First discovered by \emph{ROSAT} as a very soft X-ray transient source which increased dramatically in flux over a few days and then declined by a factor of 80 two years later,
\citet{1996A&A...309L..35B} classified this source as an X-ray TDE. \citet{2004ApJ...604..572H} followed up this source using \emph{Chandra} and found the emission has decreased such that it is consistent with the basal star burst emission of the host Galaxy. \citet{2002ApJ...576..753L} modelled the emission from this event and speculated that it arose from either the partial stripping of a low mass main sequence star or the disruption of a brown dwarf or a giant planet. However, even though the TDE scenario is currently favoured in the literature (see e.g., \citealt{2004ApJ...604..572H}), \citet{2003ApJ...592...42G, 2004ApJ...601.1159G} found using the Hubble Space Telescope (\emph{HST}) narrow emission lines in the inner nucleus of the host, indicating that there is a low level of prior non-stellar photoionisation powered by accretion. This raises questions on the TDE origin of the flare, making it more possible that the emission arises from AGN activity. Similar to \citet{1996A&A...309L..35B} and \citet{2004ApJ...604..572H} we detect X-ray emission arising from the position of the source using \emph{ROSAT} and \emph{Chandra} which then decays. Using a follow up \emph{Chandra} observation of this source taken in 2007, we also detect X-ray emission from the source consistent with the late time emission derived by \citet{2004ApJ...604..572H}. However, unlike \citet{2004ApJ...604..572H} we derive an X-ray luminosity for the peak of the X-ray flare that is an order of magnitude less than that derived by \citet{2004ApJ...604..572H}. This difference most likely arises from using different energy ranges, and regions to derive the source counts from this event.

\subsection{OGLE16aaa} Discovered by the OGLE-IV survey at the centre of a Galaxy which shows evidence of some weak ongoing star formation and AGN emission, this transient exhibited a long rise, slow decline and broad He and H spectral features similar to those of other optical/UV TDEs \citep{2017MNRAS.465L.114W}. \emph{Swift} observations taken around the same time of optical flare detected no X-ray emission, as noted in \citep{2017MNRAS.465L.114W}. However further follow up observations using \emph{Swift} allowed us to detect significant X-ray emission from this event, which then decayed and faded over a few months.

\subsection{PGC1185375} Using the \emph{Swift} BAT, \citet{2016A&A...586A...9H} surveyed over 50000 Galaxies to search for X-ray flare emission from inactive Galaxies that could potentially arises from a TDE. From their analysis they found nine X-ray TDE candidates arising from hosts which show no evidence of AGN emission with PGC1185375 being one of these candidates. This X-ray flare had a duration of 41 days and was found offset from the host Galaxy.  However, due to its location on the detector the increased PSF dramatically decreases their ability to properly localise the source making it possible for this source to be coincident with the centre of the host.  \citet{2016A&A...586A...9H} also rule out a contribution from an AGN. There is very limited data available for this source, and as such we are only able to put upperlimit constraints on the X-ray emission. We find no recurring X-ray flare emission from the observations we analyse of the source.

\subsection{PGC1190358} Another X-ray TDE candidate suggested by \citet{2016A&A...586A...9H}. The flare lasted for 108 days and peaked 40 days before the flare was undetectable by the \emph{Swift} BAT. The flare was coincident with the centre of its host, however the uncertainty on this position is large. From peak to non detection the flux decreased by factor of 34. Similar to PGC1185375, there is very little archival X-ray data available for this source and as such we are only able to derive upperlimits to the X-ray emission.

\subsection{Pictor A} \citet{1995ApJ...438L...1S} detected broad, transient double peaked Balmer line emission arising this radio Galaxy, which is also seen in optical TDEs. However, \citet{1997A&A...328...12P} discovered a faint radio jet connecting two radio lobes to the central nucleus of the host, while \citet{2001ApJ...547..740W} discovered using \emph{Chandra} an X-ray jet coincident with the radio jet. From our X-ray analysis, we detect significant and variable X-ray emission arising from the central source that varies over many years more indicative of AGN activity

\subsection{PS1-10jh} First discovered in the \emph{Pan-STARRS}1 survey, this UV-optical flare occurred at the centre of an inactive Galaxy and was quickly classified as an optical/UV TDE \citet{2012Natur.485..217G}. The UV/optical spectra from this source was characterised by the presence of broad [He \textsc{ii}]$\lambda$4686 emission, and the absence of a number of hydrogen lines, indicating that the disrupted star was a He-rich red giant that had its outer envelope stripped. \citet{2015MNRAS.454.2321S} determined that to disrupt such a dense object, the BH must of had a mass of $<10^{5}M_{\odot}$. However, \citet{2014ApJ...783...23G} also showed using hydrodynamical simulations that it is possible to explain the properties of this event using a main sequence star that was disrupted by a SMBH. This result was supported by photoionisation modelling by \citet{2014MNRAS.438L..36G}. After the original detection in \emph{Pan-STARRS1}, follow up observations by \emph{Chandra} detected no X-ray emission from the source, while ruling out an AGN origin\citep{2012Natur.485..217G}. We also find no X-ray emission immediately after the original UV/Optical flare. However, we detect using a \emph{XMM-Newton} slew observation, week X-ray emission ($\sim4$ source photons) arising from the position of the source a few years after the flare.

\subsection{PS1-11af} Another optical/UV TDE candidate that was discovered using \emph{Pan-STARRS}. \citet{2014ApJ...780...44C} discovered a long-lived optical transient that is coincident with the centre of its host Galaxy. UV spectroscopy revealed broad, transient absorption features arising from the source, while its luminosity and colour changed slowly through its detection. The spectra obtained also showed no other features other than the broad UV absorption component. Based on its properties \citet{2014ApJ...780...44C} classified it as an optical/UV TDE which arose from the partial disruption of a main sequence star by a $10^{6}M_{\odot}$ BH. The host also shows no evidence of AGN emission. This source is very similar to PS1-10jh. There are very limited X-ray observations of this event, and as such we are only able to constrain upperlimits to the X-ray emission of the source.

\subsection{PS1-12yp} This event was first discovered in the PanSTARRS1 survey for large amplitude transients that occur at the centre of galaxies that show no evidence of AGN \citep{2016MNRAS.463..296L}. This event was the bluest transient in their sample, and showed significant increase in its optical luminosity. The optical emission from the event then decayed similar to that of PS1-10jh, leading the authors to possibly classify it as a TDE or a SNe. There is very little X-ray coverage of this source, with only shallow \emph{XMM-slew} observations of this event. We detect no X-ray emission from the position of this source and derive only X-ray upperlimits.

\subsection{PTF-09axc} \citet{2014ApJ...793...38A} analysed archival Palomar Transient Factory (\emph{PTF}) in search for transients that have a peak magnitude between -21 and -19. These authors found six events which have similar rise times to that of optical/UV TDE PS1-10jh. PTF-09axc shows evidence of broad hydrogen features and is coincident with the centre of its host Galaxy, however its optical light curve is poorly sampled. \citet{2014ApJ...793...38A} detect very weak  [O\textsc{iii}]$\lambda$5007 emission from its host Galaxy possibly indicating the presence of a very weak AGN. Using a \emph{Swift} TOO observation of PTF-09axc, \citet{2014ApJ...793...38A} detect X-ray emission with a luminosity of $\sim10^{42}$ erg s$^{-1}$ arising from the position of the source, which they state is consistent with that expected from AGN. However, the AGN origin of this emission was not confirmed. Similar to \citet{2014ApJ...793...38A} we also detect X-ray emission arising from PTF-09axc with a luminosity of $\sim10^{42}$ erg s$^{-1}$, however the uncertainty on this value is very large. There are not many available X-ray observations of this source, however we do not detect any recurring X-ray emission.

\subsection{PTF-09djl} Another potential optical/UV TDE candidate discovered in archival \emph{PTF} data by \citet{2014ApJ...793...38A}. Similar to PTF-09axc this flare like emission is coincident with its host Galaxy and its spectrum shows evidence of broad hydrogen features. \citet{2014ApJ...793...38A} find no evidence of AGN emission lines from its host spectra and follow up \emph{Swift} observations of the source detect no X-ray emission. There are only two archival observations which overlap the position of the source, and from these we are only able to derive 3$\sigma$ upperlimits, consistent with the analysis by \citet{2014ApJ...793...38A}.

\subsection{PTF-09ge} One of the six optical/UV TDE candidates discovered in the \emph{PTF} analysis by \citet{2014ApJ...793...38A}. Spectroscopically, PTF-09ge shows evidence of He-rich features, while hydrogen is absent from this spectrum. Its light curve is very well sampled and is similar to that of PS1-10jh, while its flare like UV/optical emission is coincident with the centre of its host Galaxy. There was no evidence of AGN emission in its host spectrum, and follow up \emph{Swift} TOO observations of the source produce only 3$\sigma$ upperlimits. We also are only able to extract $3\sigma$ upperlimits from our analysis of the three archival observations which overlap the position of the source.

\subsection{PTF-10iam} Another of the PTF optical/UV TDE candidates presented by \citet{2014ApJ...793...38A}, however this flare is found off set from the centre of its host Galaxy. The spectrum of its host Galaxy shows evidence of Balmer absorption features only. The light curve of PTF-10iam rises significantly faster to peak emission compared to the other TDE candidates suggested by \citet{2014ApJ...793...38A}. More recently, \citet{2016ApJ...819...35A} suggested that this event is more likely to be a peculiar Type II supernova, or a hybrid Type Ia - Type II supernova event. We detect no X-ray emission arising from this source .

\subsection{PTF-10iya} First discovered by \citet{2012MNRAS.420.2684C} using \emph{PTF} as a short lived, luminous, UV/optical transient event coincident with the centre of its host Galaxy. Based on its host spectrum, they find no evidence of AGN activity. \emph{Swift} TOO observations of this source immediately after the PTF detection, detect significant X-ray emission arising from the source. However further observations derive only upperlimits. This is consistent with the X-ray emission we derive for PTF-10iya.

\subsection{PTF-10nuj} A PTF optical/UV TDE candidate that is found systematically offset from its host Galaxy \citep{2014ApJ...793...38A}. Its optical emission rises quickly over $\sim10$ days, but then suddenly drops off $\sim30$ days after peak. We find no X-ray emission arising from the position of the source and derive $3\sigma$ upperlimits.

\subsection{PTF-11glr} Another PTF optical/UV TDE candidate presented by \citet{2014ApJ...793...38A} that is systematically offset from its host Galaxy. Its host shows evidence of Balmer absorption features and emission lines, while its optical light curve rises and falls in a similar way to PTF-10nuj.  No other information about the source is available. From our analysis, we detect no X-ray emission from the source over three observational epochs and as such derive $3\sigma$ upperlimits.

\subsection{RBS 1032} A bright, luminous X-ray flare arising from dwarf Galaxy RBS 1032 was first discovered by \citet{2006MNRAS.371.1587G} using \emph{ROSAT}. Follow up \emph{XMM-Newton} observations by \citet{2014ApJ...792L..29M} indicated the presence of a very faint X-ray source within the 30\arcsec\, error circle of the \emph{ROSAT} position of the source. This emission is 200 times less than that detected in the \emph{ROSAT} observation and is very soft in nature.  The optical monitor of \emph{XMM} also detected a source coincident with the faint X-ray source. \citet{2006MNRAS.371.1587G} find no evidence of AGN activity from the host. \citet{2014ApJ...792L..29M} classifies RBS 1032 as an X-ray TDE, however the suggestion by \citet{2006MNRAS.371.1587G} that the observed emission arises from an intermediate mass BH binary cannot be ruled out. Unlike \citet{2014ApJ...792L..29M}, we find no evidence for a faint X-ray source arising from the position of RBS 1032 using \emph{XMM-Newton} and as such derive 3$\sigma$ upperlimits from the source. This most likely arises using different background, where our background incorporates the basal emission from the host Galaxy, in addition to our more stringent requirement for emission to be classified as a detection.

\subsection{RX J1242-11A} This event was first discovered by \citet{1999A&A...349L..45K}  using \emph{ROSAT} as a large increase in the detected X-ray flux from the inactive Galaxy pair RX J1242-1119. The inactive nature of the host was confirmed by \citet{2003ApJ...592...42G, 2004ApJ...601.1159G} using \emph{HST}. Due to its soft spectrum and lack of evidence for AGN activity, \citet{1999A&A...349L..45K} classified this source as an X-ray TDE. Follow up observations using \emph{Chandra} and \emph{XMM-Newton} showed that the emission from this source dropped by a factor 200 compared to its \emph{ROSAT} detection, and was still consistent with that of a TDE \citep{2004ApJ...603L..17K, 2004ApJ...604..572H}. We similarly find that the X-ray emission of this source has decreased by a factor of $\sim200$ compared to its \emph{ROSAT} detection. Follow up observations of this source taken nearly 15 years later show no evidence of recurring X-ray emission.

\subsection{RX J1420+53} By comparing the emission from a pointed \emph{ROSAT} observation of this host with the emission detected in the RASS, \citet{2000A&A...362L..25G} discovered an X-ray transient source which displayed a variation in its flux by a factor of 150. Based on optical observations of the source \citet{2000A&A...362L..25G} classified the host as inactive, and tentatively classified this emission as an X-ray TDE. Since the original discovery, there has been follow-up \emph{Chandra} observations of the source. From our analysis, we detect no X-ray emission arising from the position of the source, $\sim$15 years after the original detection.

\subsection{RX J1624+75} Another X-ray transient event discovered using \emph{ROSAT}. Originally detected as bright X-ray emission in the RASS, \citet{1999A&A...350L..31G} found that within 1.5 years after the original discovery, that the emission from this source had faded. They took optical spectra of the host and determined that the host is most likely inactive, however they also detected a weak signature of [N\textsc{ii}]$\lambda$6584 emission which could imply weak AGN activity. However \citet{1999A&A...350L..31G} classify the observed X-ray flare as an X-ray TDE rather than AGN activity based on the timescale in which the X-rays were detected and then disappeared. Follow up \emph{Chandra} observations of the source by \citet{2004ApJ...604..572H} confirmed that the X-ray emission from this source had decreased by a factor $>1000$ and was consistent with that of X-ray TDE. Follow up observations of the host by \citet{2003ApJ...592...42G, 2004ApJ...601.1159G} using \emph{HST} confirmed that this source is an inactive Galaxy. Our analysis of the X-ray emission from the host is consistent with that derived by \citet{1999A&A...350L..31G, 2004ApJ...604..572H}.

\subsection{SDSS J0159} First analysed by \citet{2015ApJ...800..144L}, these authors performed detailed follow-up optical observations of the SDSS Galaxy SDSS J015957.64$+$003310.4. This host seems to be transitioning from a Type 1 broad-line AGN to a Type 1.9 AGN, showing weak broad H$\alpha$ emission lines over a 10 year period. They found that over this period, the optical flux from the source decreased by a factor of 6, while its H$\alpha$ weakened and became broader. Serendipitous \emph{Chandra} and \emph{XMM-Newton} observations of the source also found that the 2.0-10.0 keV emission in a high state which then decreased by an order of magnitude. \citet{2015ApJ...800..144L} attributed this change in both the optical and X-ray emission as the dimming of the AGN continuum, however they caveat that this type of change in the properties of this source is quite rare. \citet{2015MNRAS.452...69M} re-analysed archival data of the source and argued that the properties of its light curve and emission is consistent with that of an optical TDE whose accretion energy is reprocessed by dense, large scale height material \citep{2014ApJ...783...23G}. \citet{2015MNRAS.452...69M} also suggested that under certain assumptions, this TDE would be one of the most luminous non-beamed TDE discovered so far. From our analysis, we also detected X-ray emission from the host using both \emph{XMM} and \emph{Chandra}, which then decreases by an order of magnitude between these two observations. However, we find that this detected emission is less than order of magnitude greater than its \emph{ROSAT} upperlimits, which is not commonly seen in TDEs.

\subsection{SDSS J0748} \citet{2011ApJ...740...85W, 2012ApJ...749..115W} analysed the SDSS spectrum of Galaxy SDSS J074820.67$+$471214.3 and found evidence of strong high-ionization coronal lines such as [Fe \textsc{x}]$\lambda$6376 and [Ar \textsc{xiv}]$\lambda$4414, as well as very broad line emission, which can be interpreted as the blue-shifted He \textsc{ii} and Balmer lines. They also found that the source brightened in the g-band by 0.2 magnitudes between the photometric and spectroscopic observations of the source. Follow up optical observations of the source four years later had shown that these lines had weakened significantly, while line ratios ruled out the presence of AGN activity. Based on the detection of these coronal lines and broad line emission, \citet{2008ApJ...678L..13K} that suggested these lines represent a ``light echo'' of a flare, while \citet{2011ApJ...740...85W, 2012ApJ...749..115W} suggested that this source is an optical TDE. \citet{2016arXiv160505145D} recently presented a study of the mid-infrared emission from this source and discovered significant emission arising from this source many years after the original detection. They classified this late time mid-infrared as an general signature of a TDE occuring in a gas-rich environment. There is very limited X-ray data available of this object, with only a \emph{ROSAT} and \emph{XMM-Newton} slew observation of the source. Both observations detect no X-ray emission from the position of the TDE and thus three sigma upperlimits are derived.

\subsection{SDSS J0938} \citet{2012ApJ...749..115W} performed a survey of SDSS Galaxies to search for strong coronal lines from [Fe \textsc{x}]$\lambda$6376 up to [Fe \textsc{xiv}]$\lambda$5304 which could represent a ``light echo'' of a flare arising from a TDE. SDSS J093801.64$+$135317.0 is one of the seven host Galaxies that they selected which show evidence of these strong coronal lines and they suggested that these lines arise from a TDE. However, follow up spectroscopic observations using the Multi-Mirror Telescope (MMT) of SDSS J0938 by \citet{2013ApJ...774...46Y} shows that these coronal lines are super imposed over narrower, low ionisation lines that arise from star forming regions. As such \citet{2013ApJ...774...46Y} suggest that the strong coronal lines that were detected by \citet{2012ApJ...749..115W} most likely arise from an obscured AGN, however the TDE scenario cannot be ruled out. We detect no X-ray emission arising from this host and derive only X-ray upperlimits.

\subsection{SDSS J0939} This candidate was suggested by \citet{2007A&A...462L..49E} to be an X-ray TDE based on its detection in the \emph{XMM-Newton} slew survey. However, even though the host Galaxy, SDSS J093922.90$+$370944.0, is found within the error circle of \emph{XMM}, the position of this source is offset from the main optical emission of the host. Based on the width of the H$\alpha$ line, the strength of the [Fe \textsc{ii}] multiplets and the ratio of [O \textsc{iii}]$\lambda$5007 to H$\beta$, \citet{2007A&A...462L..49E} classify the host as a narrow-line Seyfert 1 galaxy. We detect variable X-ray emission from the host which shows evidence of a high and a low state emission more indicative of AGN emission.

\subsection{SDSS J0952} \citet{2008ApJ...678L..13K} discovered strong, coronal Fe lines, along with broad Balmer and double peaked, narrow He $\beta$ emission lines arising from galaxy SDSS J095209.56$+$214313.3. These lines which were strong in an SDSS spectrum taken during 2005, faded significantly over a two year period. Follow up photometric observations of the source in the NUV, optical and NIR bands showed evidence of variability in these wavelengths, which \citet{2008ApJ...678L..13K} suggest arises from an large X-ray flare. Follow up observations using the Lincoln Near Earth Asteroid Research (LINEAR) survey by \citet{2016ApJ...819..151P} detected a strong UV flare 1.8 years after when the flare was expected to have started, that arose from the center of the host Galaxy. \emph{Chandra} X-ray observations taken three years after the flare revealed faint X-ray emission coincident with the host Galaxy. \citet{2016arXiv160505145D} also detected years after the original event mid-infrared emission that declines following a simple powerlaw model which they attribute to being a signature of the original TDE. We also find X-ray emission arising from the position of the source, but the emission does not decay following a powerlaw, but increases and then plateaus for two years. This type of emission is not consistent with that seen in other X-rays TDEs which usually dramatically peak and then decay following approximately a powerlaw.


\subsection{SDSS J1011} \citet{2016MNRAS.455.1691R} classified this source as a ``changing look'' quasar whose broad emission lines and continuum emission, is representative of a quasar. This source was first detected by \emph{SDSS} in 2003, and then dramatically weakened in follow-up time domain spectroscopy of the source. Even though \citet{2016MNRAS.455.1691R} attribute the changes in the spectrum to changes in the accretion rate onto the BH, they also suggest that a TDE scenario could be responsible for the changes in the observed optical spectrum. However, even though a TDE is consistent with the decay rate of the light curve, this scenario cannot explain the fact that the emission of the source stayed in a high state for many years prior to decay, or explain the strength of the observed emission lines. Unfortunately, there is very little archival X-ray data overlapping the period in which these spectral changes were observed, however from the observations we analysed we detect no X-ray emission from the position of the source and derive only X-ray upperlimits.

\subsection{SDSS J1055} Another Galaxy which shows evidence of strong coronal lines, and broad Balmer lines which could indicate that a flare from a TDE occurred \citet{2012ApJ...749..115W}.  However, \citet{2013ApJ...774...46Y} detect broad Balmer lines, Fe II and the non-stellar continuum which suggests that it is a Type I Seyfert Galaxy, while its narrow line ratios indicate that it is an AGN. We detect weak X-ray emission from the host using \emph{ROSAT}, however all follow up observations detect only X-ray upperlimits.

\subsection{SDSS J1201} Using the slew capability of \emph{XMM-Newton}, \citet{2012A&A...541A.106S} first discovered a soft X-ray flare coincident with the nucleus of the inactive Galaxy SDSS J120136.02$+$300305.5. Follow-up observations of the source revealed significant variability in the emission of the source, however the emission declined following the canonical t$^{-5/3}$ of a X-ray TDE fading significantly over nearly a year. Deep radio observations of the source revealed that no X-ray jet was launched during the event. \citet{2014ApJ...786..103L} showed that the observed variability in the X-ray light curve of SDSS J1201 can be well explained by a super-massive BH binary system undergoing variability accretion when the star was disrupted. Similar to \citet{2012A&A...541A.106S} we find flare like X-ray emission using both \emph{XMM} and \emph{Swift}, which shows evidence of variability. Prior to this X-ray flare event, we do not detect any X-ray emission from the source. Even though our X-ray light curve is consistent with that of \citet{2012A&A...541A.106S}, we derive slightly lower luminosities than these authors. This most likely arises from using a different model to reproduce the X-ray spectrum, and from using different background regions to define the spectra and source counts.

\subsection{SDSS J1241} SDSS J124134.25$+$442639.2 is another strong coronal line emitting Galaxy that was discovered by \citet{2009MNRAS.397..172G} and \citet{2012ApJ...749..115W} using \emph{SDSS} while searching for the imprint of an X-ray flare that could arise from a TDE. \citet{2013ApJ...774...46Y} performed follow-up MMT observations of this Galaxy and found that that these emission lines did not show significant variation in their strength with time, indicating that the lines most likely arise from the presence of an AGN, rather than a transient event like a TDE. We detect using archival \emph{Swift} and \emph{Chandra} observations of the source soft X-ray emission that stays at maximum for over two years, then is suddenly undetectable over a year later making it less likely to be a TDE.

\subsection{SDSS J1311} Using archival \emph{Chandra} and \emph{XMM} data, \citet{2010ApJ...722.1035M} discovered a transient X-ray event that they suggest most likely arises from a TDE. This event was coincident with the centre of its host Galaxy SDSS J131122.15$-$012345.6, which shows no evidence of strong optical emission lines that would suggest the presence of an AGN. This event was very soft, showed signs of variability and decayed following a t$^{-5/3}$ over a two year period. The temperature of this TDE (kT $\sim$ 0.12 keV) is one of the highest temperature TDEs discovered so far. Late time radio observations of the source found no evidence of radio emission indicating the formation of a jet \citep{2013ApJ...763...84B}. In our analysis, we are able to reproduce the results of \citet{2010ApJ...722.1035M}, however we are able to extend on their work and characterise the emission properties of the source over a longer base line. We find that apart from the original flare detected by \citet{2010ApJ...722.1035M}, no detectable X-ray emission arises from the source.

\subsection{SDSS J1323} This source was originally discovered by \citet{2007A&A...462L..49E} in the \emph{XMM-Newton} slew survey searching for potential TDE candidates. They found that this source varied by $>80$ compared to its \emph{ROSAT} upperlimit. Optical observations of the host taken before the burst showed no evidence of emission lines which could indicate the presence of an AGN \citep{2008A&A...489..543E}. \citet{2008A&A...489..543E} obtained follow-up \emph{XMM-Newton} and \emph{Swift} observation of the source two years after the original \emph{XMM-Newton} slew observation. They found that X-ray emission detected by  \emph{XMM-Newton} and \emph{Swift} had decreased by a factor of $\sim40$ compared to the \emph{XMM-Newton} slew observation and decayed following t$^{-5/3}$. Unlike \citet{2008A&A...489..543E} we detect X-ray emission from this source using only \emph{XMM-Newton}, while we derive only upper limits for the \emph{XMM-Newton} slew and \emph{Swift} observation. In addition, our derive luminosity is also an order of magnitude lower than \citep{2008A&A...489..543E}. The difference in the number of detected data points most likely arises from our more stringent requirement for classifying a detection, while our luminosity estimation is most likely lower due to using regions to extract the source and background spectra, as well as fitting the spectra using a different model.

\subsection{SDSS J1342} Identified by \citet{2013ApJ...774...46Y} as a potential TDE by the detection of coronal emission lines. These lines disappeared years after the original SDSS observation that lead to this object being a source of interest. More recently \citet{2016arXiv160505145D} detected mid-infrared emission from this source many years after the original flare which they state is most likely a signature of the original TDE. The X-ray data available for this source is quite limited, however a \emph{Swift} observation that overlapped the position of the source detected soft X-ray emission. This emission was comparable to the \emph{ROSAT} upperlimit we derived. However due to the limited data, we are unable to draw any conclusions about the X-ray emission from the source.

\subsection{SDSS J1350} Another potential TDE classified using its coronal emission lines, while also showing evidence of mid-infrared emission which faded many years after the original flaring event  \citet{2013ApJ...774...46Y, 2016arXiv160505145D}. No X-ray emission was detected from the position of the source.

\subsection{Swift J1112-82} First detected as an long duration $\gamma$-ray outburst using the \emph{Swift} BAT, \citet{2015MNRAS.452.4297B} obtained follow up \emph{Swift} observations of the source and found that this burst also exhibited a bright X-ray flare coincident with the centre of its host Galaxy. The X-ray emission decayed following approximately a powerlaw model, while showing significant short-term variability. The X-ray emission was also quite hard in nature, with \citet{2016PASJ...68...58K} also detecting hard X-ray emission from this source using 37 months of \emph{MAXI} observations. Using the Fermi-LAT $\gamma$-ray satellite, \citet{2016ApJ...825...47P} searched for $\gamma$ emission arising from the position of the source, however they were only able to derive upperlimits to the source. Simultaneous \emph{Swift} UVOT observations of the source detected no UV/optical emission arising from the position of the flare, and follow up observations using \emph{Gemini} detected a weak point-like source that decayed quickly. Optical spectra taken of the source reveal a single, weak emission line arising from [O\textsc{ii}]$\lambda$3727 but no other additional emission lines or continuum emission was detected. \citet{2015MNRAS.452.4297B} rule out an AGN origin for this flare and suggest that Swift J1112-82 is a non-thermal (relativistic) X-ray TDE similar to that of Swift J1644+57. This flare fades quickly and disappears $\sim$100 days after the original discovery. From \emph{ROSAT} and \emph{XMM-Newton} slew observations of the source, we detect no other X-ray emission arising from the source. The luminosity derived by \citet{2015MNRAS.452.4297B} is a few orders of magnitude larger than what we derive. The reason behind this is because we derive the luminosity in the 0.3-2.0 keV, while the luminosity presented by \citet{2015MNRAS.452.4297B} represents the 0.2-10.0 keV. As this source is quite hard in nature, a majority of the emission from this source falls into the hard X-ray band rather than the soft X-ray band presented in this work.


\subsection{Swift J1644+57} Similar to Swift J1112-82, Swift J1644+57 was first discovered by the \emph{Swift} BAT as a long duration $\gamma$-ray outburst \citep{2011Sci...333..203B, 2011Natur.476..421B}. Observations by the \emph{Swift} XRT indicated that this source was highly variable, long lived, had an isotropic peak 0.3-10.0 keV luminosity that exceeded $10^{48}$ erg/s, and decayed following approximately a $t^{-5/3}$ powerlaw law, ruling out the possibility that this source was a GRB. Using follow-up optical observations of the position of the source, \citet{2011Sci...333..199L} detected an optical counterpart which they associated to be the host Galaxy of this source. \citet{2011Sci...333..199L}  triggered \emph{Chandra} TOO observations of the source and was able to confirmed its highly variable nature and that this source was coincident with the nucleus of its host Galaxy. \citet{2011Sci...333..199L} was also able to rule out the presence of permanent AGN activity. Following the first month of evolution of this source, \citet{2011Natur.476..425Z} detected a radio transient coincident with the centre of the host Galaxy, which was suggested to be a collimated relativistic outflow. The unique properties of this source quickly lead to the conclusion that Swift J1644+57 is a highly beamed, non-thermal (relativistic) jetted X-ray TDE. The event is thought to arise from a $\sim10^{6}-10^{7}M_{\odot}$ BH, while the observed X-rays arise from internal dissipation from the inner part of the jet, while the radio emission arises from an expanding shock front \citep[e.g., see ][]{2011Sci...333..203B, 2011Natur.476..421B,2011Sci...333..199L,2011Natur.476..425Z,2011arXiv1104.2528B}. 

Swift J1644+57 is one of the most data rich X-ray TDEs detected. This is best shown by \citet{2016ApJ...817..103M} who present a complete analysis of the all available \emph{Swift} and \emph{Chandra} data of the object, 507 days after the first trigger, while \citet{2016ApJ...819...51L} and \citet{2016ApJ...816L..10C} analyse the late-time \emph{XMM}, \emph{Swift} and \emph{Chandra} emission from the source. Even though we have merged observations of Swift J1644+57 that have a similar MJD, we find the general trend obtained by \citet[e.g.,][]{2011Sci...333..203B, 2011Natur.476..421B,2016ApJ...817..103M}. However, as we derive the luminosity of the source over 0.3-2.0 keV energy range, our maximum luminosity is lower than that derived in the literature. The literature on Swift J1644+57 focuses predominantly on the rich \emph{Swift} data of this source, however we also take advantage of the number of \emph{XMM-Newton} observations available to derive the X-ray emission from the source. We find that the \emph{XMM} data shows the strong variability seen in the \emph{Swift} data, while also providing a constraint on the X-ray emission immediately before the flare. \emph{ROSAT} observations of the source show that there is no variable X-ray emission arising from the source, however between the \emph{ROSAT} data and the original trigger there is nearly 20 years in which there is no X-ray coverage of the source.

\subsection{Swift J2058+05} Soon after the discovery by Swift J1644+57, the \emph{Swift} BAT discovered another X-ray transient source, Swift J2058+05, which showed many similarities with Swift J1644+57 \citep{2011ATel.3384....1K, 2012ApJ...753...77C}. \citet{2012ApJ...753...77C} performed a detailed multi-wavelength follow-up campaign of the source and discovered a long-lived, very luminous X-ray transient coincident with the centre of an inactive Galaxy. The X-ray emission from this source decays following a simple powerlaw, and then drops off rapidly much like Swift J1644+57 \citep{2015ApJ...805...68P}. However the source does not show such dramatic variability as that seen in the Swift J1644+57 light curve. In addition, \citet{2012ApJ...753...77C} also revealed a radio counterpart to this flare  which they associate with a jet like outflow. The properties of this source led \citet{2012ApJ...753...77C} to conclude that Swift J2058+05 was another non-thermal (relativistic) jetted X-ray TDE. Due to the relatively faint optical emission arising from the host Galaxy, \citet{2012ApJ...753...77C} was not able to rule out the presence of an AGN even though the observations suggested that the host did not harbour an AGN. As an alternative to the TDE scenario, \citet{2012ApJ...753...77C} also suggested that if this source arose from an AGN, then the discovery of this object would represent a new mode of variability in AGN.

The light curve we derive from our analysis is very similar to that derived by \citet{2012ApJ...753...77C} and \citet{2015ApJ...805...68P}. However, our luminosities are lower due to us focusing on the soft X-rays rather than the full energy band that is analysed in the literature. We also take advantage of available \emph{ROSAT} and \emph{XMM-Newton} slew observations of the source and determine that prior to the flare, there was no X-ray emission arising from the source. However, similar to Swift J1644+57 there is no X-ray data available for Swift J2058+05 between $\sim1990-2008$.

\subsection{TDE2, VV-2} Using archival \emph{SDSS} data, \citet{2011ApJ...741...73V} discovered optical TDE, TDE2, coincident with the nucleus of its host Galaxy. Using follow up UV and optical observations \citet{2011ApJ...741...73V} were able to rule out the presence of an AGN, or its origin as a supernova explosion or another transient phenomenon. This event had an optical blackbody temperature of 1.82$\times10^{4}$ K and a peak magnitude of $M_{g} = -20.4$. The UV and optical light curve of TDE2 decayed following the standard $t^{-5/3}$, while \citet{2014ApJ...792...53V} found that the light curve has a very similar decay rate as that of PS1-10jh, however it is much more luminous. There is very limited X-ray data about this object, with no X-ray observations taken around the same time the original flare was detected. We find that there is no X-ray emission from this source.

\subsection{Wings (A1795)}\label{wingsappendix} Using the large number of archival X-ray observations of Galaxy cluster Abell 1795, \citet{2013MNRAS.435.1904M} and \citet{2014ApJ...781...59D} discovered a very luminous, soft X-ray flare located significantly off centre from the core of cluster. However, the transient flare was found to be coincident with the centre of its inactive host, a dwarf Galaxy called WINGS J134849.88+263557.5 (Wings J1348). The flare lasted for more than five years and decayed following the canonical TDE powerlaw decay rate, making it a strong candidate for being a TDE. Archival observations using the Extreme Ultraviolet Explorer (EUVE) seemed to suggest a strong correlation between the detected X-ray flare in this work and a EUVE transient observed by \citet{1999ApJ...526..592B}. Based on this observation, \citet{2013MNRAS.435.1904M} and \citet{2014ApJ...781...59D} derived a BH mass of $<10^{6}M_{\odot}$. Follow up optical observations of the host using \emph{Gemini} by \citet{2014MNRAS.444..866M} revealed that the host Galaxy is an extremely low mass Galaxy making it one of the smallest galaxies to host a BH. Based on their optical spectrum that they obtained of the host, \citet{2014MNRAS.444..866M} infer the presence of weak or temporary nuclear activity based on broad [O III] emission. 

The X-ray emission arising from Wings J1348 is located in a very complicated region. The Galaxy cluster is very bright in X-rays and diffuse X-rays from this host dominate the background emission of Wings J1348. We attempted to remove the contribution of the Galaxy cluster from our analysis by defining a background region that immediately surrounds the position of the possible X-ray TDE. In addition, we used source and background regions similar to the PSF size of each instrument rather that using the region sizes specified in Section \ref{dataanalysis}. By specifying smaller source and background regions, we minimise the contamination from Galaxy cluster emission. However, due to the complicated nature of the X-ray emission from this source, our results differ in some respects to that of \citet{2013MNRAS.435.1904M} and \citet{2014ApJ...781...59D}. 

In our analysis, we are able to reproduce the X-ray lightcurve that \citet{2013MNRAS.435.1904M} derived from the \emph{Chandra} data of Wings. Compared to \citet{2014ApJ...781...59D} our luminosity values are much higher due to using a smaller energy range to define our luminosity. Compared to \citet{2013MNRAS.435.1904M} we derived an upperlimit to the X-ray flux using \emph{XMM-Newton} rather than an X-ray detection. This discrepancy most likely arises from the use of different source and background regions. In addition to the \emph{Chandra} and \emph{XMM-Newton} observations, we find that there are multiple pointed \emph{ROSAT} observations and a RASS observation that overlap the position of the transient which cover the early 1990s, in addition to a \emph{Swift} observation that was taken in $\sim$2006. Neither \citet{2013MNRAS.435.1904M} or \citet{2014ApJ...781...59D} analysed the X-ray emission from these \emph{ROSAT} or \emph{Swift} observations. Using these observations, we are able to derive X-ray upperlimits to the emission from the source.

\subsection{XMM SL1 J0740-85} During the \emph{XMM-Newton} slew survey,  \citet{2016arXiv161001788S} detected a bright X-ray flare from the centre of quiescent galaxy, 2MASX $07400785-8539307$. The event was detected in both the full X-ray energy band and the UV band using \emph{XMM-Newton} and \emph{Swift}, showed signs of X-ray variability, and decayed by a factor of $>$70 and $>$12 in X-rays and UV respectively. \citet{2016arXiv161003861A} followed up this source in radio using ATCA and detected weak radio emission consistent with a non-relativistic outflow, similar to that seen in ASASSN-14li \citep{2016ApJ...819L..25A}. Using CITO and LCO optical observations of the host Galaxy, \citet{2016arXiv161001788S} determined that the host shows no evidence of AGN or current star formation activity indicating that the host of this event is a post-starburst Galaxy. As the galaxy showed no signs of previous AGN activity, this lead  \citet{2016arXiv161001788S} to classify XMM SL1 J0740-85 as a TDE. From our analysis, we are able to produce similar results to that of \citet{2016arXiv161001788S}.

\input{lightcurves.tex}

\section{Powerlaw Model fits}\label{powerlawmodels}

\begin{figure*}[h]
	\begin{center}
		\includegraphics[width=0.3\textwidth]{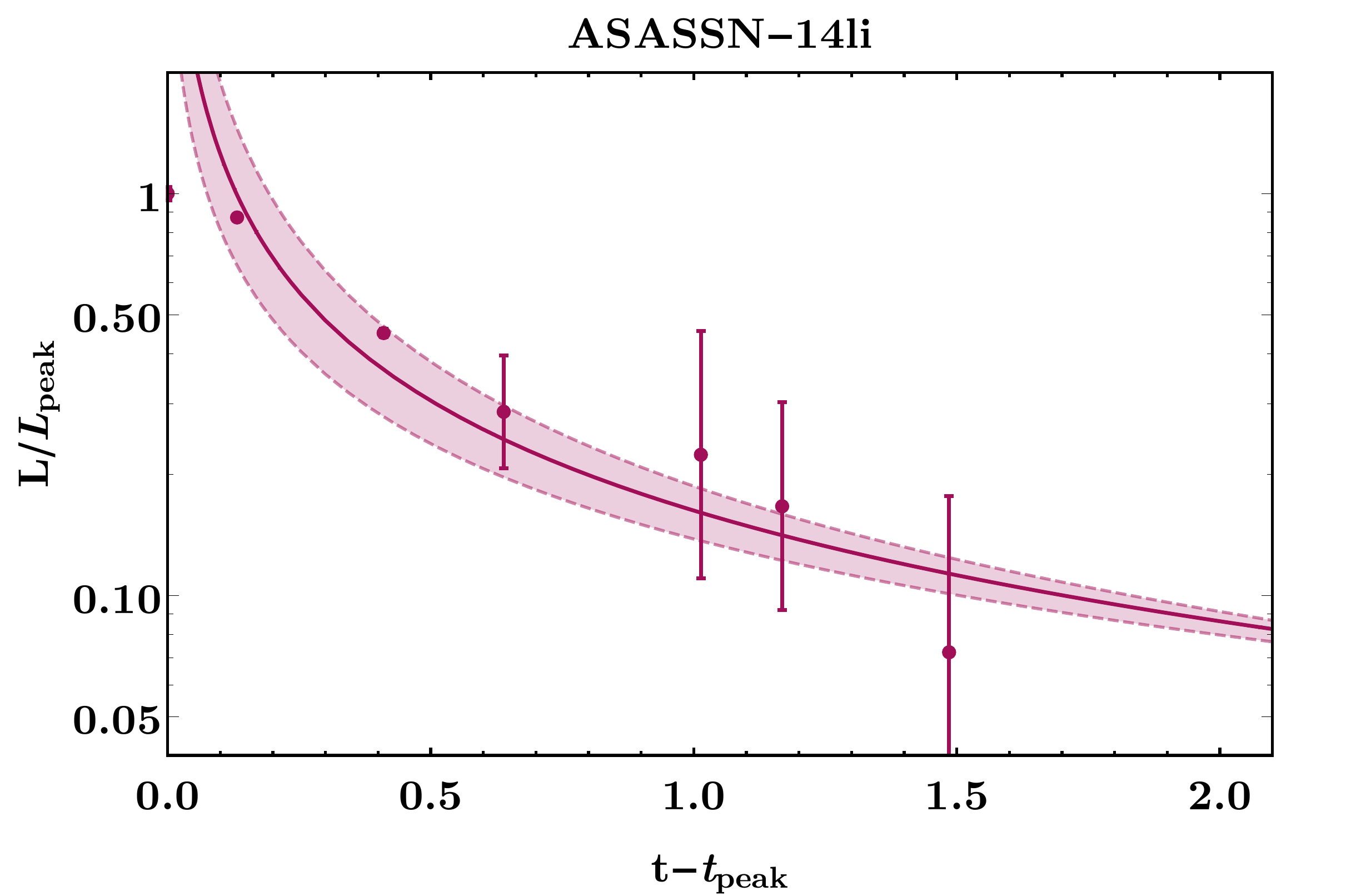}
		\includegraphics[width=0.3\textwidth]{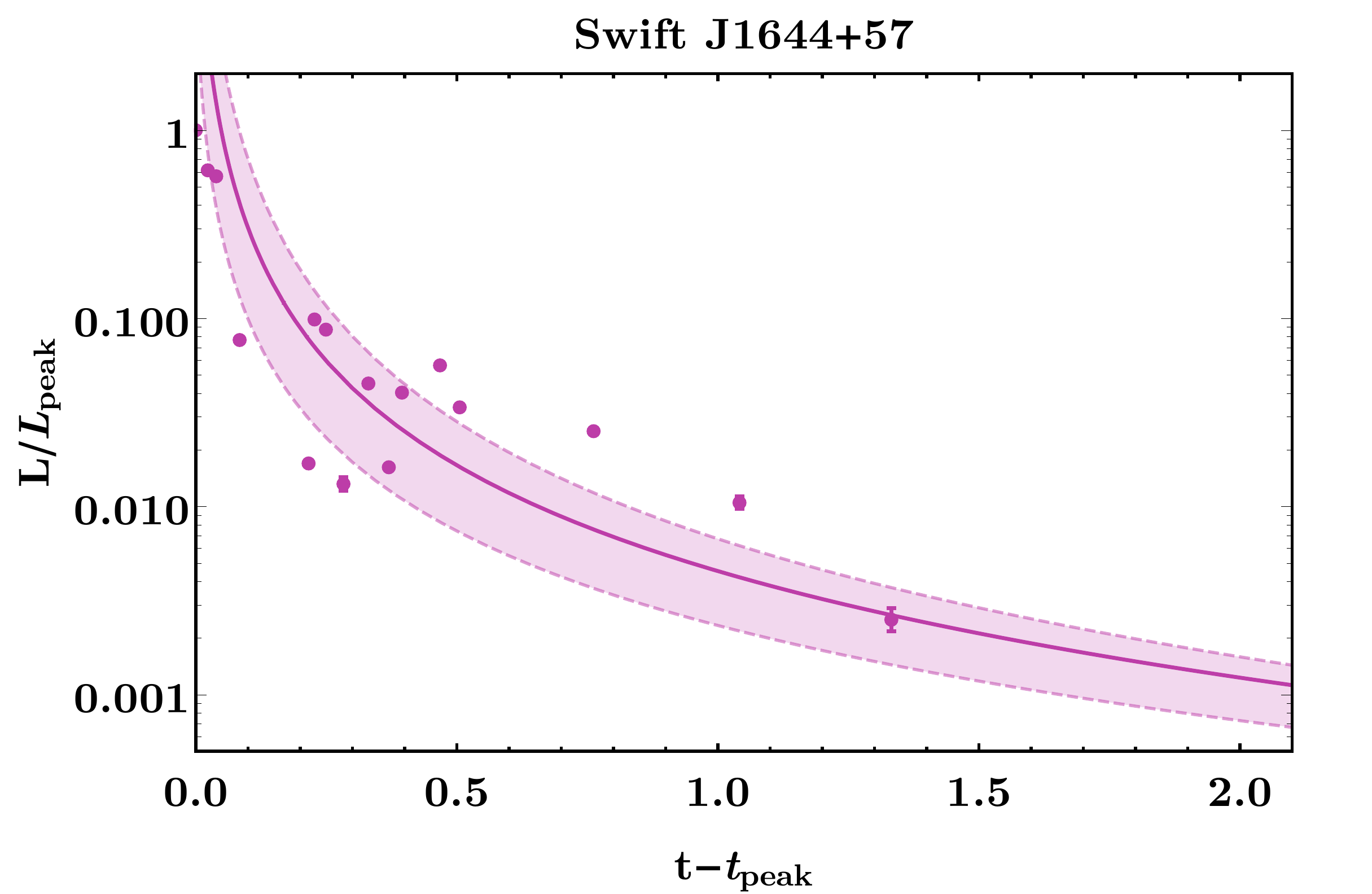}
		\includegraphics[width=0.3\textwidth]{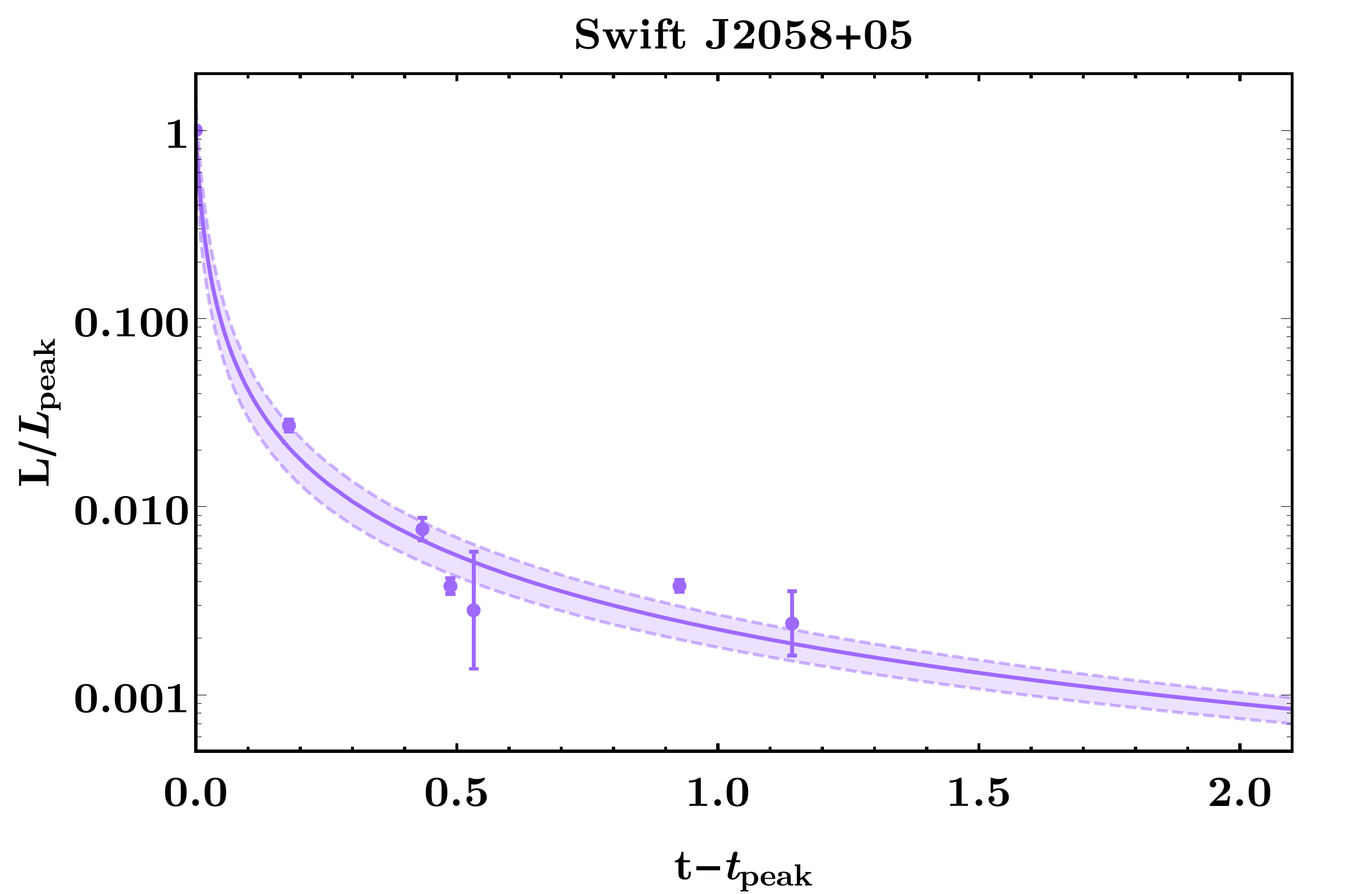}
				\includegraphics[width=0.3\textwidth]{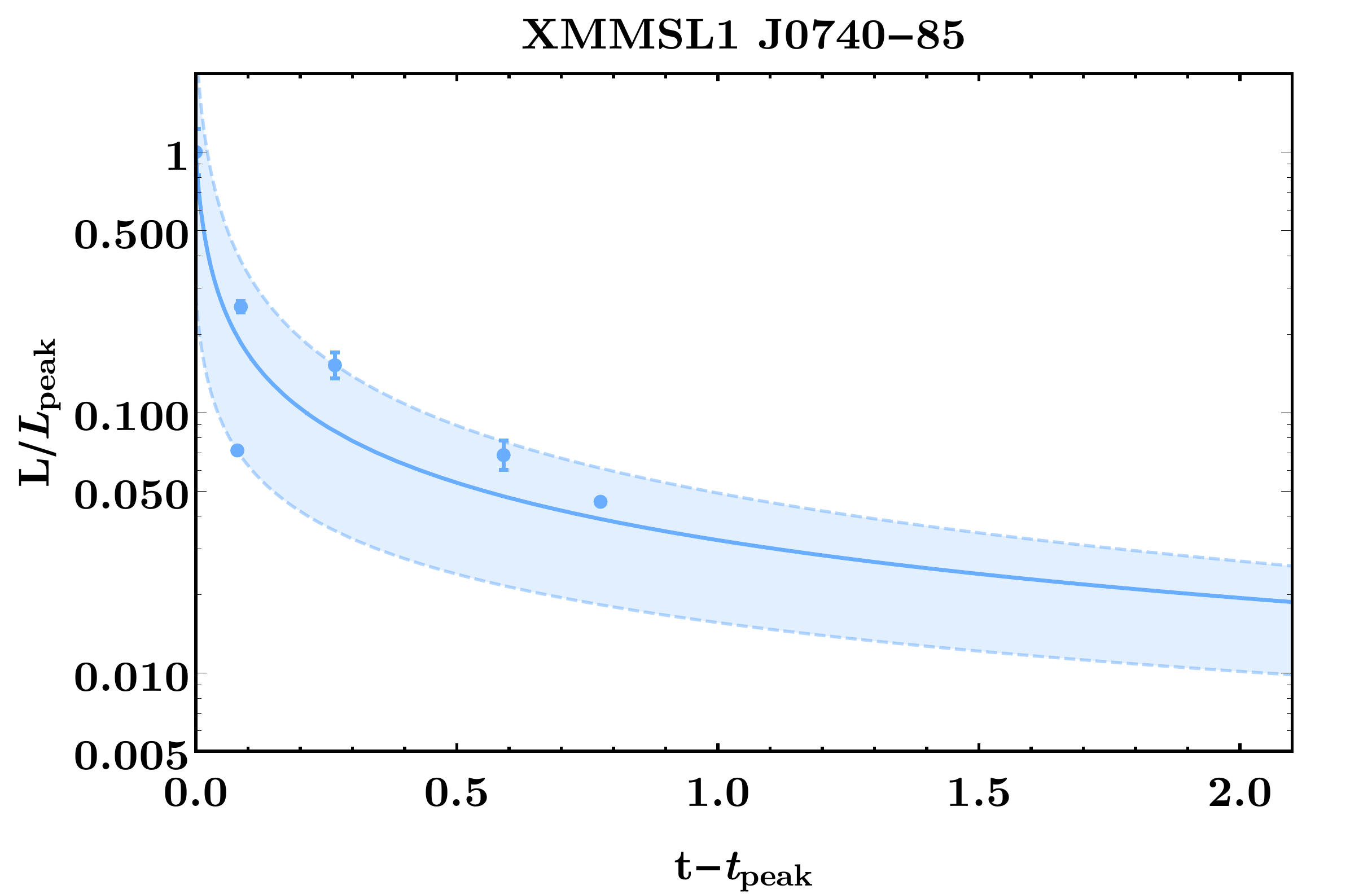}
		\includegraphics[width=0.3\textwidth]{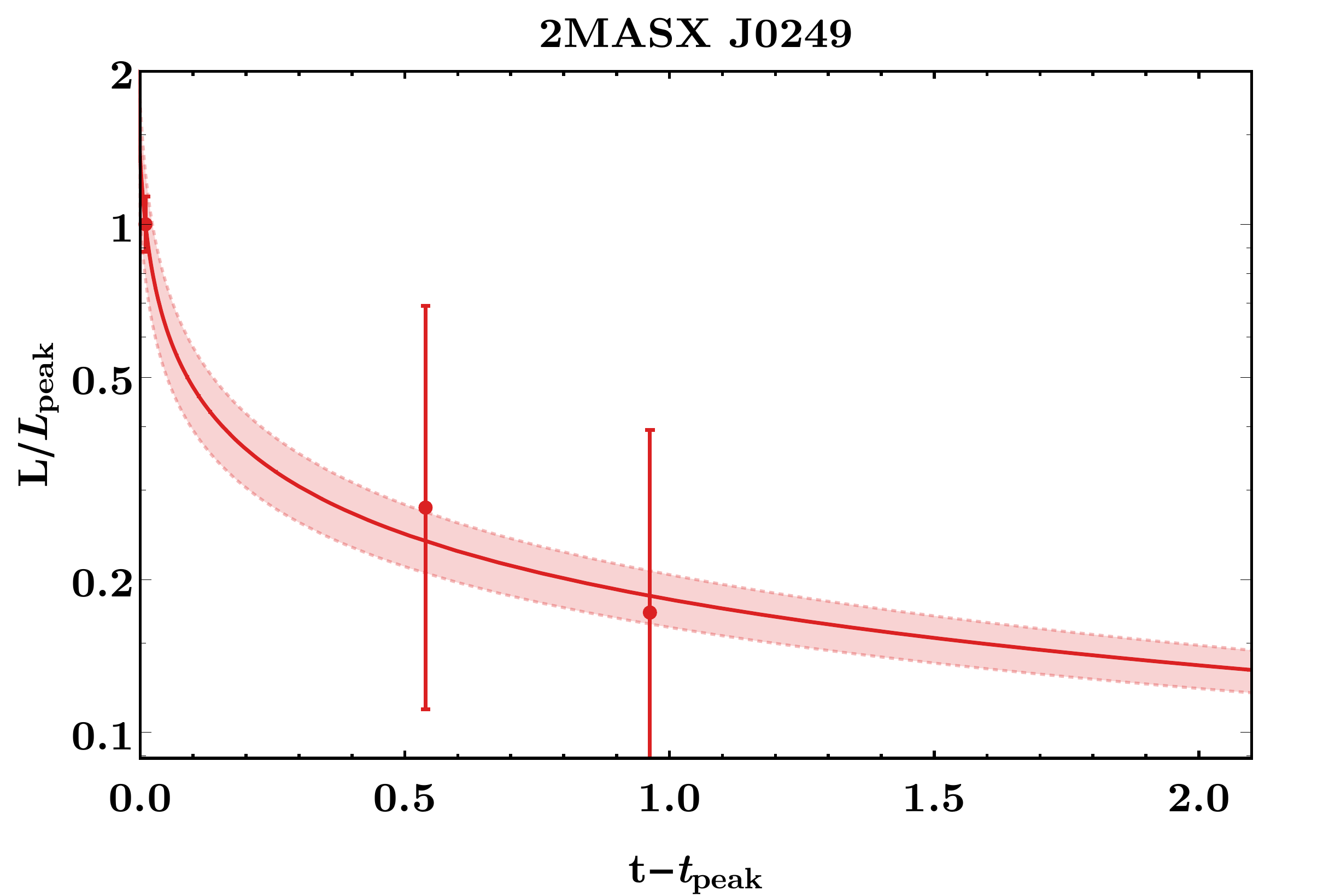}
		\includegraphics[width=0.3\textwidth]{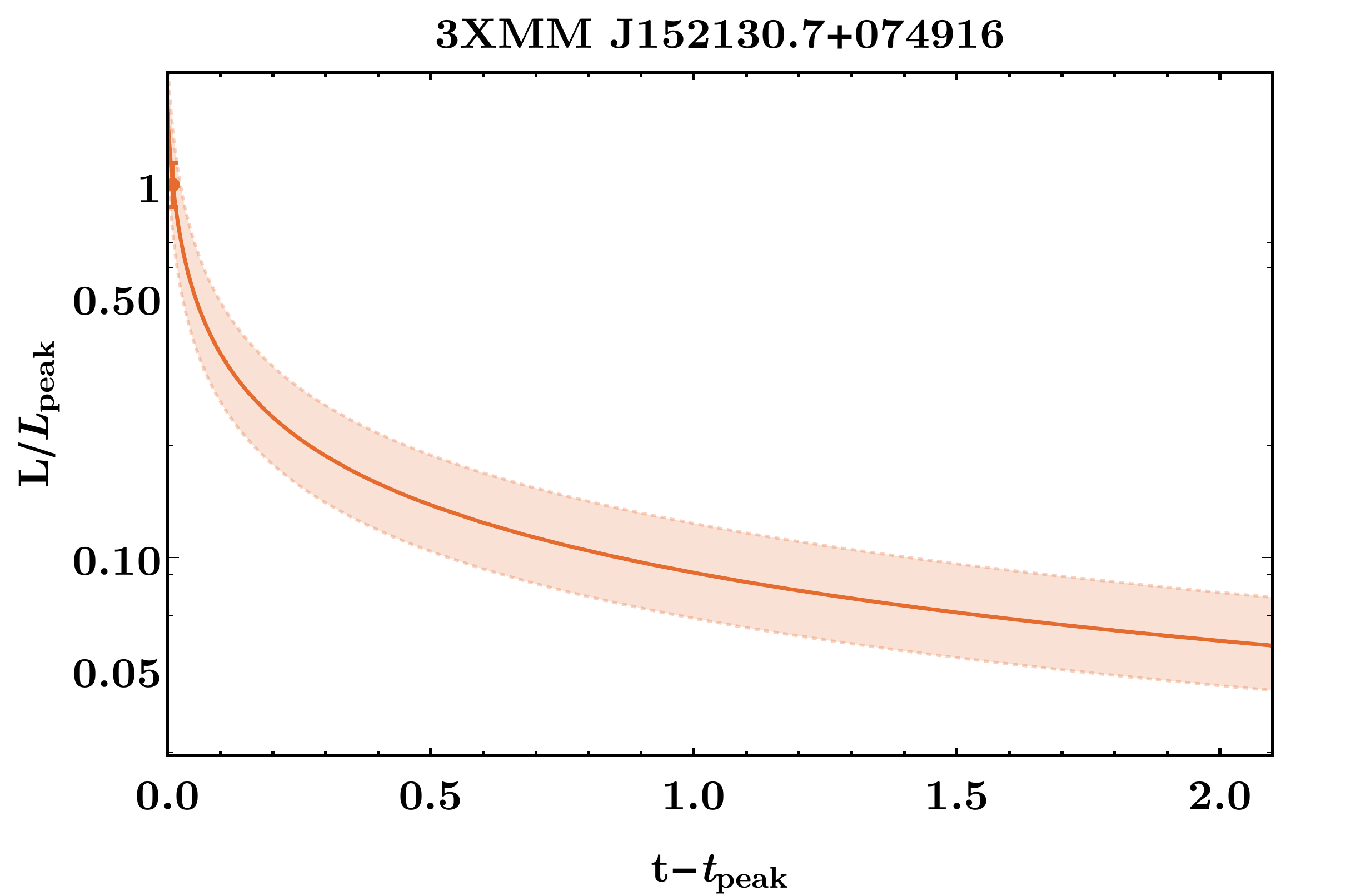}
		\includegraphics[width=0.3\textwidth]{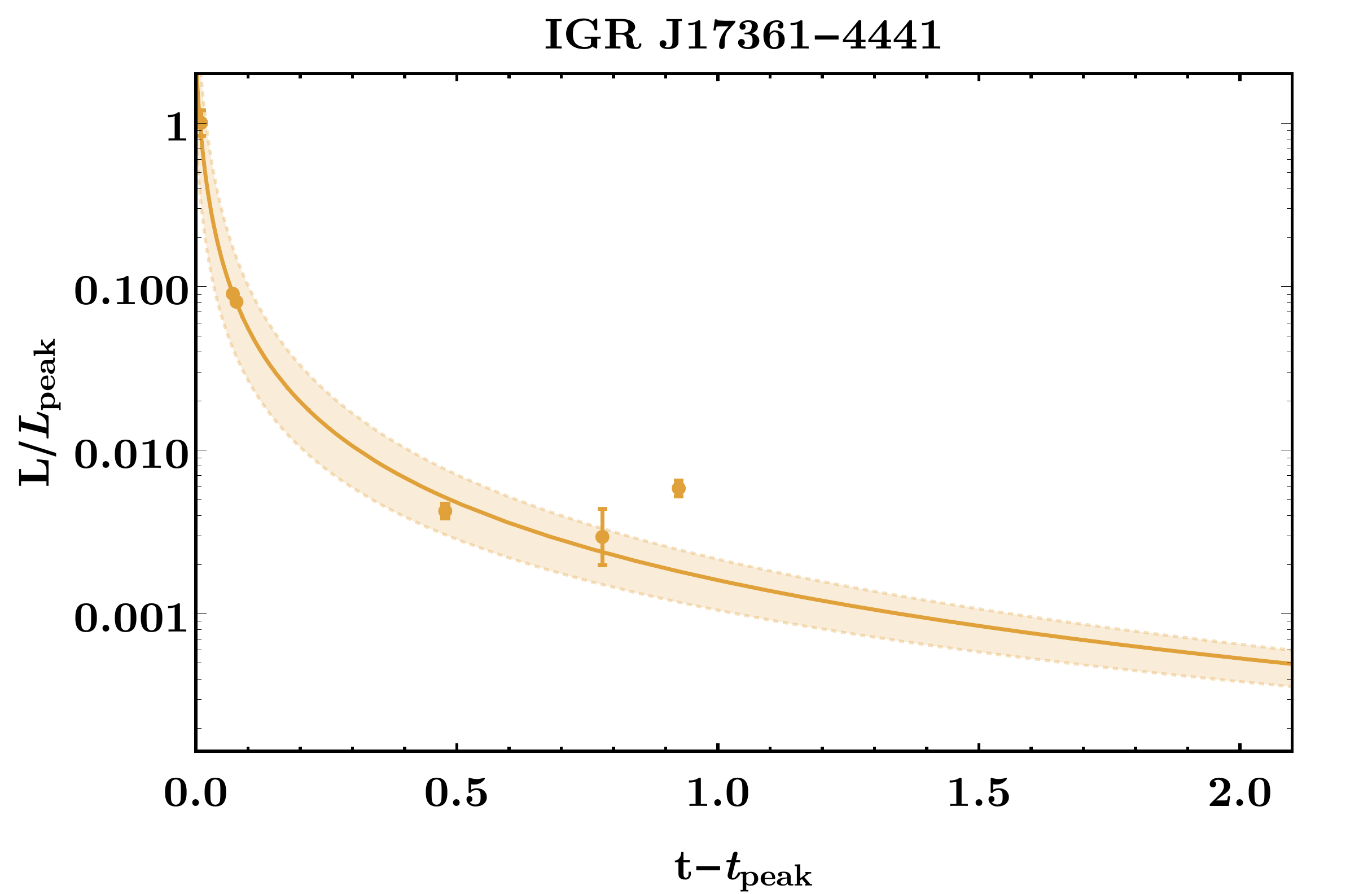}
		\includegraphics[width=0.3\textwidth]{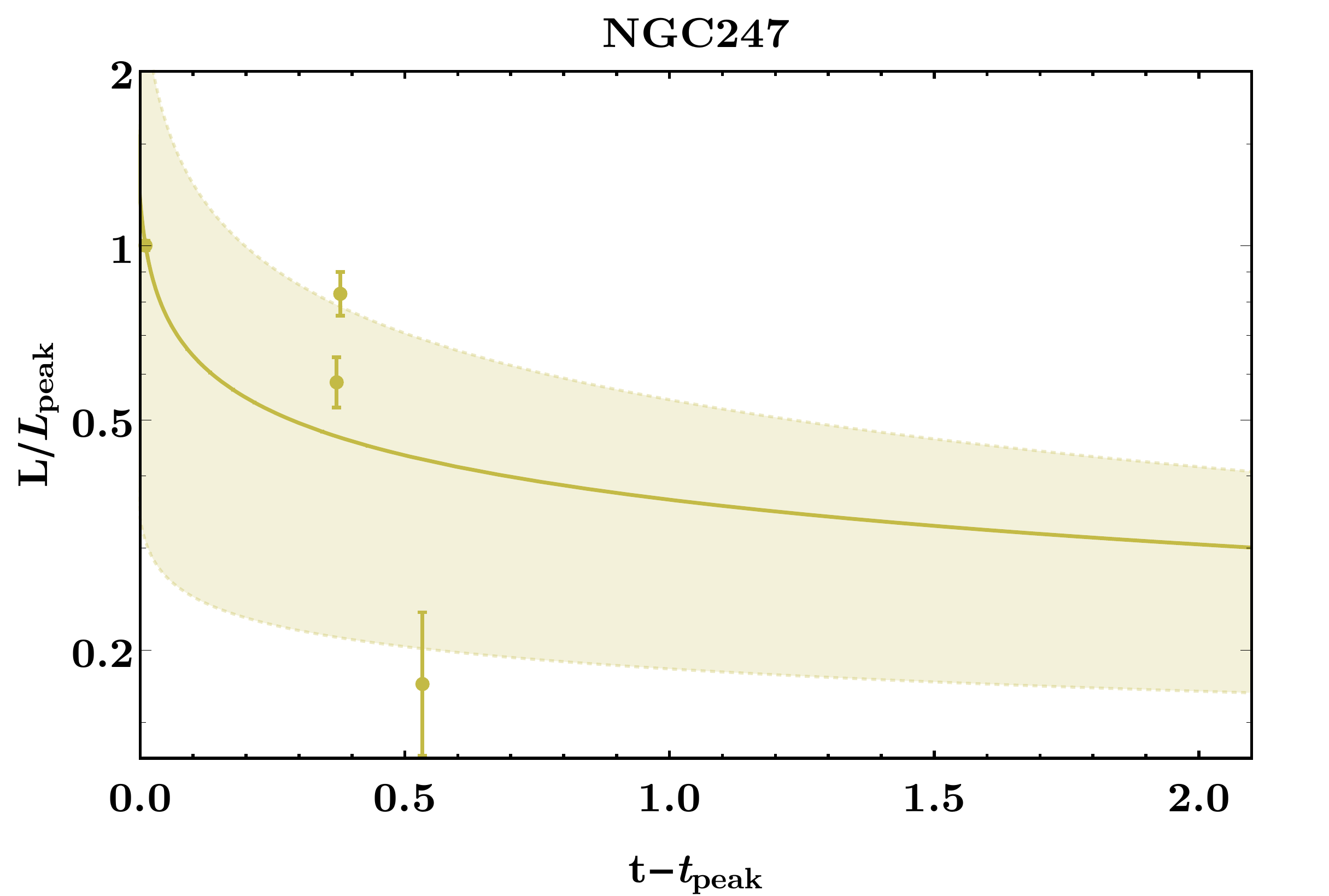}
		\includegraphics[width=0.3\textwidth]{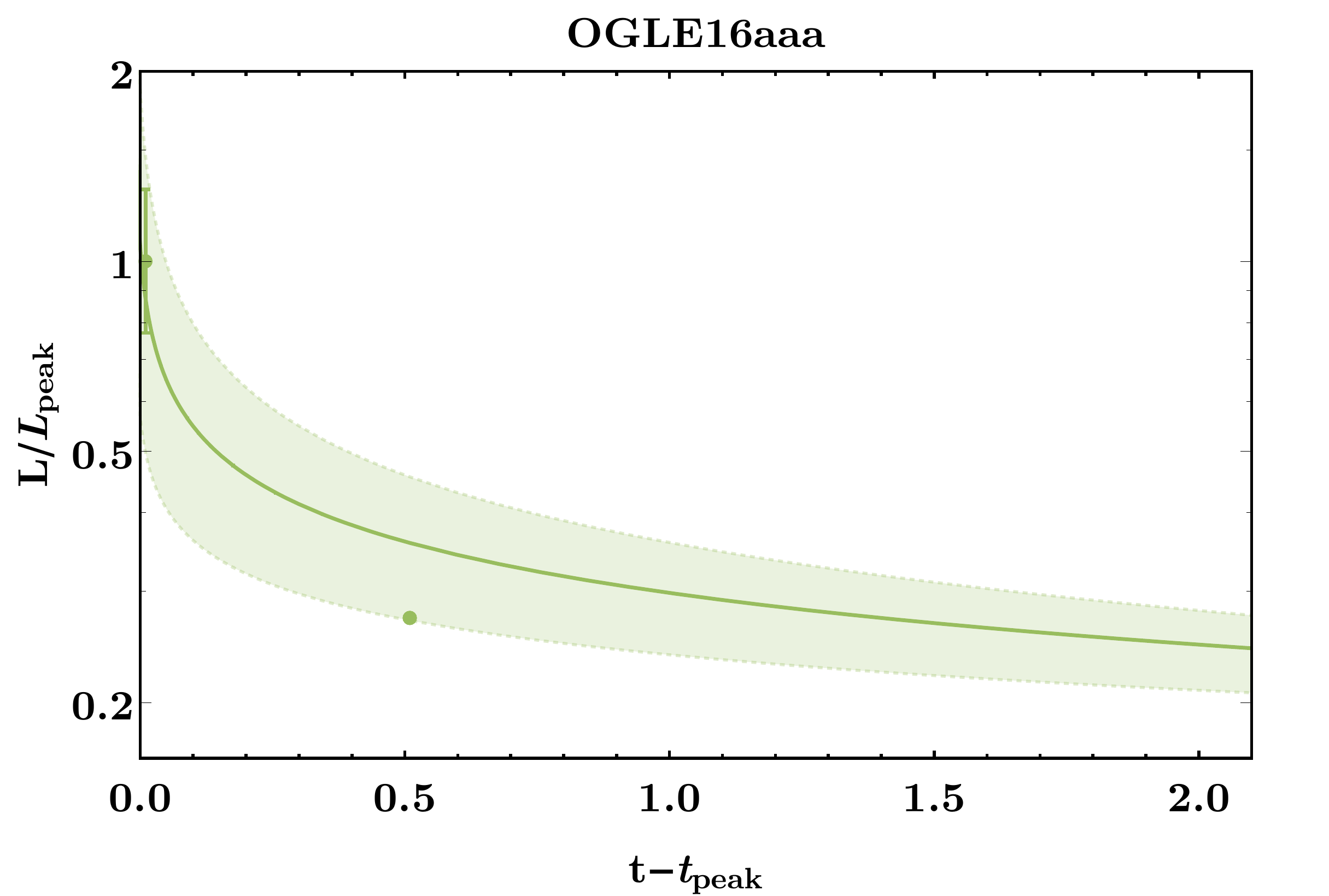}
		\includegraphics[width=0.3\textwidth]{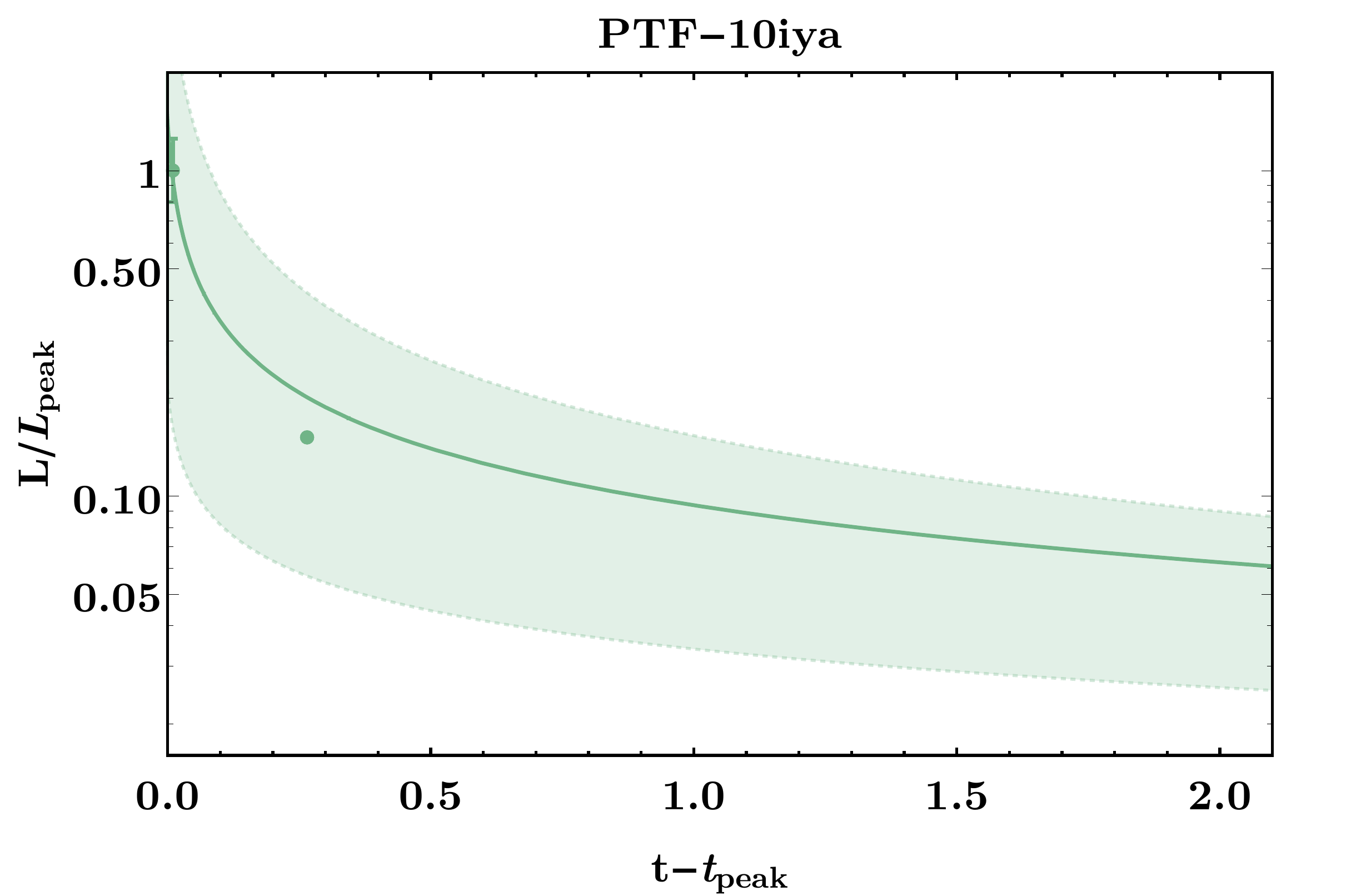}
		\includegraphics[width=0.3\textwidth]{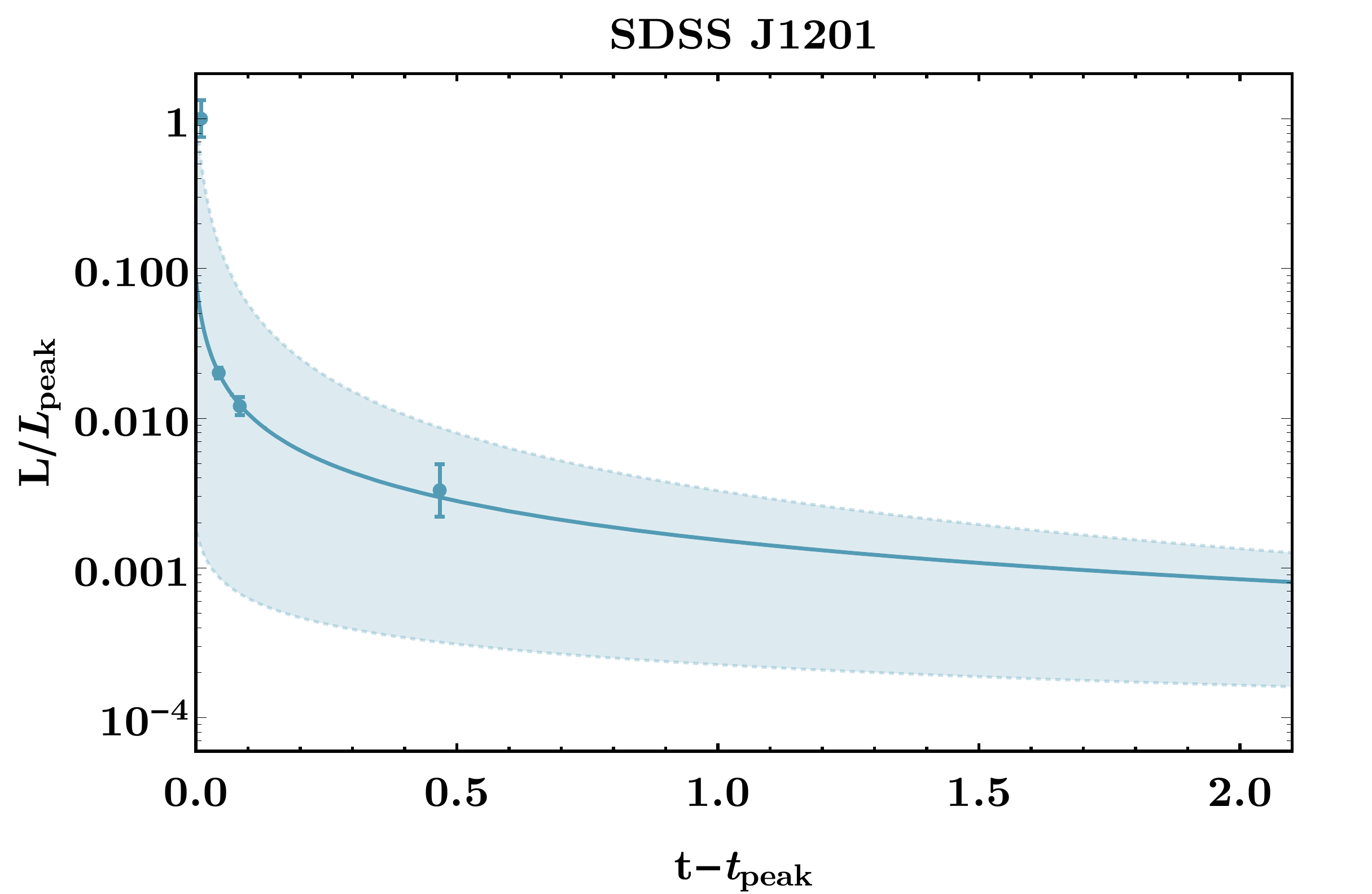}
		\includegraphics[width=0.3\textwidth]{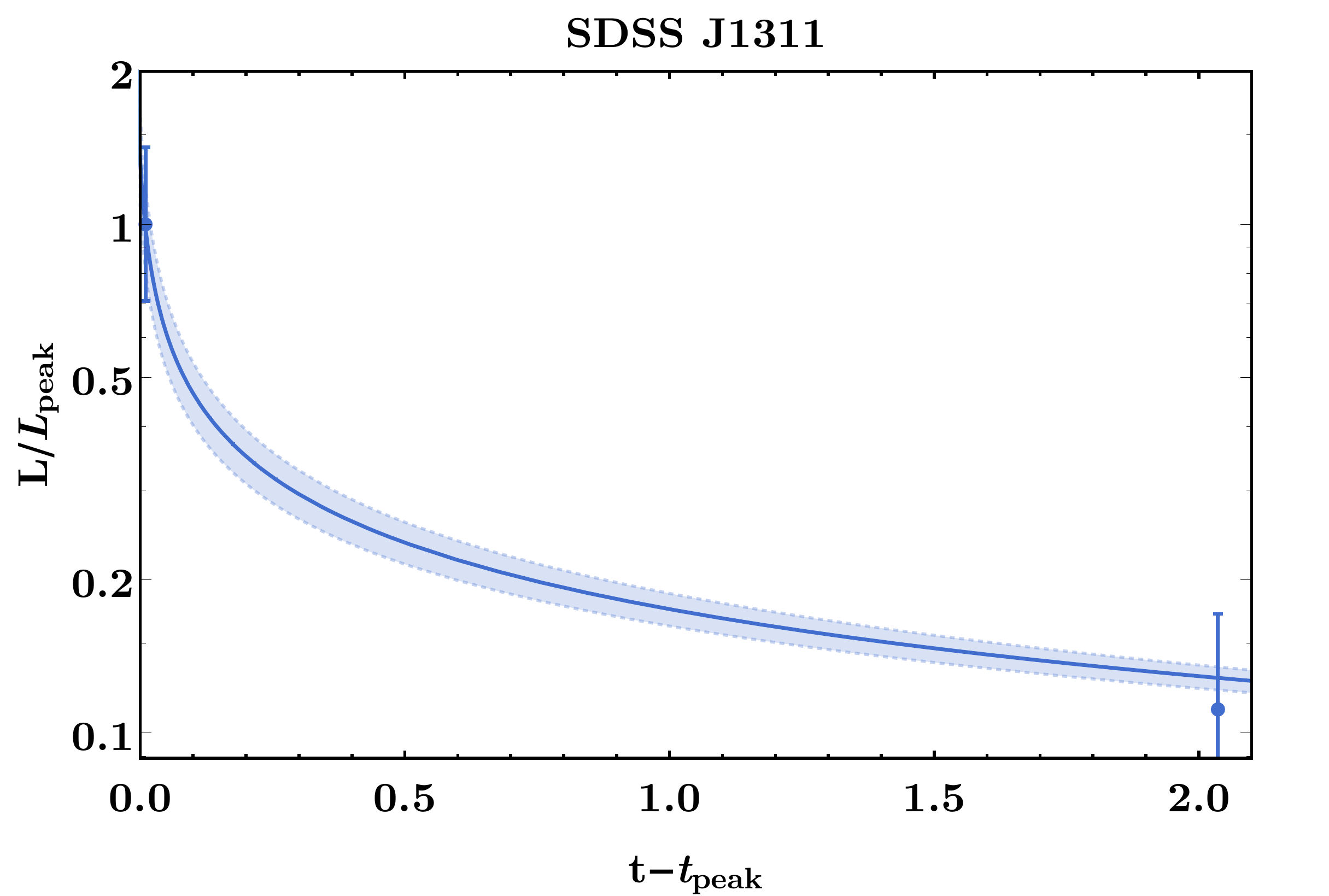}
			\includegraphics[width=0.3\textwidth]{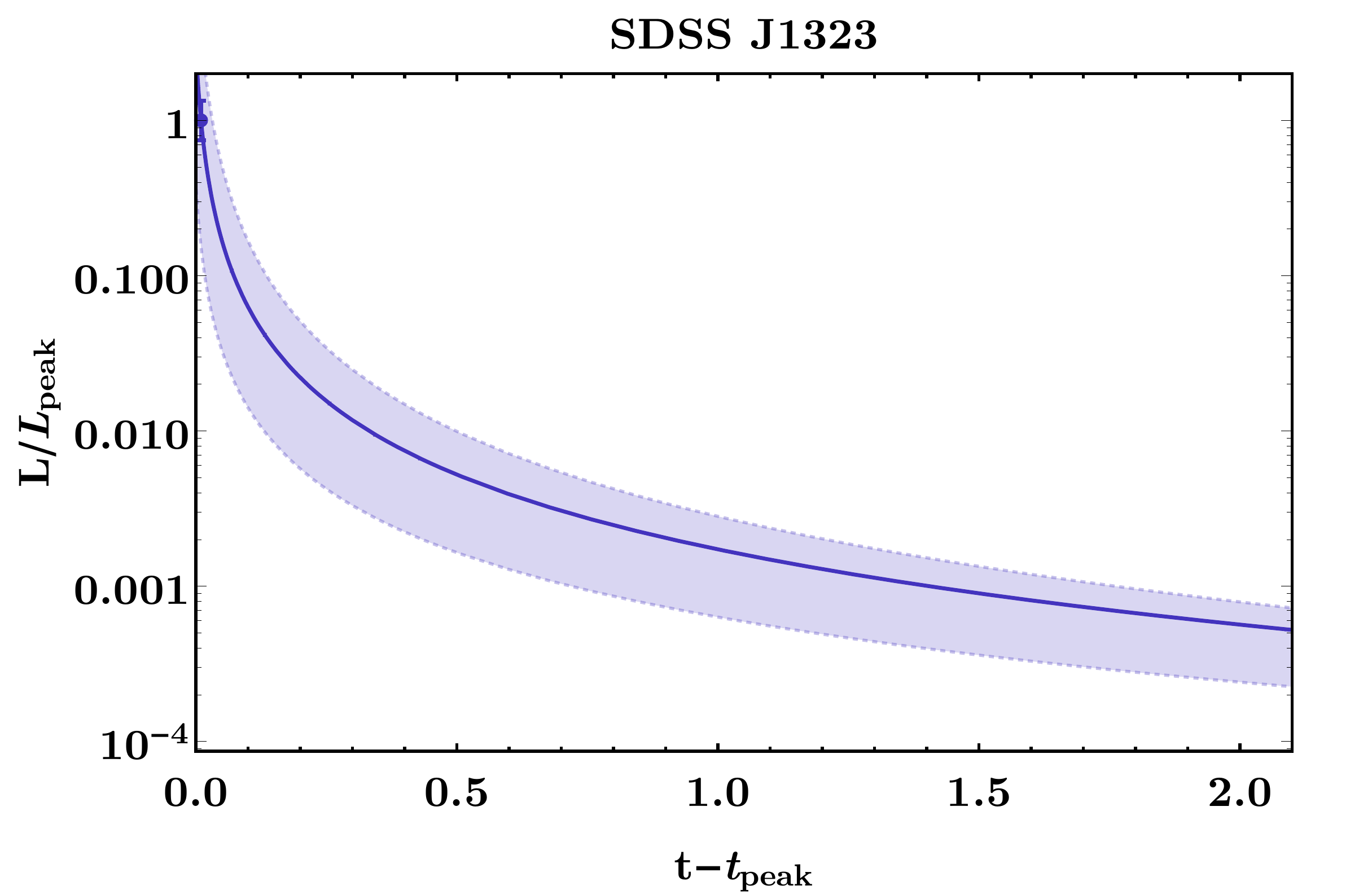}
		
		\caption{The best fit models (with their uncertainties) of the X-ray light curves of each of our \textit{X-ray TDE} and \textit{likely TDE} sample, that we obtained using a simple powerlaw $a\times(t-t_{peak})^{n}$, where $a$ is the normalisation, $t_{max}$ is the time when the maximum X-ray luminosity was measured and $n$ is the powerlaw index. For simplicity, we normalised both the X-ray luminosity and time of observation of each data set by the maximum (peak) X-ray luminosity and the time that this maximum luminosity was detected. \label{gspwlfigures}}
	\end{center}
\end{figure*}

In Section \ref{pwldecay} we model the full X-ray light-curves of our TDE sample listed in Table \ref{goldsilvers}. To do this we assume a simple powerlaw model defined by $a(t-t_{peak})^{n}$, where $a$ is the normalisation, and $n$ is the powerlaw index. Here we let the normalisation and powerlaw index be free parameters. In Table \ref{gspwlfits} we have listed our best fit powerlaw models and their one sigma uncertainties. Using these values, we generated the Gaussians seen in Figure \ref{pwlindex}, where the best fit value defines the peak of the Gaussian and the uncertainty defines the width. In Figure \ref{gspwlfigures} we have plotted the best fit power models and their uncertainties for each of the TDE candidates we consider.

\input{powerlawmodelfits.tex}

\section{Individual spectral energy distributions of each X-ray TDE.}\label{individalsed}

\begin{figure*}[th]
	\begin{center}
		\includegraphics[width=0.30\textwidth]{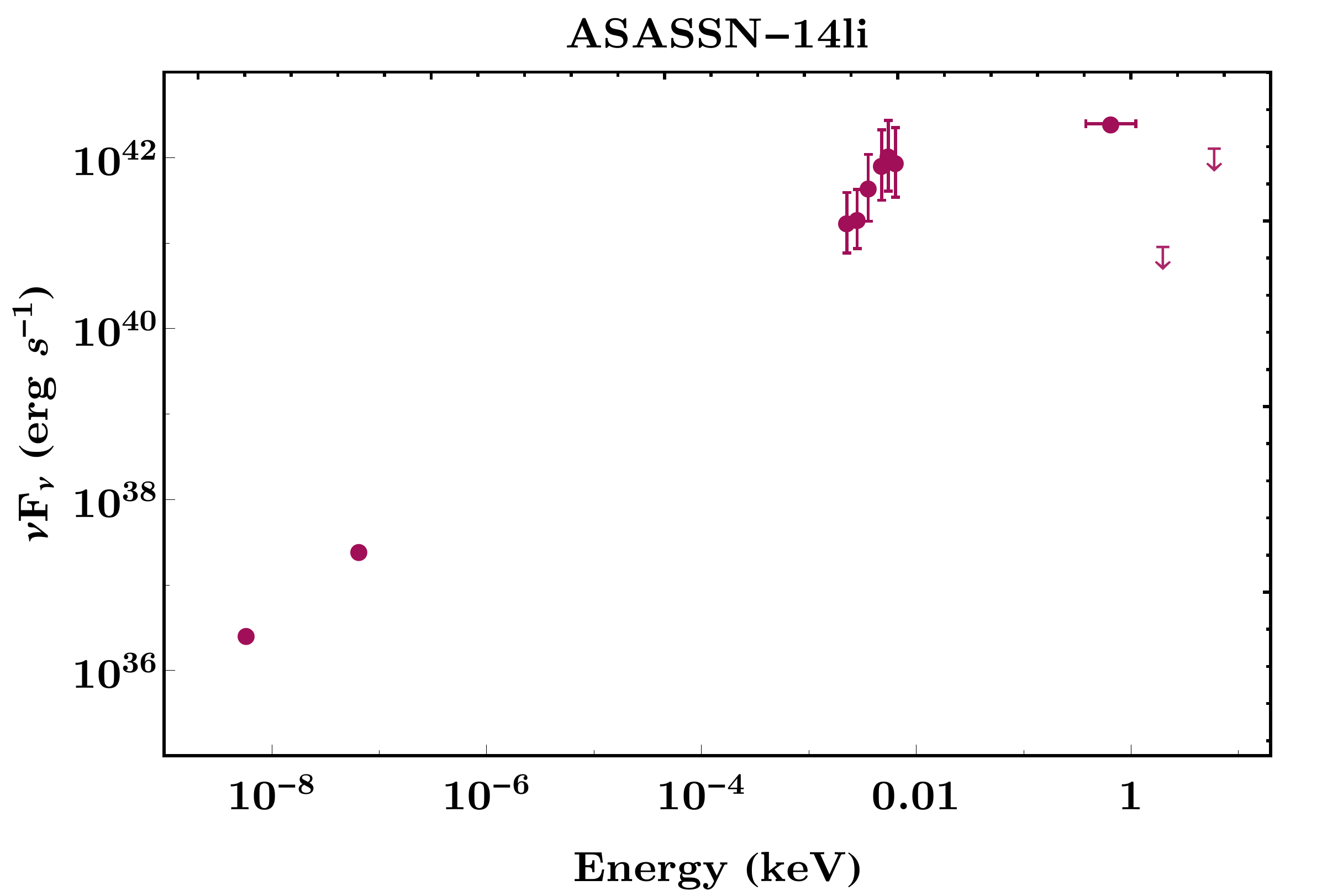}
		\includegraphics[width=0.30\textwidth]{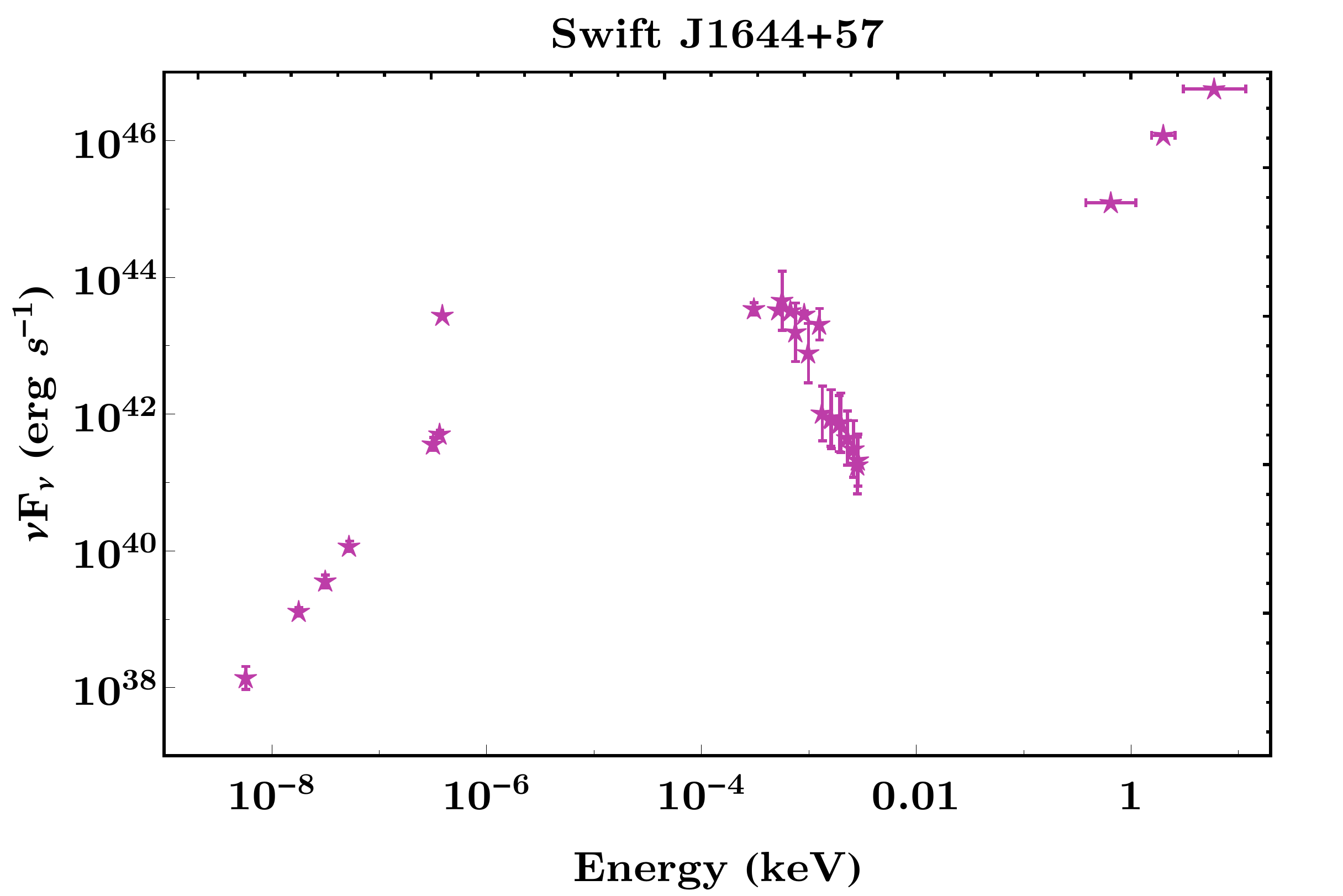}
		\includegraphics[width=0.30\textwidth]{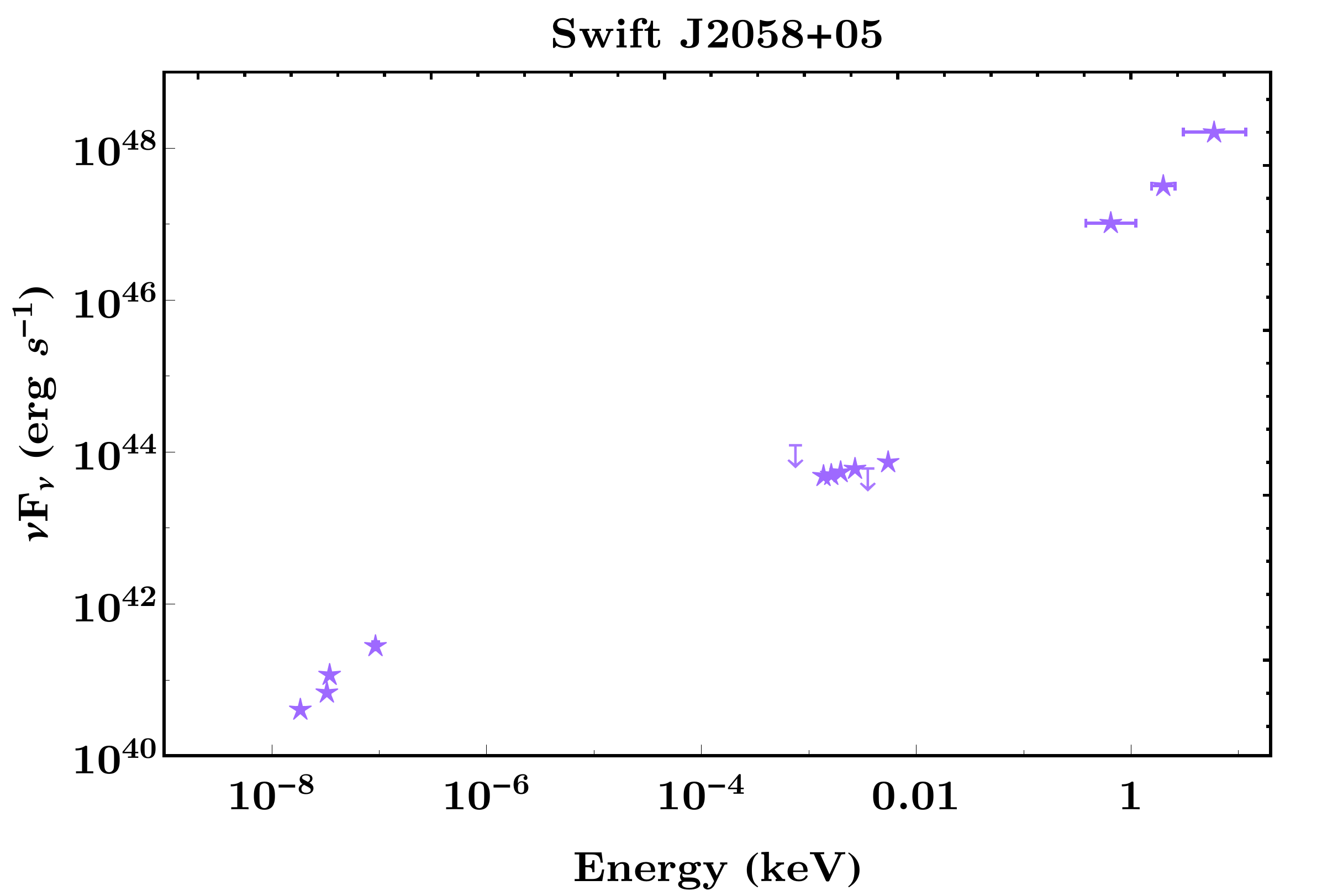}
			\includegraphics[width=0.30\textwidth]{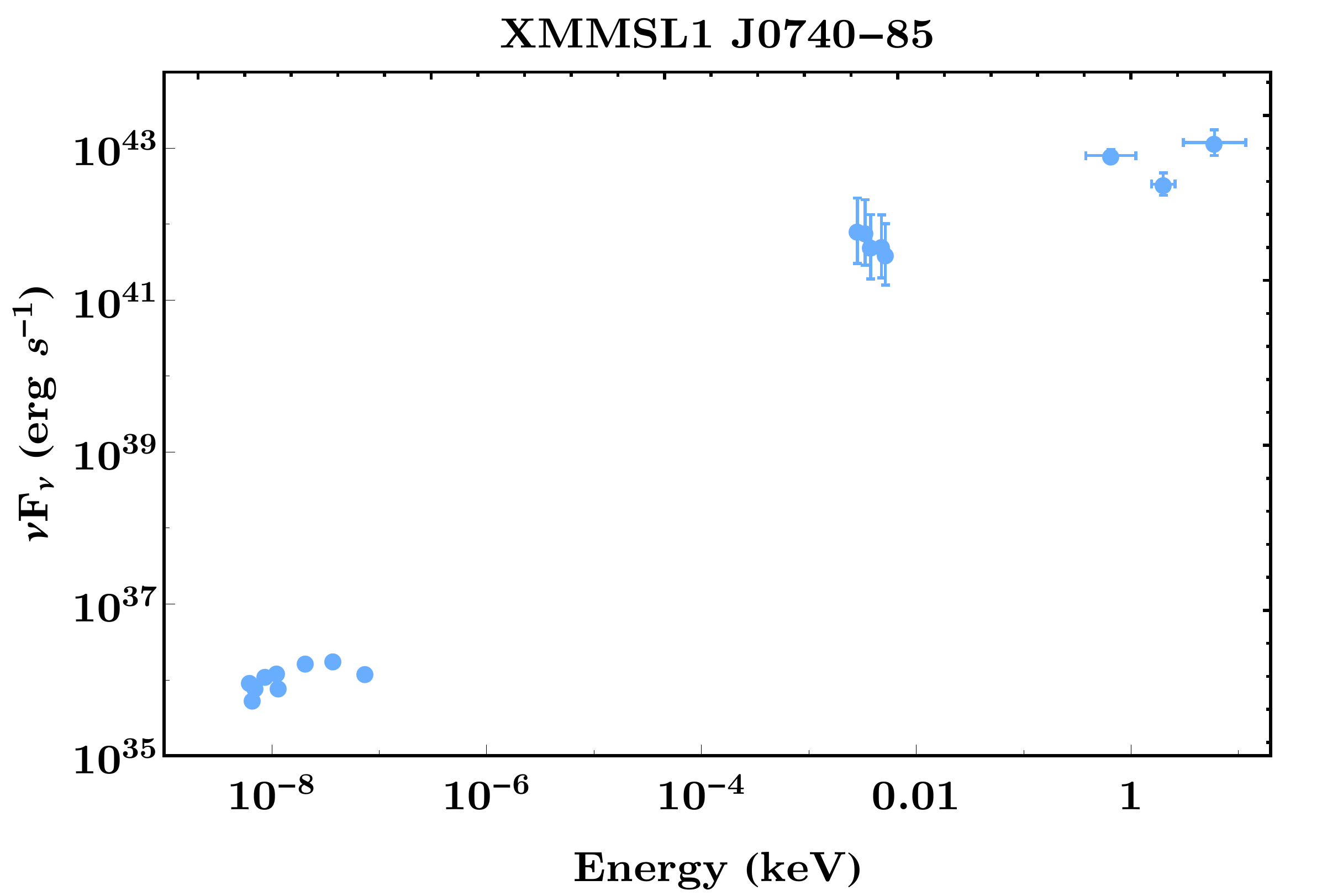}
		\includegraphics[width=0.30\textwidth]{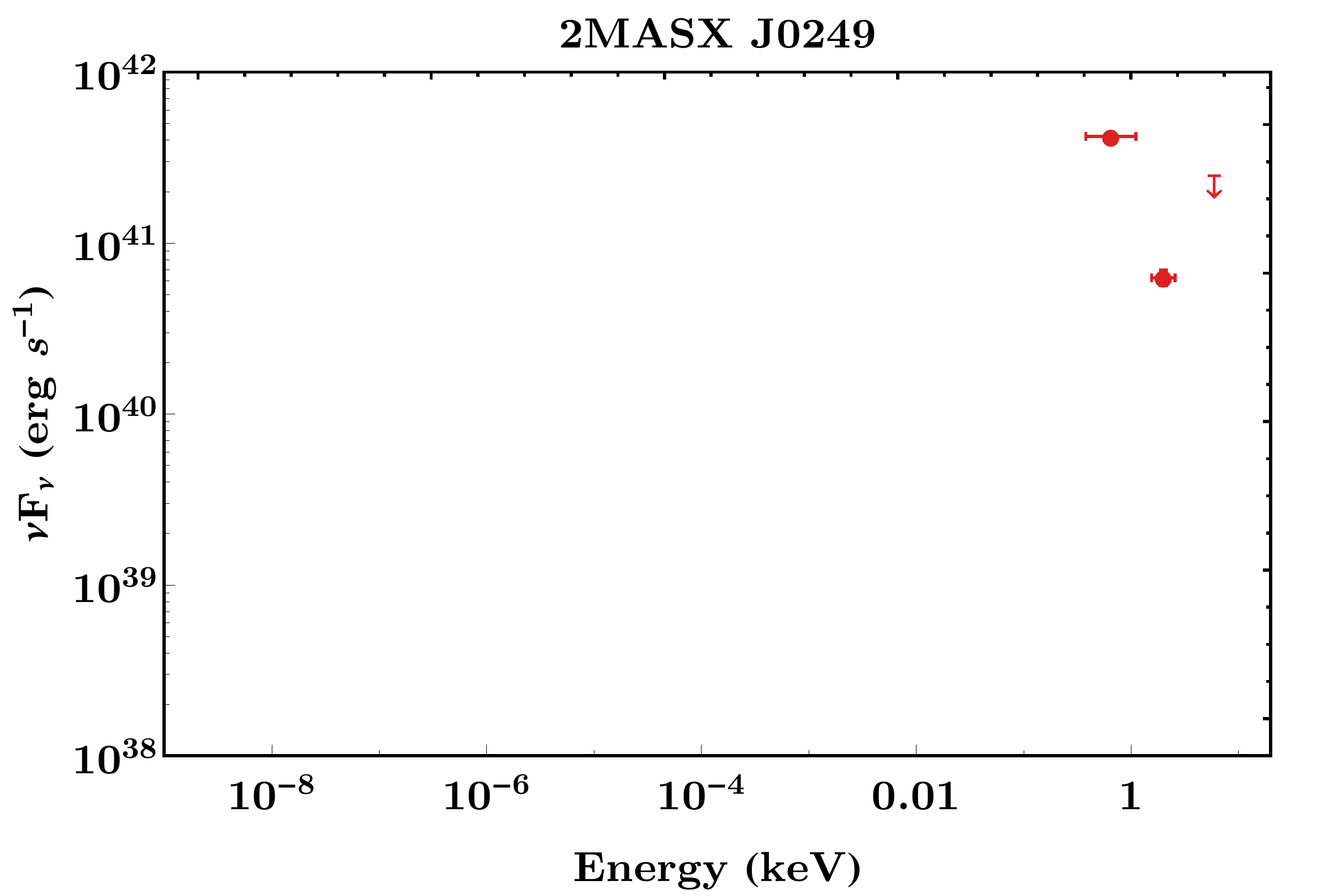}
		\includegraphics[width=0.30\textwidth]{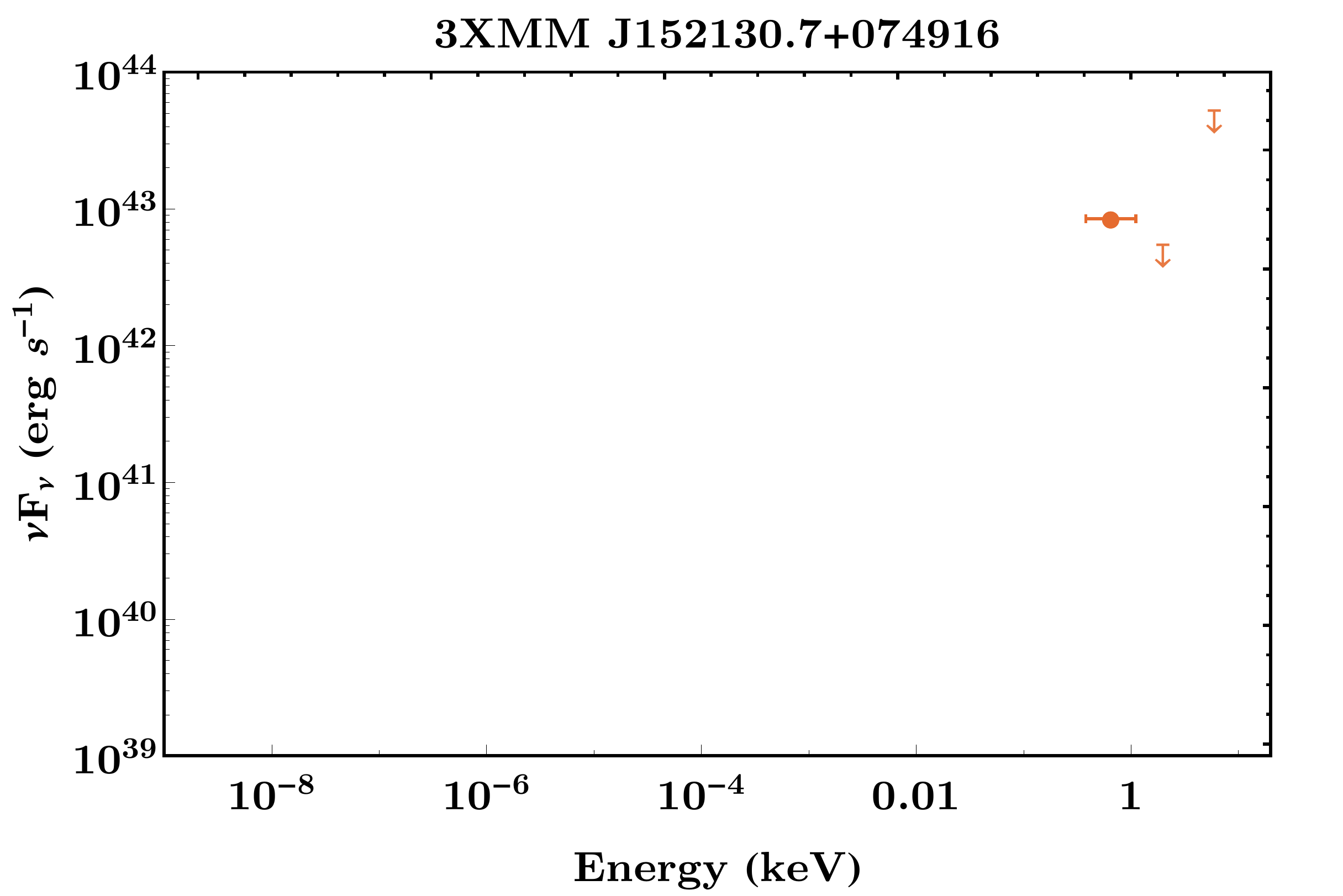}
		\includegraphics[width=0.30\textwidth]{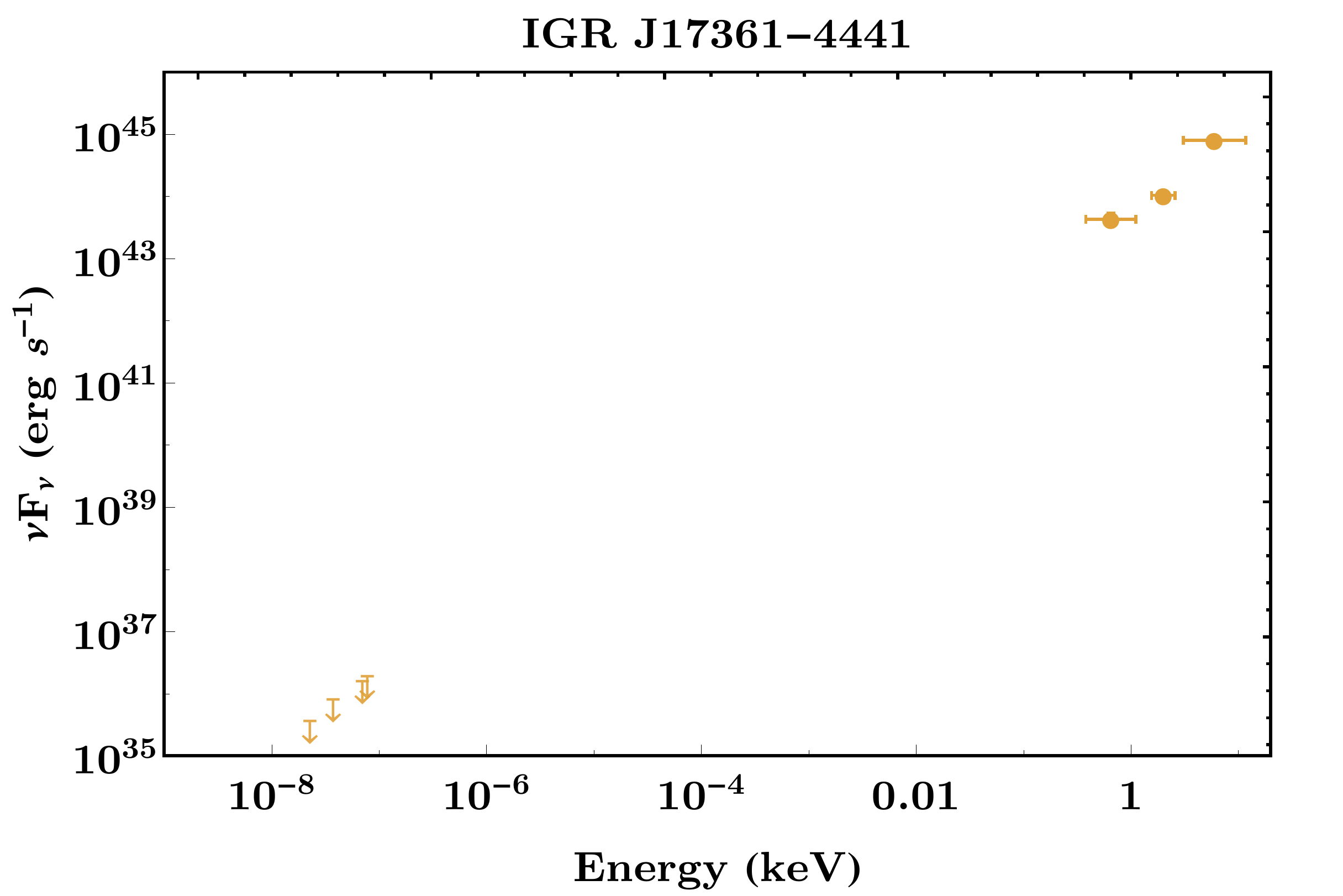}
		\includegraphics[width=0.30\textwidth]{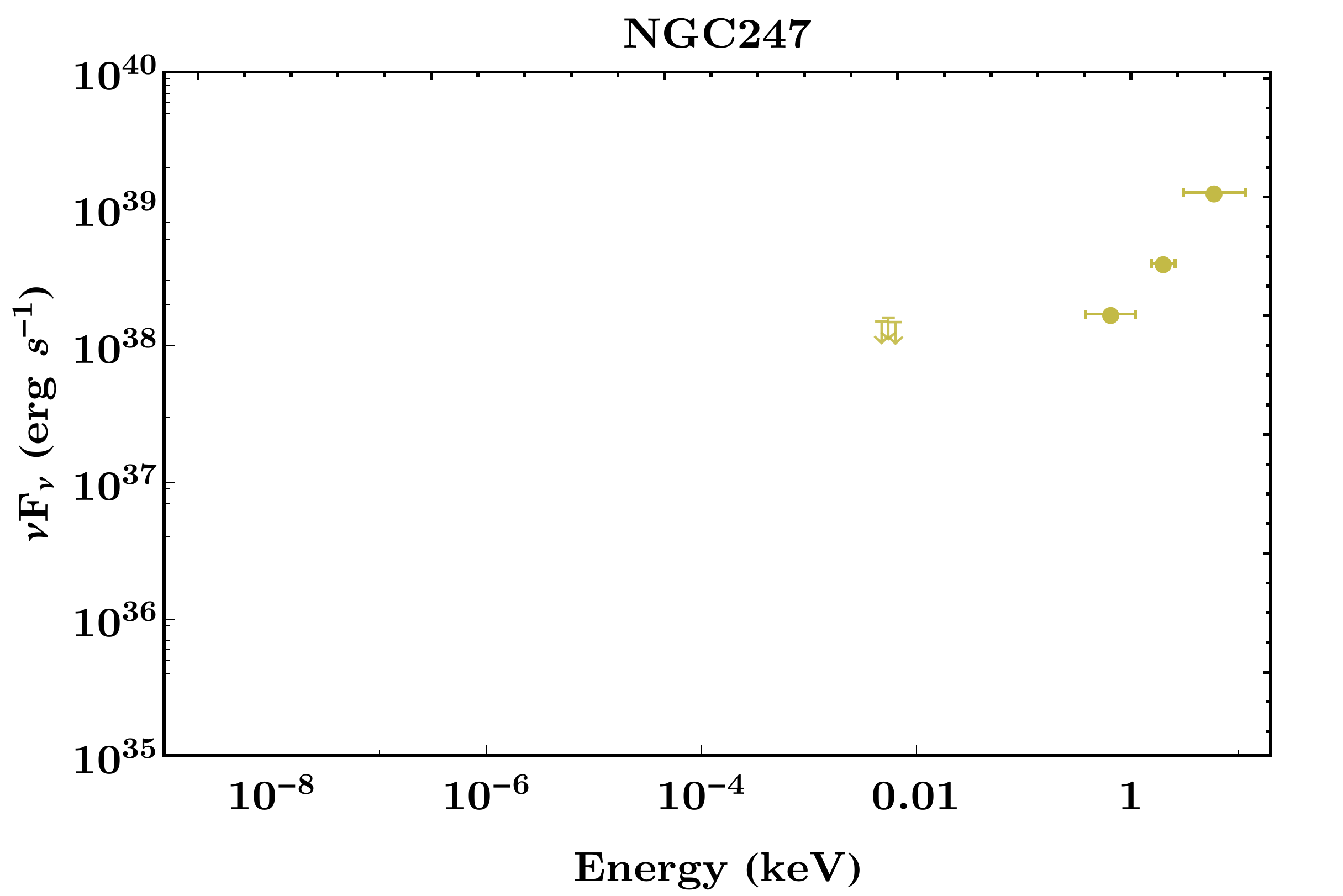}
		\includegraphics[width=0.30\textwidth]{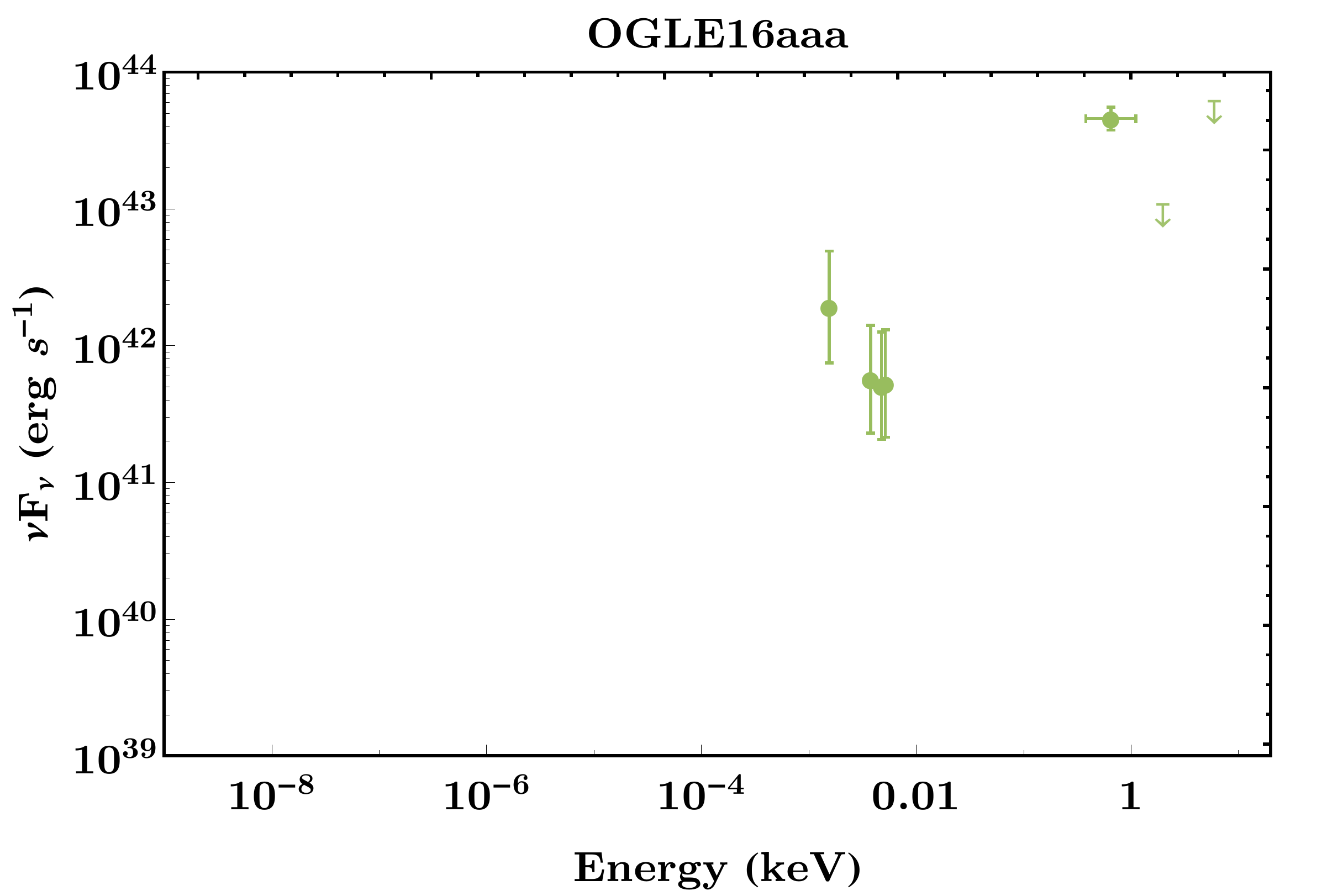}
		\includegraphics[width=0.30\textwidth]{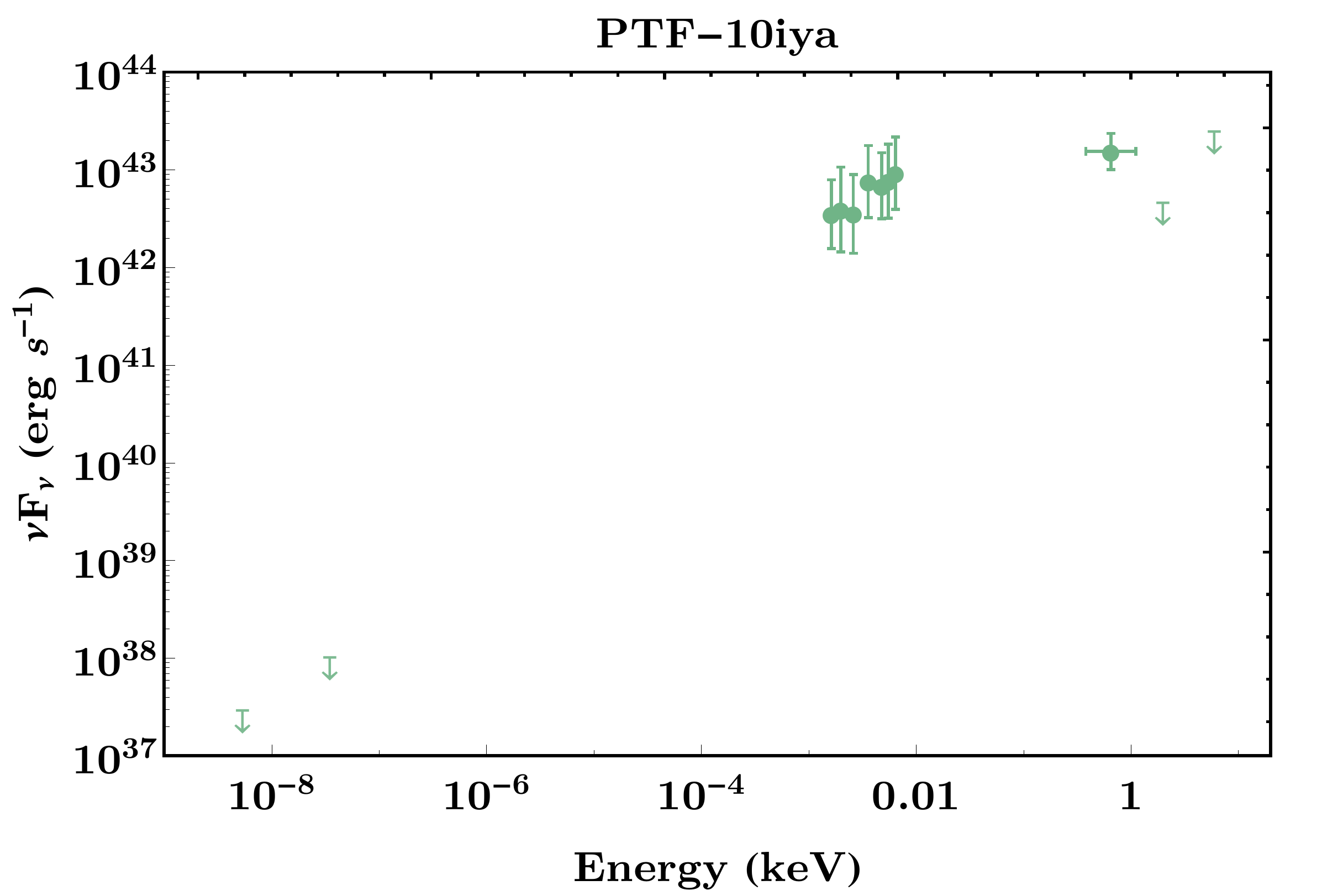}
		\includegraphics[width=0.30\textwidth]{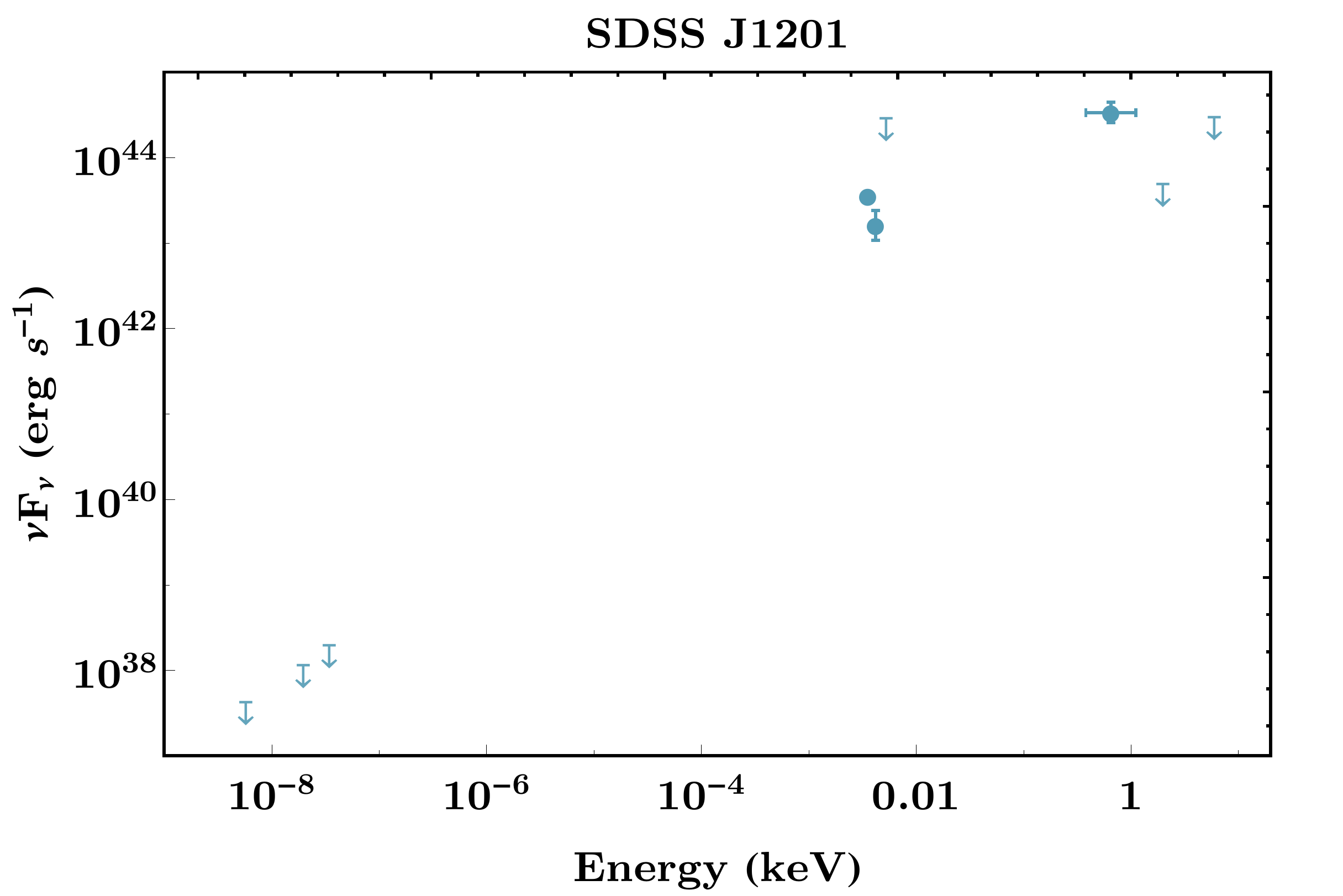}
		\includegraphics[width=0.30\textwidth]{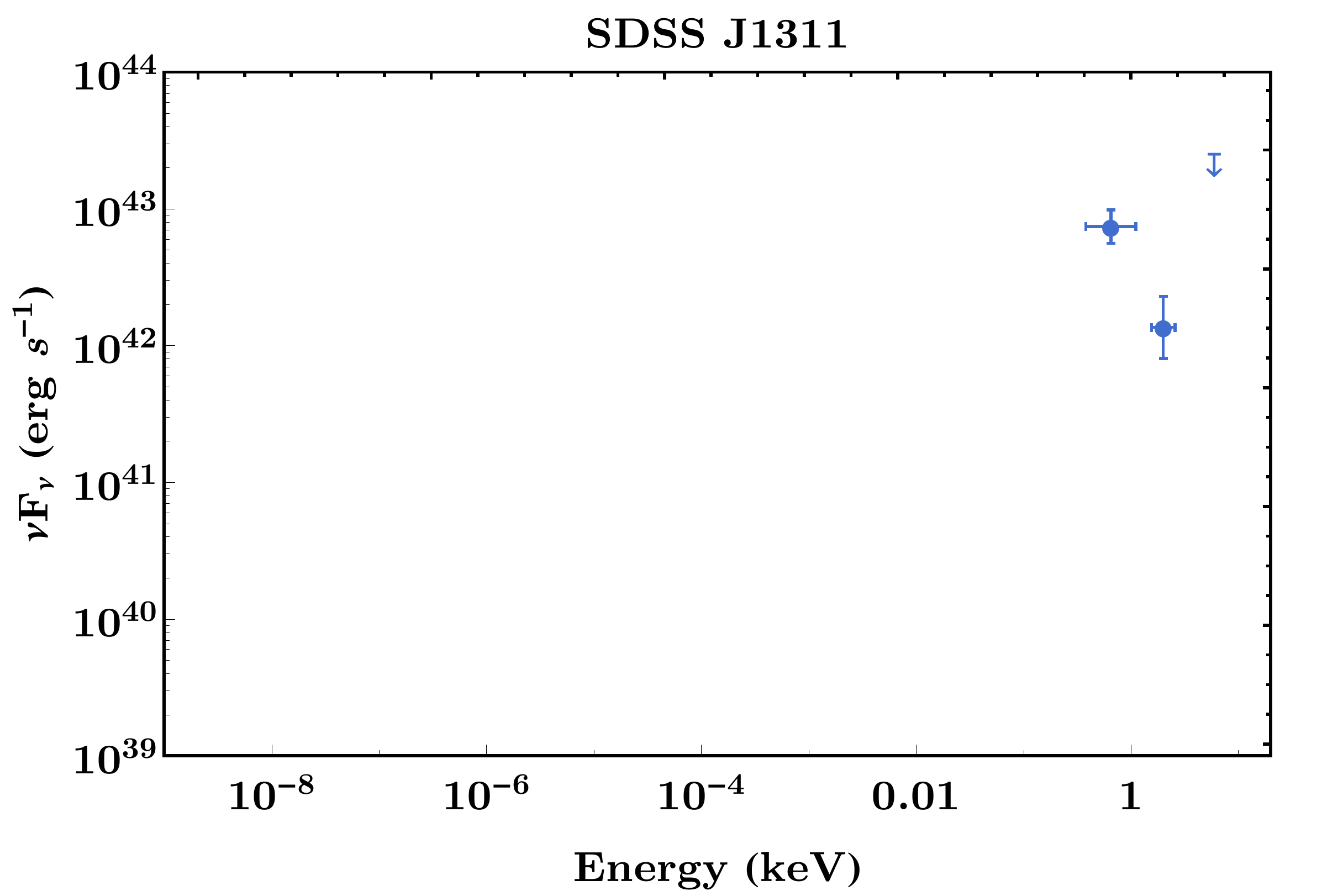}
		\includegraphics[width=0.30\textwidth]{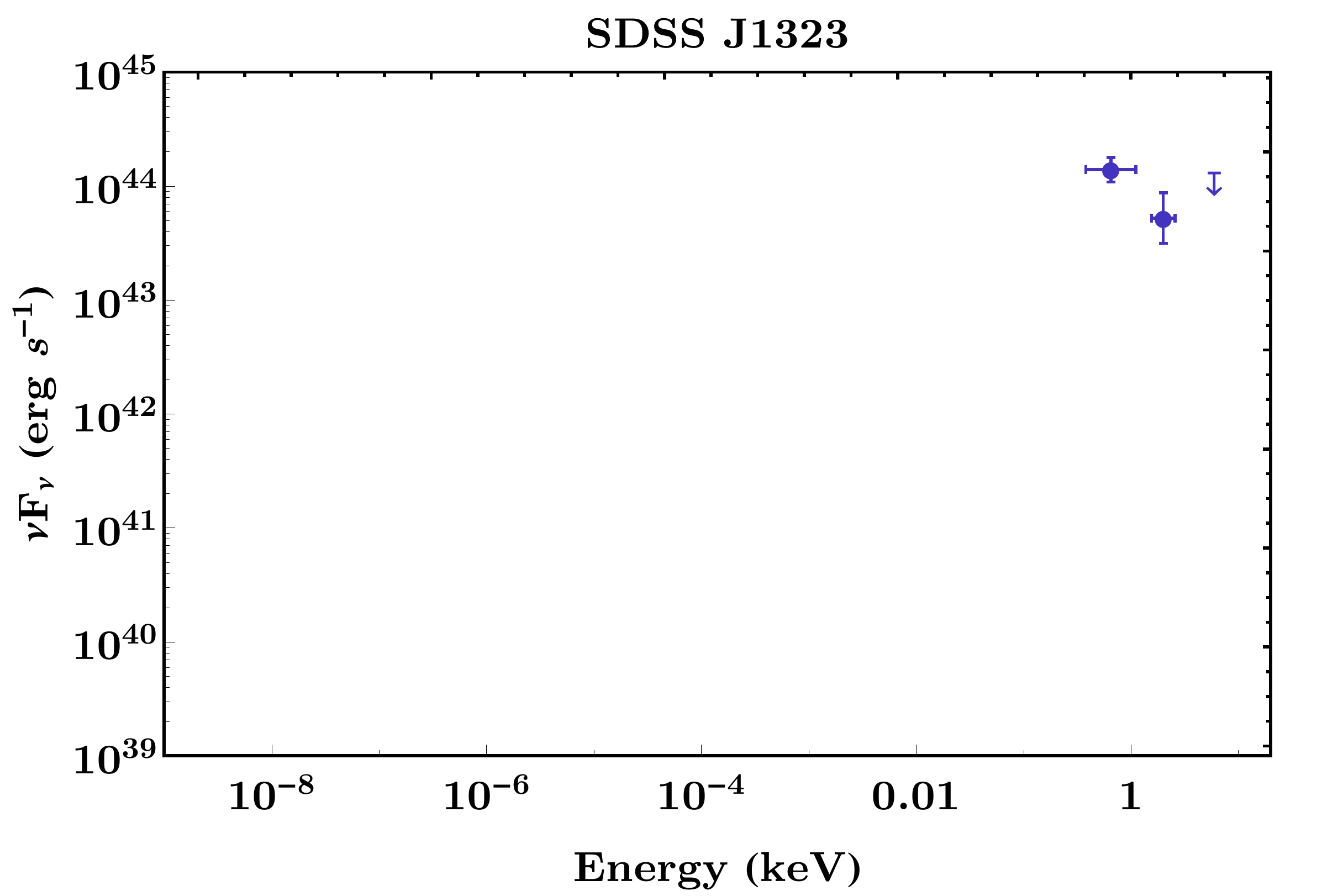}
		\caption{The individual spectral energy distribution of each of our \textit{X-ray TDE} and \textit{likely TDE} candidates in $\nu F_{\nu}$ (erg s$^{-1}$) vs. Energy (keV). The $\nu F_{\nu}$ values in the X-ray energies were derived from the number of counts detected at peak in the soft, medium and hard energy bands, and converted these into their corresponding $\nu F_{\nu}$ to characterise the X-ray emission of these sources. For sources in which optical, UV and/or radio data was taken simultaneously or close to ($t_{optical}-t_{peak} < +/-0.5$ year) when the X-ray luminosity was maximum, we have also plotted the corresponding $\nu F_{\nu}$ in this energy band derived from the literature values.
			 \label{individualnuFnu}}
	\end{center}
\end{figure*}

To derive the $\nu F_{\nu}$ spectral energy distributions for each of the TDE candidates we consider we took the soft, medium and hard X-ray counts rates for each as detected at peak. We then covered these counts in to fluxes $F_{\nu}$, taking into account the effective area of each instrument. To get $\nu F_{\nu}$, we then multiplied these fluxes by the energy band of interest. In the optical/UV and radio energy bands, we took the fluxes or magnitudes from the literature and converted these into $\nu F_{\nu}$. We selected only measurements in these bands that were taken around approximately the same time as the original TDE flare was detected. In Figure \ref{individualnuFnu} we have plotted the individual SEDs for each of our TDE sample, which are also overlaid with each other in Figure \ref{nuFnu}.

For ASASSN-14li, we took radio data from \citet{2016Sci...351...62V}, while the optical/UV data of this event was taken from \citet{2016MNRAS.455.2918H}. The radio and optical/UV data for Swift J1644+57 was taken from \citet{2011Sci...333..203B}.  For  Swift J2058+05, we used the optical/UV data for this event from \citet{2012ApJ...753...77C}, while its radio data was taken from \citet{2015ApJ...805...68P}. Optical/UV data for NGC 247, SDSS J1201, and PTF-10iya was taken from \citet{2015ApJ...807..185F}, \citet{2012A&A...541A.106S}, \citet{2012MNRAS.420.2684C} and \citet{2015MNRAS.452.4297B} respectively, while radio data for IGR J17361-4441 was taken from \citet{2011ATel.3617....1F} respectively. For XMM SL1 J0740-58, we used the optical/UV data derived using \emph{Swift} in \citet{2016arXiv161001788S}, and the radio data from \citet{2016arXiv161003861A}.

\section{Table of results obtained from this analysis}

In Tables \ref{rosatcounts}--\ref{swiftfits}, we have listed all the results of our X-ray analysis from each TDE candidate listed in Table \ref{tdes}. For each instrument, we have created three tables. Tables \ref{rosatcounts}--\ref{swiftcounts} contains the details of the observations we used, source and background counts in the full instrument energy band, whether we classified these as a detection or not and the derive count rate we used for our analysis. In Tables \ref{rosathms}--\ref{swifthms} we listed the counts we extracted in the soft, medium and hard energy bands for each event. In Tables \ref{rosatfits}--\ref{swiftfits}, we have list the parameters of the absorbed powerlaw model that best describe the emission from the event, and the corresponding 0.3-2.0 keV flux and luminosity we derived from these models. These tables will be made available to download from \url{https://tde.space}.

In Table \ref{t90l90values}, we have listed the derived $T_{90}$ and $L_{90}$ values used in this work to produce Figures~\ref{t90s} and ~\ref{bhmass}. Here uncertainties are listed are one sigma uncertainties.

In Table \ref{nhmnhgalcvalue} we have listed the ratio of the measured $N_{\rm H}$ derived from modelling the X-ray spectrum of the observations for which we could extract an X-ray spectrum, and the corresponding Galactic $N_{\rm H}$ as derived from the Leiden/Argentine/Bonn (LAB) Survey of Galactic H$\textsc{i}$ \citep{2005A&A...440..775K}.  Here we have ignored observations in which we were unable to extract an X-ray spectrum and instead assumed the Galactic $N_{\rm H}$ towards the source of interest. To derive $N_{\rm H}$ we assume \citet{2000ApJ...542..914W} solar abundances. This ratio was used to produce Figure~\ref{nhvshr}.

In Table \ref{hrvalues} we have listed the hardness ratios (HRs) derived from the Tables \ref{rosathms}--\ref{swifthms}. Here we have listed only the HRs for observations of the \textit{X-ray TDE} and \textit{likely X-ray TDE} candidates for which we used to produce Figure~\ref{nhvari}, ~\ref{hr} and ~\ref{nhvshr}. We do not list observations for which we derive an upperlimit. 

In Table \ref{xtooratio}, we have listed the integrated optical/UV (0.002-0.1 keV) and X-ray (0.3-10 keV) luminosities derived for the events which have both optical/UV and X-ray data (see the individual SEDs listed in Figure~\ref{individualnuFnu}). These values were used to produce Figure~\ref{opticaltoxray}.

\input{rosat_table_counts.tex}
\input{chandra_table_counts.tex}
\input{xmm_table_counts.tex}

\input{xmm_slew_table_counts.tex}

\input{swift_table_counts.tex}

\input{rosat_table_hms.tex}
\input{chandra_table_hms.tex}
\input{xmm_table_hms.tex}
\input{slew_table_hms.tex}
\input{swift_table_hms.tex}

\input{rosat_table_fits.tex}
\input{chandra_table_fits.tex}
\input{xmm_table_fits.tex}
\input{xmm_slew_table_fits.tex}
\input{swift_table_fits.tex}

\input{t90l90.tex}

\input{nh_nhgalc_values.tex}

\input{hrvalues.tex}

\input{opticaluvxrayratio.tex}
\end{appendix}

%% file: lightcurves.tex
\begin{figure*}[!th]
	\begin{center}
		\includegraphics[width=0.31\textwidth]{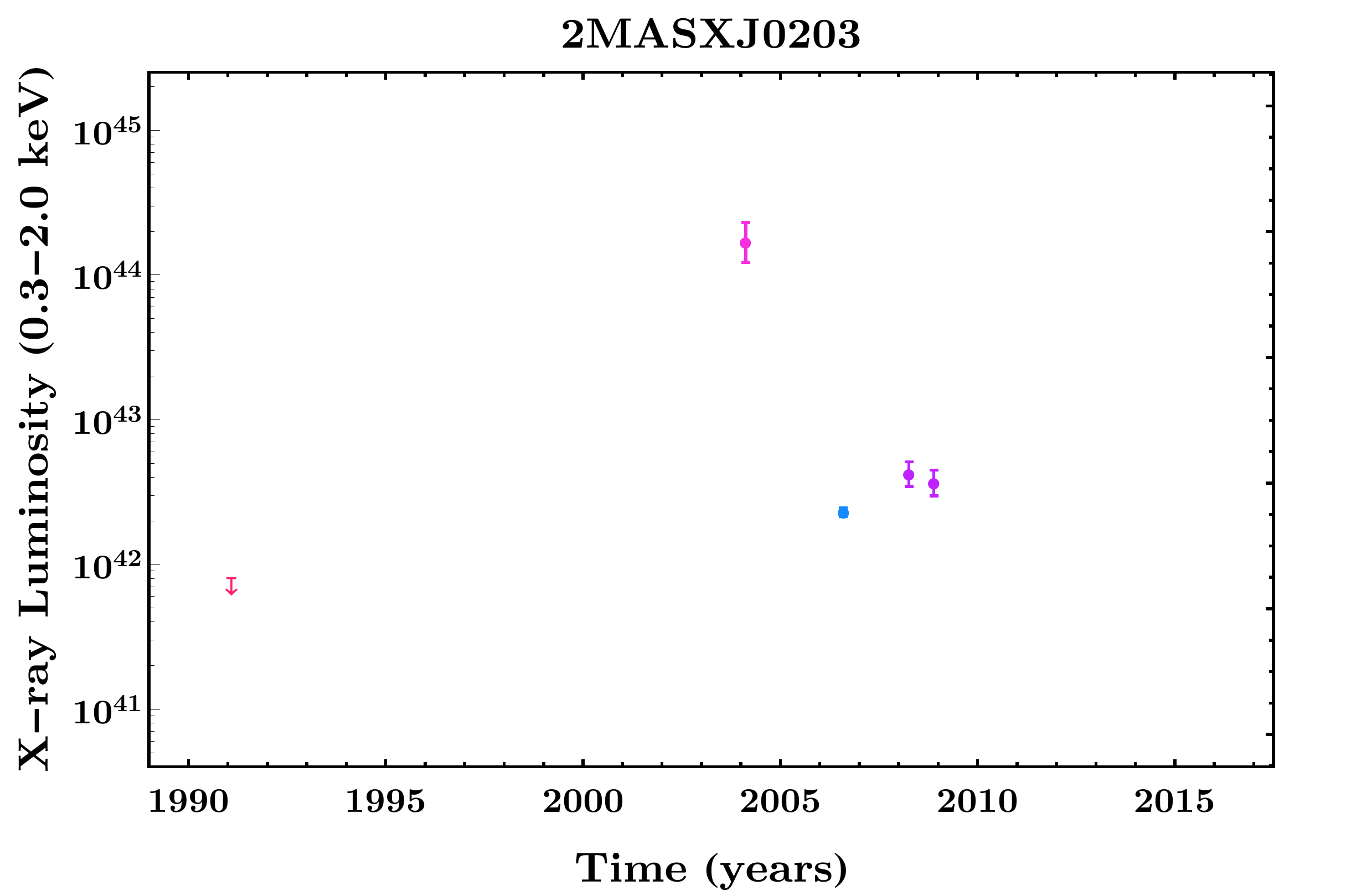}
		\includegraphics[width=0.31\textwidth]{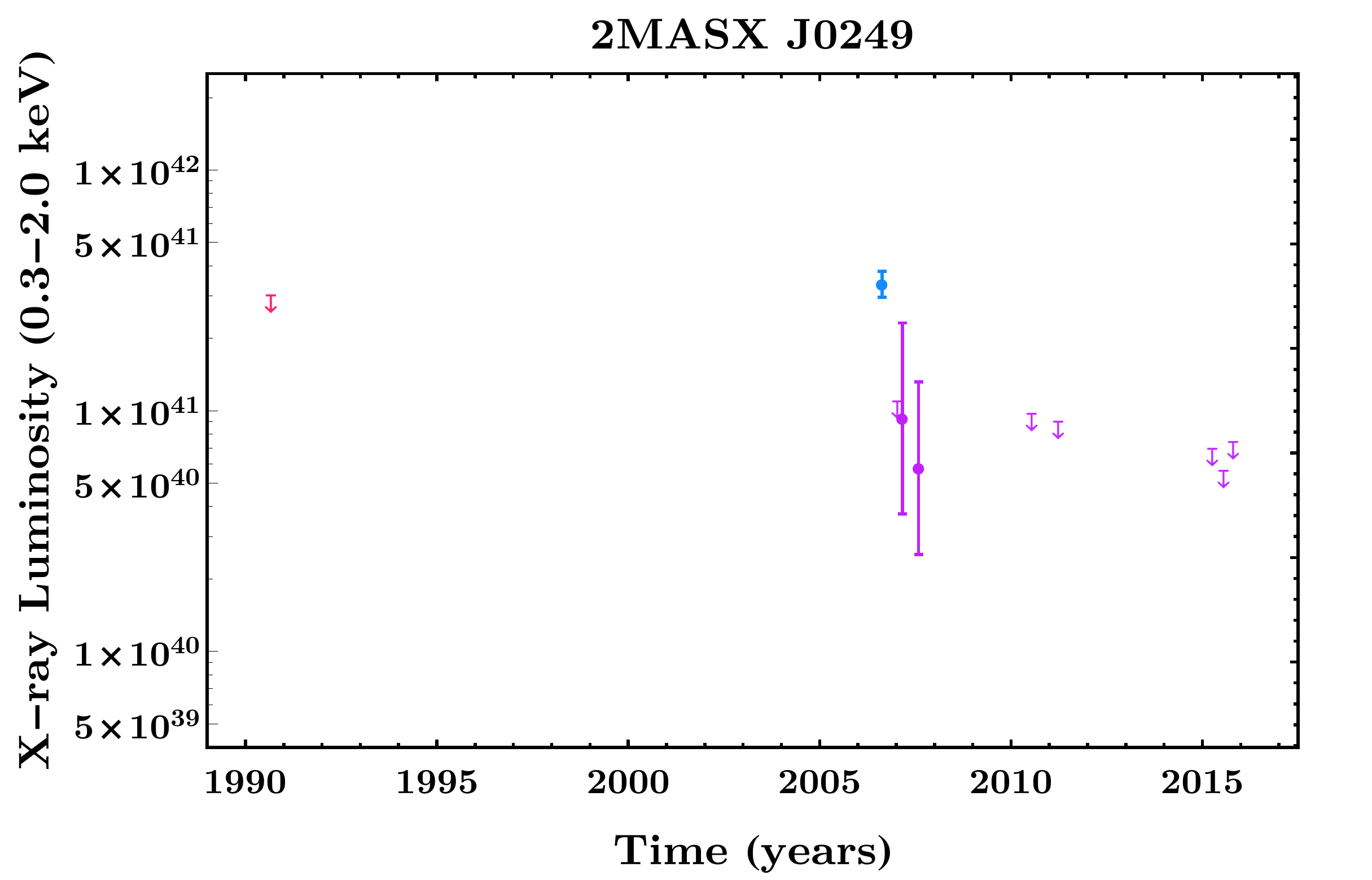}
		\includegraphics[width=0.31\textwidth]{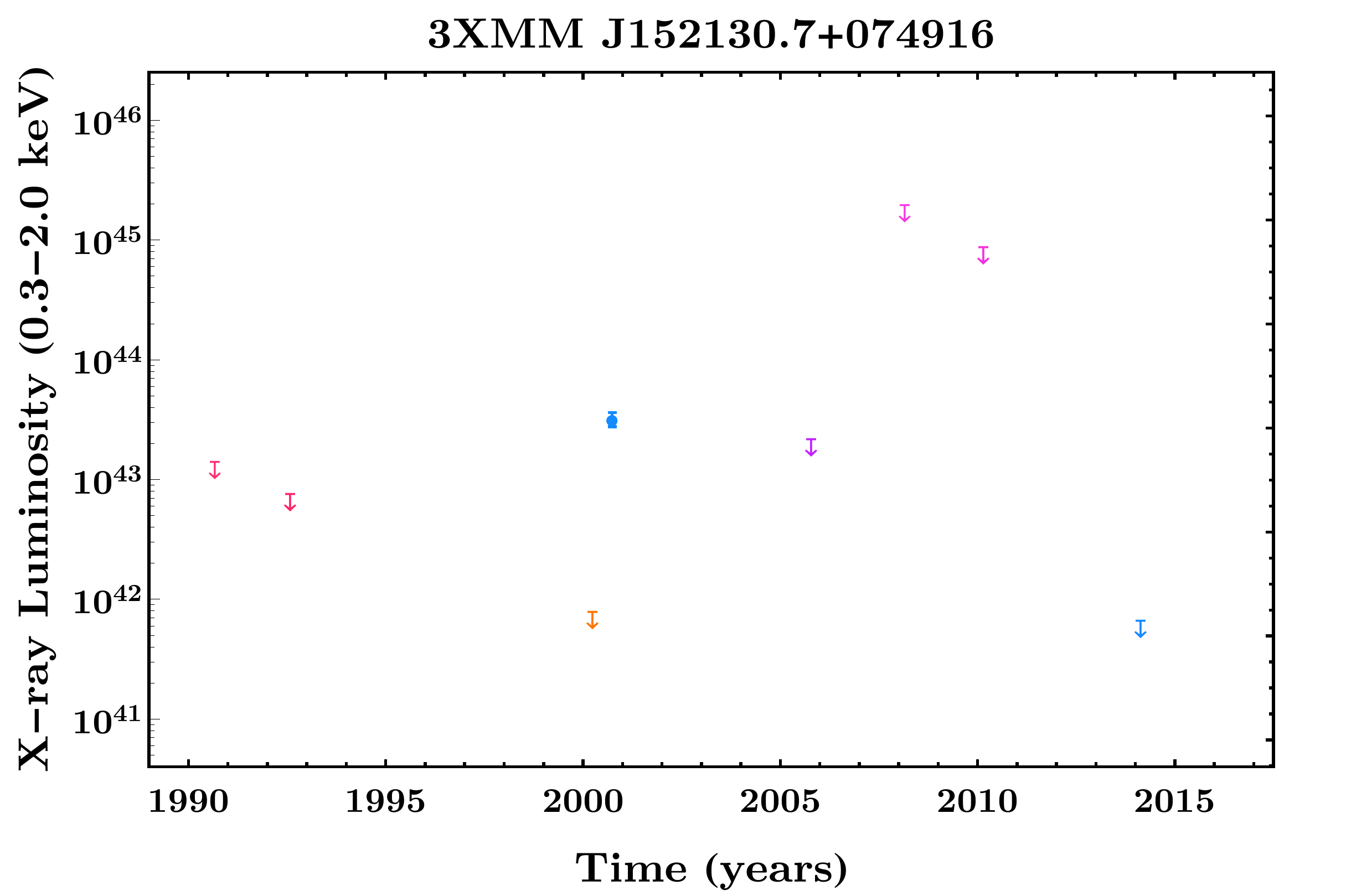}
		\includegraphics[width=0.31\textwidth]{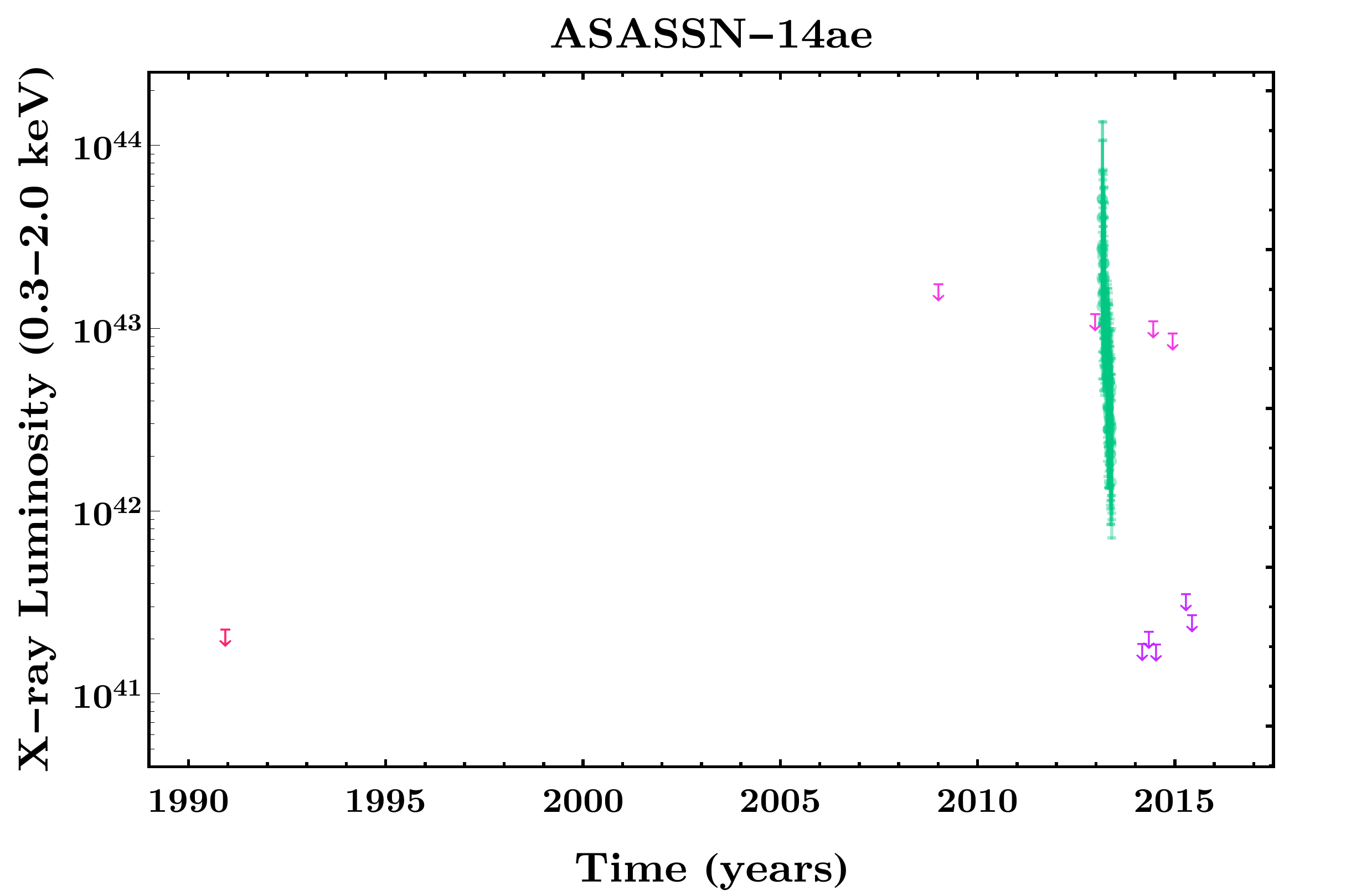}
		\includegraphics[width=0.31\textwidth]{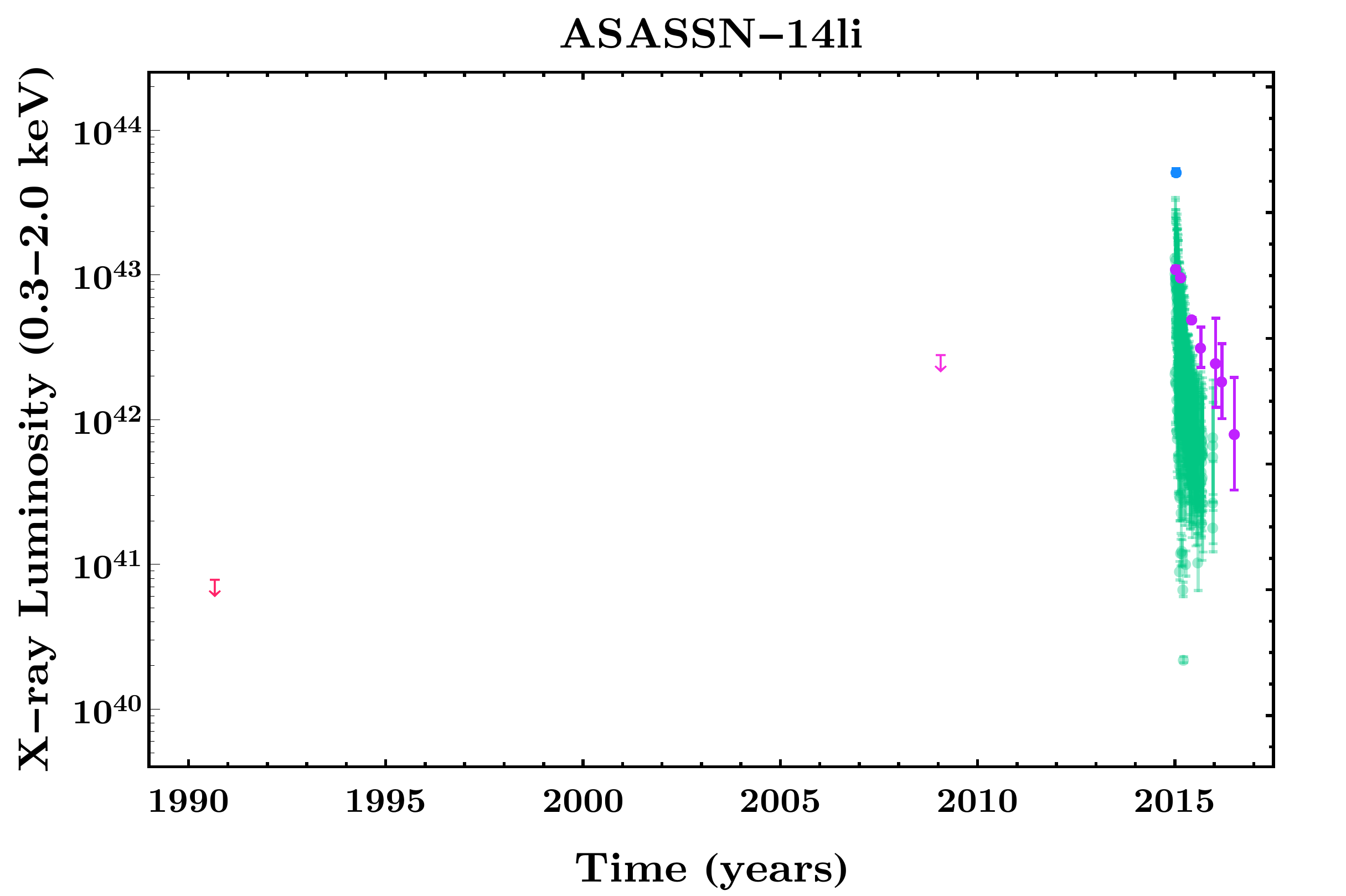}
		\includegraphics[width=0.31\textwidth]{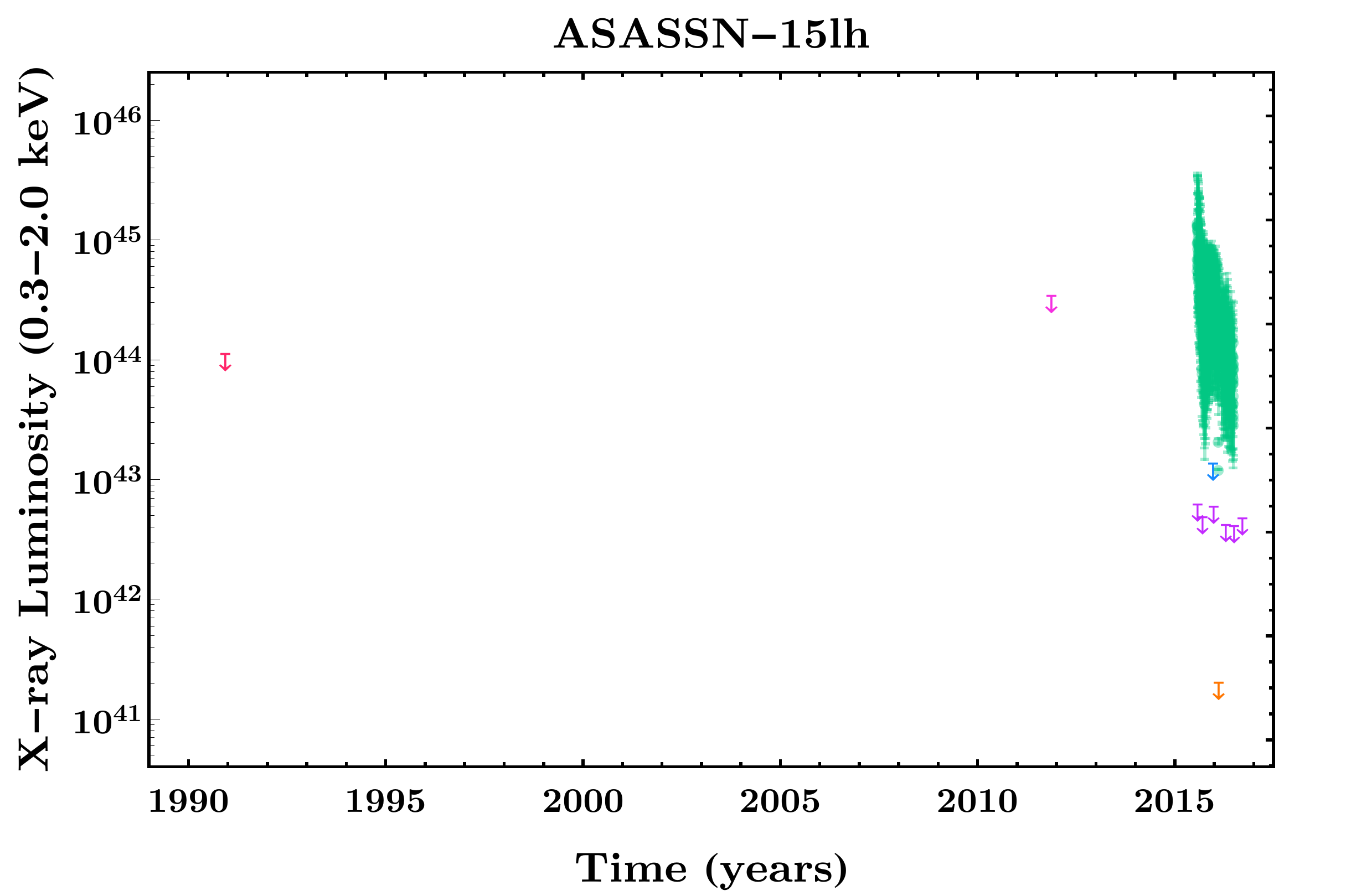}
		\includegraphics[width=0.31\textwidth]{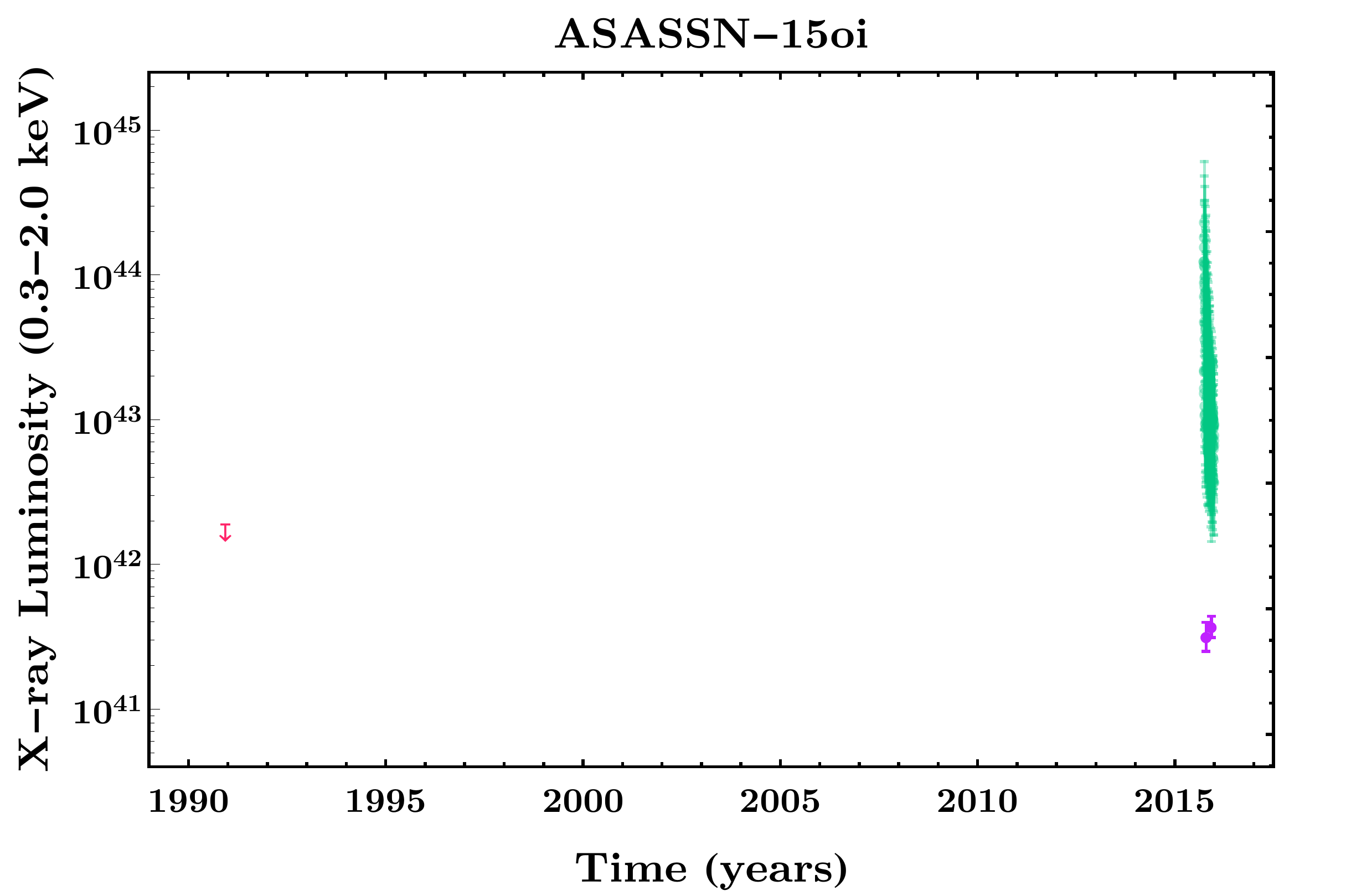}
		\includegraphics[width=0.31\textwidth]{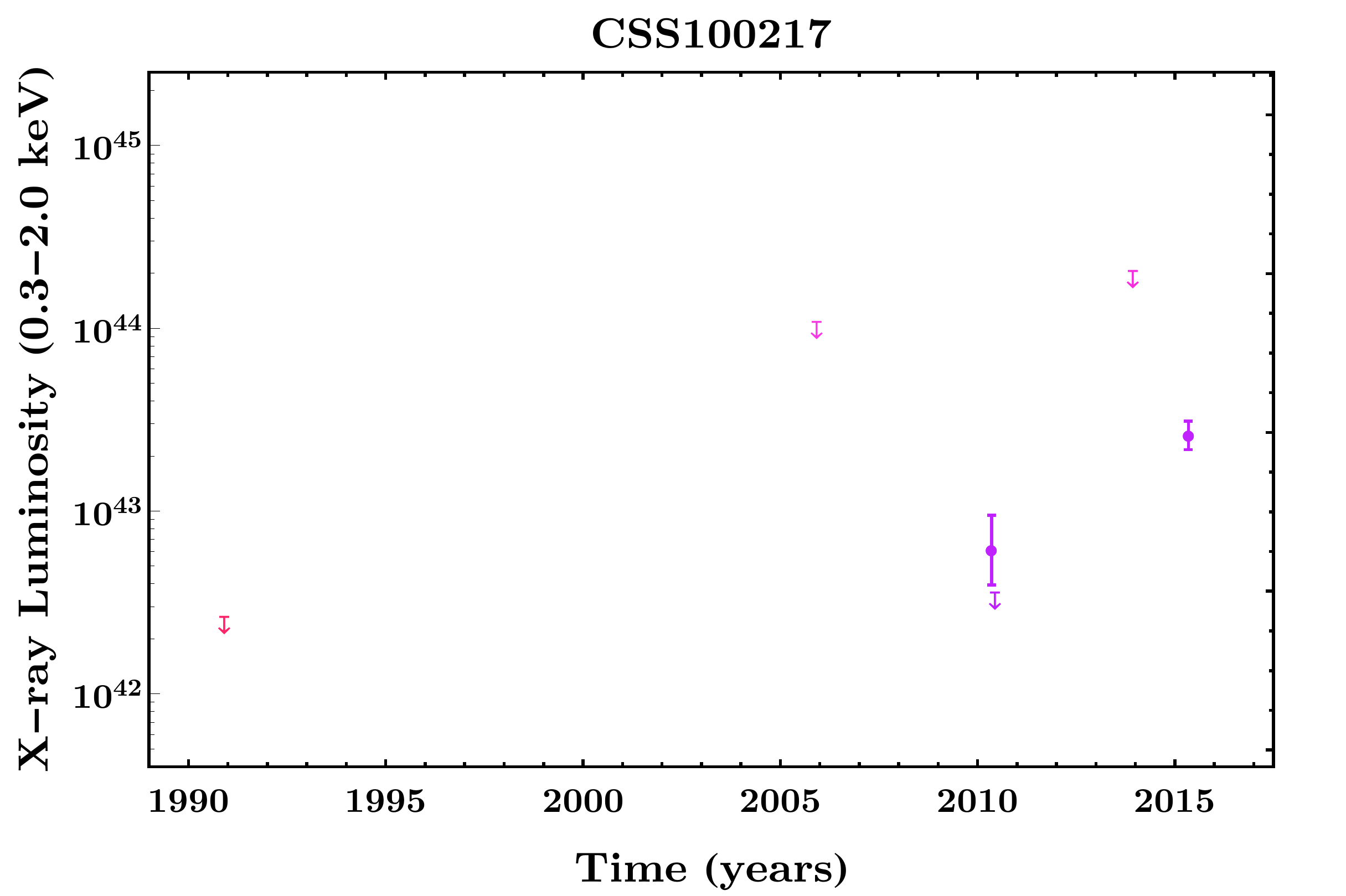}
		\includegraphics[width=0.31\textwidth]{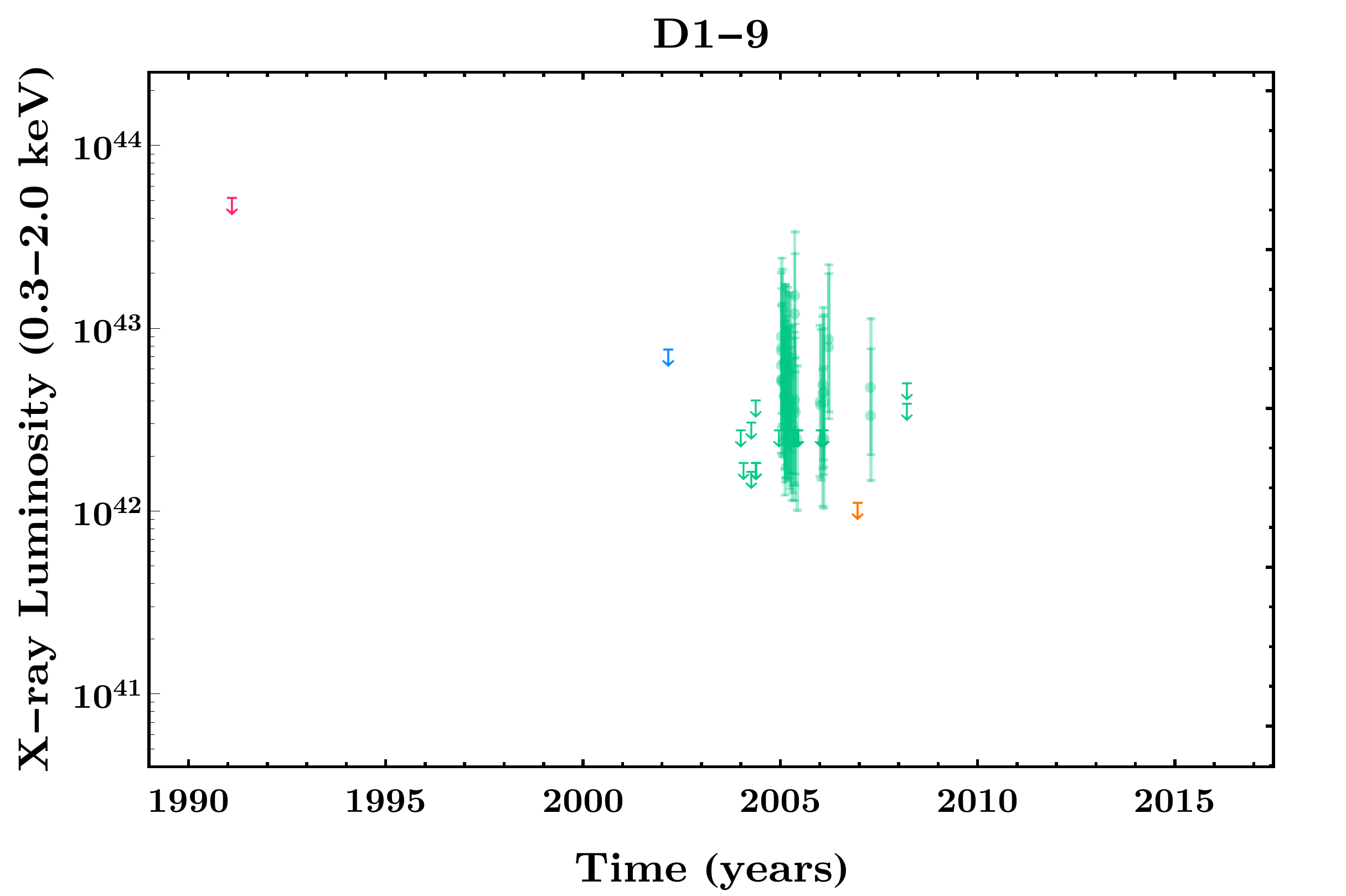}
		\includegraphics[width=0.31\textwidth]{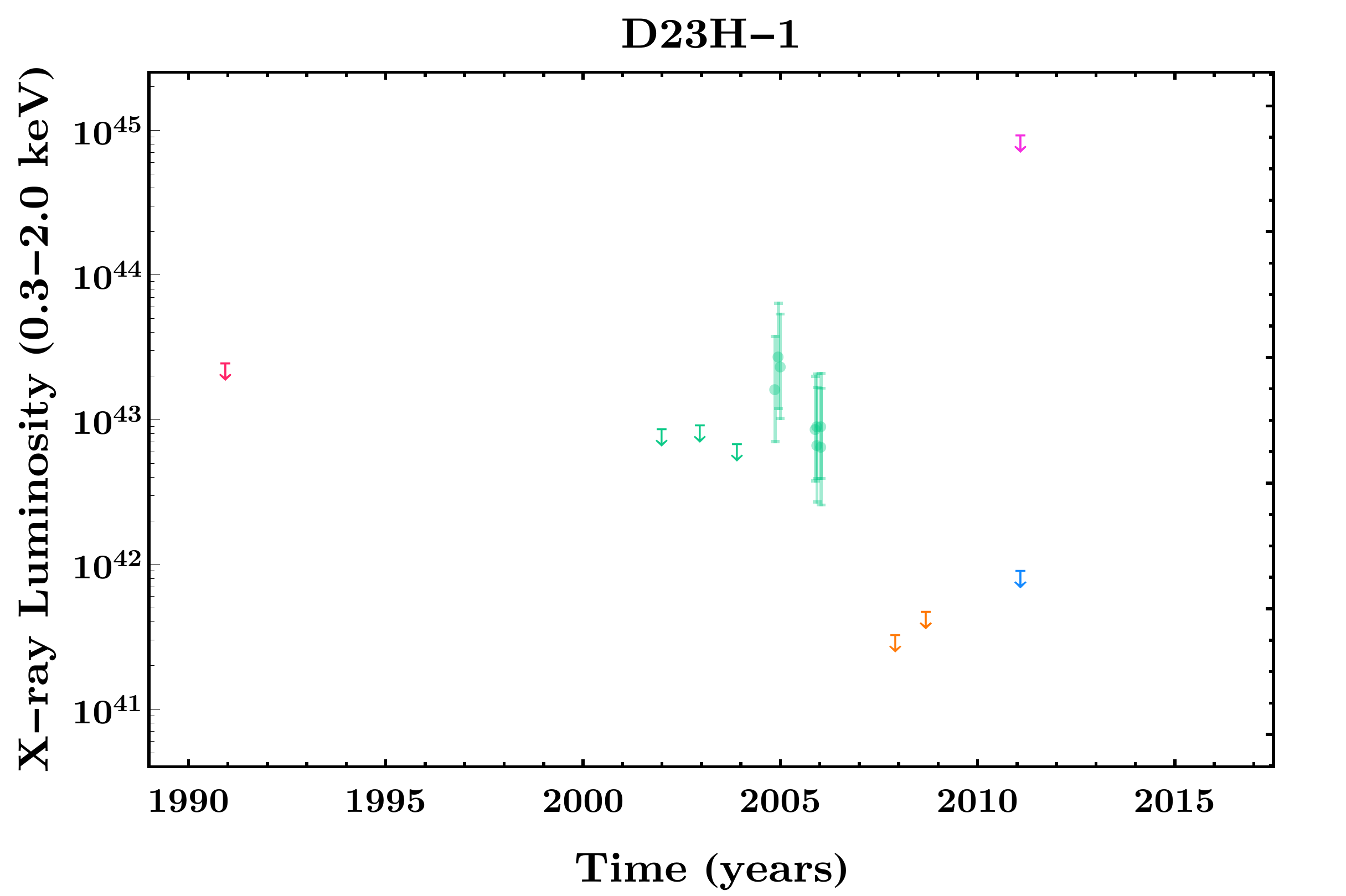}
		\includegraphics[width=0.31\textwidth]{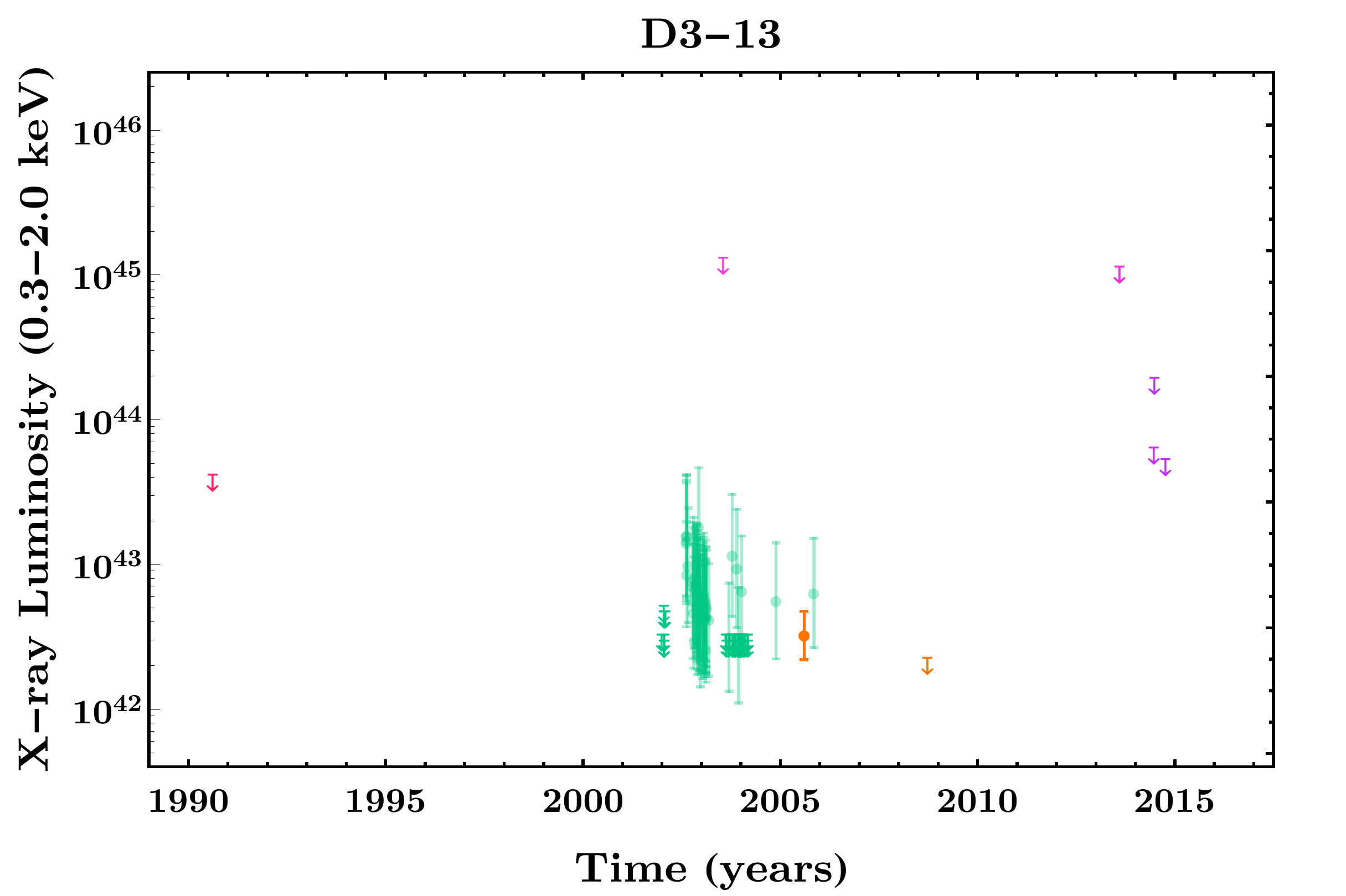}
		\includegraphics[width=0.31\textwidth]{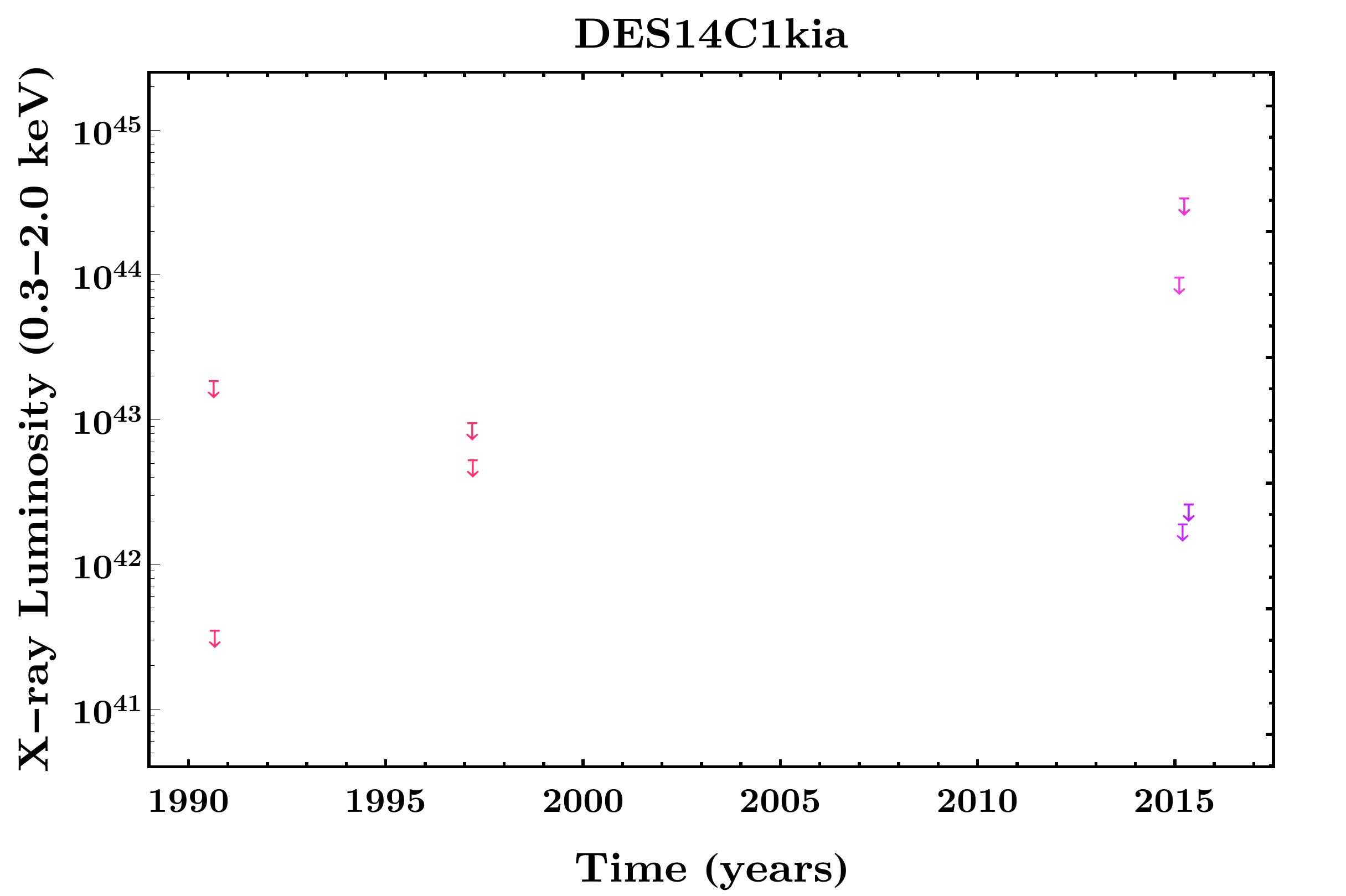}
		\includegraphics[width=0.31\textwidth]{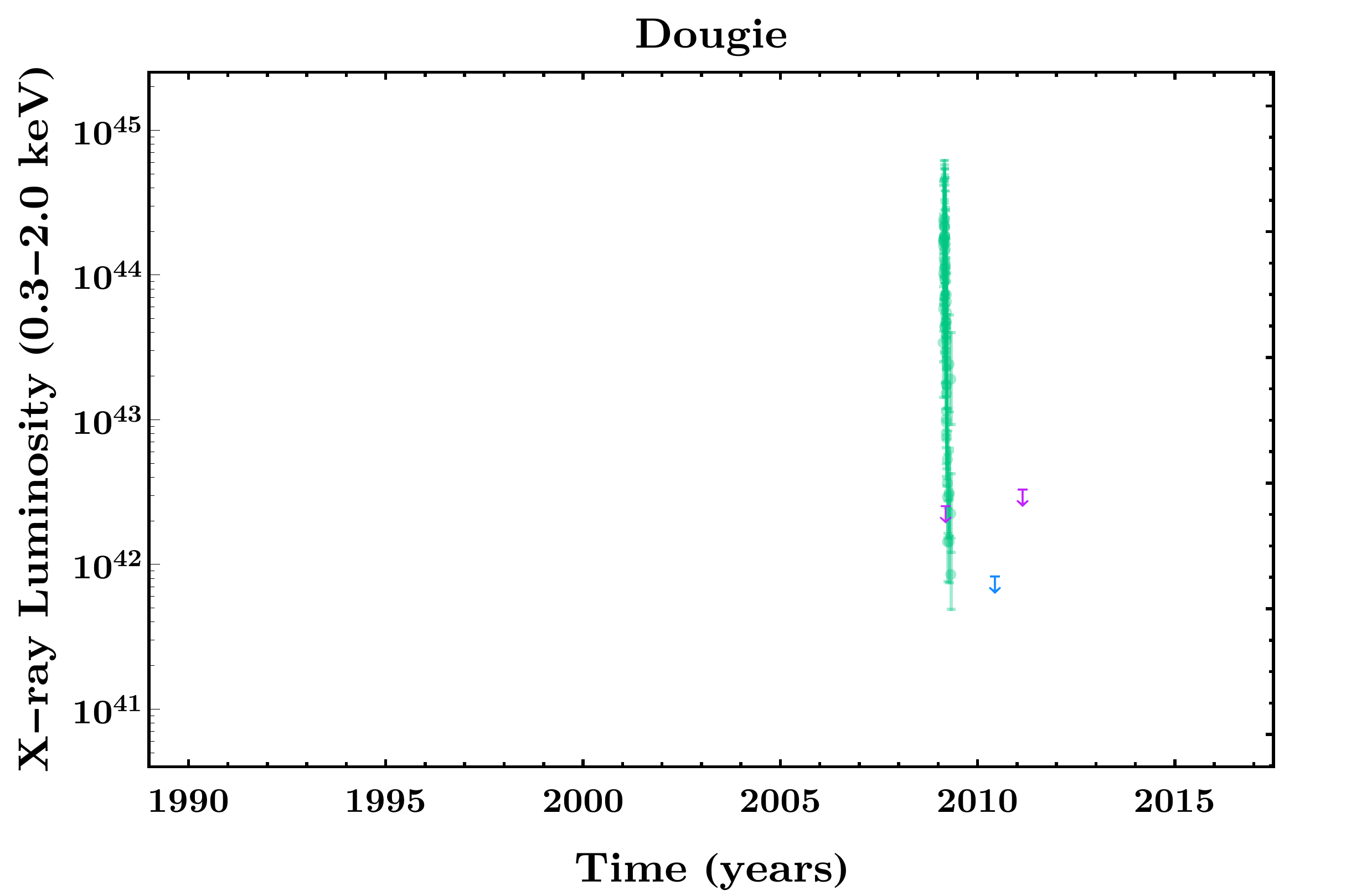}
		\includegraphics[width=0.31\textwidth]{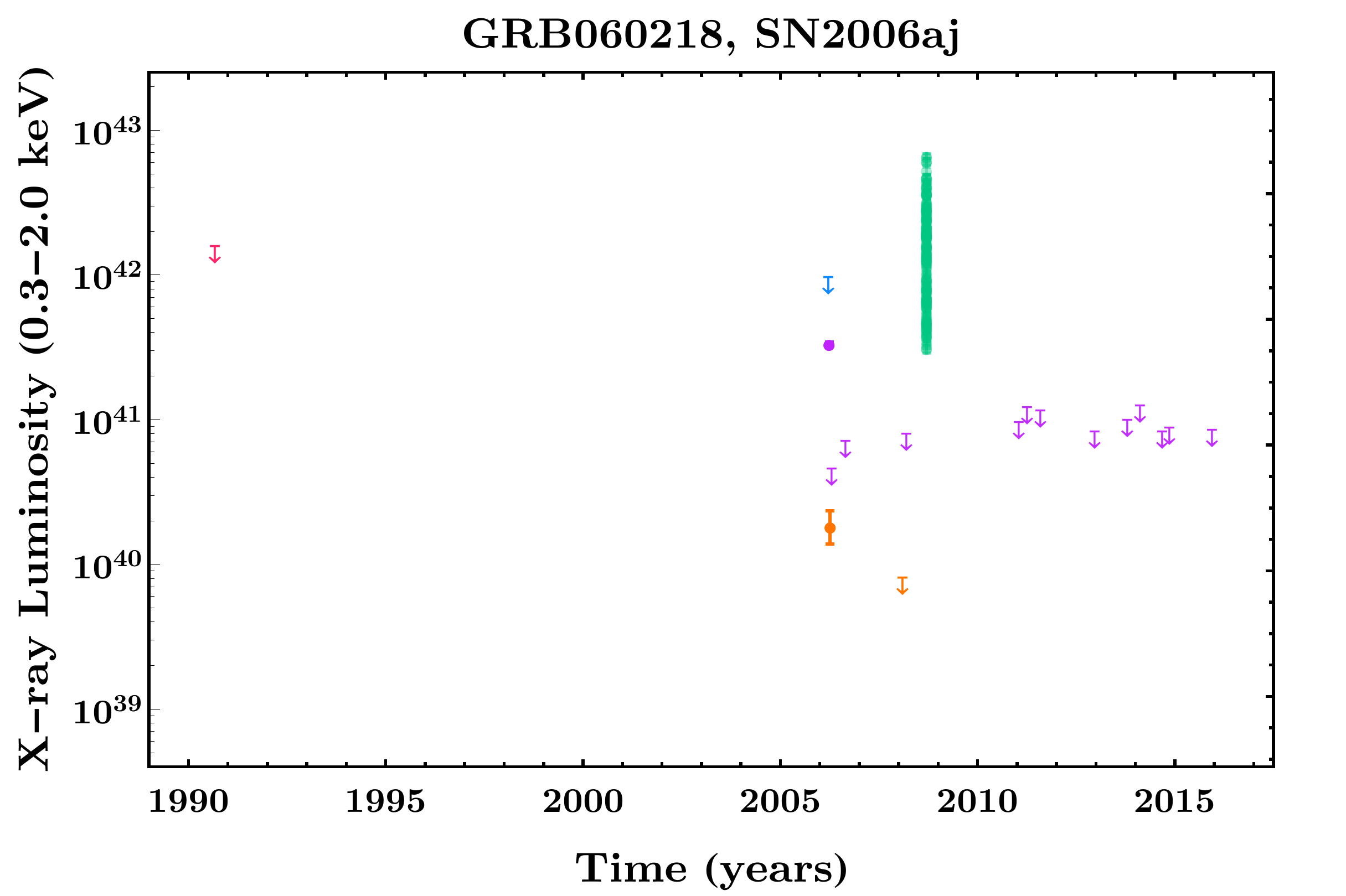}
		\includegraphics[width=0.31\textwidth]{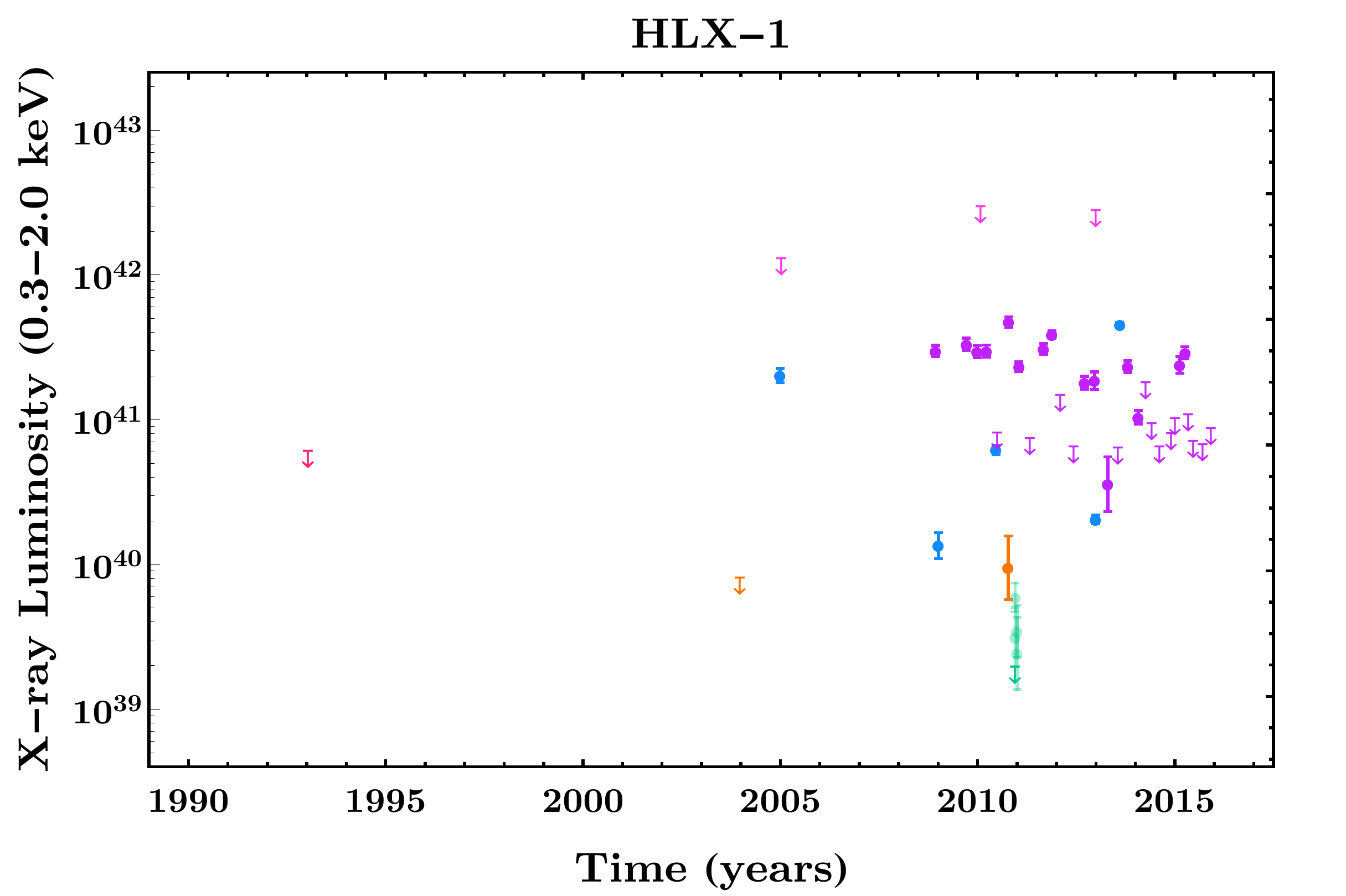}
		\includegraphics[width=0.31\textwidth]{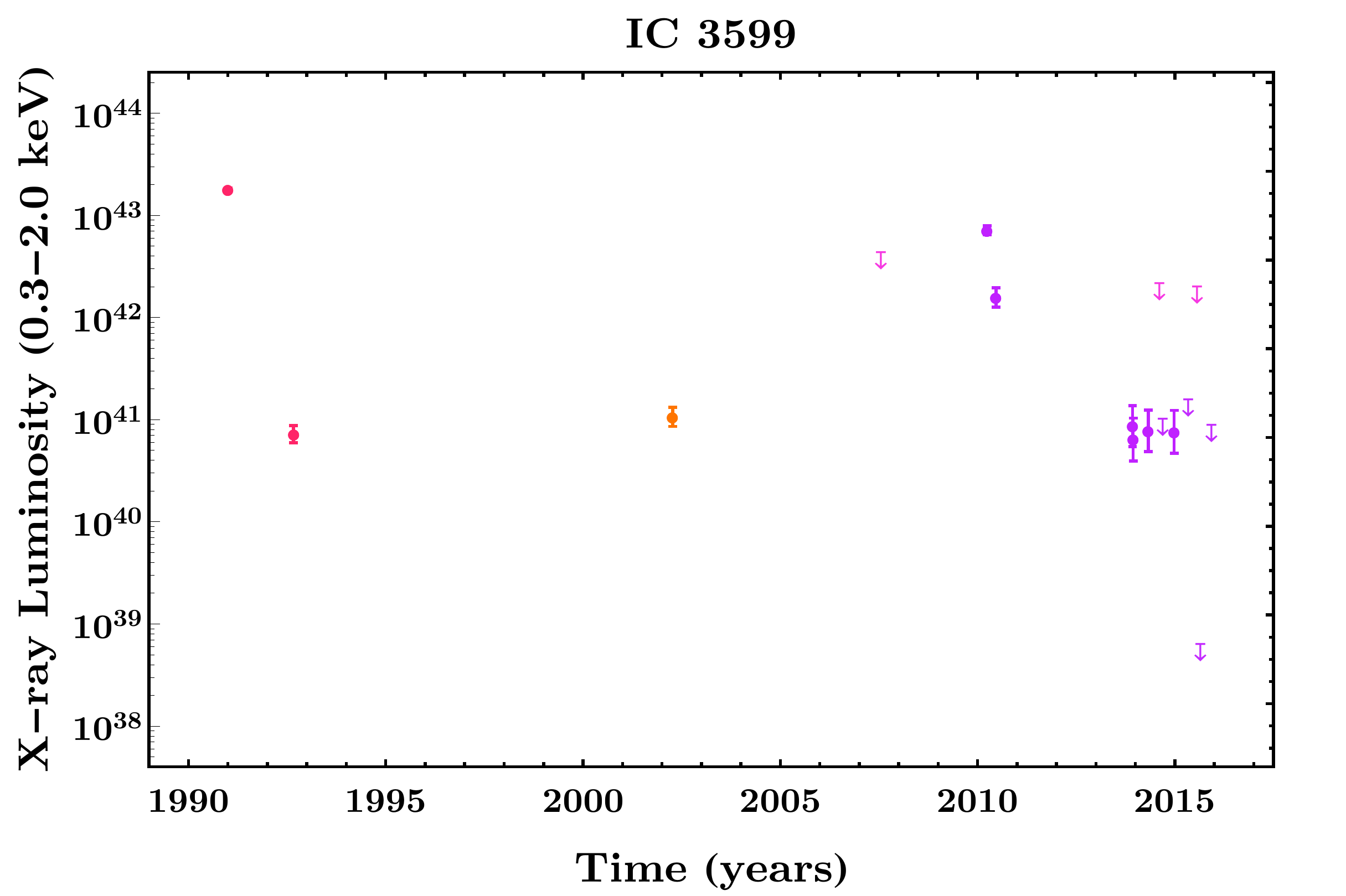}
				\includegraphics[width=0.31\textwidth]{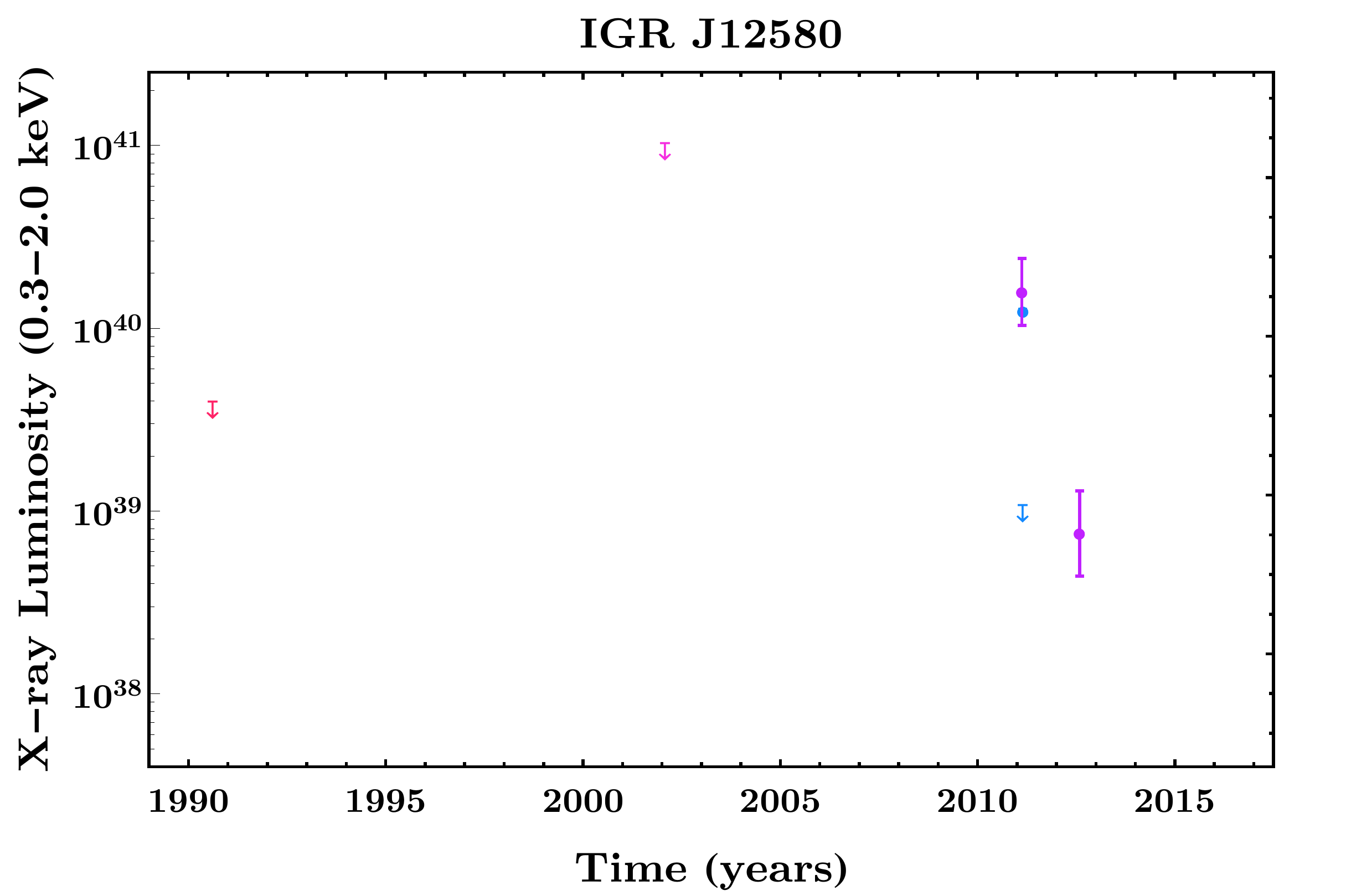}
		\includegraphics[width=0.10\textwidth]{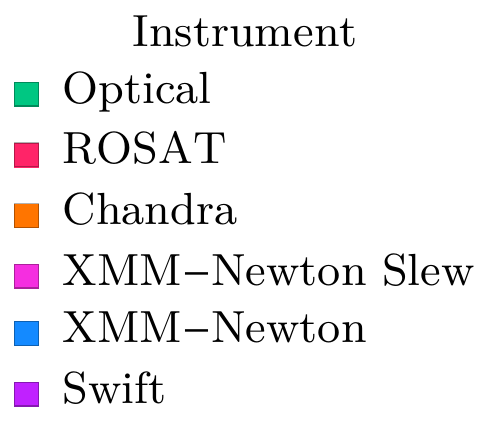}
		\caption{The X-ray/optical light curves of all TDE candidates listed in Table \ref{tdes}. The X-ray data which makes up these plots is derived from this work, while the optical data was taken from the literature. Here we have also coloured the data points based on the X-ray instrument in which we obtained this constraint. The exception is the optical data which is plotted as one colour, even though data points were taken by different instruments. Listed in the last panel in this figure is the colour key for each instrument. \label{lightcurves1}}
	\end{center}
\end{figure*}

\begin{figure*}[!th]
	\begin{center}
		\includegraphics[width=0.31\textwidth]{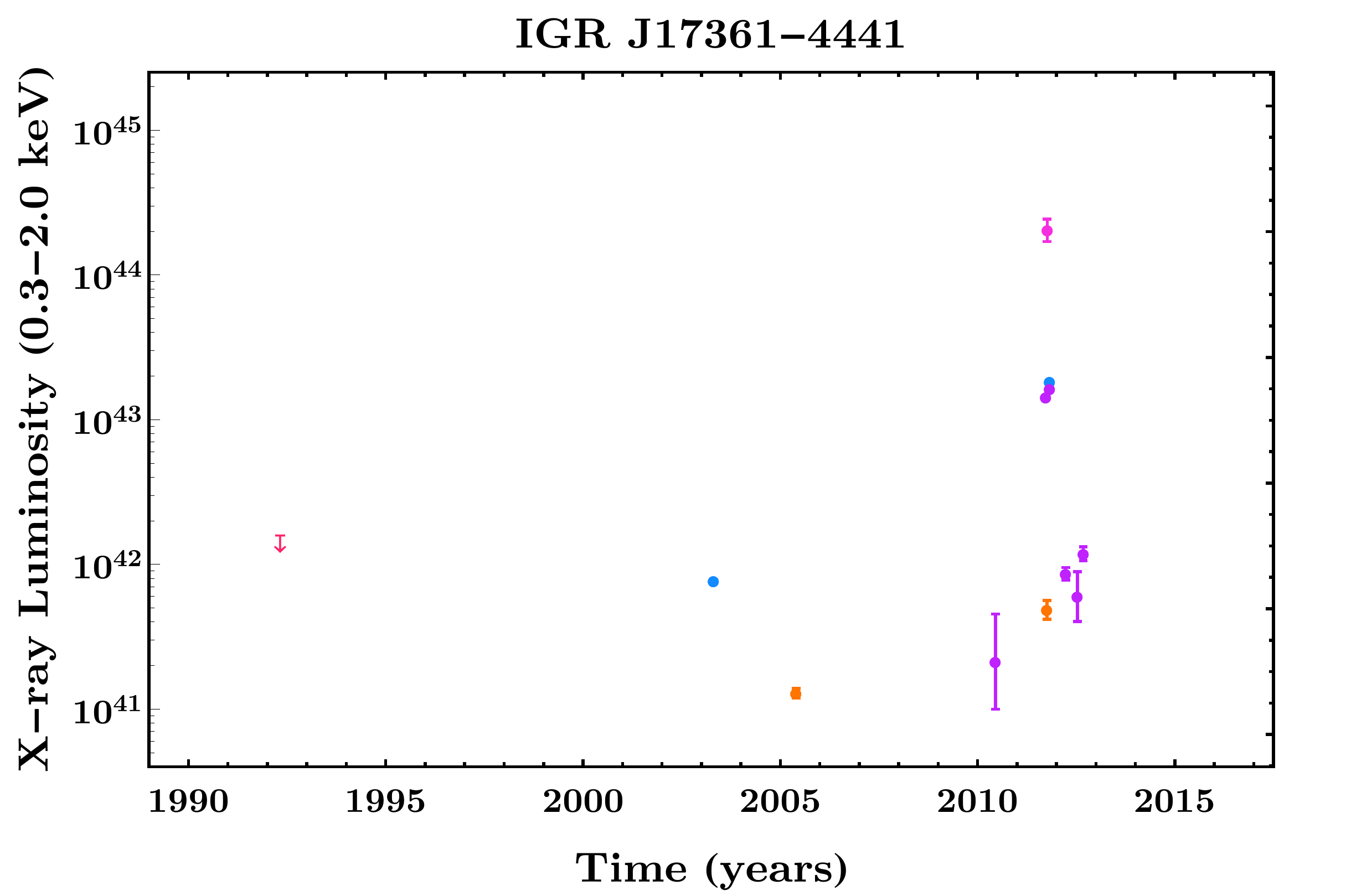}
		\includegraphics[width=0.31\textwidth]{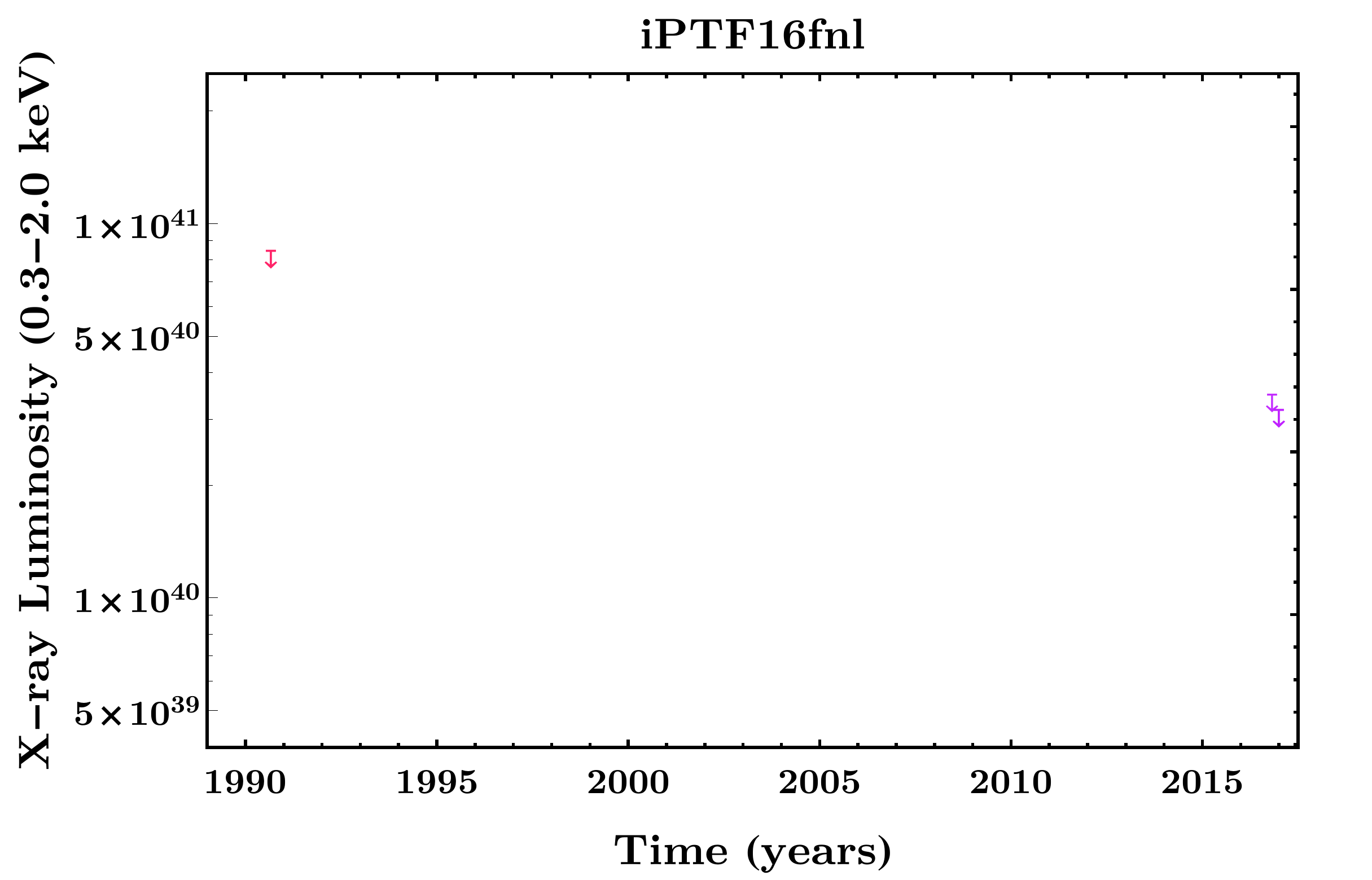}
		\includegraphics[width=0.31\textwidth]{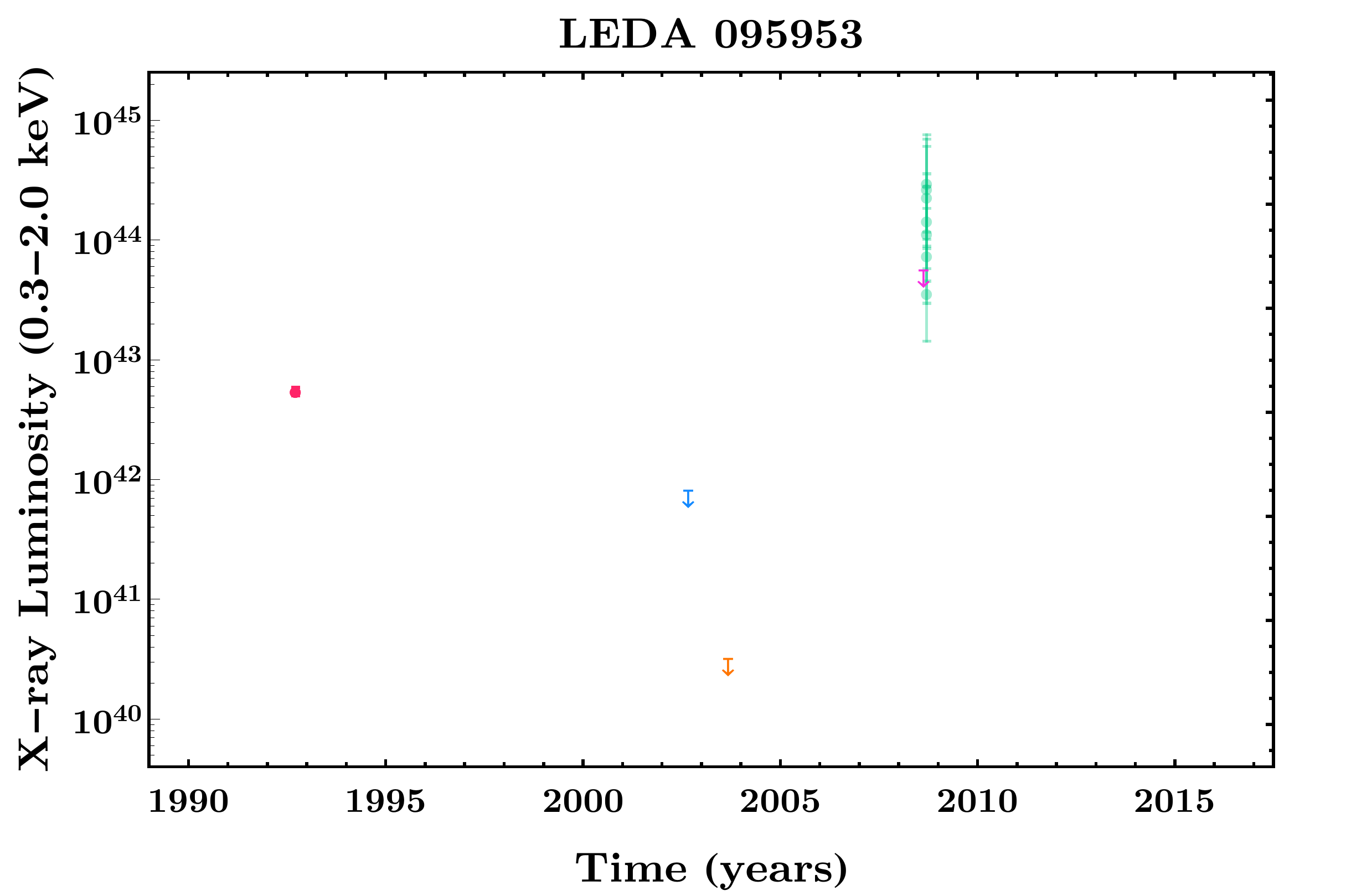}
		\includegraphics[width=0.31\textwidth]{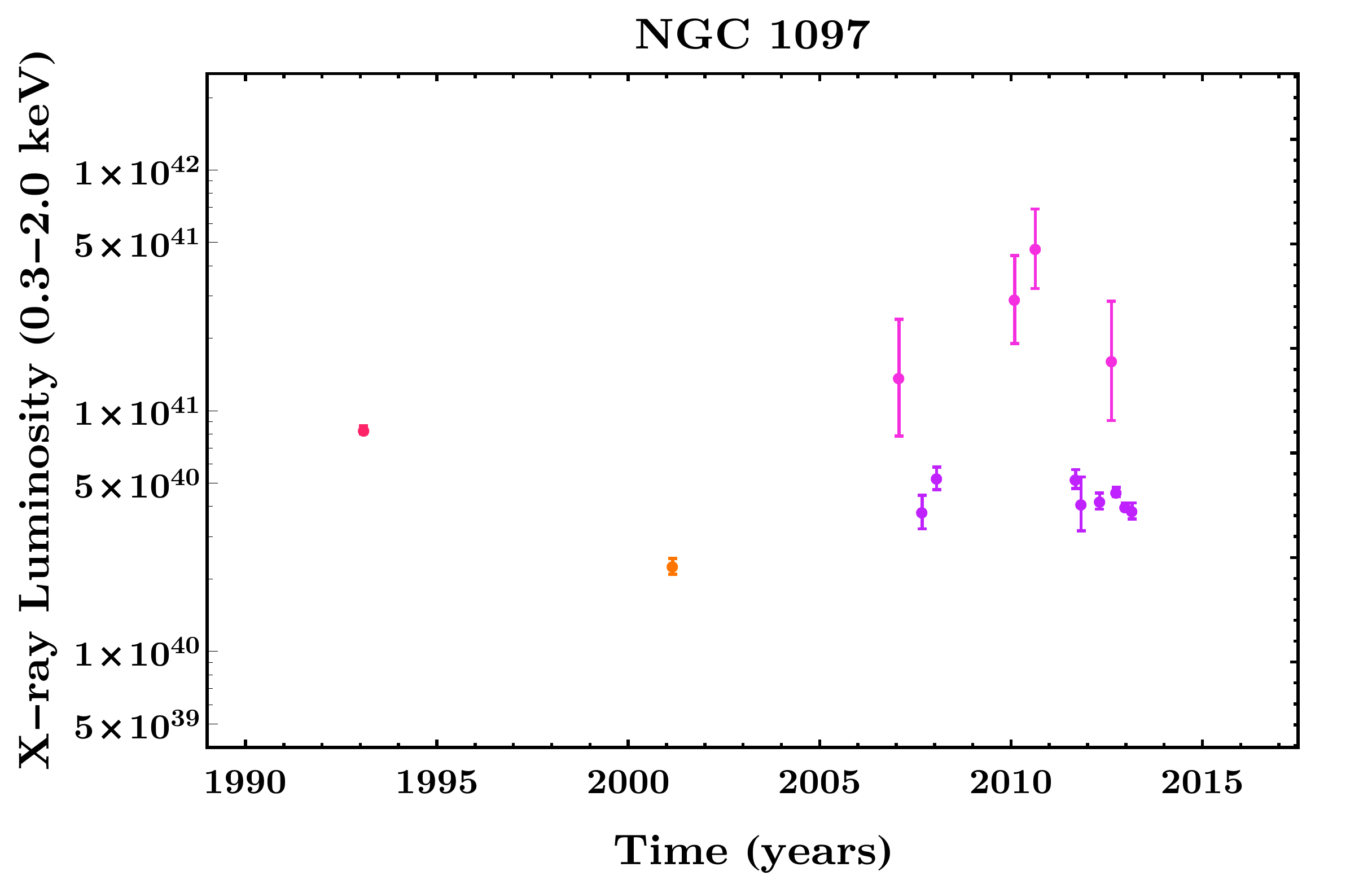}
		\includegraphics[width=0.31\textwidth]{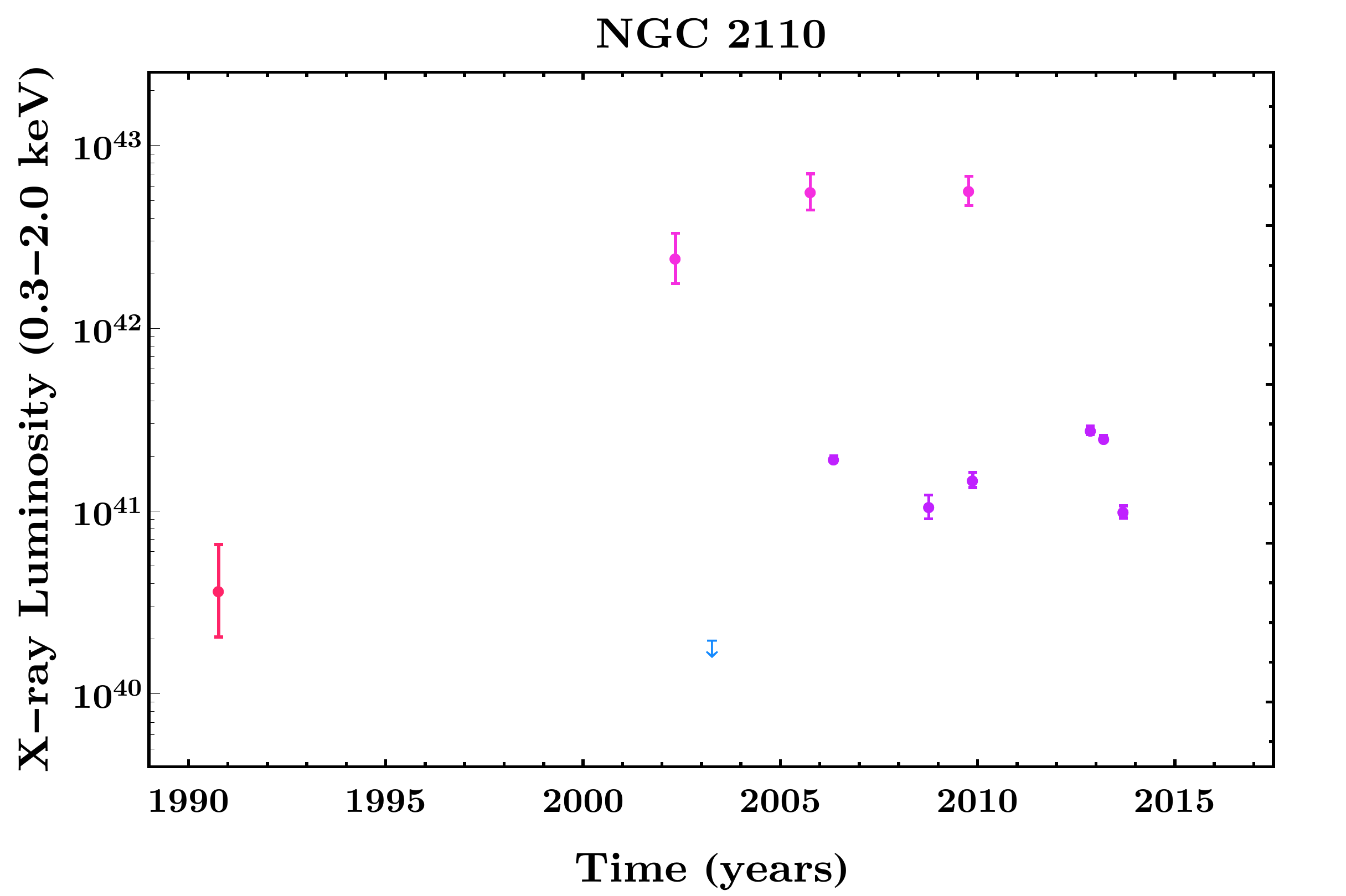}
		\includegraphics[width=0.31\textwidth]{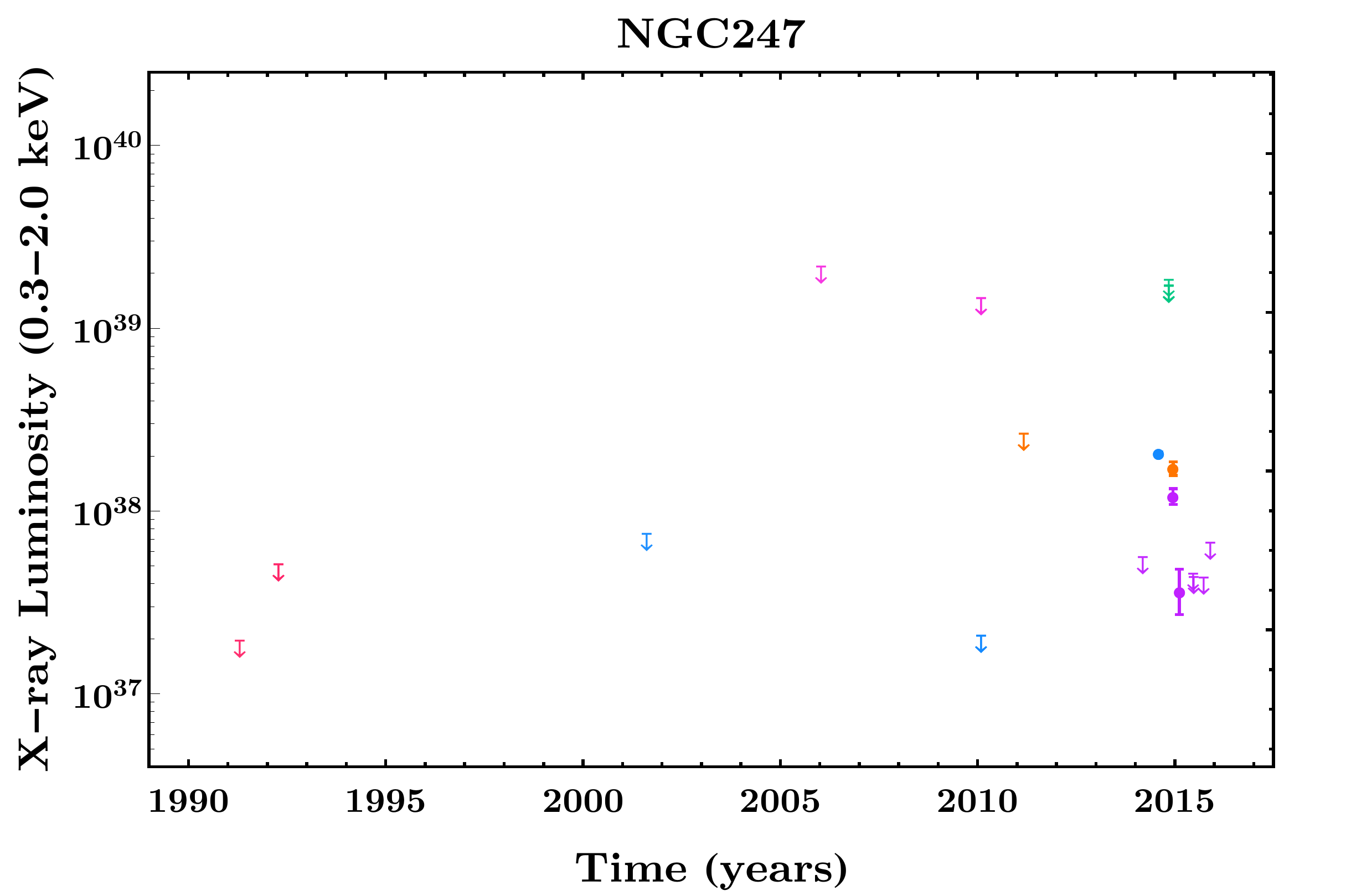}
		\includegraphics[width=0.31\textwidth]{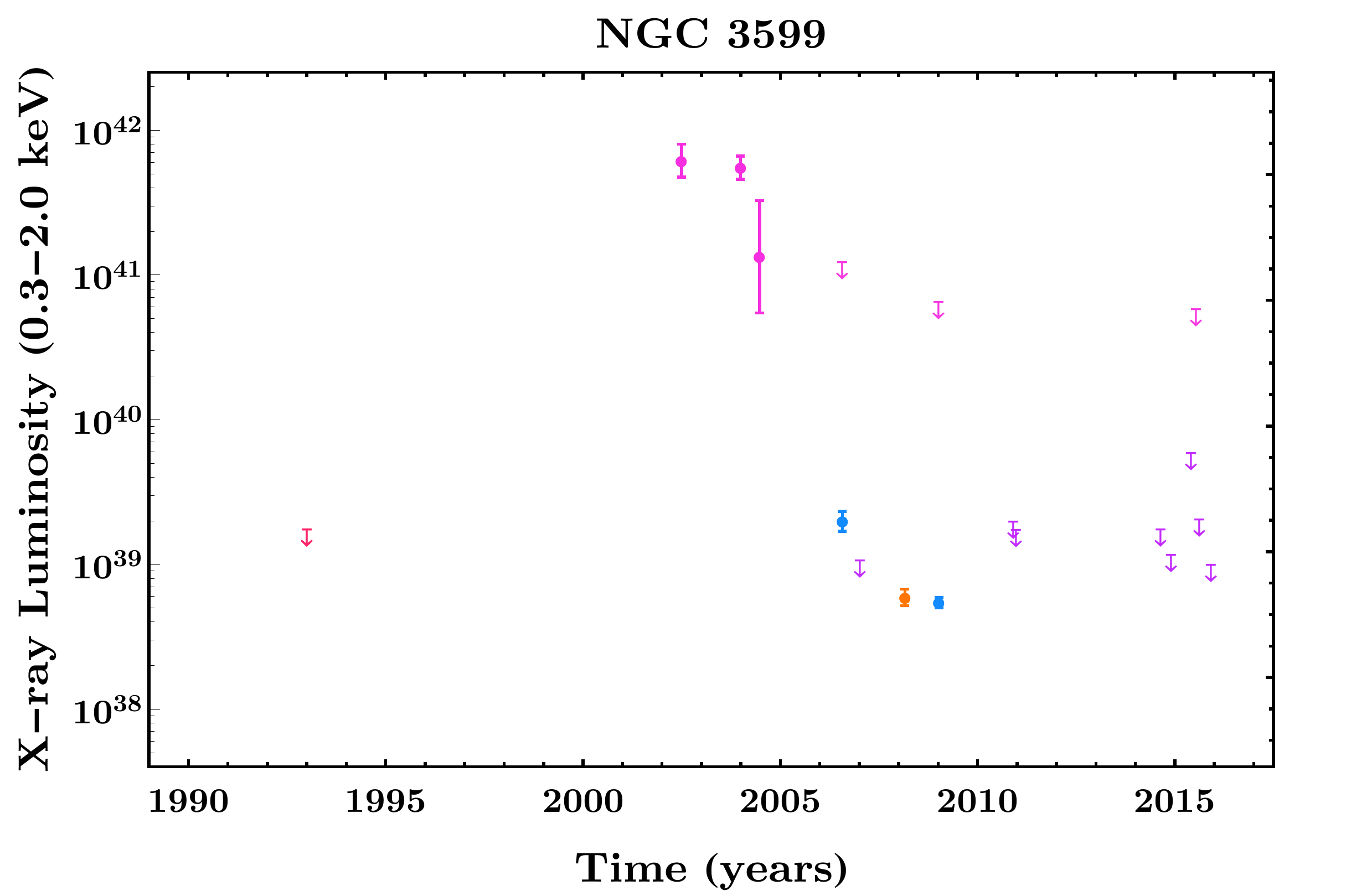}
		\includegraphics[width=0.31\textwidth]{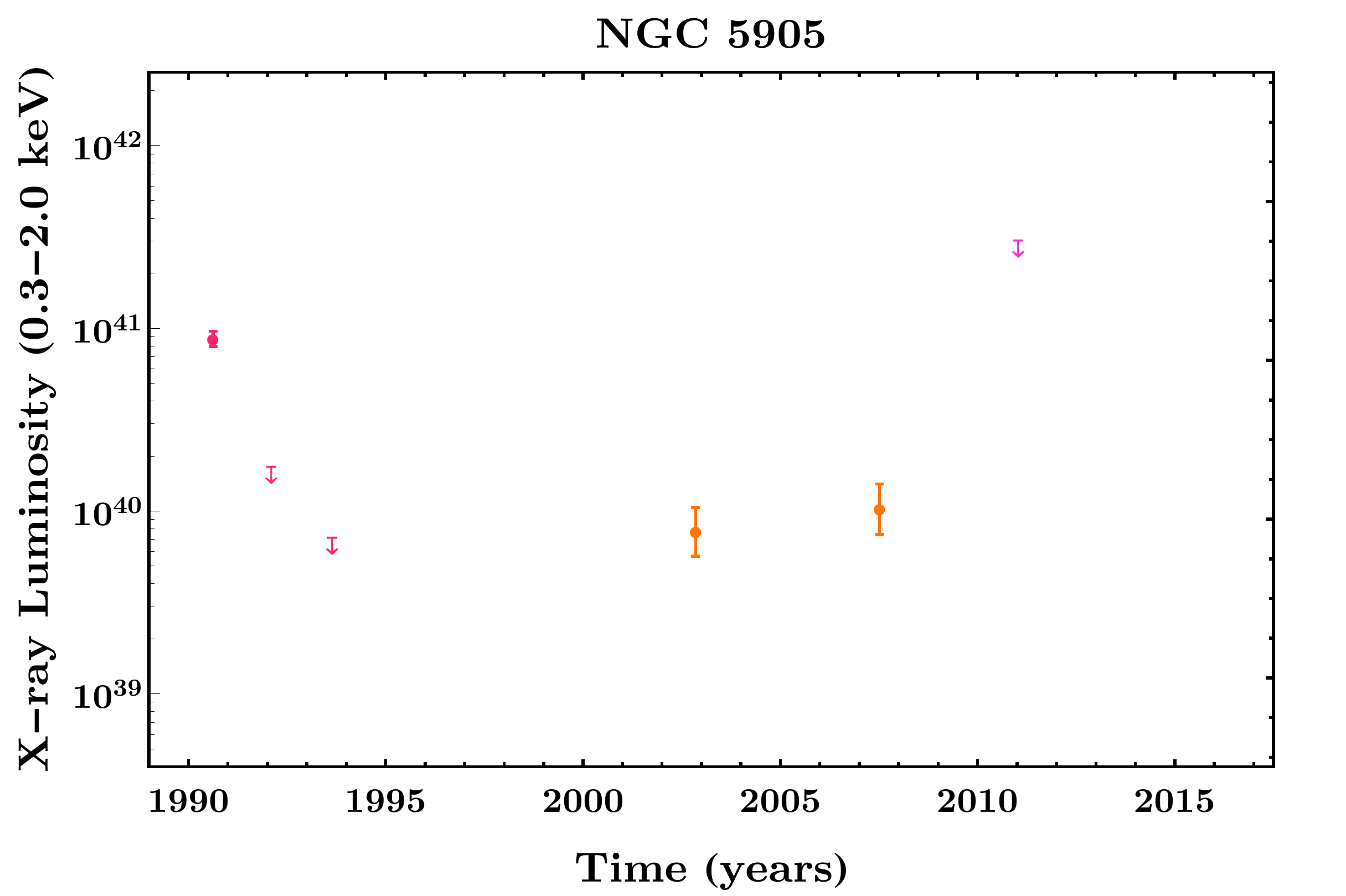}
		\includegraphics[width=0.31\textwidth]{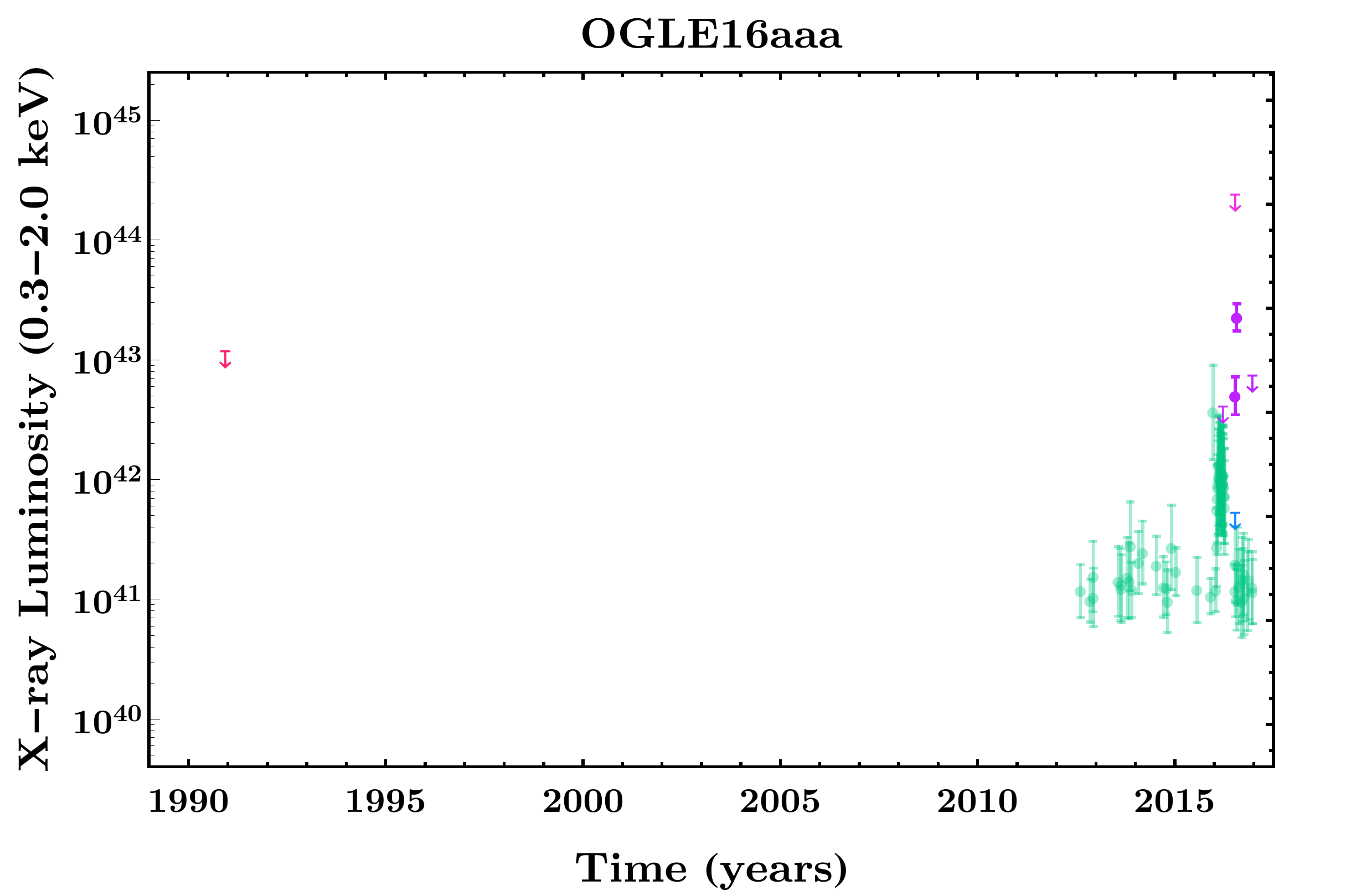}
		\includegraphics[width=0.31\textwidth]{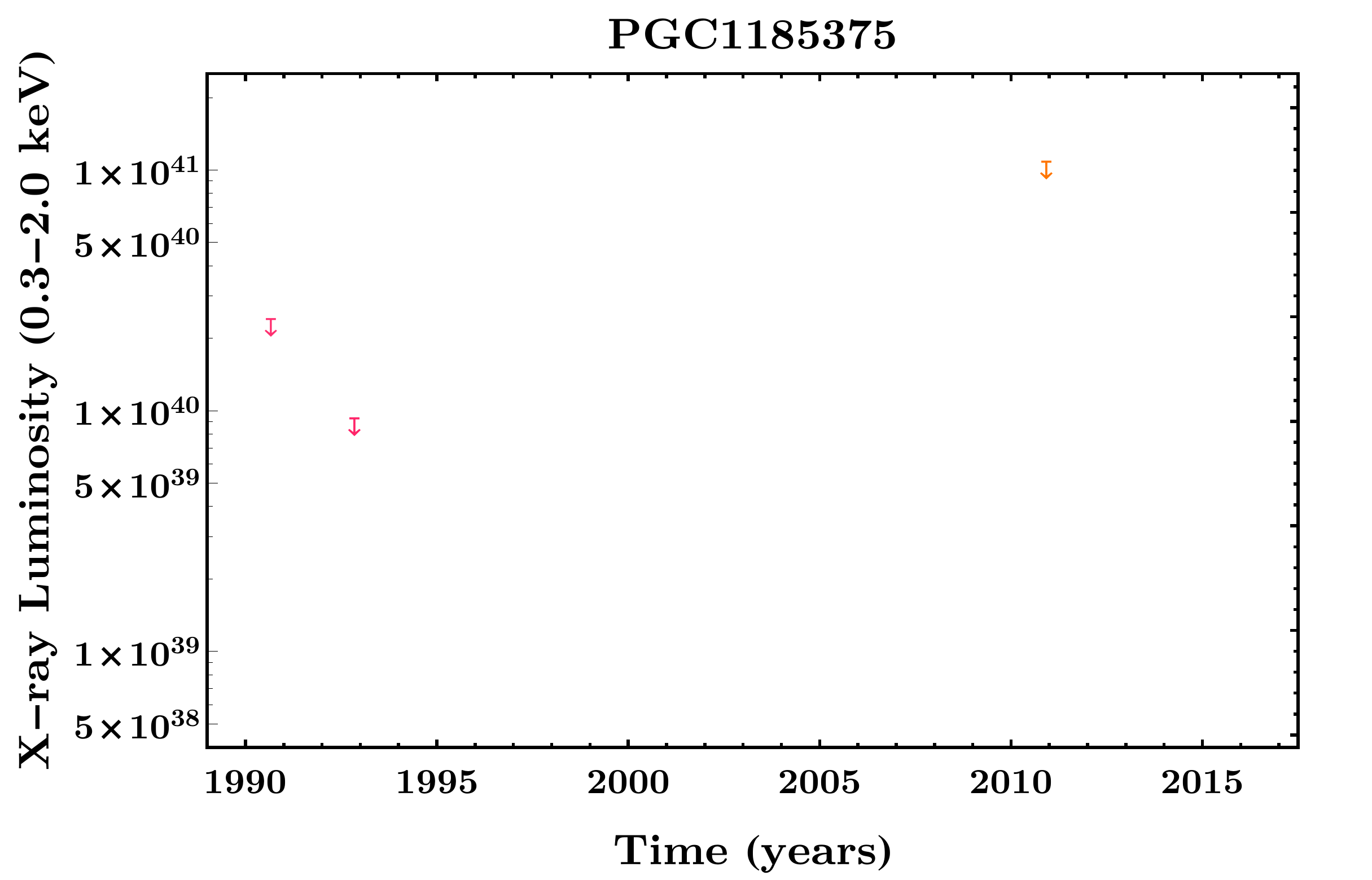}
		\includegraphics[width=0.31\textwidth]{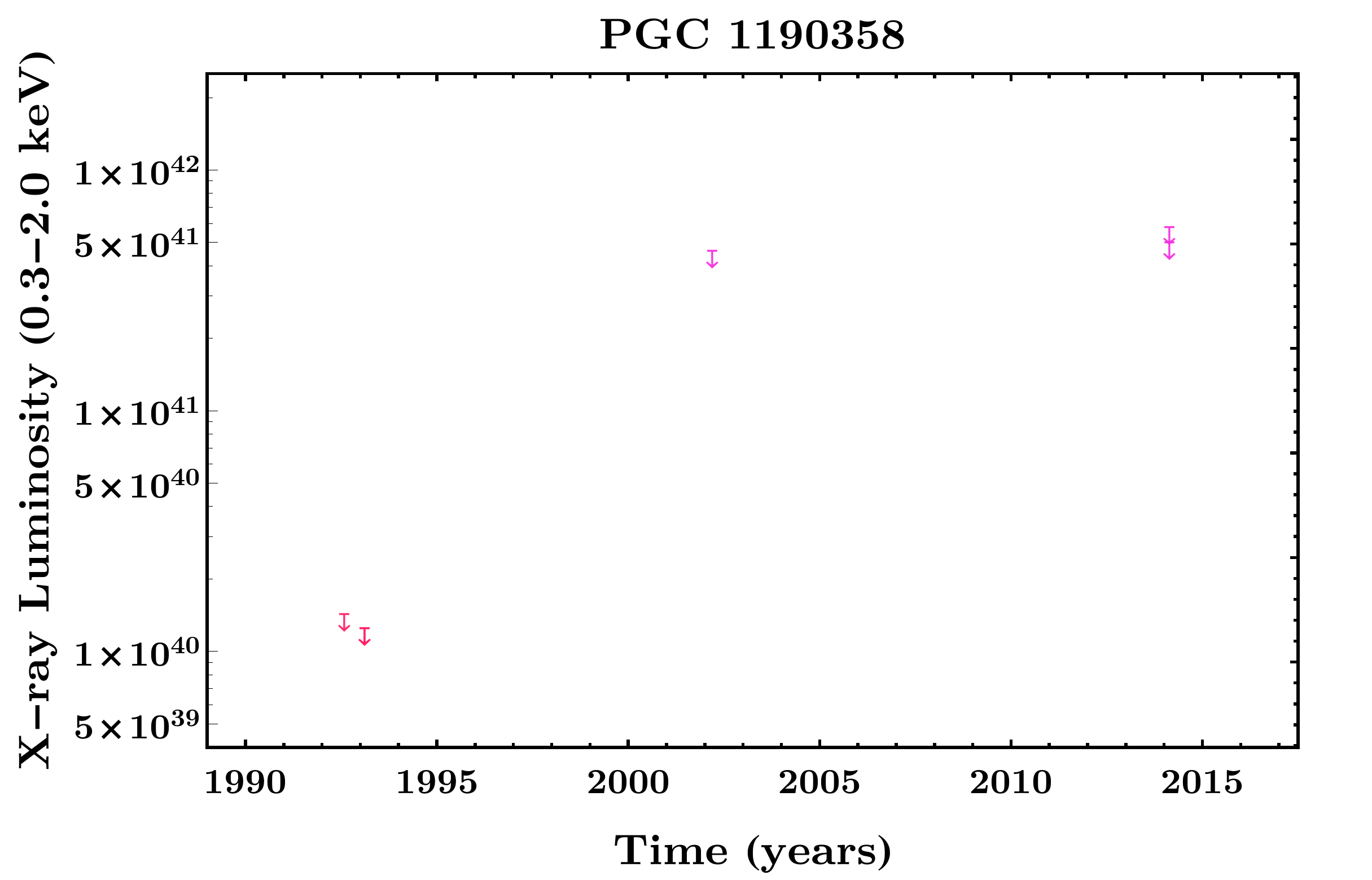}
		\includegraphics[width=0.31\textwidth]{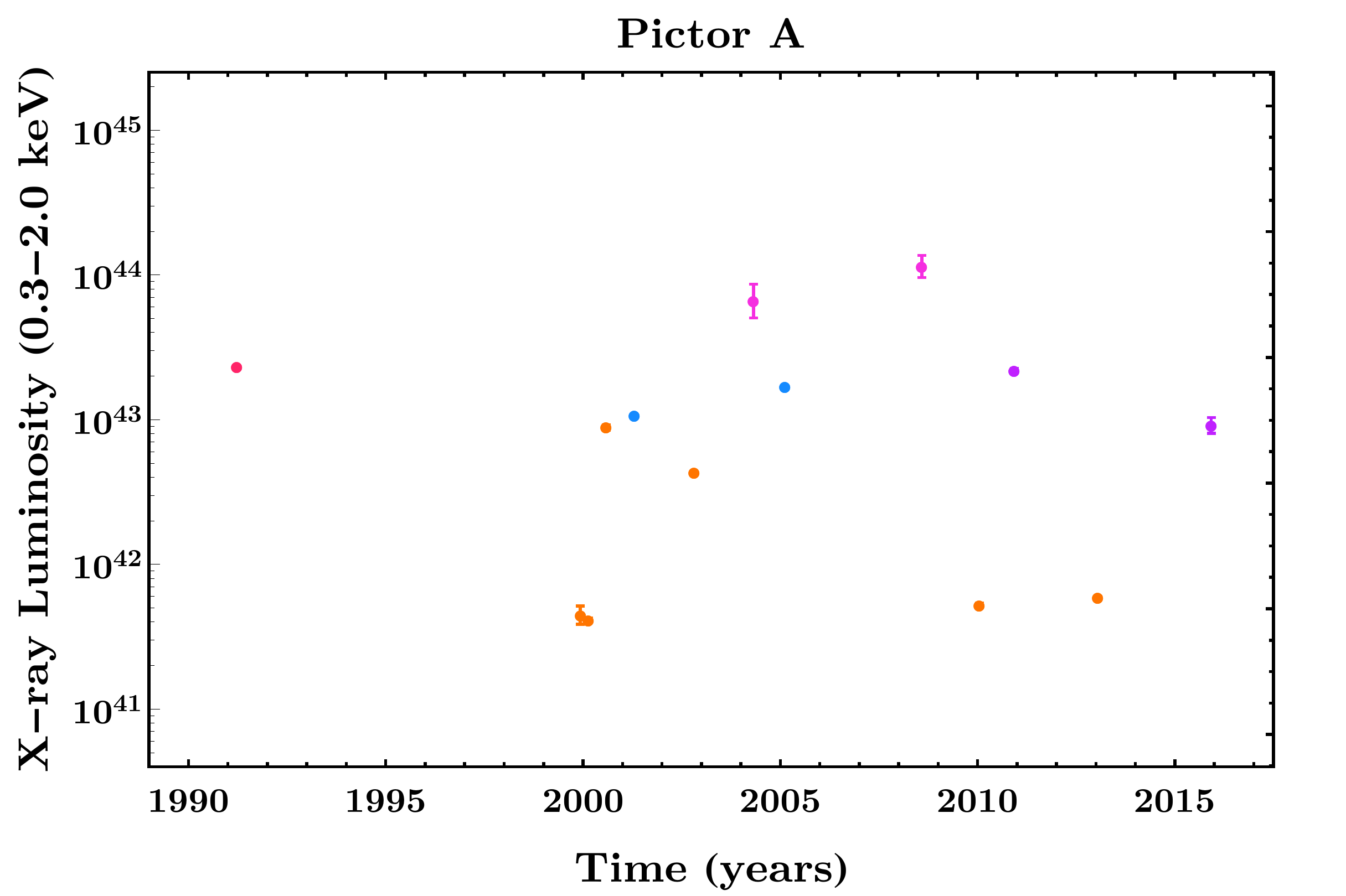}
		\includegraphics[width=0.31\textwidth]{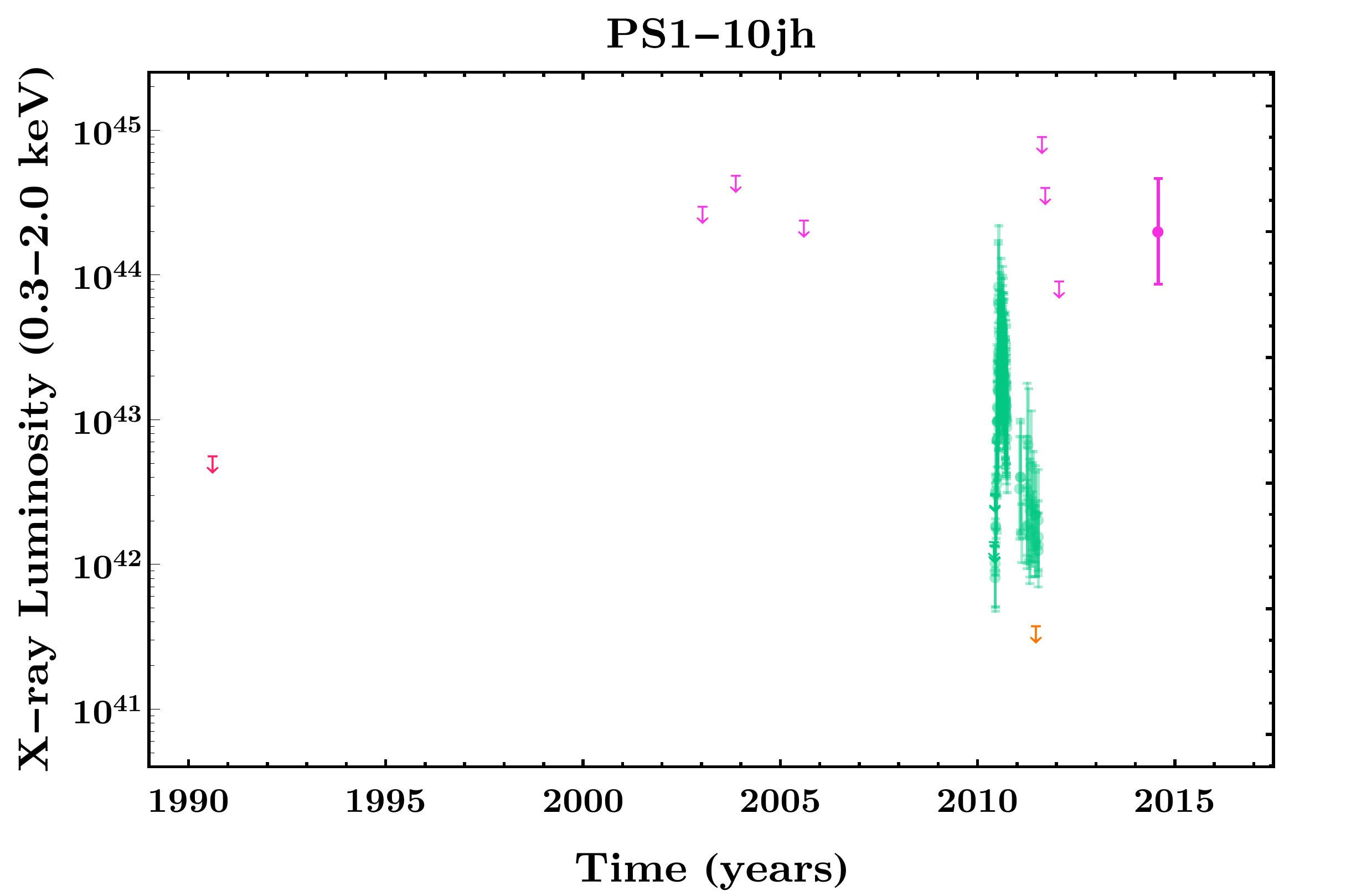}
		\includegraphics[width=0.31\textwidth]{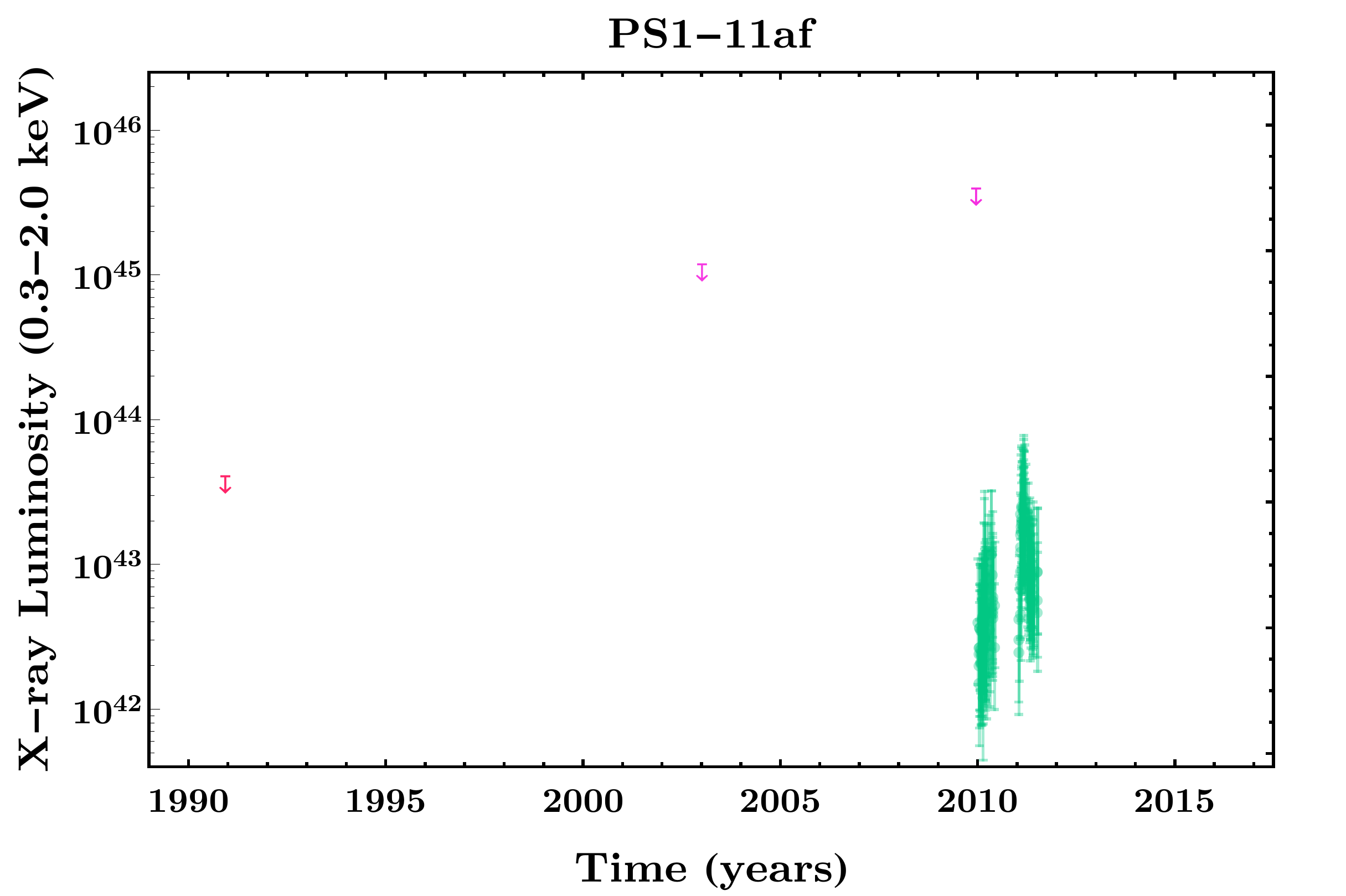}
		\includegraphics[width=0.31\textwidth]{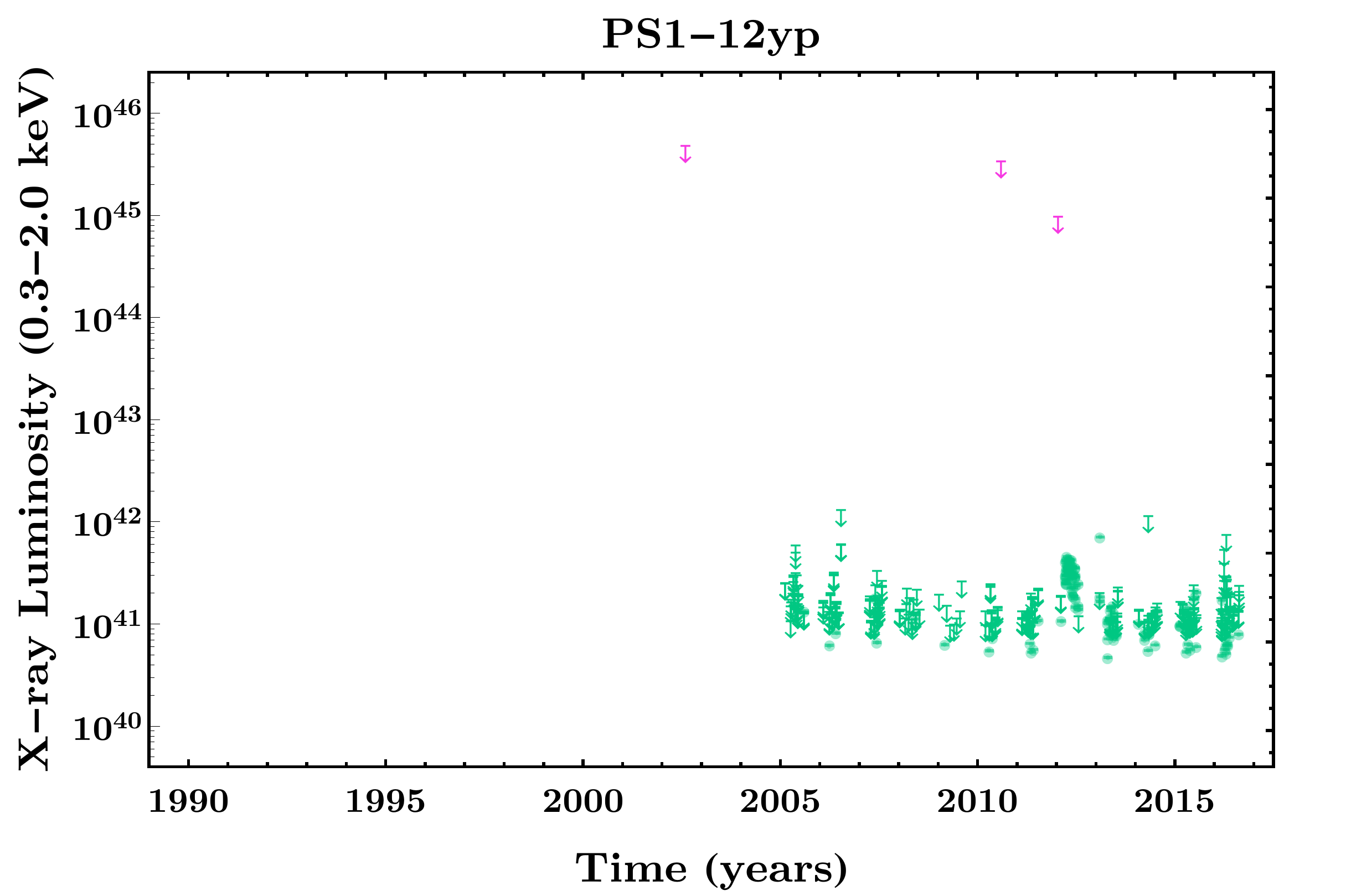}
		\includegraphics[width=0.31\textwidth]{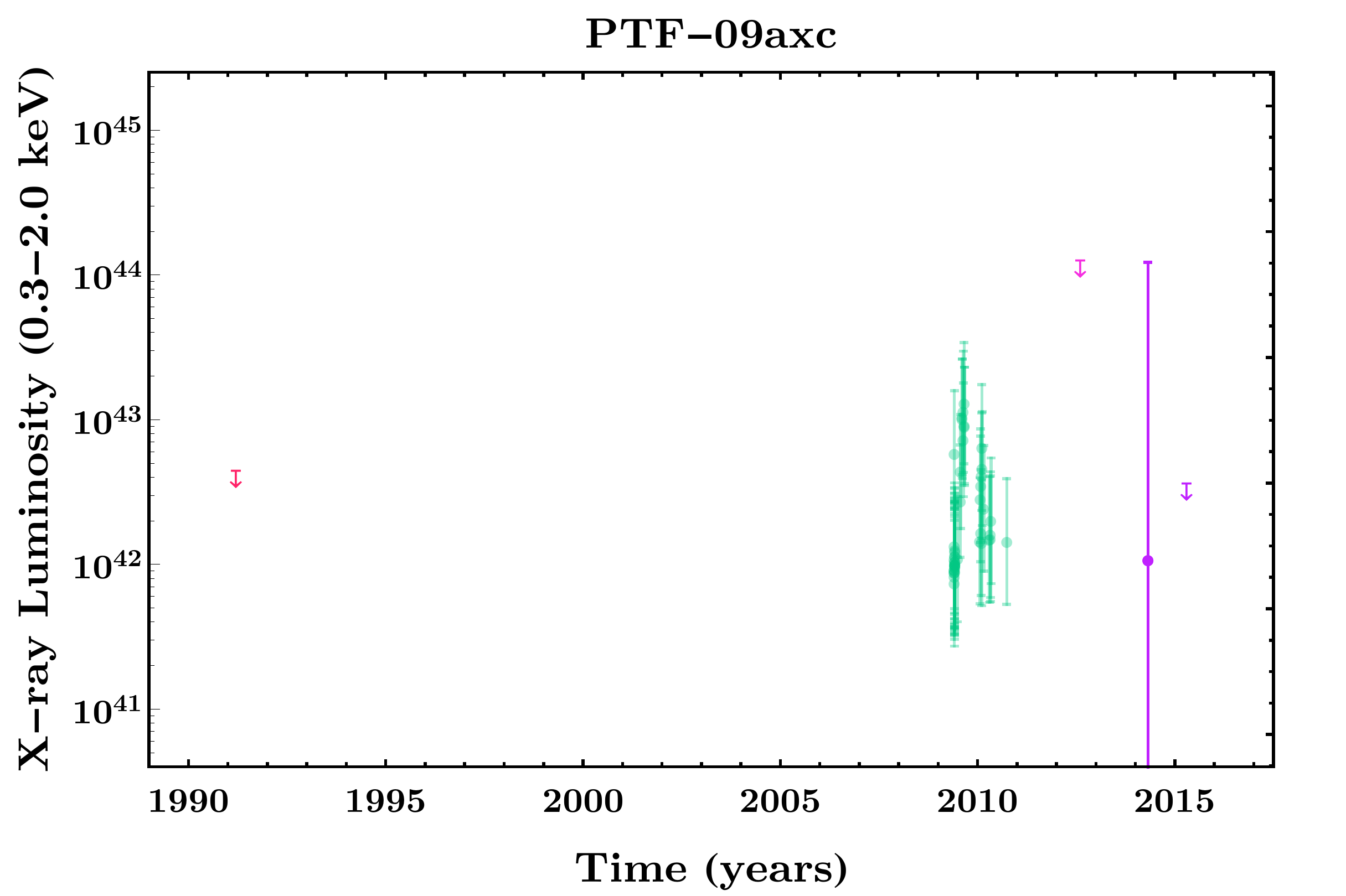}
		\includegraphics[width=0.31\textwidth]{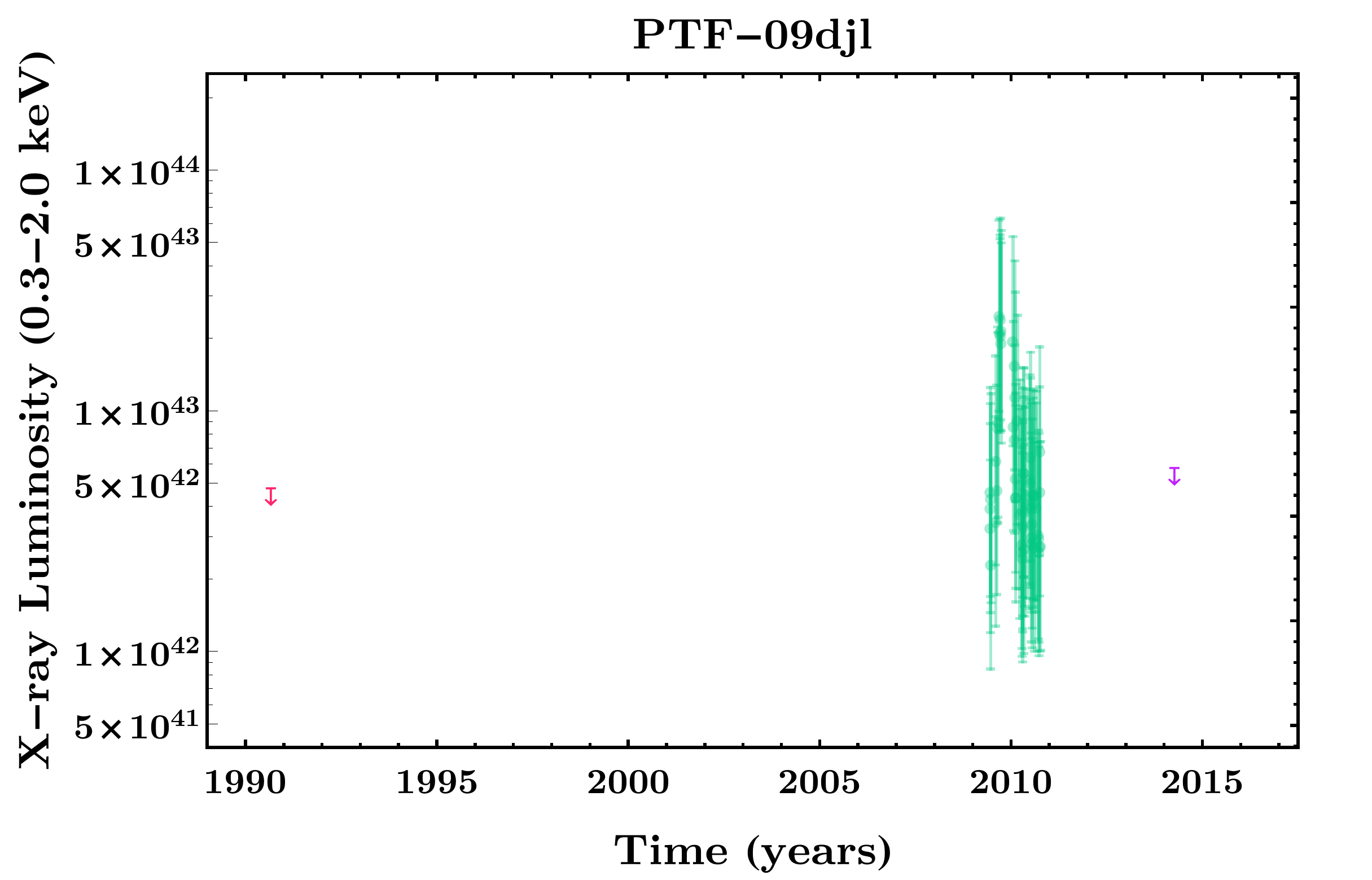}
		\includegraphics[width=0.10\textwidth]{lightcurve_key.pdf}
		\caption{Light Curves for all TDE candidates listed in Table 1 continued. Similar to that of Figure~\ref{lightcurves1}. \label{lightcurves2}}
	\end{center}
\end{figure*}

\begin{figure*}
	\begin{center}
			\includegraphics[width=0.31\textwidth]{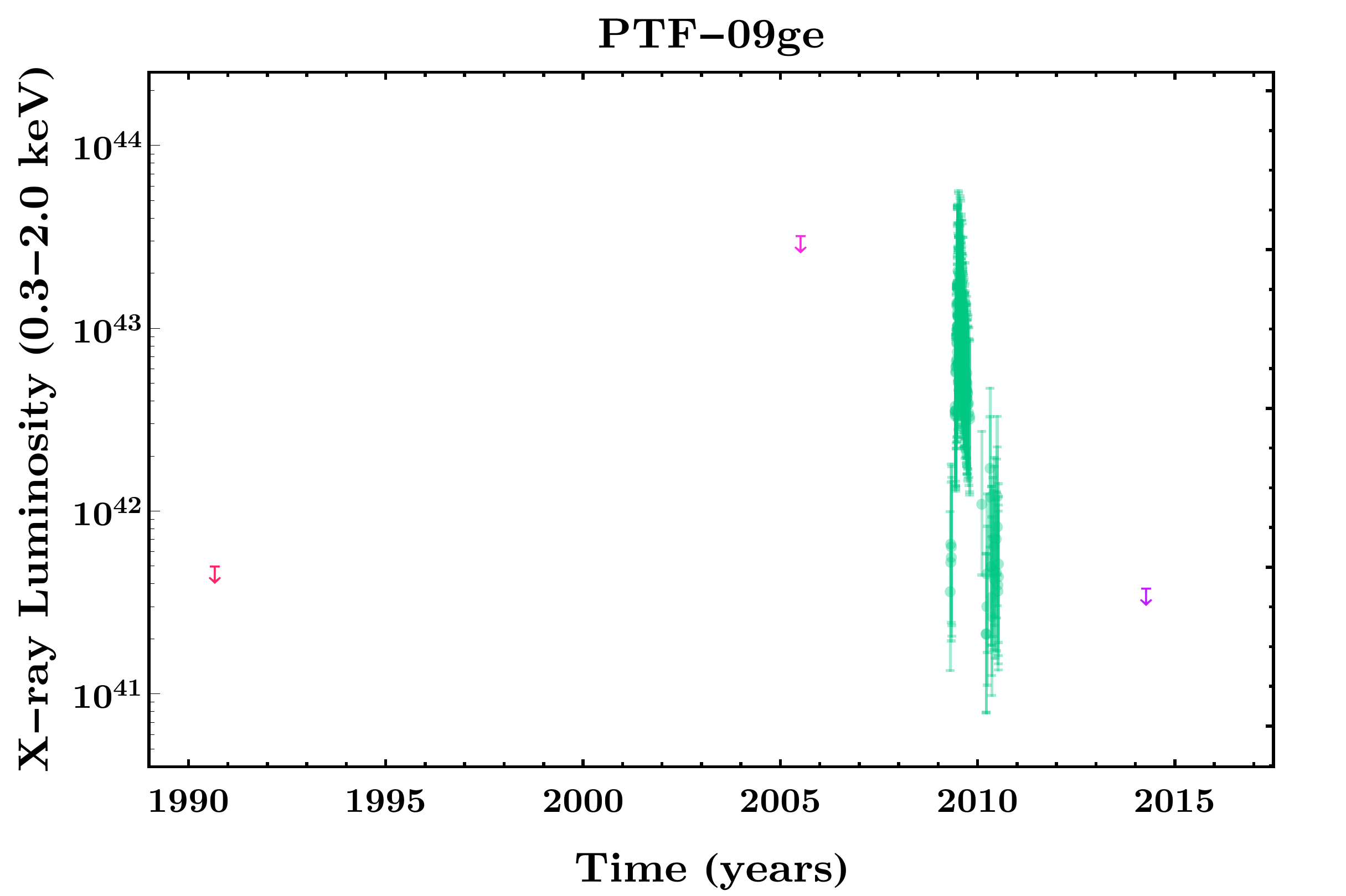}
		\includegraphics[width=0.31\textwidth]{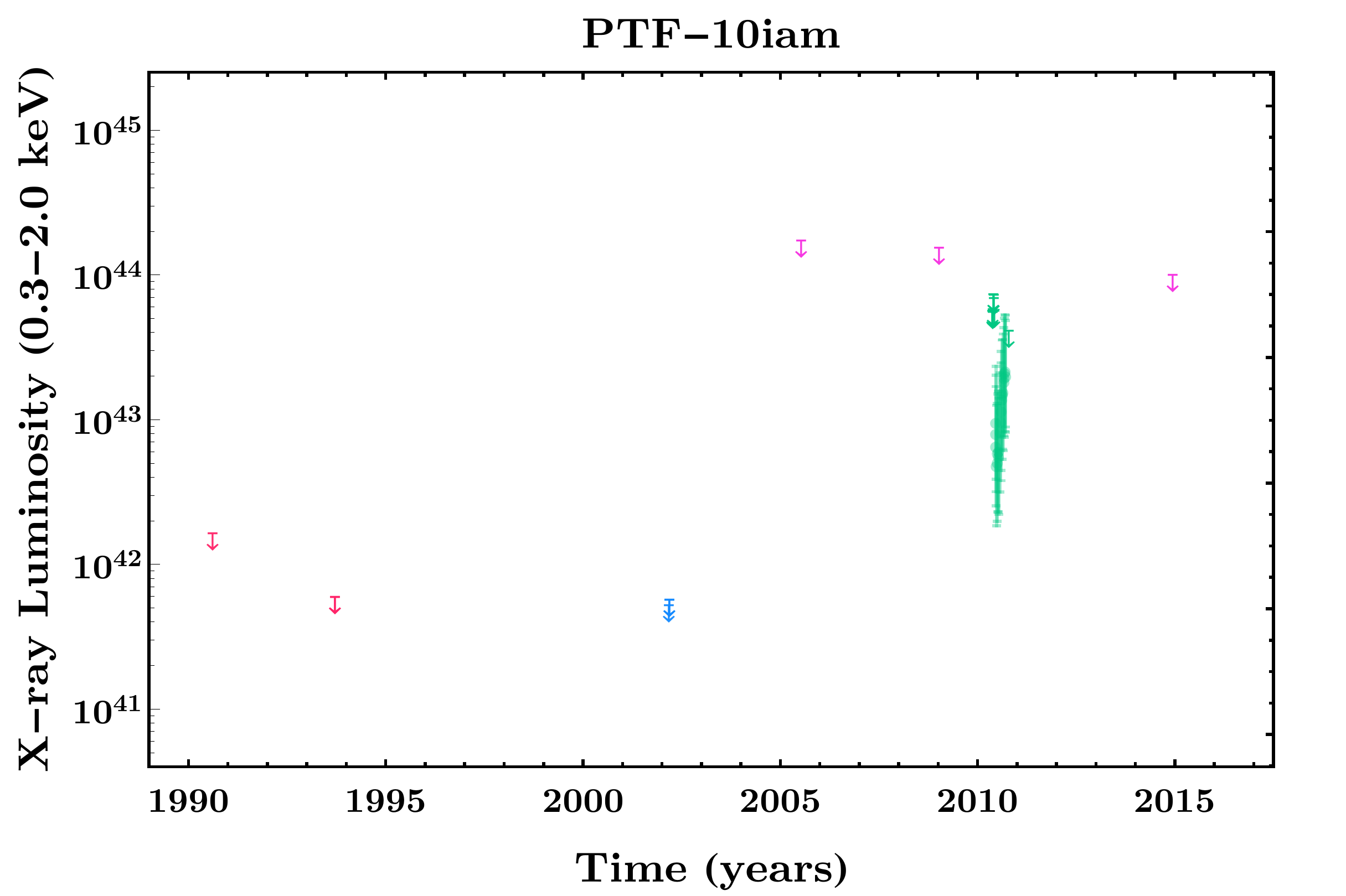}
		\includegraphics[width=0.31\textwidth]{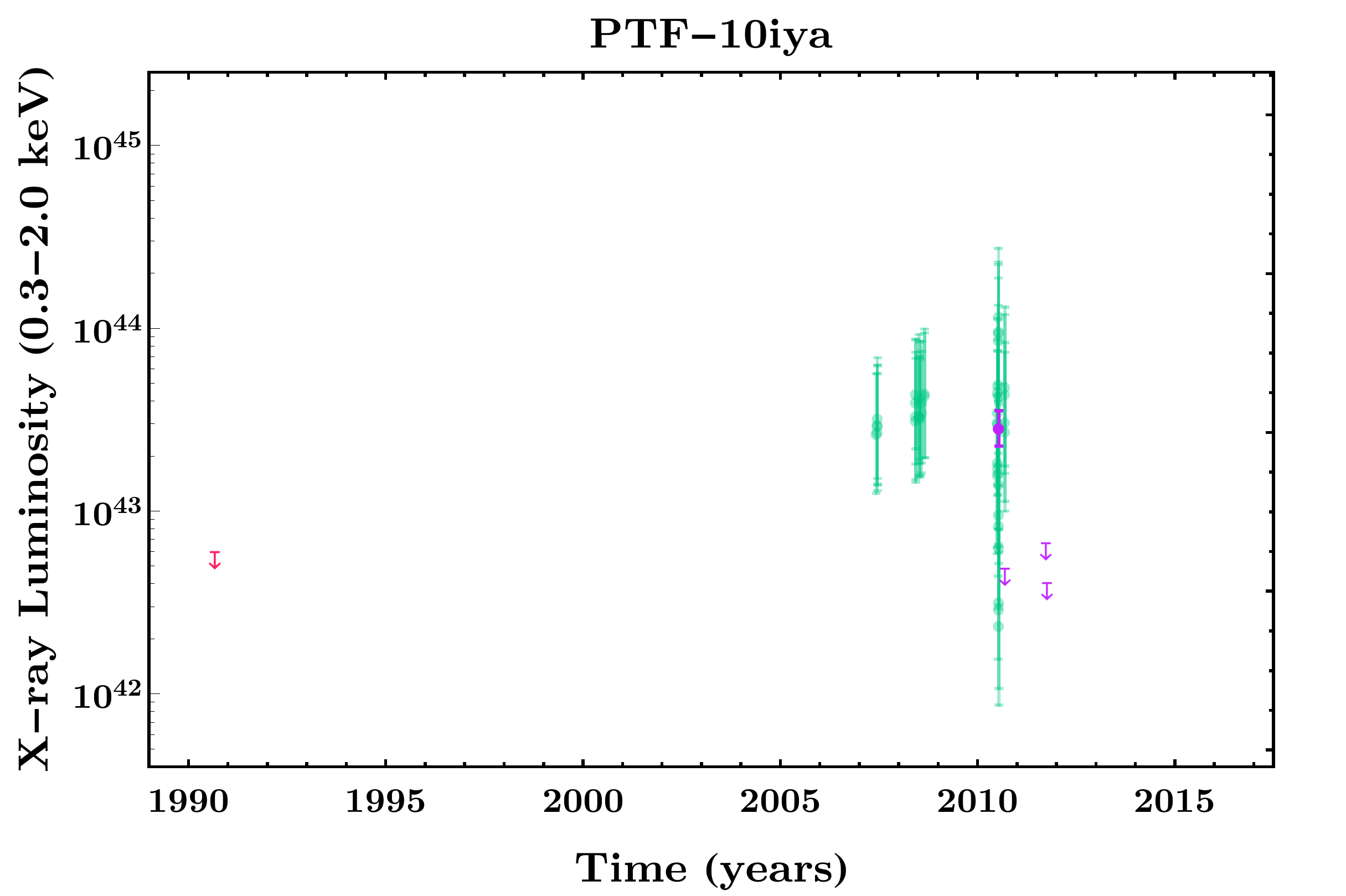}
		\includegraphics[width=0.31\textwidth]{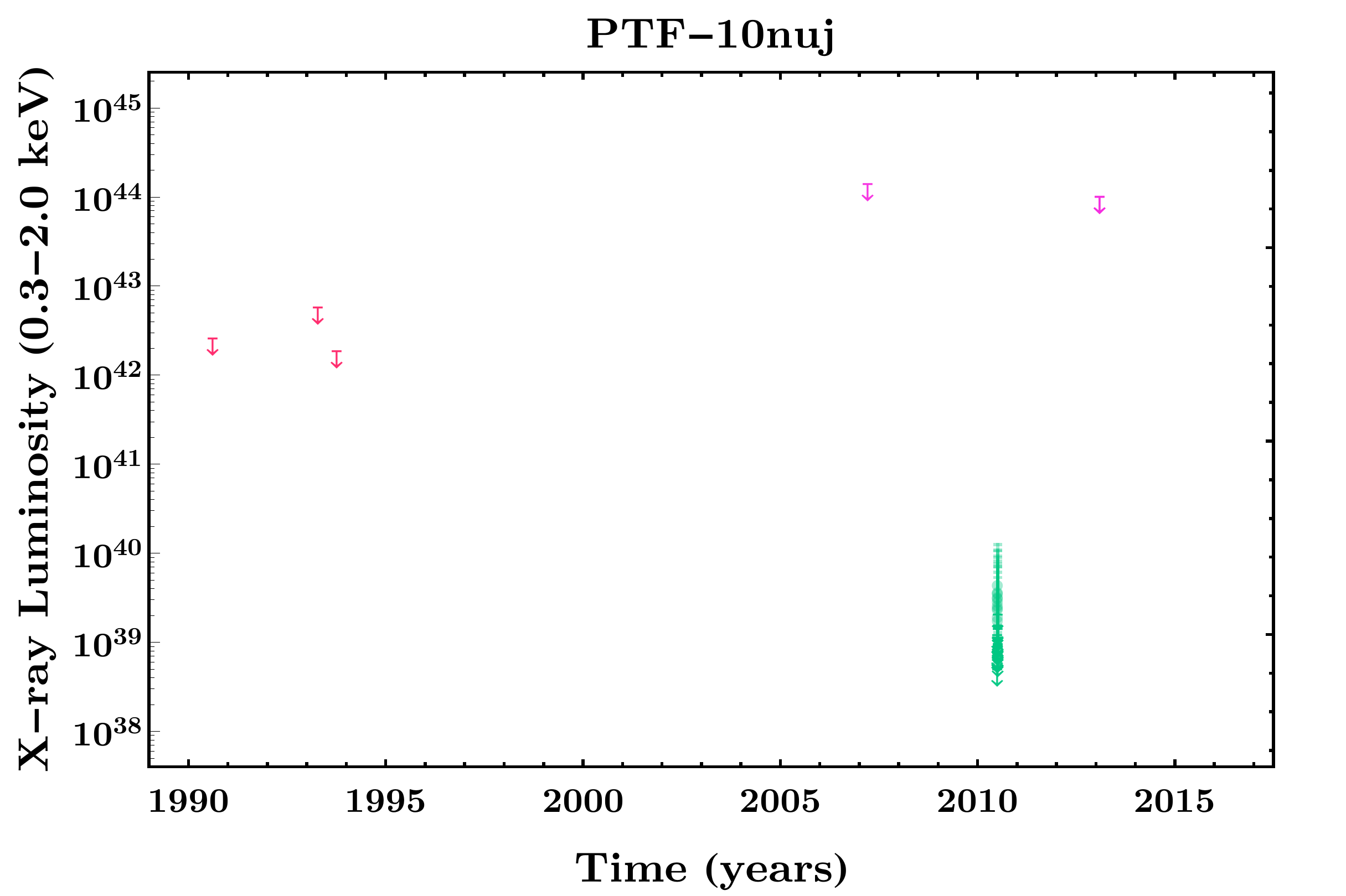}
		\includegraphics[width=0.31\textwidth]{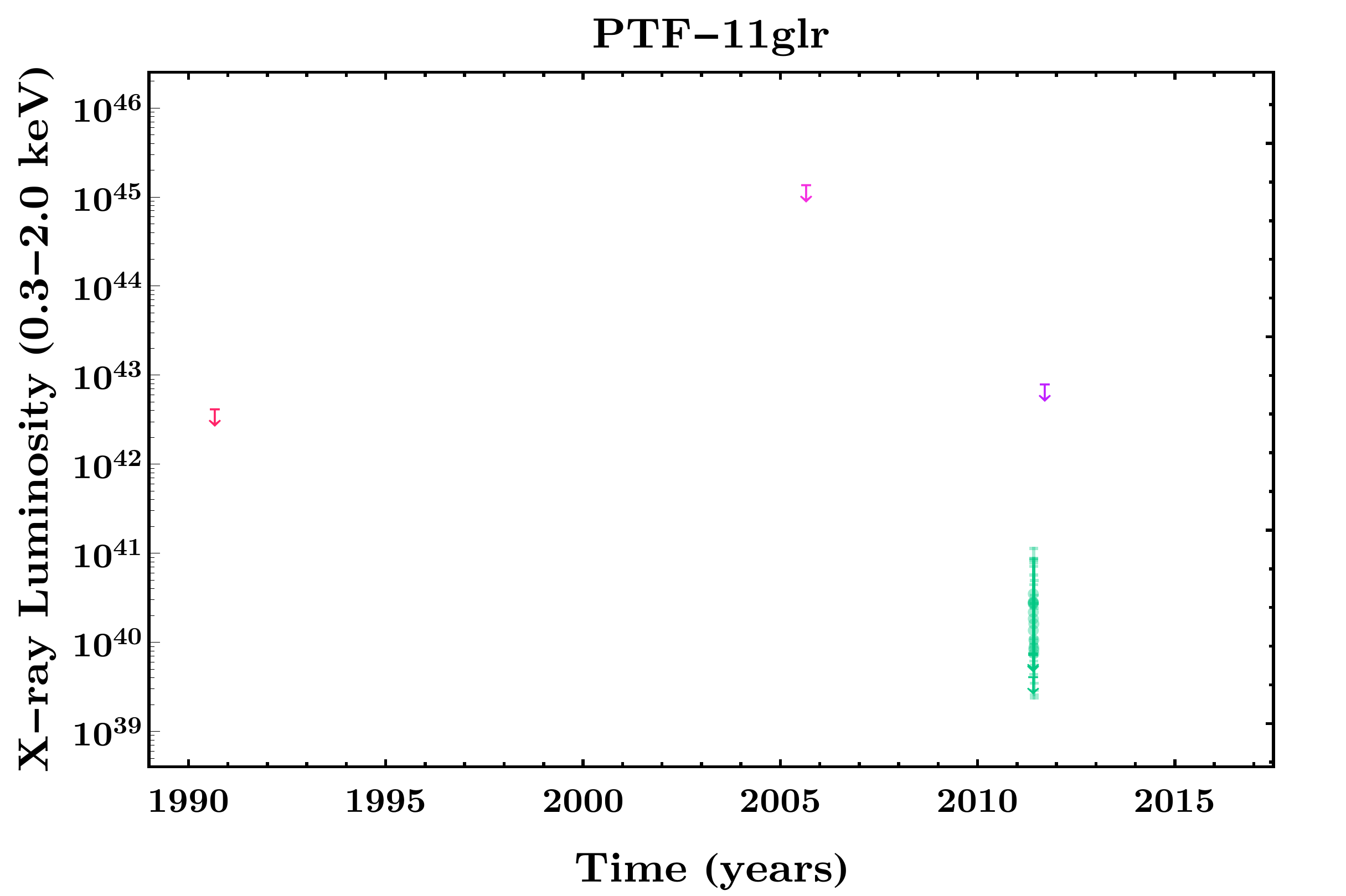}
		\includegraphics[width=0.31\textwidth]{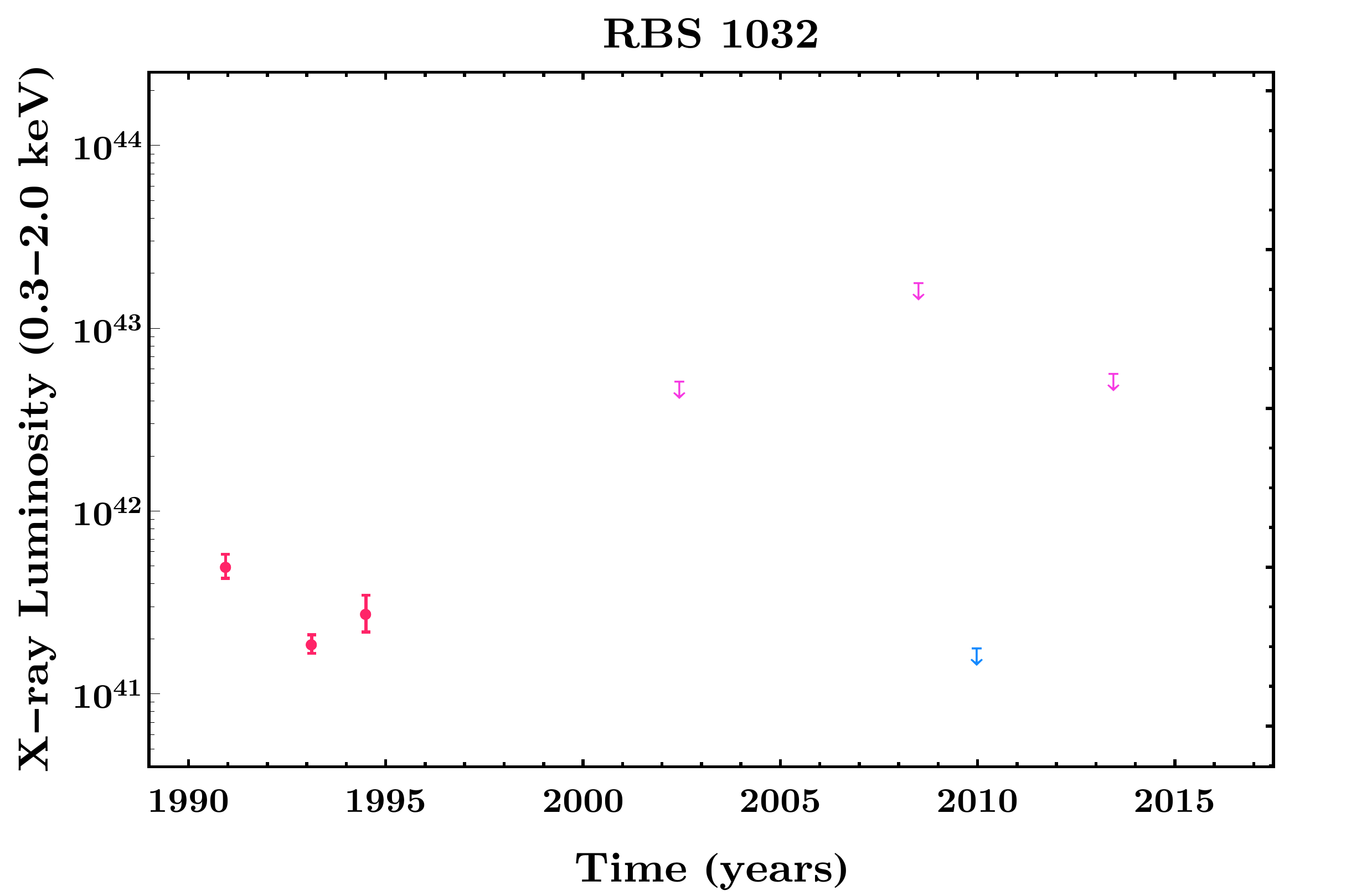}
		\includegraphics[width=0.31\textwidth]{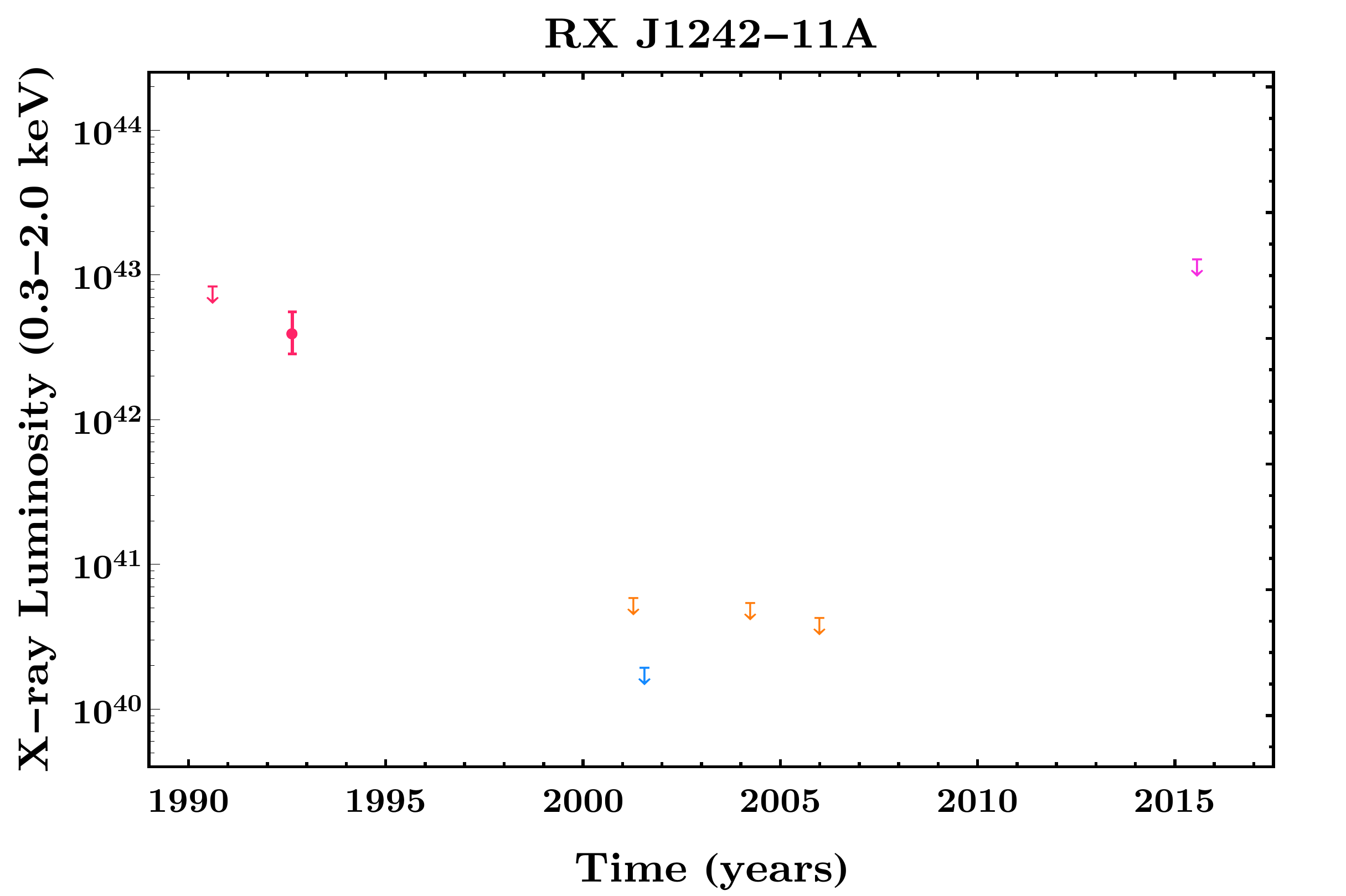}
		\includegraphics[width=0.31\textwidth]{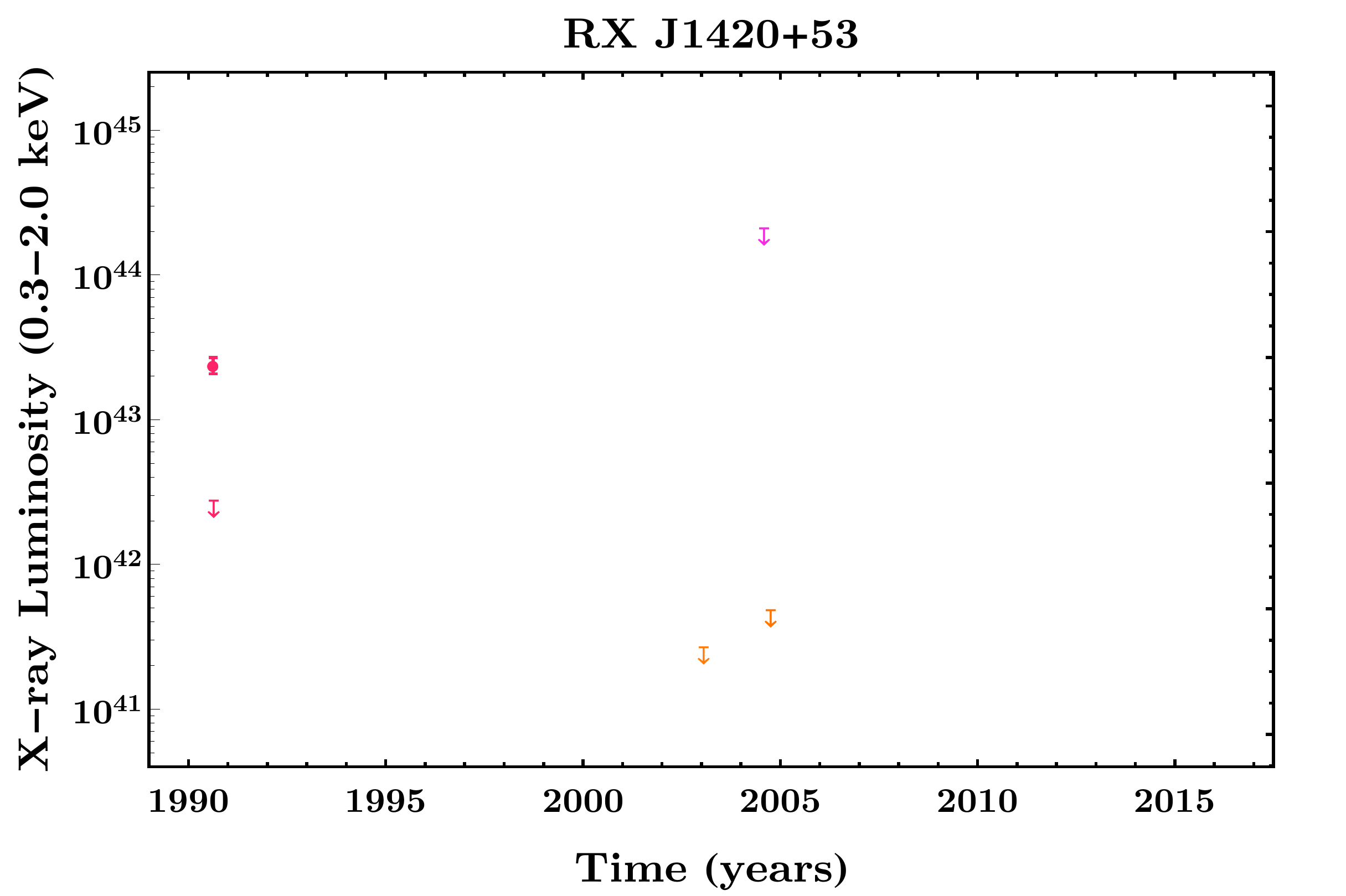}
		\includegraphics[width=0.31\textwidth]{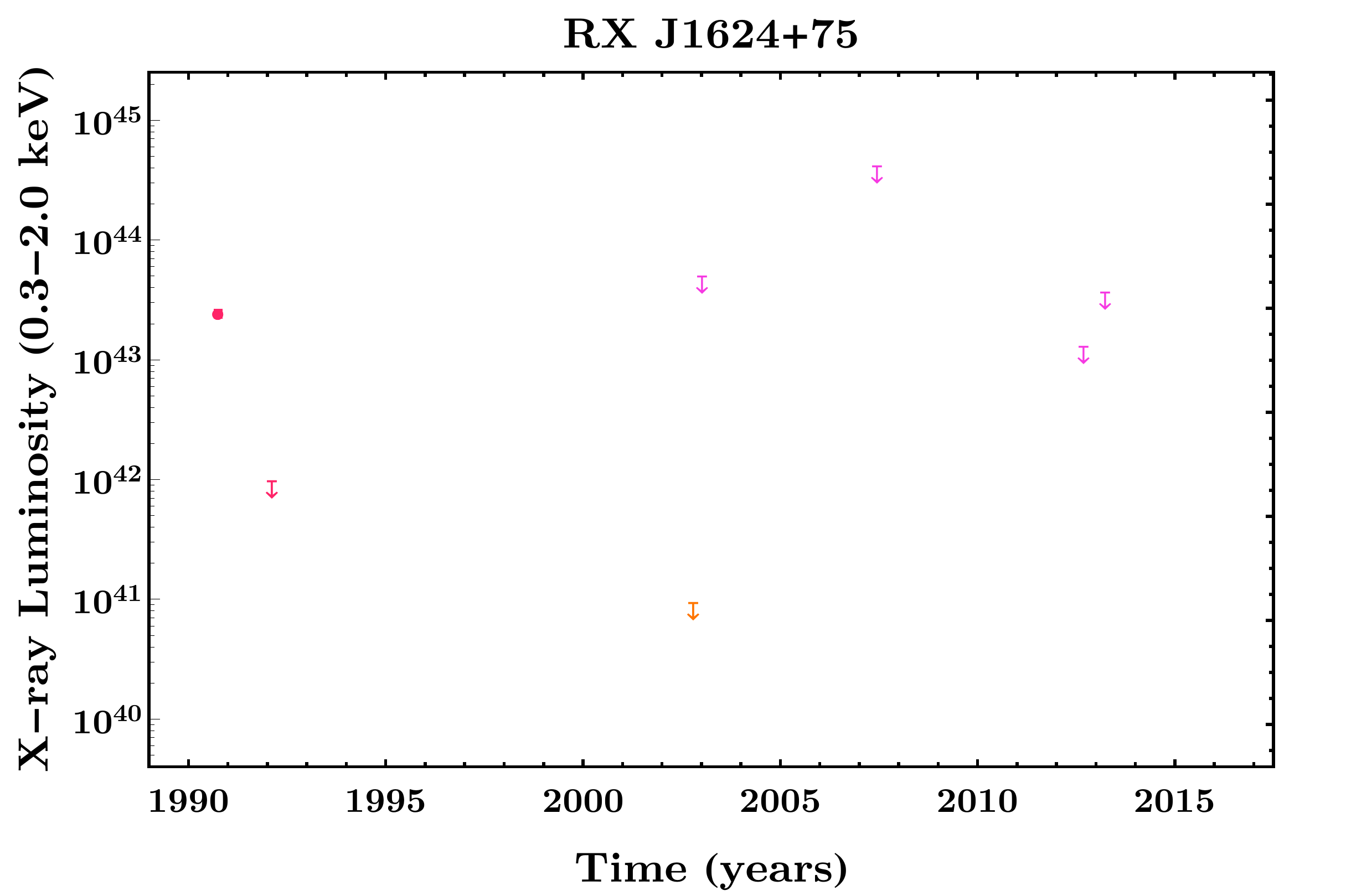}
		\includegraphics[width=0.31\textwidth]{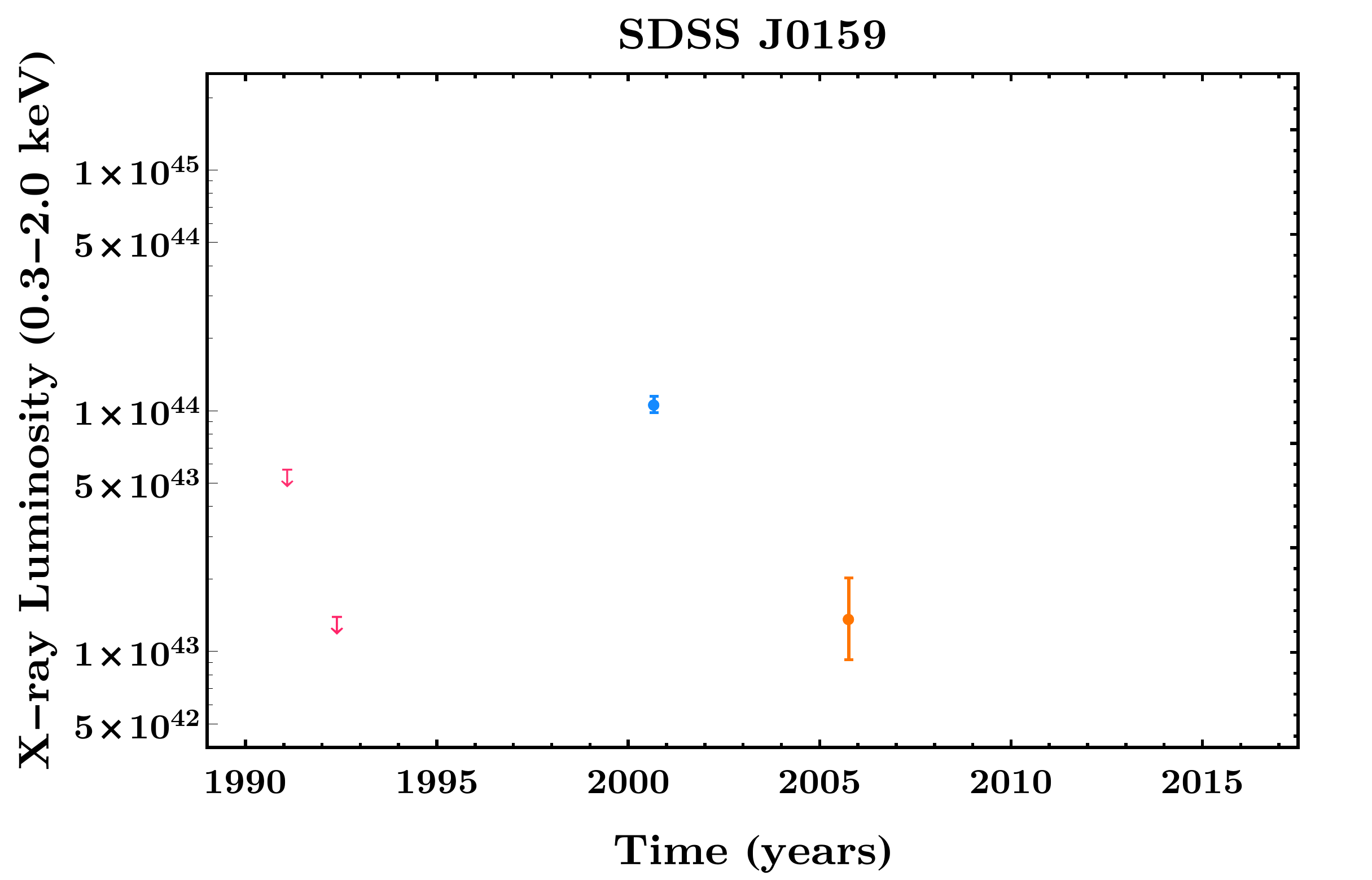}
		\includegraphics[width=0.31\textwidth]{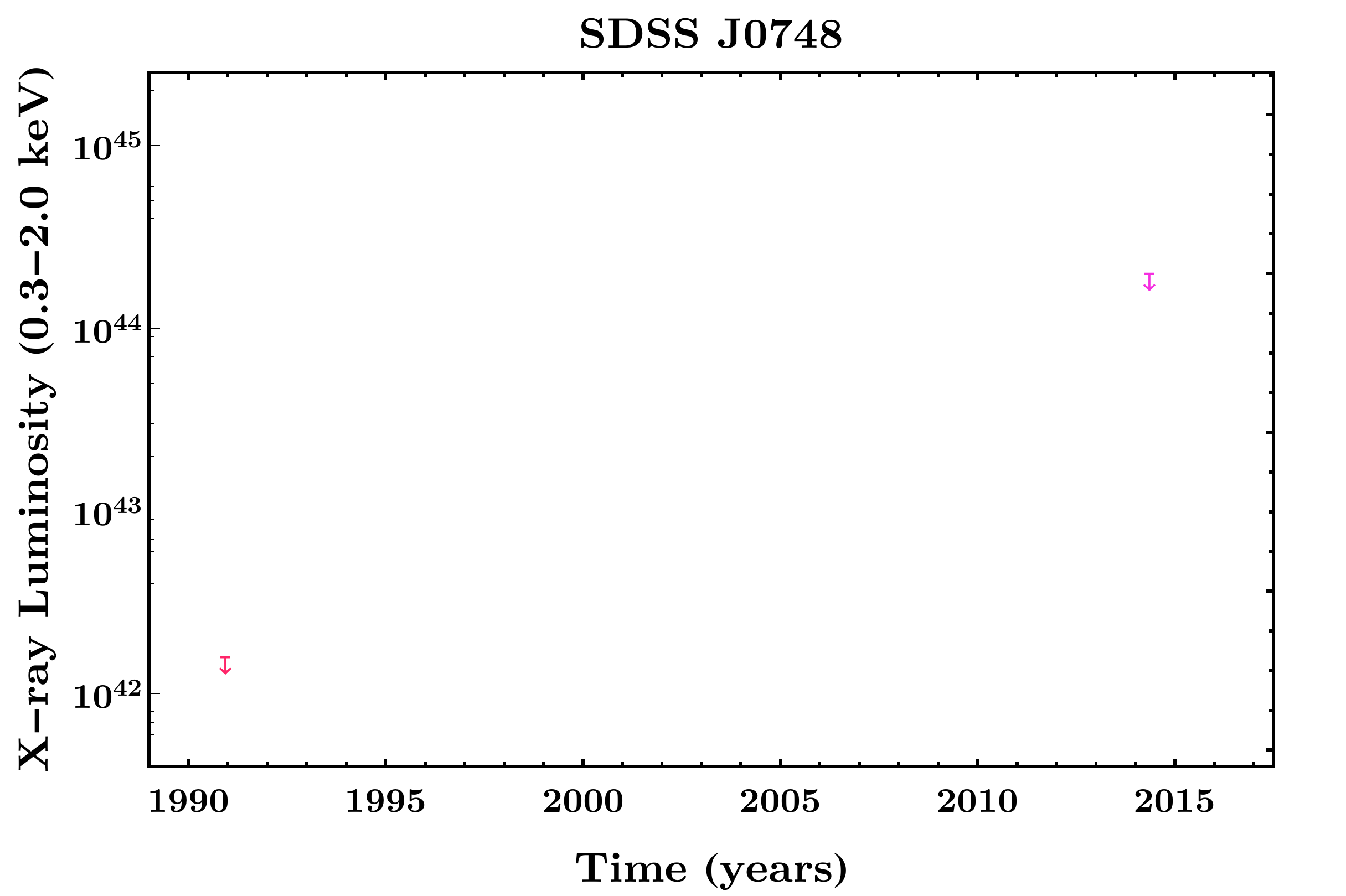}
		\includegraphics[width=0.31\textwidth]{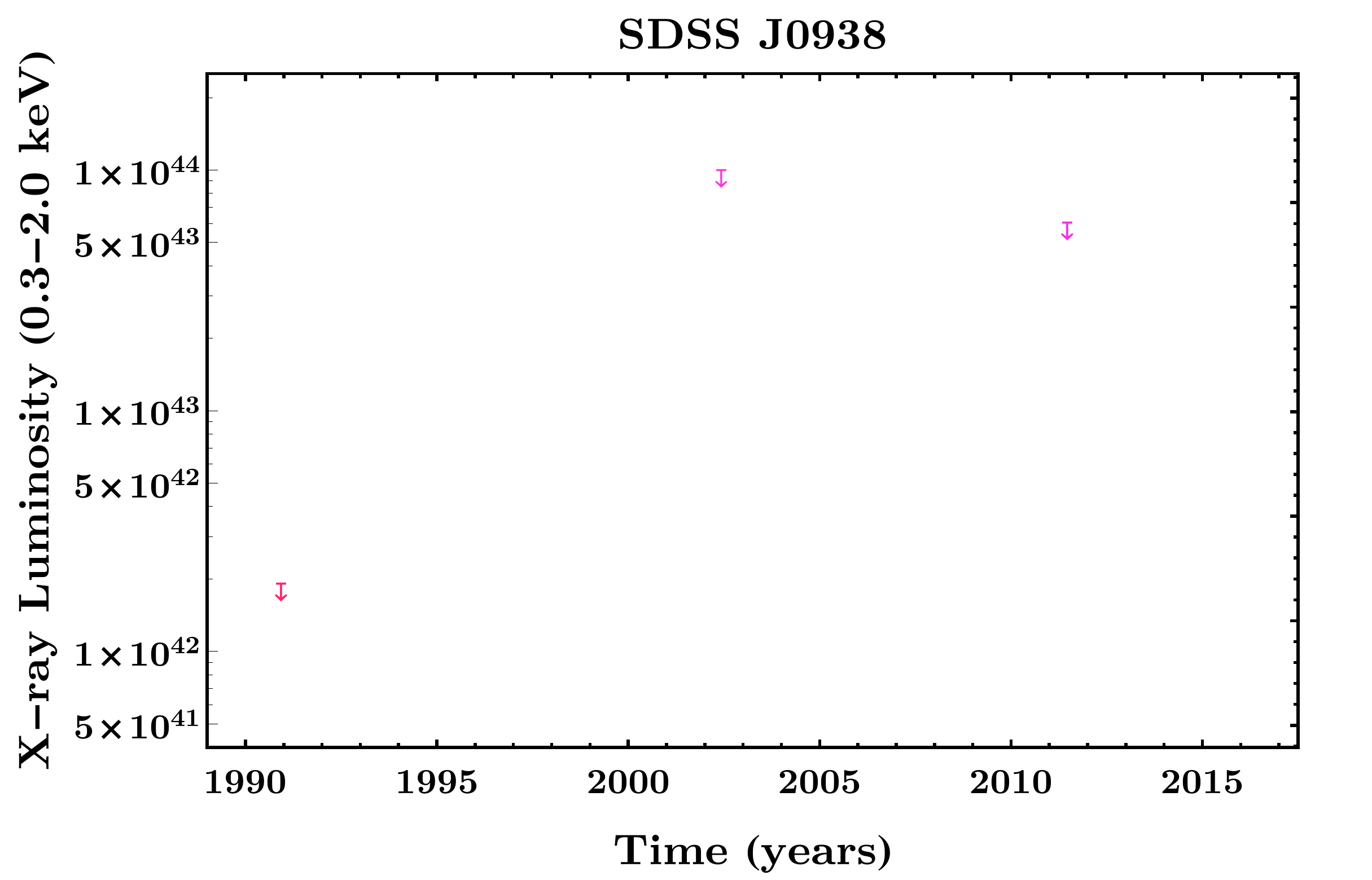}
		\includegraphics[width=0.31\textwidth]{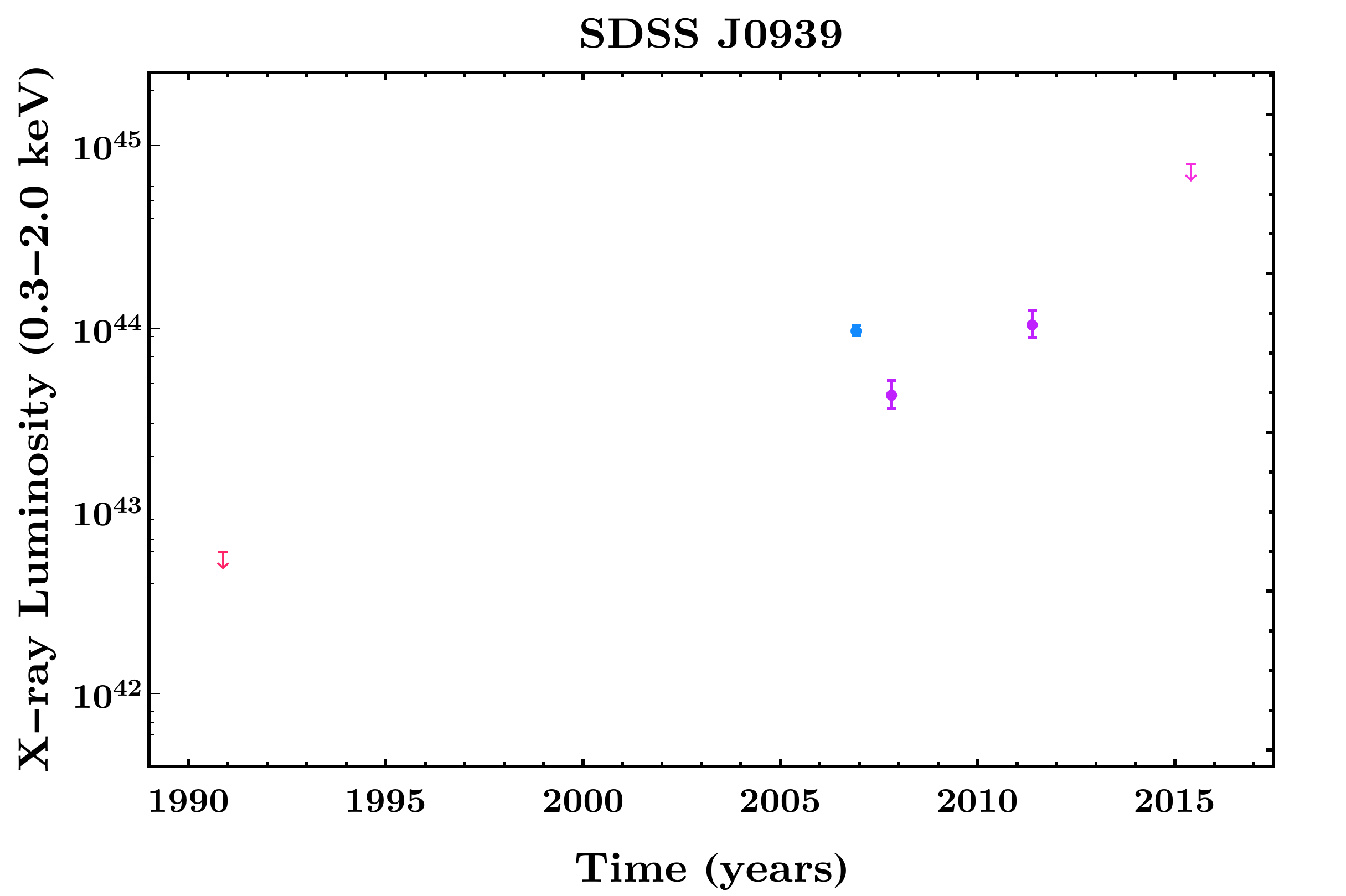}
		\includegraphics[width=0.31\textwidth]{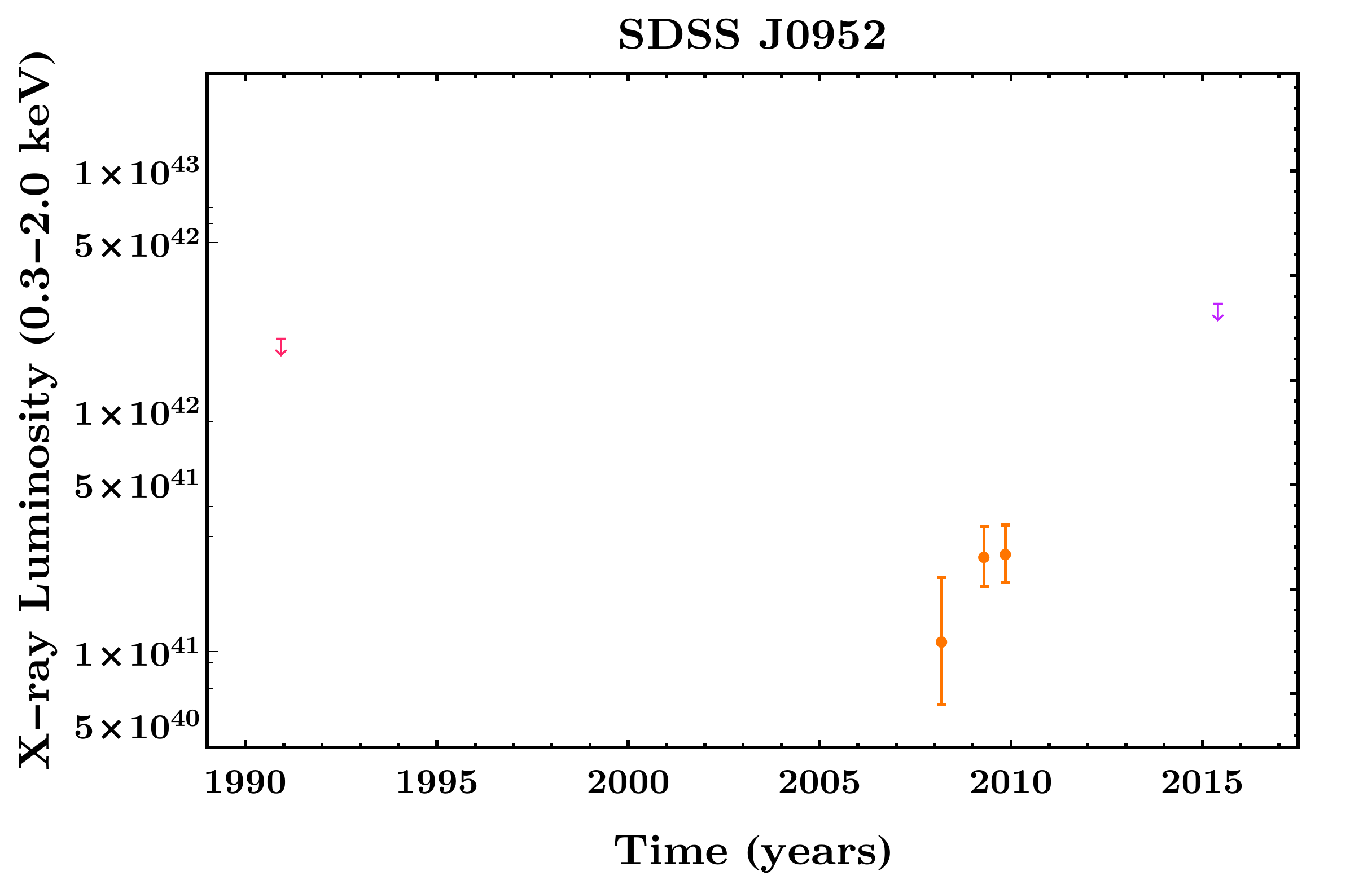}
		\includegraphics[width=0.31\textwidth]{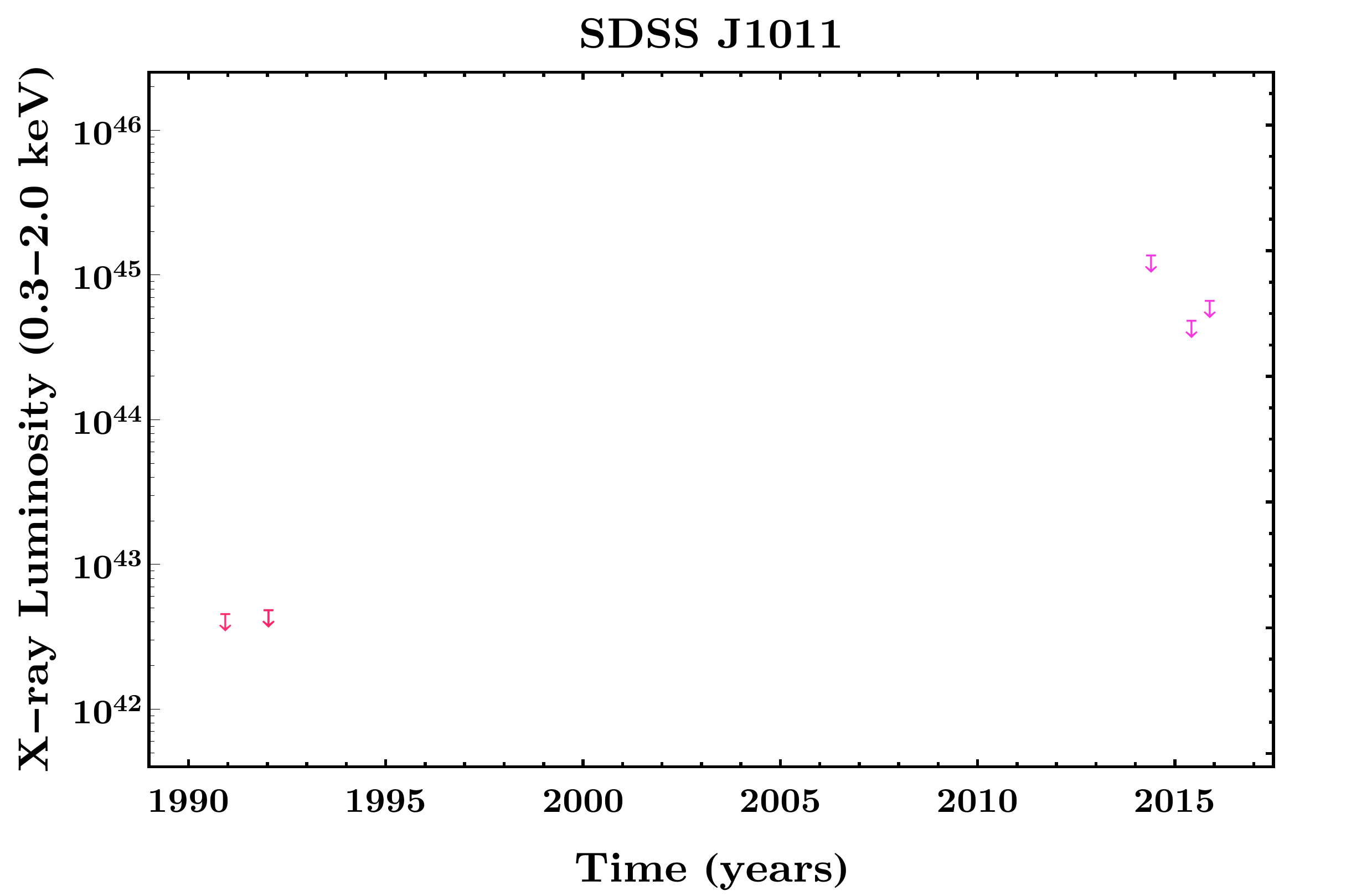}
		\includegraphics[width=0.31\textwidth]{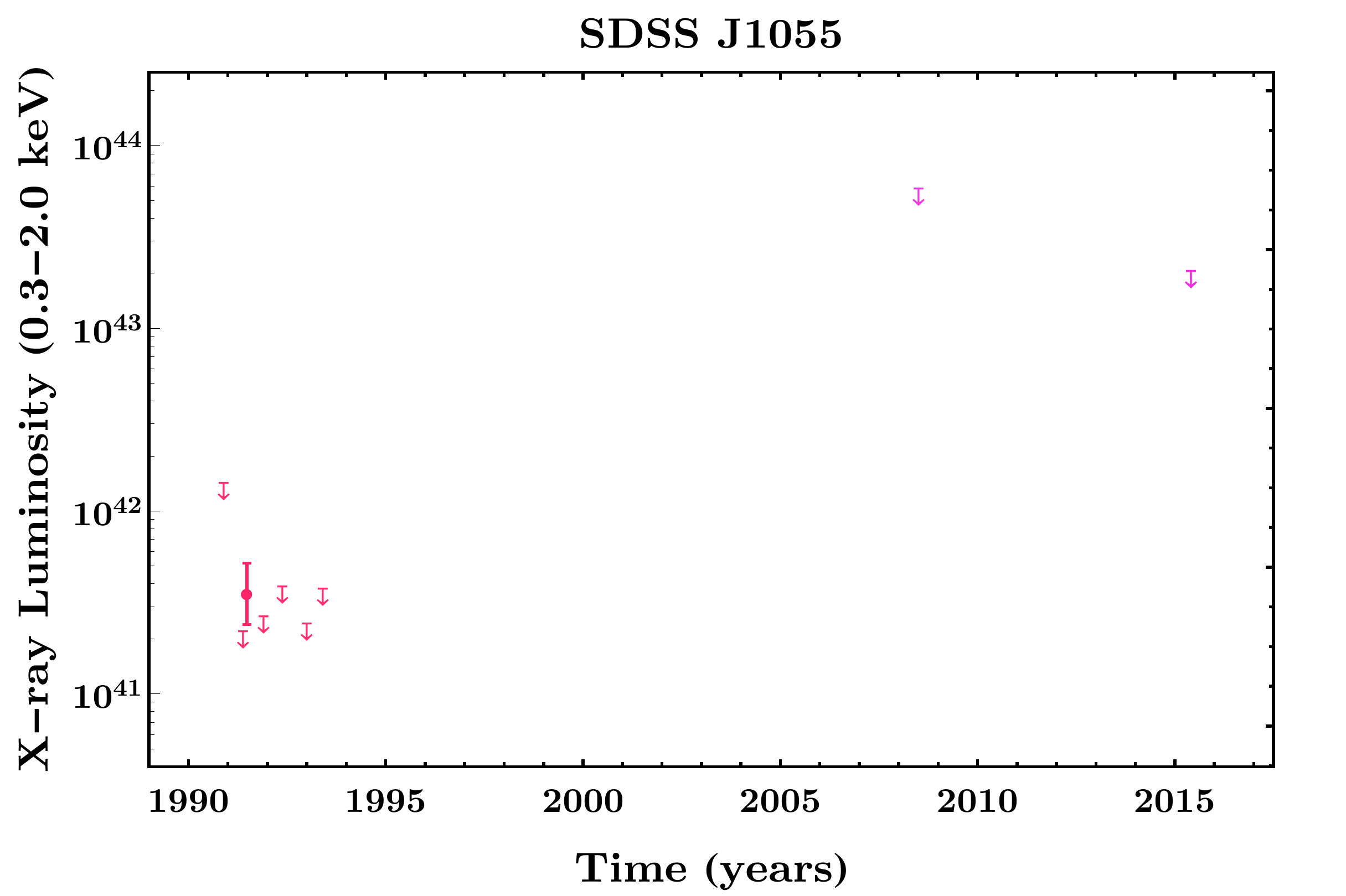}
		\includegraphics[width=0.31\textwidth]{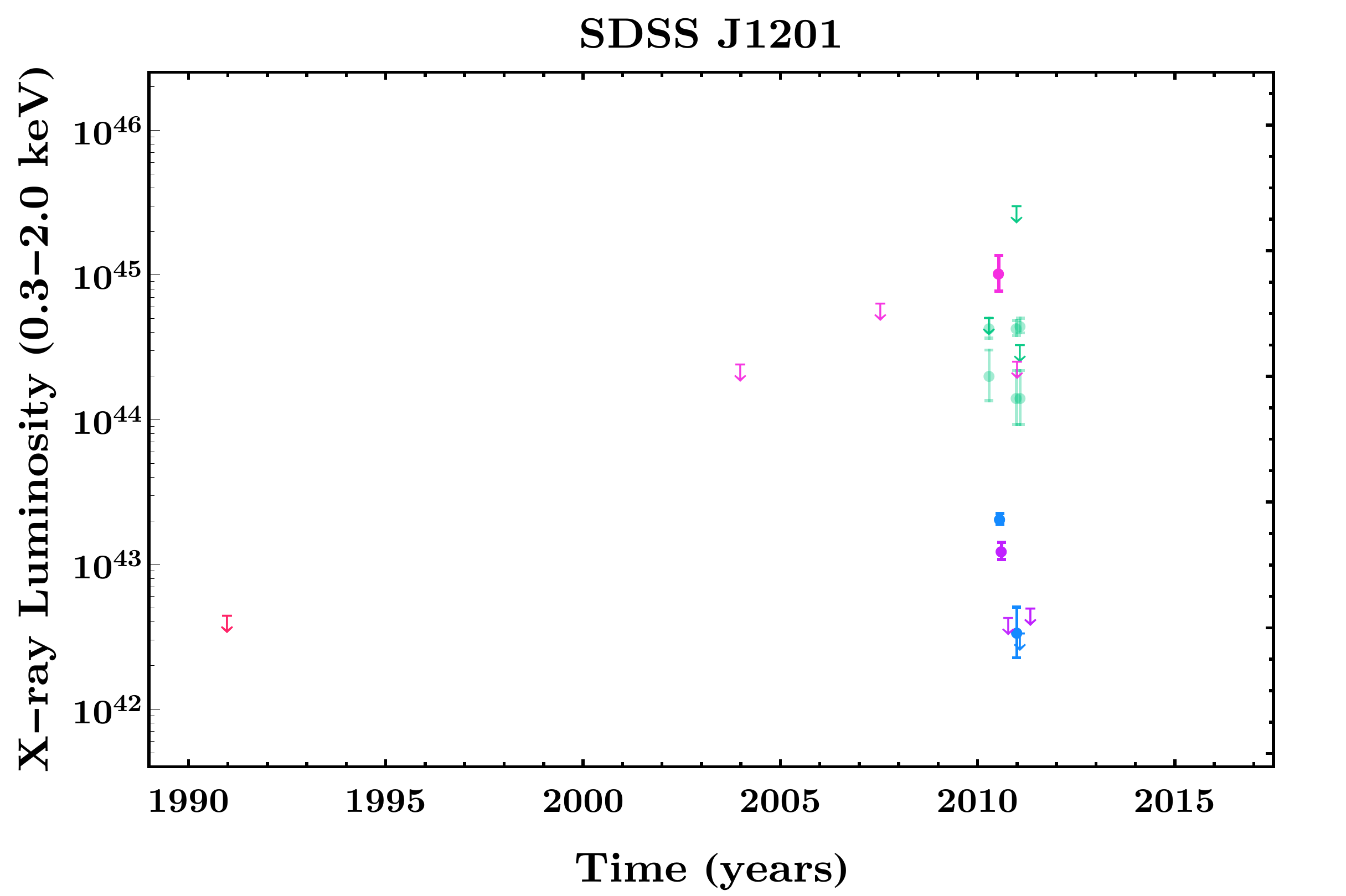}
		\includegraphics[width=0.10\textwidth]{lightcurve_key.pdf}
		
		\caption{Light Curves for all TDE candidates listed in Table 1 continued. Similar to that of Figure~\ref{lightcurves1}. \label{lightcurve3}}
		
	\end{center}
\end{figure*}

\begin{figure*}
	\begin{center}
			\includegraphics[width=0.31\textwidth]{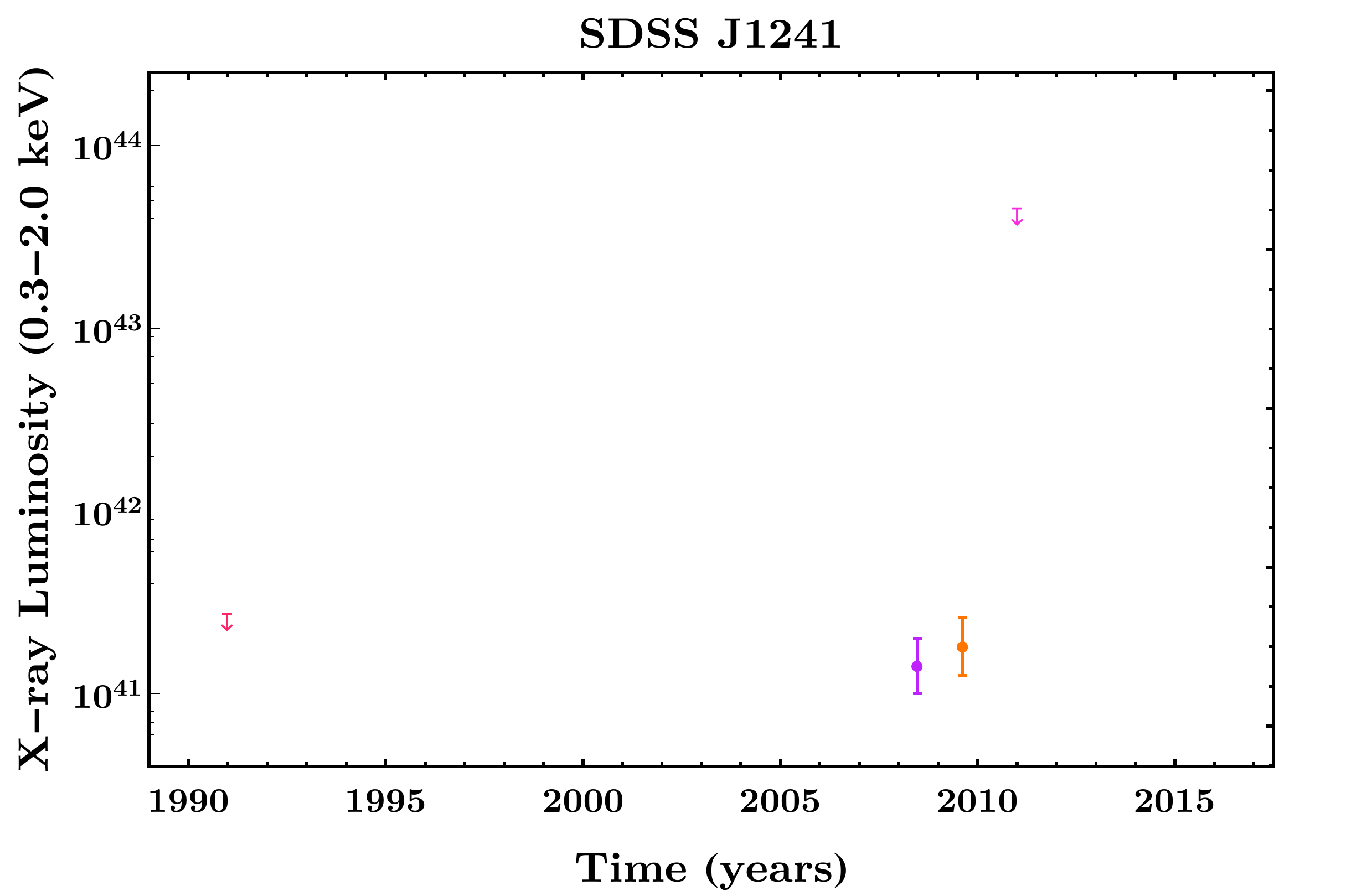}
		\includegraphics[width=0.31\textwidth]{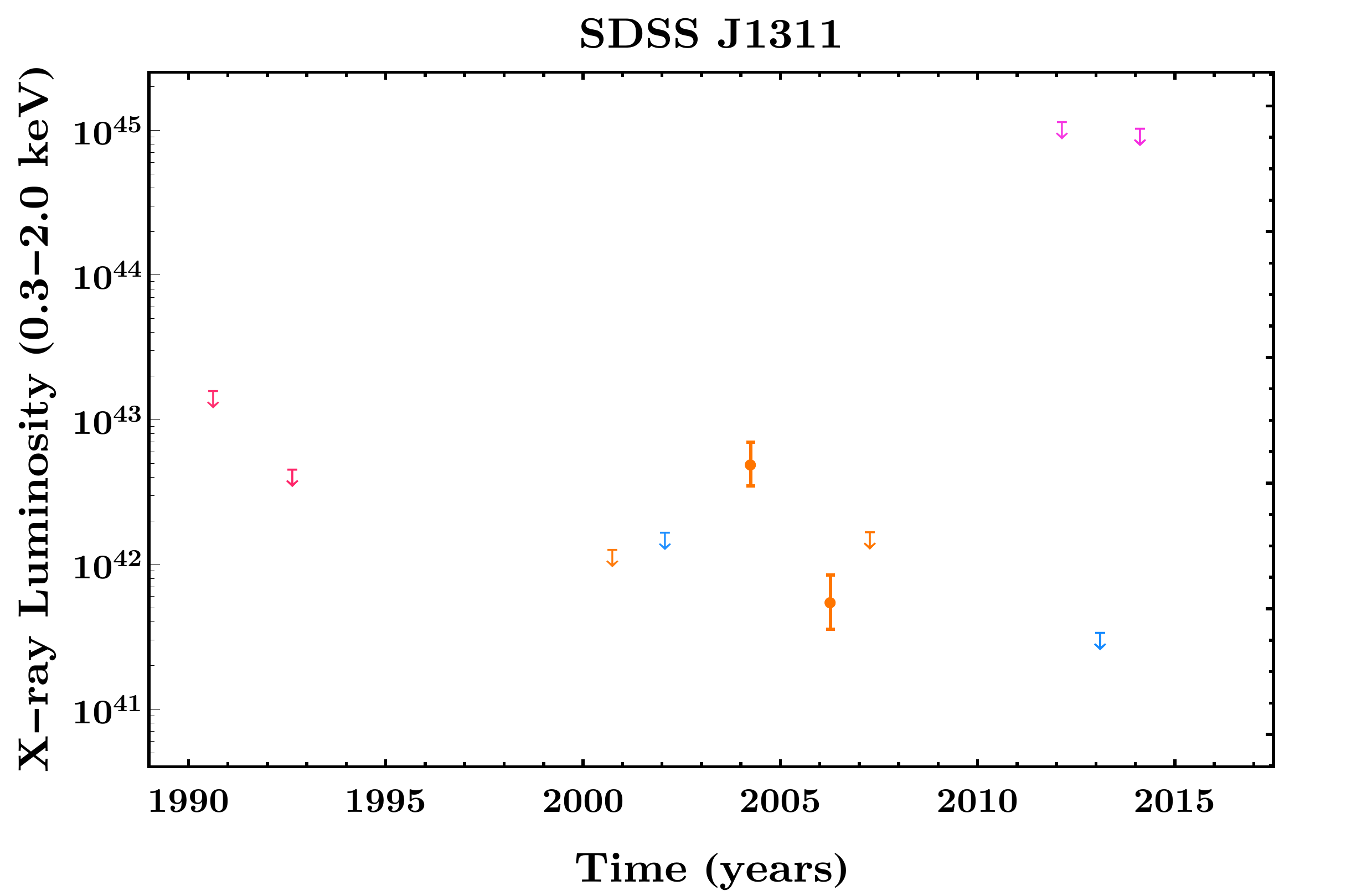}
		\includegraphics[width=0.31\textwidth]{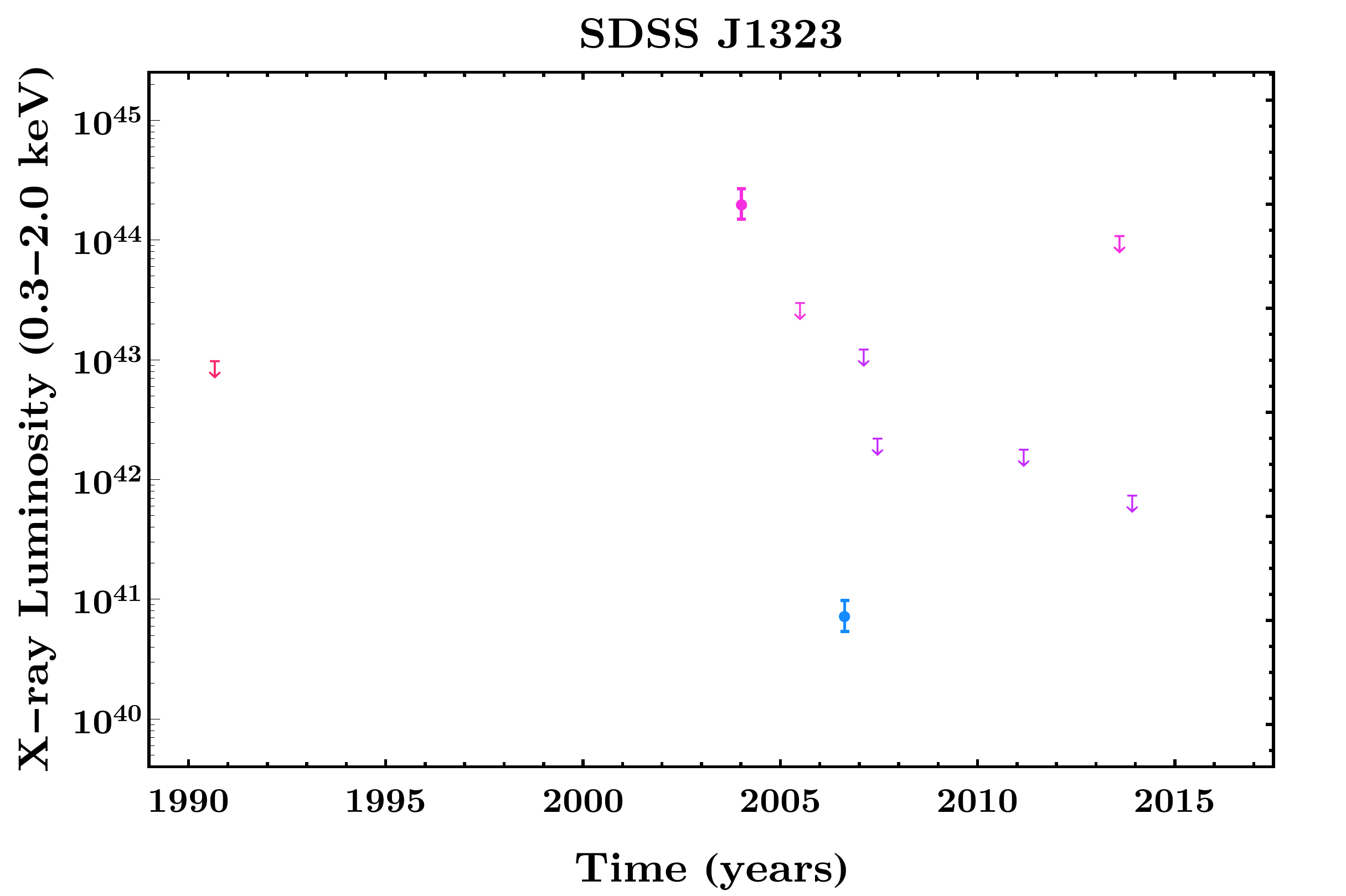}
		\includegraphics[width=0.31\textwidth]{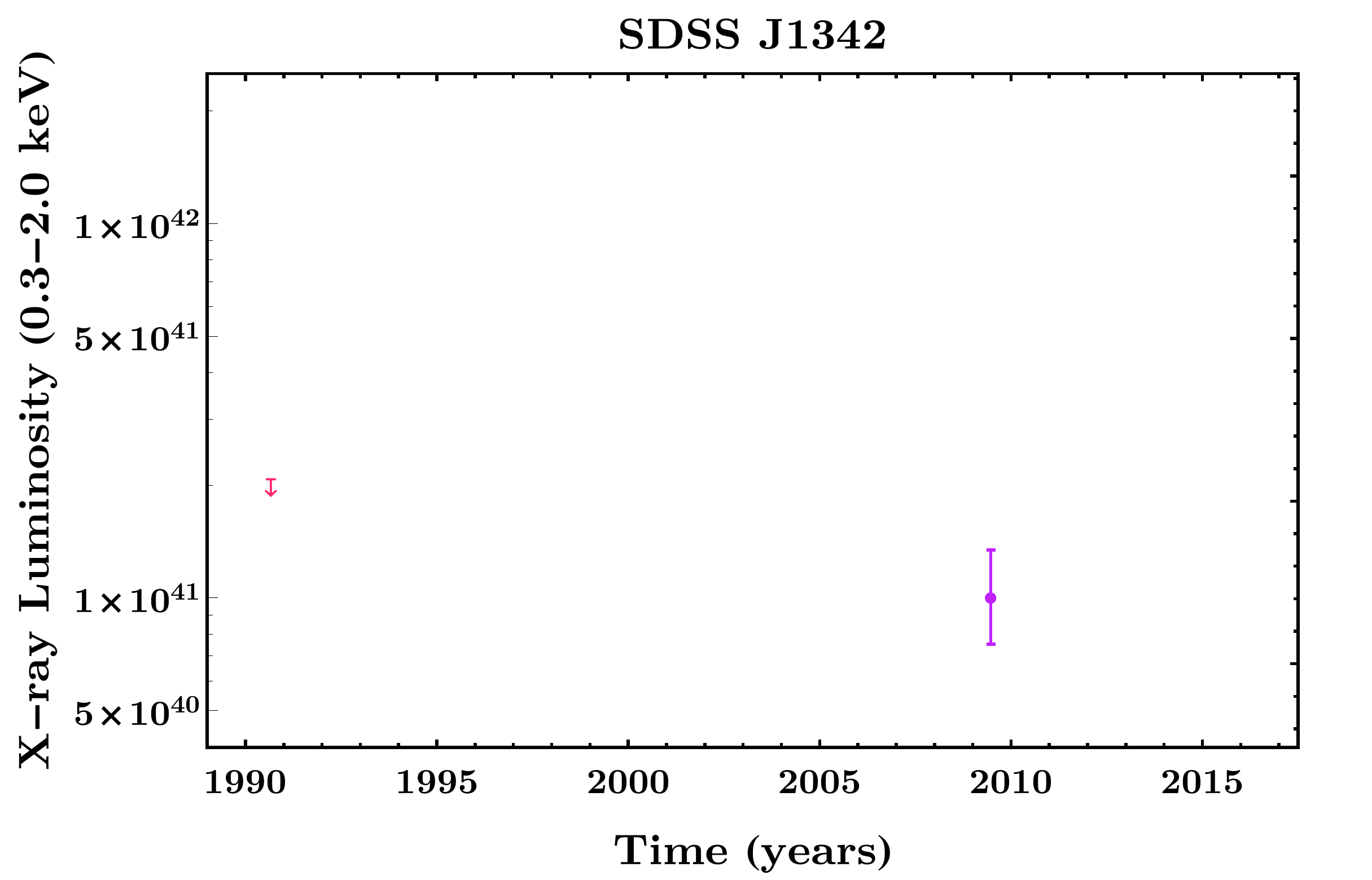}
		\includegraphics[width=0.31\textwidth]{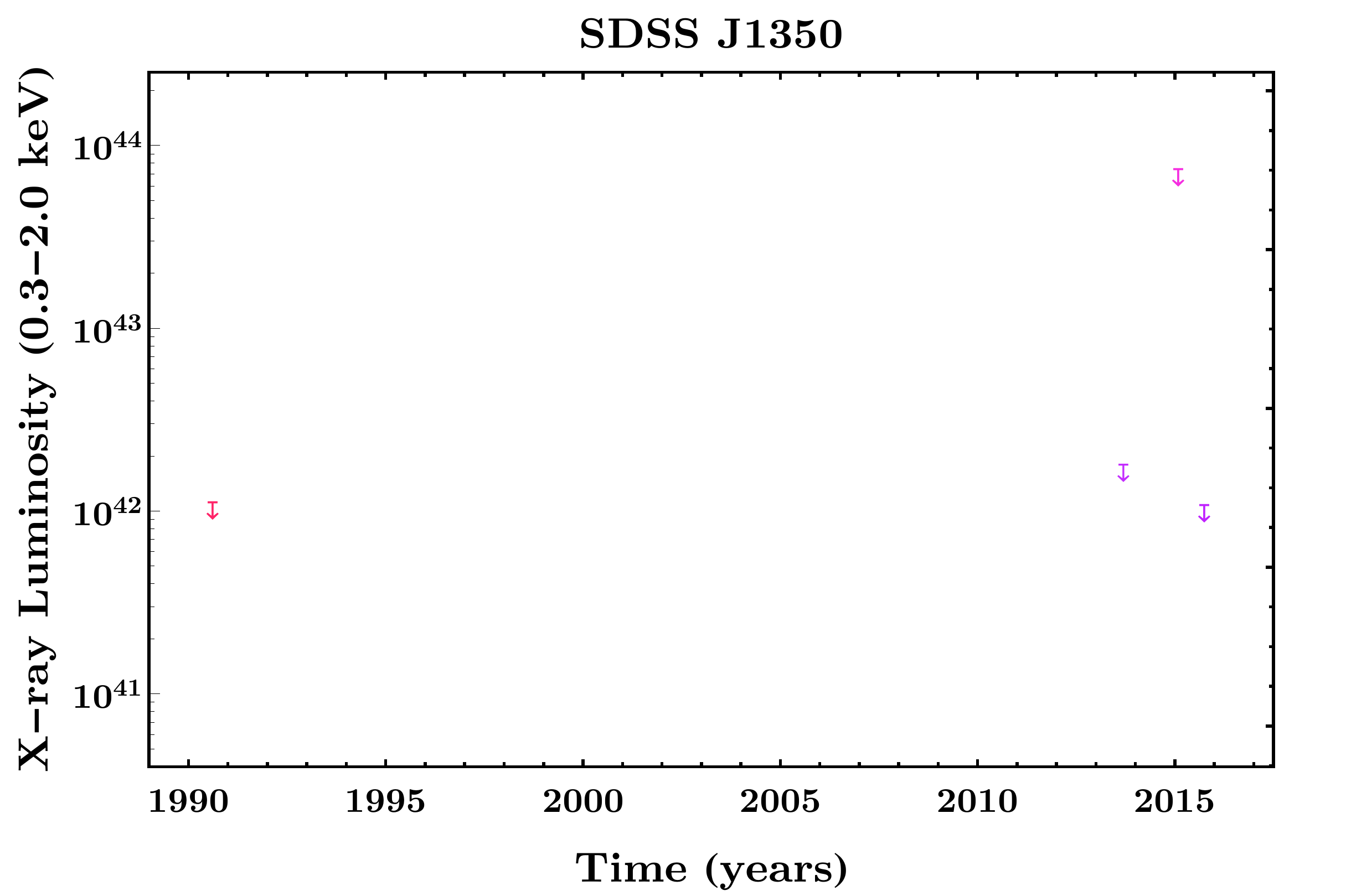}
		\includegraphics[width=0.31\textwidth]{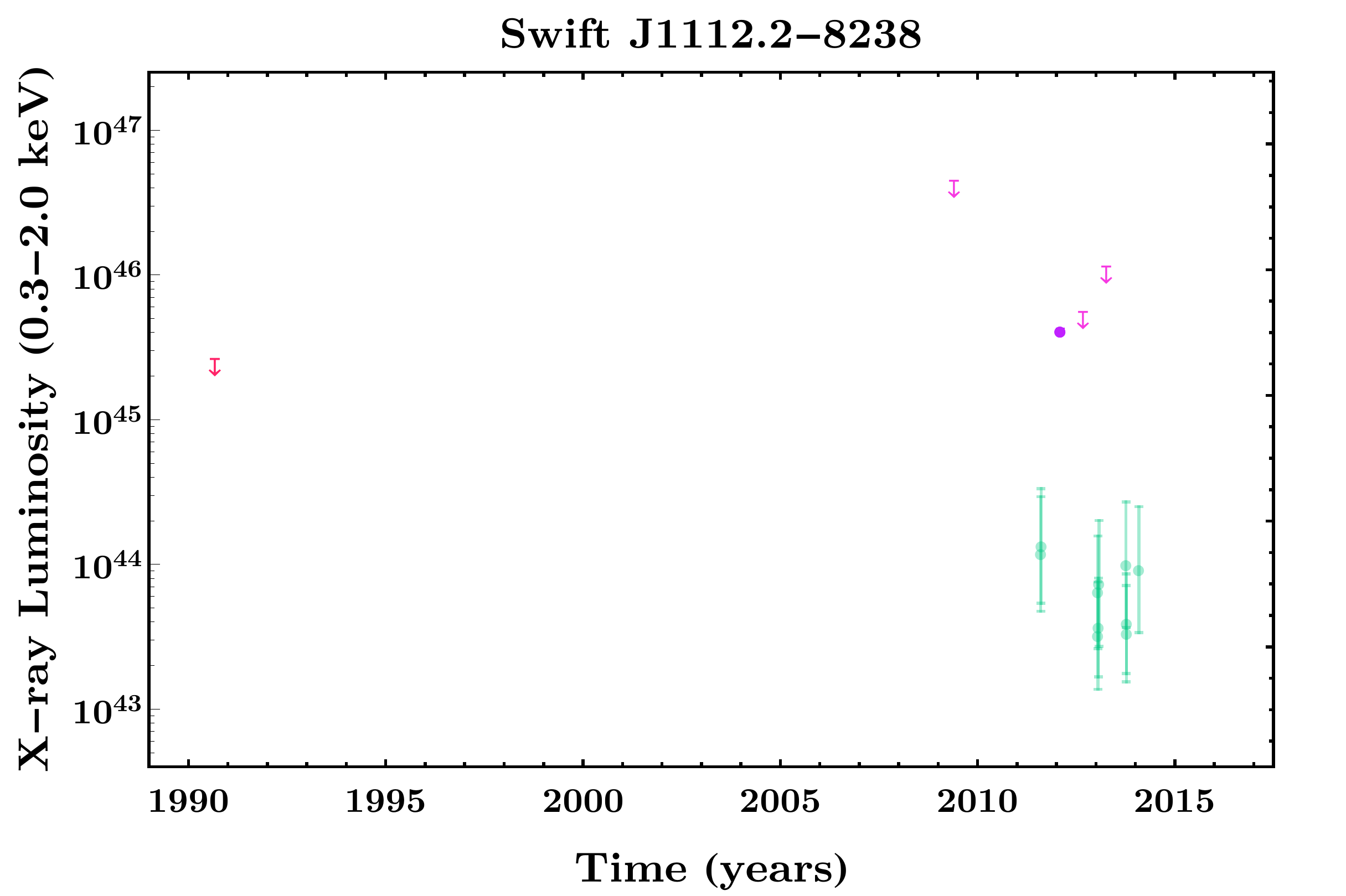}
			\includegraphics[width=0.31\textwidth]{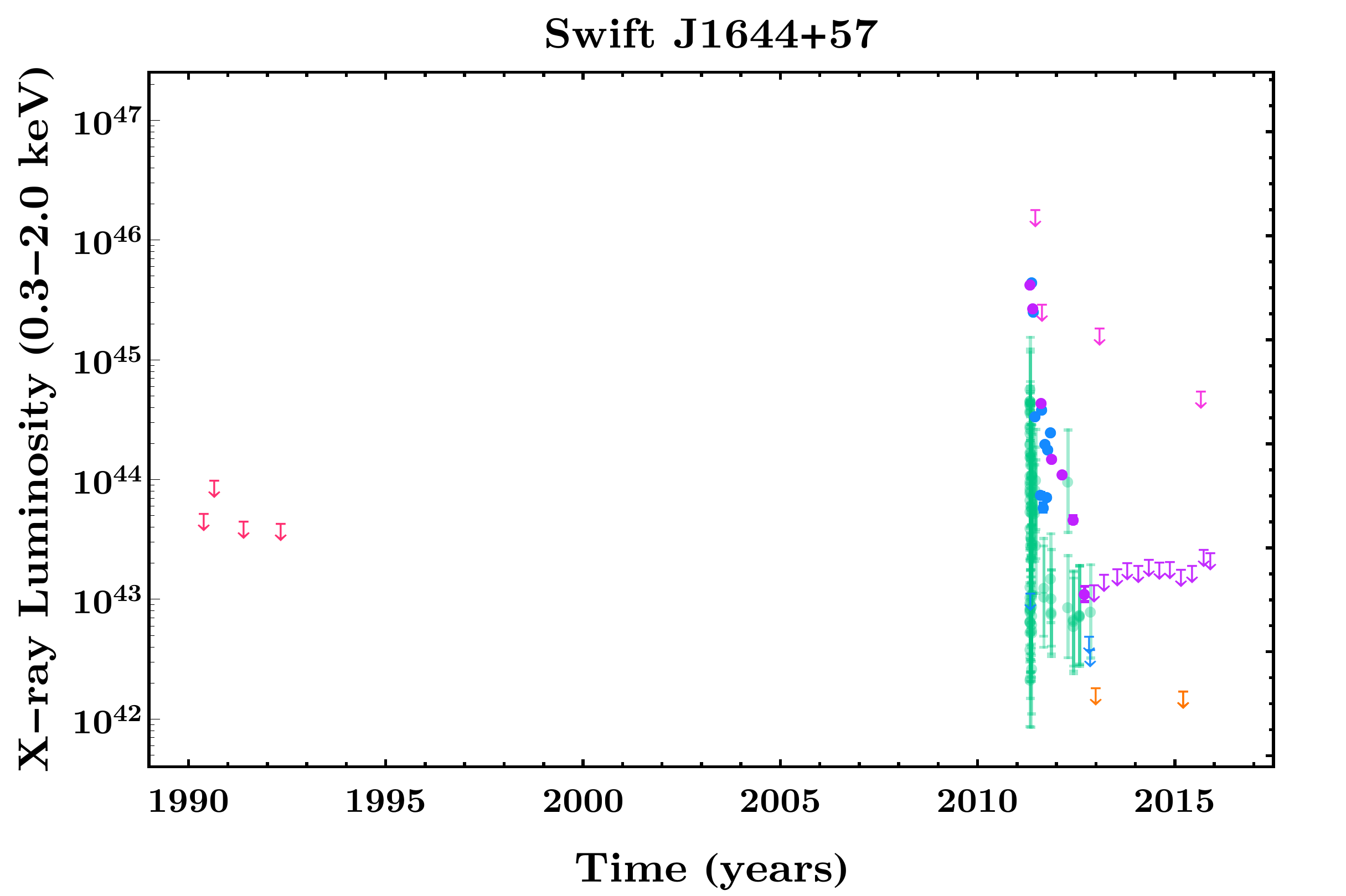}
				\includegraphics[width=0.31\textwidth]{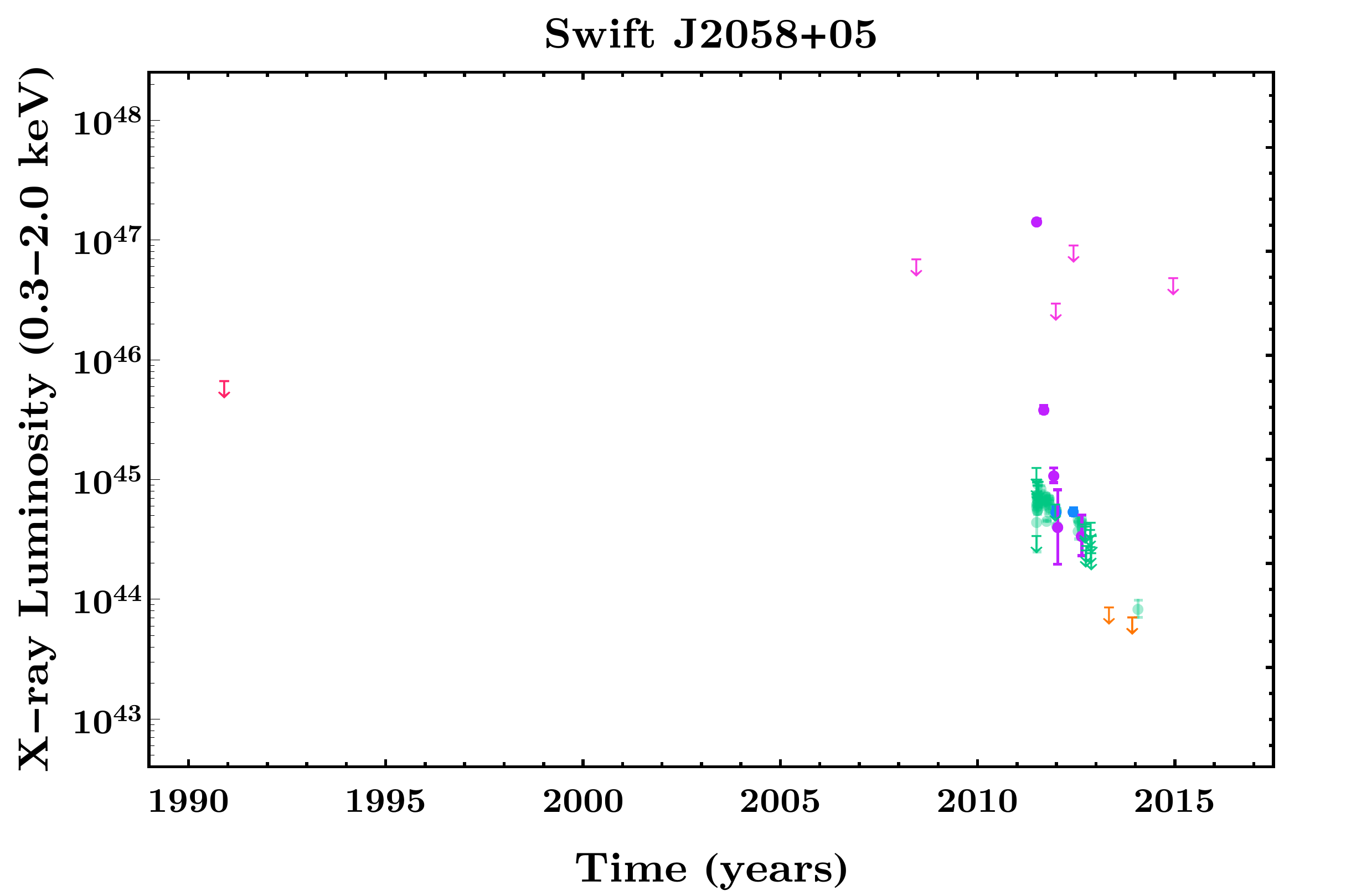}
			\includegraphics[width=0.31\textwidth]{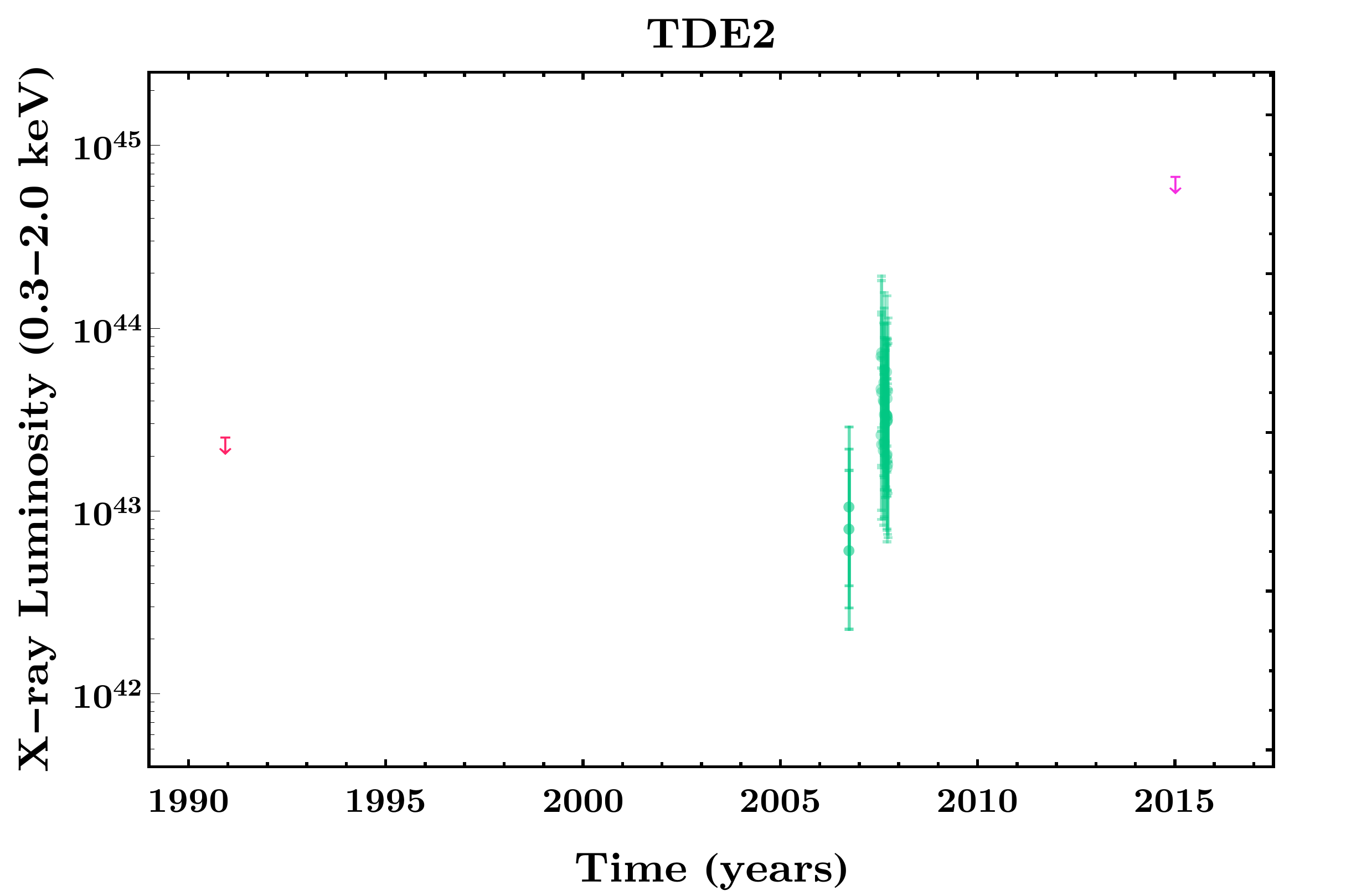}
			\includegraphics[width=0.31\textwidth]{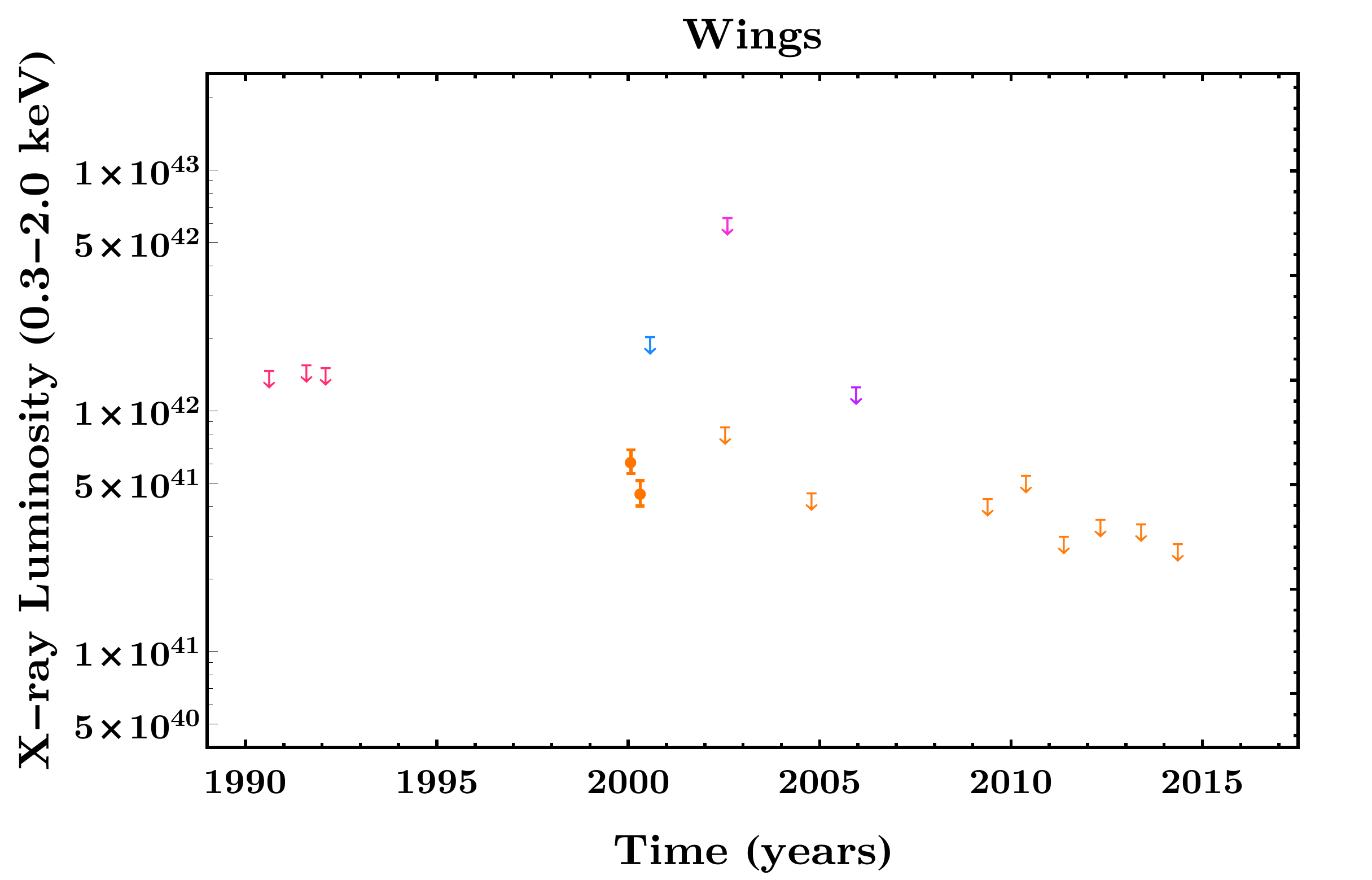}
					\includegraphics[width=0.31\textwidth]{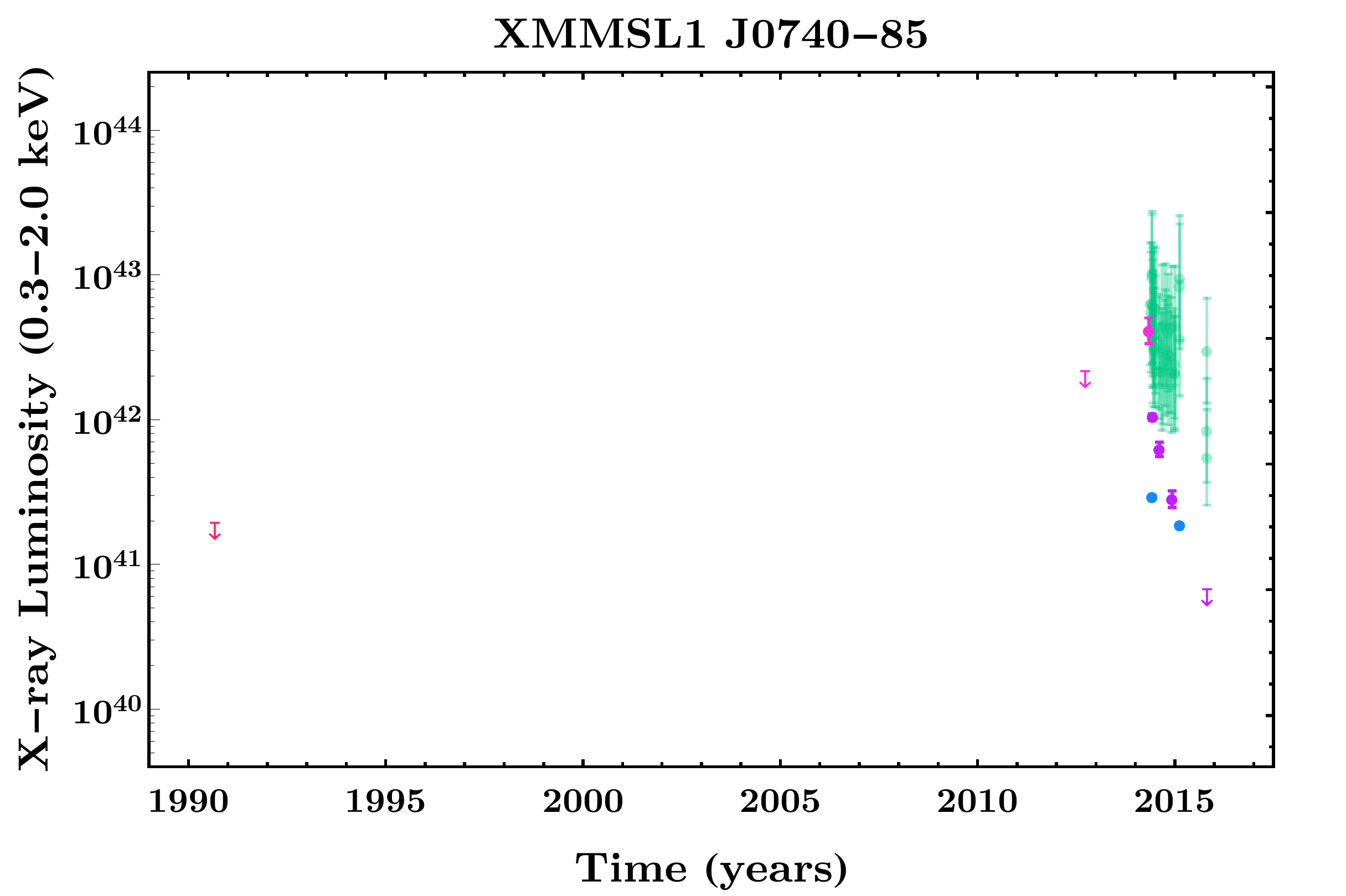}
		\includegraphics[width=0.10\textwidth]{lightcurve_key.pdf}
		\caption{Light Curves for all TDE candidates listed in Table 1 continued. Similar to that in Figure~\ref{lightcurves1}. \label{lightcurves4}}
	\end{center}
\end{figure*}

%% file: powerlawmodelfits.tex
\begin{table}[t]
	\begin{center}
		\caption{The best fit powerlaw models as derived from fitting the full X-ray light curve of the \textit{X-ray TDE} and \textit{likely X-ray TDE} candidates. \label{gspwlfits}}
	\begin{tabular}{lllll}
		\hline
Name & Power law index \\
\hline\hline
ASASSN-14li & $0.92\pm0.12$ \\
Swift J1644+57 & $1.89\pm0.20$ \\ 
Swift J2058+05 &$1.32\pm0.06$ \\
XMM SL1 J0740-85&$0.74\pm0.10$ \\
\hline
2MASX J049 & $0.42\pm0.03$  \\
3XMM J152130.7+074916 &$0.61\pm0.01$\\
IGR J17361-4441 & $1.60\pm0.14$\\ 
NGC247 & $0.26\pm0.10$ \\
OGLE16aa &$0.27\pm0.10$\\
PTF-10iya  & $0.61\pm0.20$ \\
SDSS J1201 & $0.88\pm0.40$ \\
SDSS J1311 &$0.44\pm0.03$ \\
SDSS J1323 &$0.62\pm0.22$ \\
	\hline
\end{tabular}
\end{center}
\end{table}

%% file: rosat_table_counts.tex
\newcolumntype{R}{>{\centering\arraybackslash}m{2.0cm}}
             \clearpage
\LongTables              
\begin{deluxetable*}{lRccccccccccccc}              
 \tabletypesize{\scriptsize}         
	\tablecolumns{11}               
 \tablewidth{0pt}              
 \tablecaption{X-ray properties, as derived using \emph{ROSAT}, for TDE candidates which had \emph{ROSAT pointed or RASS} observations overlapping the position of the source. \label{rosatcounts}}              
	\tablehead{ \colhead{Name} &  \colhead{ObsID} & \colhead{ObsID}& \colhead{MJD} &  \colhead{Exposure} &  \colhead{Source} &  \colhead{Bkg} &   \colhead{$\mathcal{P}_{fluct}^{d}$}&  \colhead{$\mathcal{D}^{d}$}&  \colhead{U or D$^{e}$}&  \colhead{Count Rate$^{f,g}$} \\
		&	 & \colhead{label$^a$}&	 &	 \colhead{time (s)} &	\colhead{Counts$^b$} &	\colhead{Counts$^c$} &	   &	  &	&	 \colhead{(ct/s)} 
	}          
 \startdata              

2MASXJ0203	&	rs931806n00	&	r1	&	48258	&	400	&	$	0.00	\pm	1.1	$	&	$	3.40	\pm	0.5	$	&	1.0	&	0.0	&	U	&	$	(	2.63		)	\times 10^{	-2	}	$	\\
2MASX J0249	&	rs931808n00	&	r2	&	48100	&	237	&	$	0.00	\pm	1.5	$	&	$	2.22	\pm	0.4	$	&	1.0	&	0.0	&	U	&	$	(	3.32		)	\times 10^{	-2	}	$	\\
3XMM J152130.7+074916	&	rp800128n00 +rp800376n00	&	r3	&	48800	&	14058	&	$	129.32	\pm	21.0	$	&	$	290.68	\pm	4.4	$	&	1.0	&	0.0	&	U	&	$	(	2.86		)	\times 10^{	-2	}	$	\\
3XMM J152130.7+074916	&	rs931641n00	&	r4	&	48100	&	408	&	$	7.73	\pm	4.2	$	&	$	9.27	\pm	0.8	$	&	0.8	&	0.2	&	U	&	$	(	5.31		)	\times 10^{	-2	}	$	\\
ASASSN-14ae	&	rs931126n00	&	r5	&	48200	&	323	&	$	0.73	\pm	2.1	$	&	$	3.27	\pm	0.5	$	&	1.0	&	0.0	&	U	&	$	(	3.17		)	\times 10^{	-2	}	$	\\
ASASSN-14li	&	rs931434n00	&	r6	&	48100	&	324	&	$	2.27	\pm	3.1	$	&	$	6.73	\pm	0.7	$	&	1.0	&	0.0	&	U	&	$	(	5.27		)	\times 10^{	-2	}	$	\\
ASASSN-15lh	&	rs932821n00	&	r7	&	48200	&	396	&	$	0.80	\pm	2.3	$	&	$	4.20	\pm	0.5	$	&	1.0	&	0.0	&	U	&	$	(	3.07		)	\times 10^{	-2	}	$	\\
ASASSN-15oi	&	rs932250n00	&	r8	&	48200	&	324	&	$	0.00	\pm	2.1	$	&	$	5.40	\pm	0.6	$	&	1.0	&	0.0	&	U	&	$	(	4.49		)	\times 10^{	-2	}	$	\\
\enddata              
 \tablecomments{All uncertainties correspond to the 90\% confidence level and observations that have similar MJD are combined.}  
\tablenotetext{a}{Due to the large number of observations used in this analysis, we have listed a label for which we can reference in other Tables which observation ID (ObsID) we are referring to.  }         
 \tablenotetext{b}{Obtained from a 100 arcsecond circular region surrounding the position of the TDE candidate. These values listed have \textbf{not} been corrected for encircled energy fraction. The exception to this is for TDE candidate \textit{Wings} as discussed in Appendix~\ref{wingsappendix}}              
 \tablenotetext{c}{Obtained from a 400 arcsecond circular region surrounding the position of the TDE which has been scaled such that the equivalent 100 arcsecond background count rate is presented. These values listed have \textbf{not} been corrected for encircled energy fraction. The exception to this is for TDE candidate \textit{Wings} as discussed in Appendix~\ref{wingsappendix}}              
 \tablenotetext{d}{Calculated assuming Poisson statistics}              
 \tablenotetext{e}{Here U indicates an upperlimit, while D corresponds to a data point as classified using Poission statsitics (see previous two columns).}              
 \tablenotetext{f}{The count rate \textbf{has} been corrected for encircled energy fraction.}              
\tablenotetext{g}{Values which do not have any uncertainties are upperlimits. These upperlimits are derived assuming that we would detect a signal if it is $3\sigma$ above background.}             
\end{deluxetable*}              
\clearpage              

%% file: chandra_table_counts.tex
\newcolumntype{R}{>{\raggedright\arraybackslash}m{2.2cm}} 
\clearpage
\LongTables
\begin{deluxetable*}{lRccccccccccccc}
 \tabletypesize{\scriptsize}              
 \tablecolumns{11}               
 \tablewidth{0pt}              
 \tablecaption{X-ray properties, as derived using \emph{Chandra}, for TDE candidates which had \emph{Chandra} observations overlapping the position of the source. \label{chandracounts}}              
	\tablehead{ \colhead{Name} &  \colhead{ObsID} & \colhead{ObsID}& \colhead{MJD} &  \colhead{Exposure} &  \colhead{Source Counts$^b$} &  \colhead{Bkg Counts$^c$} &   \colhead{$\mathcal{P}_{fluct}^{d}$}&  \colhead{$\mathcal{D}^{d}$}&  \colhead{U or D$^{e}$}&  \colhead{Count Rate$^{f,g}$} \\
		&	 & \colhead{label$^a$}&	 &	 \colhead{time (s)} &	 &	 &	   &	  &	&	 \colhead{(ct/s)} 
	}                  
 \startdata  
3XMM J152130.7+074916	&	900	&	c1	&	51600	&	58100	&	$	0.49	\pm	1.7	$	&	$	2.51	\pm	0.2	$	&	1.0	&	0.0	&	U	&	$	(	1.39		)	\times 10^{	-4	}	$	\\
ASASSN-15lh	&	17879+17880	&	c2	&	57400	&	19800	&	$	0.00	\pm	1.7	$	&	$	0.21	\pm	0.1	$	&	0.2	&	0.8	&	U	&	$	(	8.97		)	\times 10^{	-5	}	$	\\
D1-9	&	6864	&	c3	&	54051	&	29741	&	$	2.98	\pm	2.5	$	&	$	3.02	\pm	0.2	$	&	0.8	&	0.2	&	U	&	$	(	3.08		)	\times 10^{	-4	}	$	\\
D23H-1	&	8601	&	c4	&	54683	&	9060	&	$	1.34	\pm	1.4	$	&	$	0.66	\pm	0.1	$	&	0.5	&	0.5	&	U	&	$	(	3.79		)	\times 10^{	-4	}	$	\\
D23H-1	&	7867+7868+9719	&	c5	&	54400	&	59850	&	$	4.48	\pm	3.3	$	&	$	6.52	\pm	0.3	$	&	0.9	&	0.1	&	U	&	$	(	2.63		)	\times 10^{	-4	}	$	\\
D3-13	&	5848+5847+5850 +6216	&	c6	&	53556	&	184025	&	$	17.63	\pm	4.8	$	&	$	5.37	\pm	0.2	$	&	0.0	&	1.0	&	D	&	$	(	1.06	\pm	0.3	)	\times 10^{	-4	}	$	\\
D3-13	&	9727+9729+9730 +9881+9454+9455 +9456+9457+9460 +9733+9878+9879 +9880	&	c7	&	54700	&	592580	&	$	5.70	\pm	5.1	$	&	$	20.30	\pm	0.5	$	&	1.0	&	0.0	&	U	&	$	(	6.34		)	\times 10^{	-5	}	$	\\
GRB060218, SN2006aj	&	7604	&	c8	&	54471	&	19096	&	$	1.75	\pm	2.2	$	&	$	3.26	\pm	0.2	$	&	1.0	&	0.0	&	U	&	$	(	5.04		)	\times 10^{	-4	}	$	\\
\enddata
 \tablecomments{All uncertainties correspond to the 90\% confidence level and observations that have similar MJD are combined.}

 		\tablenotetext{a}{Due to the large number of observations used in this analysis, we have listed a label for which we can reference in other Tables which observation ID (ObsID) we are referring to.  }
 		\tablenotetext{b}{Obtained from a 2 arcsecond circular region surrounding the position of the TDE candidate. These values listed have \textbf{not} been corrected for encircled energy fraction.}              
 		\tablenotetext{c}{Obtained from a 20 arcsecond circular region surrounding the position of the TDE which has been scaled such that the equivalent 2 arcsecond background count rate is presented. These values listed have \textbf{not} been corrected for encircled energy fraction. }              
 		\tablenotetext{d}{Calculated assuming Poisson statistics}              
 		\tablenotetext{e}{Here U indicates an upperlimit, while D corresponds to a data point as classified using Poission statsitics (see previous two columns).}              
 		\tablenotetext{f}{The count rate \textbf{has} been corrected for encircled energy fraction.}              
 		\tablenotetext{g}{Values which do not have any uncertainties are upperlimits. These upperlimits are derived assuming that we would detect a signal if it is $3\sigma$ above background.}  
 
\end{deluxetable*}
\clearpage

%% file: xmm_table_counts.tex
\newcolumntype{R}{>{\raggedright\arraybackslash}m{2.0cm}}
             \clearpage
\LongTables              
\begin{deluxetable*}{lRccccccccccccc}              
 \tabletypesize{\scriptsize}              
	\tablecolumns{11}          
 \tablewidth{0pt}              
 \tablecaption{X-ray properties, as derived using \emph{XMM-Newton}, for TDE candidates which had \emph{XMM-Newton pointed} observations overlapping the position of the source. \label{xmmcounts}}              
	\tablehead{ \colhead{Name} &  \colhead{ObsID} & \colhead{ObsID}& \colhead{MJD} &  \colhead{Exposure} &  \colhead{Source Counts$^b$} &  \colhead{Bkg Counts$^c$} &   \colhead{$\mathcal{P}_{fluct}^{d}$}&  \colhead{$\mathcal{D}^{d}$}&  \colhead{U or D$^{e}$}&  \colhead{Count Rate$^{f,g}$} \\
		&	 & \colhead{label$^a$}&	 &	 \colhead{time (s)} &	 &	 &	   &	  &	&	 \colhead{(ct/s)} 
	}                     
 \startdata              

2MASXJ0203	&	0411980201	&	x1	&	53919	&	9528	&	$	1678.00	\pm	42.6	$	&	$	127.87	\pm	2.9	$	&	0.0	&	1.0	&	D	&	$	(	2.20	\pm	0.1	)	\times 10^{	-1	}	$	\\
2MASX J0249	&	0411980401	&	x2	&	53931	&	9760	&	$	1396.00	\pm	39.5	$	&	$	154.00	\pm	3.2	$	&	0.0	&	1.0	&	D	&	$	(	1.68		)	\times 10^{	-1	}	$	\\
3XMM J152130.7+074916	&	0109930101, 0109930201	&	x3	&	51779	&	50830	&	$	1503.27	\pm	59.3	$	&	$	1493.73	\pm	10.5	$	&	0.0	&	1.0	&	D	&	$	(	3.48	\pm	0.1	)	\times 10^{	-2	}	$	\\
3XMM J152130.7+074916	&	0723801501	&	x4	&	56677	&	114200	&	$	367.45	\pm	37.0	$	&	$	945.55	\pm	7.7	$	&	1.0	&	0.0	&	U	&	$	(	1.14		)	\times 10^{	-2	}	$	\\
ASASSN-14li	&	0722480201	&	x5	&	57000	&	33834	&	$	21781.06	\pm	154.3	$	&	$	1896.94	\pm	11.2	$	&	0.0	&	1.0	&	D	&	$	(	7.57	\pm	0.1	)	\times 10^{	-1	}	$	\\
ASASSN-15lh	&	0770581201	&	x6	&	57345	&	6430	&	$	19.73	\pm	18.4	$	&	$	313.27	\pm	2.7	$	&	1.0	&	0.0	&	U	&	$	(	6.70		)	\times 10^{	-2	}	$	\\
D1-9	&	0112681001, 0112681101	&	x7	&	52305	&	34035	&	$	0.00	\pm	26.6	$	&	$	675.78	\pm	8.2	$	&	1.0	&	0.0	&	U	&	$	(	2.77		)	\times 10^{	-2	}	$	\\
D23H-1	&	0652401401	&	x8	&	55560	&	35660	&	$	15.53	\pm	16.7	$	&	$	247.47	\pm	4.1	$	&	1.0	&	0.0	&	U	&	$	(	9.72		)	\times 10^{	-3	}	$	\\
\enddata              
 \tablecomments{All uncertainties correspond to the 90\% confidence level and observations that have similar MJD are combined.}  
 
 		\tablenotetext{a}{Due to the large number of observations used in this analysis, we have listed a label for which we can reference in other Tables which observation ID (ObsID) we are referring to.  }
 		\tablenotetext{b}{Obtained from a 30 arcsecond circular region surrounding the position of the TDE candidate. These values listed have \textbf{not} been corrected for encircled energy fraction. The exception to this is for TDE candidate \textit{Wings} as discussed in Appendix~\ref{wingsappendix}}              
 		\tablenotetext{c}{Obtained from a 120 arcsecond circular region surrounding the position of the TDE which has been scaled such that the equivalent 30 arcsecond background count rate is presented. These values listed have \textbf{not} been corrected for encircled energy fraction. The exception to this is for TDE candidate \textit{Wings} as discussed in Appendix~\ref{wingsappendix}}              
 		\tablenotetext{d}{Calculated assuming Poisson statistics}              
 		\tablenotetext{e}{Here U indicates an upperlimit, while D corresponds to a data point as classified using Poission statsitics (see previous two columns).}              
 		\tablenotetext{f}{The count rate \textbf{has} been corrected for encircled energy fraction.}              
 		\tablenotetext{g}{Values which do not have any uncertainties are upperlimits. These upperlimits are derived assuming that we would detect a signal if it is $3\sigma$ above background.}  
\end{deluxetable*}              
\clearpage

%% file: xmm_slew_table_counts.tex
\newcolumntype{R}{>{\raggedright\arraybackslash}m{2.0cm}}
         \clearpage
\LongTables              
\begin{deluxetable*}{lRccccccccccccc}              
 \tabletypesize{\scriptsize}                  
	\tablecolumns{11}              
	\tablewidth{0pt}              
	\tablecaption{X-ray properties, as derived using \emph{XMM-Newton slew}, for TDE candidates which had \emph{XMM-Newton slew} observations overlapping the position of the source. \label{slewcounts}}              
	\tablehead{ \colhead{Name} &  \colhead{ObsID} & \colhead{ObsID}& \colhead{MJD} &  \colhead{Exposure} &  \colhead{Source Counts$^b$} &  \colhead{Bkg Counts$^c$} &   \colhead{$\mathcal{P}_{fluct}^{d}$}&  \colhead{$\mathcal{D}^{d}$}&  \colhead{U or D$^{e}$}&  \colhead{Count Rate$^{f,g}$} \\
		&	 & \colhead{label$^a$}&	 &	 \colhead{time (s)} &	 &	 &	   &	  &	&	 \colhead{(ct/s)} 
	}                    
	\startdata              

2MASXJ0203	&	9100500004	&	xs1	&	53018	&	5.00	&	$	20.33	\pm	4.6	$	&	$	0.67	\pm	0.2	$	&	0.0	&	1.0	&	D	&	$	(	4.52	\pm	1.0	)	\times 10^{	0	}	$	\\
3XMM J152130.7+074916	&	9148900002	&	xs2	&	54491	&	5.83	&	$	0.00	\pm	3.4	$	&	$	12.37	\pm	0.9	$	&	1.0	&	0.0	&	U	&	$	(	4.36		)	\times 10^{	0	}	$	\\
3XMM J152130.7+074916	&	9185400002	&	xs3	&	55220	&	3.40	&	$	0.00	\pm	1.1	$	&	$	1.87	\pm	0.4	$	&	1.0	&	0.0	&	U	&	$	(	1.95		)	\times 10^{	0	}	$	\\
ASASSN-14ae	&	9164500004	&	xs4	&	54803	&	5.13	&	$	0.00	\pm	1.0	$	&	$	1.07	\pm	0.3	$	&	1.0	&	0.0	&	U	&	$	(	9.02		)	\times 10^{	-1	}	$	\\
ASASSN-14ae	&	9237200005	&	xs5	&	56253	&	5.27	&	$	0.00	\pm	0.2	$	&	$	0.60	\pm	0.2	$	&	1.0	&	0.0	&	U	&	$	(	6.17		)	\times 10^{	-1	}	$	\\
ASASSN-14ae	&	9264300002	&	xs6	&	56792	&	8.80	&	$	0.80	\pm	1.4	$	&	$	1.20	\pm	0.3	$	&	1.0	&	0.0	&	U	&	$	(	5.66		)	\times 10^{	-1	}	$	\\
ASASSN-14ae	&	9273100003	&	xs7	&	56969	&	9.90	&	$	0.00	\pm	0.3	$	&	$	1.13	\pm	0.3	$	&	1.0	&	0.0	&	U	&	$	(	4.86		)	\times 10^{	-1	}	$	\\
ASASSN-14li	&	9165300002	&	xs8	&	54819	&	6.42	&	$	0.00	\pm	0.2	$	&	$	0.94	\pm	0.2	$	&	1.0	&	0.0	&	U	&	$	(	6.64		)	\times 10^{	-1	}	$	\\
\enddata             
 \tablecomments{All uncertainties correspond to the 90\% confidence level and observations that have similar MJD are combined.}      
 		\tablenotetext{a}{Due to the large number of observations used in this analysis, we have listed a label for which we can reference in other Tables which observation ID (ObsID) we are referring to.  }
 		\tablenotetext{b}{Obtained from a 50 arcsecond circular region surrounding the position of the TDE candidate. These values listed have \textbf{not} been corrected for encircled energy fraction. The exception to this is for TDE candidate \textit{Wings} as discussed in Appendix~\ref{wingsappendix}}              
 		\tablenotetext{c}{Obtained from a 200 arcsecond circular region surrounding the position of the TDE which has been scaled such that the equivalent 50 arcsecond background count rate is presented. These values listed have \textbf{not} been corrected for encircled energy fraction. The exception to this is for TDE candidate \textit{Wings} as discussed in Appendix~\ref{wingsappendix}}              
 		\tablenotetext{d}{Calculated assuming Poisson statistics}              
 		\tablenotetext{e}{Here U indicates an upperlimit, while D corresponds to a data point as classified using Poission statsitics (see previous two columns).}              
 		\tablenotetext{f}{The count rate \textbf{has} been corrected for encircled energy fraction.}              
 		\tablenotetext{g}{Values which do not have any uncertainties are upperlimits. These upperlimits are derived assuming that we would detect a signal if it is $3\sigma$ above background.}            
 \end{deluxetable*}             
 \clearpage

%% file: swift_table_counts.tex
\newcolumntype{R}{>{\raggedright\arraybackslash}m{2.2cm}}
\clearpage
   \LongTables              
\begin{deluxetable*}{lRccccccccccccc}              
 \tabletypesize{\scriptsize}               
	\tablecolumns{11}               
	\tablewidth{0pt}              
	\tablecaption{X-ray properties, as derived using \emph{Swift XRT}, for TDE candidates which had \emph{Swift XRT} observations overlapping the position of the source. \label{swiftcounts}}              
	\tablehead{ \colhead{Name} &  \colhead{ObsID} & \colhead{ObsID}& \colhead{MJD} &  \colhead{Exposure} &  \colhead{Source} &  \colhead{Bkg} &   \colhead{$\mathcal{P}_{fluct}^{d}$}&  \colhead{$\mathcal{D}^{d}$}&  \colhead{U or D$^{e}$}&  \colhead{Count Rate$^{f,g}$} \\
		&	 & \colhead{label$^a$}&	 &	 \colhead{time (s)} &\colhead{Counts$^b$}	 &	\colhead{ Counts$^c$} &	   &	  &	&	 \colhead{(ct/s)} 
	}                  
	\startdata               
2MASXJ0203	&	00035746001, 00040304001	&	sw1	&	54756	&	5621	&	$	175.34	\pm	13.5	$	&	$	7.66	\pm	0.7	$	&	0.0	&	1.0	&	D	&	$	(	3.47	\pm	0.3	)	\times 10^{	-2	}	$	\\
2MASXJ0203	&	00035746002	&	sw2	&	54527	&	3634	&	$	118.86	\pm	11.2	$	&	$	5.14	\pm	0.6	$	&	0.0	&	1.0	&	D	&	$	(	3.63	\pm	0.3	)	\times 10^{	-2	}	$	\\
2MASX J0249	&	00035748002, 00035748003	&	sw3	&	54076	&	829	&	$	2.80	\pm	2.0	$	&	$	1.20	\pm	0.3	$	&	0.3	&	0.7	&	U	&	$	(	6.01		)	\times 10^{	-3	}	$	\\
2MASX J0249	&	00035748004	&	sw4	&	54127	&	617	&	$	3.34	\pm	2.0	$	&	$	0.67	\pm	0.2	$	&	0.0	&	1.0	&	D	&	$	(	5.60	\pm	3.6	)	\times 10^{	-3	}	$	\\
2MASX J0249	&	00035748005	&	sw5	&	54279	&	3042	&	$	27.26	\pm	5.6	$	&	$	3.74	\pm	0.5	$	&	0.0	&	1.0	&	D	&	$	(	3.49	\pm	2.0	)	\times 10^{	-3	}	$	\\
2MASX J0249	&	00040306001	&	sw6	&	55361	&	689	&	$	2.27	\pm	1.7	$	&	$	0.74	\pm	0.2	$	&	0.2	&	0.8	&	U	&	$	(	5.33		)	\times 10^{	-3	}	$	\\
2MASX J0249	&	00040306002--003	&	sw7	&	55618	&	1456	&	$	3.87	\pm	2.5	$	&	$	2.13	\pm	0.4	$	&	0.4	&	0.6	&	U	&	$	(	4.97		)	\times 10^{	-3	}	$	\\
2MASX J0249	&	00040306004--005	&	sw8	&	57086	&	2812	&	$0.00	\pm	1.8	$	&	$	3.81	\pm	0.5	$	&	1.0	&	0.0	&	U	&	$	(	3.82		)	\times 10^{	-3	}	$	\\
\enddata              
	\tablecomments{All uncertainties correspond to the 90\% confidence level and observations that have similar MJD are combined.}              
		\tablenotetext{a}{Due to the large number of observations used in this analysis, we have listed a label for which we can reference in other Tables which observation ID (ObsID) we are referring to.  }
	\tablenotetext{b}{Obtained from a 50 arcsecond circular region surrounding the position of the TDE candidate. These values listed have \textbf{not} been corrected for encircled energy fraction. The exception to this is for TDE candidate \textit{Wings} as discussed in Appendix~\ref{wingsappendix}}              
	\tablenotetext{c}{Obtained from a 200 arcsecond circular region surrounding the position of the TDE which has been scaled such that the equivalent 50 arcsecond background count rate is presented. These values listed have \textbf{not} been corrected for encircled energy fraction. The exception to this is for TDE candidate \textit{Wings} as discussed in Appendix~\ref{wingsappendix}}              
	\tablenotetext{d}{Calculated assuming Poisson statistics}              
	\tablenotetext{e}{Here U indicates an upperlimit, while D corresponds to a data point as classified using Poission statsitics (see previous two columns).}              
	\tablenotetext{f}{The count rate \textbf{has} been corrected for encircled energy fraction.}              
	\tablenotetext{g}{Values which do not have any uncertainties are upperlimits. These upperlimits are derived assuming that we would detect a signal if it is $3\sigma$ above background.}         
\end{deluxetable*}              
\clearpage

%% file: rosat_table_hms.tex
\clearpage              
\LongTables              
\begin{deluxetable*}{lcccccccccccccc}              
 \tabletypesize{\footnotesize}              
 \tablecolumns{5}               
 \tablewidth{0pt}              
 \tablecaption{The soft, medium and hard counts derived for the TDE candidates which have \emph{ROSAT pointed and RASS} observations overlapping the position of the event.\label{rosathms}}              
	\tablehead{ \colhead{Name} &  \colhead{ObsID} &   \colhead{Soft Counts} &  \colhead{Medium Counts} &  \colhead{Hard Counts } \\
		&  \colhead{label}  &  \colhead{0.3-1.0 keV} &  \colhead{1.0-2.0 keV}  &  \colhead{2.0-10.0keV}
	}            
 \startdata              

2MASXJ0203	&	r1	&	$	4.31		$	&	$	2.96		$	&	$	0.84		$	\\
2MASX J0249	&	r2	&	$	4.51		$	&	$	0.99		$	&	$	1.00		$	\\
3XMM J152130.7+074916	&	r3	&	$	178.38		$	&	$	85.86		$	&	$	7.82		$	\\
3XMM J152130.7+074916	&	r4	&	$	10.93		$	&	$	4.90		$	&	$	1.54		$	\\
ASASSN-14ae	&	r5	&	$	4.31		$	&	$	2.14		$	&	$	1.00		$	\\
ASASSN-14li	&	r6	&	$	6.52		$	&	$	3.44		$	&	$	1.00		$	\\
ASASSN-15lh	&	r7	&	$	5.82		$	&	$	2.05		$	&	$	1.00		$	\\
ASASSN-15oi	&	r8	&	$	10.65		$	&	$	1.81		$	&	$	1.00		$	\\
\enddata              
 \tablecomments{All uncertainties correspond to the 90\% confidence level. The soft, medium and hard count rates have been corrected for encircled energy fraction. Values which do not have any uncertainties are upperlimits. These upperlimits are derived assuming that we would detect a signal if it is $3\sigma$ above background in that energy band.}              

\end{deluxetable*}              
\clearpage              

%% file: chandra_table_hms.tex
\clearpage              
\LongTables              
\begin{deluxetable*}{lcccccccccccccc}              
 \tabletypesize{\footnotesize}     
 \tablecolumns{7}               
 \tablewidth{0pt}              
 \tablecaption{The soft, medium and hard counts derived for the TDE candidates which have \emph{Chandra} observations overlapping the position of the event. \label{chandrahms}}              
	\tablehead{ \colhead{Name} &  \colhead{ObsID} &   \colhead{Soft Counts} &  \colhead{Medium Counts} &  \colhead{Hard Counts } \\
		&  \colhead{label}  &  \colhead{0.3-1.0 keV} &  \colhead{1.0-2.0 keV}  &  \colhead{2.0-10.0keV}
	}                      
 \startdata              
   
3XMM J152130.7+074916	&	c1	&	$	2.30		$	&	$	2.90		$	&	$	2.95		$	\\
ASASSN-15lh	&	c2	&	$	1.00		$	&	$	1.18		$	&	$	3.48		$	\\
D1-9	&	c3	&	$	1.46		$	&	$	1.66		$	&	$	4.82		$	\\
D23H-1	&	c4	&	$	1.82		$	&	$	0.89		$	&	$	1.68		$	\\
D23H-1	&	c5	&	$	2.34		$	&	$	2.09		$	&	$	8.35		$	\\
D3-13	&	c6	&	$	16.23	\pm	4.3	$	&	$	2.65		$	&	$	7.23		$	\\
D3-13	&	c7	&	$	5.26		$	&	$	5.77		$	&	$	19.10		$	\\
GRB060218, SN2006aj	&	c8	&	$	1.44		$	&	$	1.08		$	&	$	4.12		$	\\
\enddata              
 \tablecomments{All uncertainties correspond to the 90\% confidence level. The soft, medium and hard count rates have been corrected for encircled energy fraction. Values which do not have any uncertainties are upperlimits. These upperlimits are derived assuming that we would detect a signal if it is $3\sigma$ above background in that energy band.}              

\end{deluxetable*}              
\clearpage              

%% file: xmm_table_hms.tex
\clearpage              
\LongTables              
\begin{deluxetable*}{lcccccccccccccc}              
 \tabletypesize{\footnotesize}     
 \tablecolumns{7}               
 \tablewidth{0pt}              
 \tablecaption{The soft, medium and hard counts derived for the TDE candidates which have \emph{XMM-Newton pointed} observations overlapping the position of the event. \label{xmmhms}}              
	\tablehead{ \colhead{Name} &  \colhead{ObsID} &   \colhead{Soft Counts} &  \colhead{Medium Counts} &  \colhead{Hard Counts } \\
		&  \colhead{label}  &  \colhead{0.3-1.0 keV} &  \colhead{1.0-2.0 keV}  &  \colhead{2.0-10.0keV}
	}                      
 \startdata              
   
2MASXJ0203	&	x1	&	$	958.67	\pm	35.3	$	&	$	665.08	\pm	29.7	$	&	$	380.07	\pm	21.2	$	\\
2MASX J0249	&	x2	&	$	1610.75	\pm	45.6	$	&	$	151.58	\pm	15.6	$	&	$	109.24		$	\\
3XMM J152130.7+074916	&	x3	&	$	1366.57	\pm	46.2	$	&	$	447.25		$	&	$	927.39		$	\\
3XMM J152130.7+074916	&	x4	&	$	371.45		$	&	$	303.55		$	&	$	562.47		$	\\
ASASSN-14li	&	x5	&	$	26914.80	\pm	185.1	$	&	$	432.16		$	&	$	1331.46		$	\\
ASASSN-15lh	&	x6	&	$	65.30		$	&	$	87.20		$	&	$	248.00		$	\\
D1-9	&	x7	&	$	136.06		$	&	$	217.48		$	&	$	430.19		$	\\
D23H-1	&	x8	&	$	61.32		$	&	$	84.56		$	&	$	207.33		$	\\
\enddata              
 \tablecomments{All uncertainties correspond to the 90\% confidence level. The soft, medium and hard count rates have been corrected for encircled energy fraction. Values which do not have any uncertainties are upperlimits. These upperlimits are derived assuming that we would detect a signal if it is $3\sigma$ above background in that energy band.}              

\end{deluxetable*}              
\clearpage              

%% file: slew_table_hms.tex
\clearpage              
\LongTables              
\begin{deluxetable*}{lcccccccccccccc}              
 \tabletypesize{\footnotesize}               
 \tablecolumns{7}               
 \tablewidth{0pt}              
 \tablecaption{The soft, medium and hard counts derived for the TDE candidates which have \emph{XMM-Newton slew} observations overlapping the position of the event.\label{slewhms}}              
	\tablehead{ \colhead{Name} &  \colhead{ObsID} &   \colhead{Soft Counts} &  \colhead{Medium Counts} &  \colhead{Hard Counts } \\
		&  \colhead{label}  &  \colhead{0.3-1.0 keV} &  \colhead{1.0-2.0 keV}  &  \colhead{2.0-10.0keV}
	}                      
 \startdata              
   
2MASXJ0203	&	xs1	&	$	10.89	\pm	3.5	$	&	$	4.37	\pm	2.2	$	&	$	1.23		$	\\
3XMM J152130.7+074916	&	xs2	&	$	7.18		$	&	$	6.44		$	&	$	14.16		$	\\
3XMM J152130.7+074916	&	xs3	&	$		\pm		$	&	$		\pm		$	&	$		\pm		$	\\
ASASSN-14ae	&	xs4	&	$	2.29		$	&	$	1.36		$	&	$	1.54		$	\\
ASASSN-14ae	&	xs5	&	$	0.94		$	&	$	0.94		$	&	$	1.54		$	\\
ASASSN-14ae	&	xs6	&	$	2.80		$	&	$	0.94		$	&	$	1.54		$	\\
ASASSN-14ae	&	xs7	&	$	1.36		$	&	$	0.94		$	&	$	1.82		$	\\
ASASSN-14li	&	xs8	&	$	1.59		$	&	$	0.87		$	&	$	1.43		$	\\
\enddata              
 \tablecomments{All uncertainties correspond to the 90\% confidence level. The soft, medium and hard count rates have been corrected for encircled energy fraction. Values which do not have any uncertainties are upperlimits. These upperlimits are derived assuming that we would detect a signal if it is $3\sigma$ above background in that energy band.}              

\end{deluxetable*}              
\clearpage              

%% file: swift_table_hms.tex
\clearpage              
\LongTables              
\begin{deluxetable*}{lcccccccccccccc}              
 \tabletypesize{\footnotesize}            
 \tablecolumns{7}               
 \tablewidth{0pt}              
 \tablecaption{The soft, medium and hard counts derived for the TDE candidates which have \emph{Swift XRT} observations overlapping the position of the event.\label{swifthms}}              
	\tablehead{ \colhead{Name} &  \colhead{ObsID} &   \colhead{Soft Counts} &  \colhead{Medium Counts} &  \colhead{Hard Counts } \\
		&  \colhead{label}  &  \colhead{0.3-1.0 keV} &  \colhead{1.0-2.0 keV}  &  \colhead{2.0-10.0keV}
	}                     
 \startdata

2MASXJ0203	&	sw1	&	$	78.22	\pm	9.5	$	&	$	60.15	\pm	8.3	$	&	$	41.47	\pm	6.6	$	\\
2MASXJ0203	&	sw2	&	$	42.89	\pm	7.0	$	&	$	49.55	\pm	7.5	$	&	$	29.86	\pm	5.7	$	\\
2MASX J0249	&	sw3	&	$	2.96	\pm	1.9	$	&	$	2.01		$	&	$	2.51		$	\\
2MASX J0249	&	sw4	&	$	4.15	\pm	2.2	$	&	$	1.00		$	&	$	1.54		$	\\
2MASX J0249	&	sw5	&	$	31.03	\pm	6.0	$	&	$	3.67		$	&	$	5.25		$	\\
2MASX J0249	&	sw6	&	$	3.18	\pm	1.9	$	&	$	1.72		$	&	$	2.07		$	\\
2MASX J0249	&	sw7	&	$	3.63	\pm	2.2	$	&	$	1.71		$	&	$	4.00		$	\\
2MASX J0249	&	sw8	&	$	5.50		$	&	$	3.03		$	&	$	4.95		$	\\
\enddata              
 \tablecomments{All uncertainties correspond to the 90\% confidence level. The soft, medium and hard count rates have been corrected for encircled energy fraction. Values which do not have any uncertainties are upperlimits. These upperlimits are derived assuming that we would detect a signal if it is $3\sigma$ above background in that energy band.}              

\end{deluxetable*}              
\clearpage

%% file: rosat_table_fits.tex
\clearpage              
\LongTables              
\begin{deluxetable*}{lcccccccccccccc}              
\tabletypesize{\footnotesize}
\tablecolumns{7}              
 \tablewidth{0pt}              
 \tablecaption{The best fit, or assumed parameters for our absorbed powerlaw model that was used to derive the 0.3-2.0 keV X-ray flux and luminosity of each TDE candidate from the \emph{ROSAT pointed or RASS} observations overlapping the position of the source. \label{rosatfits}}              
\tablehead{ \colhead{Name} &  \colhead{ObsID} &   \colhead{$N_{H}^a$} &  \colhead{$\Gamma$$^{b}$} &  \colhead{$\chi^{2}$ (dof) }&  \colhead{X-ray flux$^{c}$}&  \colhead{X-ray Luminosity$^{c}$} \\
 &  \colhead{label}  &  \colhead{($10^{22}$ cm$^{-2}$)} &   &  &  \colhead{ (erg/s/cm$^{2}$) }&   \colhead{(erg/s) }
}              
 \startdata              
2MASXJ0203	&	r1	&	$	0.02		$	&	$	4.50		$	&		&	$	(	7.55		)	\times 10^{	-14	}	$	&	$	(	6.86		)	\times 10^{	41	}	$	\\
2MASX J0249	&	r2	&	$	0.44	_{-0.14	}	^{+	0.15	}	$	&	$	7.70		$	&		&	$	(	3.57		)	\times 10^{	-13	}	$	&	$	(	2.73		)	\times 10^{	41	}	$	\\
3XMM J152130.7+074916	&	r3	&	$	0.03		$	&	$	4.50		$	&		&	$	(	6.62		)	\times 10^{	-14	}	$	&	$	(	6.29		)	\times 10^{	42	}	$	\\
3XMM J152130.7+074916	&	r4	&	$	0.03		$	&	$	4.50		$	&		&	$	(	1.23		)	\times 10^{	-13	}	$	&	$	(	1.17		)	\times 10^{	43	}	$	\\
ASASSN-14ae	&	r5	&	$	0.02		$	&	$	4.50		$	&		&	$	(	4.50		)	\times 10^{	-14	}	$	&	$	(	1.99		)	\times 10^{	41	}	$	\\
ASASSN-14li	&	r6	&	$	0.02		$	&	$	4.50		$	&		&	$	(	7.11		)	\times 10^{	-14	}	$	&	$	(	6.70		)	\times 10^{	40	}	$	\\
ASASSN-15lh	&	r7	&	$	0.03		$	&	$	4.50		$	&		&	$	(	5.29		)	\times 10^{	-13	}	$	&	$	(	9.27		)	\times 10^{	43	}	$	\\
ASASSN-15oi	&	r8	&	$	0.06		$	&	$	4.50		$	&		&	$	(	2.95		)	\times 10^{	-13	}	$	&	$	(	1.62		)	\times 10^{	42	}	$	\\
\enddata              
 \tablecomments{All uncertainties correspond to the 90\% confidence level and observations that have similar MJD are combined.}              
  \tablenotetext{a}{$N_{H}$ is determined either directly from fitting the X-ray emitting spectrum or using the column density derived from the Leiden/Argentine/Bonn (LAB) Survey of Galactic H$\textsc{i}$ (Kalberla et al. 2005)}              
 \tablenotetext{b}{Powerlaw index was derived from either fitting X-ray spectra from either \textit{ROSAT}, \textit{XMM} or \textit{Chandra} with a powerlaw model or was assumed to be equal to 4.5. If the following $\Gamma$ has an uncertainty, the powerlaw index was derived from fitting the X-ray spectrum extracted from \textit{ROSAT}.}              
 \tablenotetext{c}{Absorbed X-ray flux and X-ray luminosity is calculated for an energy range of 0.3-2.0 keV.}           
 \tablenotetext{d}{Value frozen to that obtained from modelling the \textit{ROSAT} X-ray spectrum.}
 \tablenotetext{e}{Value set to the average of that obtained from modelling multiple \textit{ROSAT} spectra.} 
\end{deluxetable*}              
\clearpage              

%% file: chandra_table_fits.tex
\clearpage
\LongTables
\begin{deluxetable*}{lcccccccccccccc}
\tabletypesize{\footnotesize}
\tablecolumns{7} 
\tablewidth{0pt}
\tablecaption{The best fit, or assumed parameters for our absorbed powerlaw model that was used to derive the 0.3-2.0 keV X-ray flux and luminosity of each TDE candidate from the \emph{Chandra} observations overlapping the position of the source. \label{chandrafits}}
\tablehead{ \colhead{Name} &  \colhead{ObsID} &   \colhead{$N_{H}^a$} &  \colhead{$\Gamma$$^{b}$} &  \colhead{$\chi^{2}$ (dof) }&  \colhead{X-ray flux$^{c}$}&  \colhead{X-ray Luminosity$^{c}$} \\
	&  \colhead{label}  &  \colhead{($10^{22}$ cm$^{-2}$)} &   &  &  \colhead{ (erg/s/cm$^{2}$) }&   \colhead{(erg/s) }
}           
 \startdata  

3XMM J152130.7+074916	&	c1	&	$	0.03		$	&	$	4.50		$	&		&	$	(	6.83		)	\times 10^{	-15	}	$	&	$	(	6.49		)	\times 10^{	41	}	$	\\
ASASSN-15lh	&	c2	&	$	0.03		$	&	$	4.50		$	&		&	$	(	9.52		)	\times 10^{	-16	}	$	&	$	(	1.67		)	\times 10^{	41	}	$	\\
D1-9	&	c3	&	$	0.02		$	&	$	4.50		$	&		&	$	(	2.46		)	\times 10^{	-15	}	$	&	$	(	9.78		)	\times 10^{	41	}	$	\\
D23H-1	&	c4	&	$	0.04		$	&	$	4.50		$	&		&	$	(	3.89		)	\times 10^{	-15	}	$	&	$	(	4.01		)	\times 10^{	41	}	$	\\
D23H-1	&	c5	&	$	0.04		$	&	$	4.50		$	&		&	$	(	2.70		)	\times 10^{	-15	}	$	&	$	(	2.78		)	\times 10^{	41	}	$	\\
D3-13	&	c6	&	$	0.01		$	&	$	4.50		$	&		&	$	(	5.91	_{-	4.3	}	^{+	7.5	}	)	\times 10^{	-15	}	$	&	$	(	3.22	_{-	2.4	}	^{+	4.1	}	)	\times 10^{	42	}	$	\\
D3-13	&	c7	&	$	0.01		$	&	$	4.50		$	&		&	$	(	3.54		)	\times 10^{	-15	}	$	&	$	(	1.93		)	\times 10^{	42	}	$	\\
GRB060218, SN2006aj	&	c8	&	$	0.11		$	&	$	2.38		$	&		&	$	(	2.71		)	\times 10^{	-15	}	$	&	$	(	6.93		)	\times 10^{	39	}	$	\\
\enddata              
 \tablecomments{All uncertainties correspond to the 90\% confidence level and observations that have similar MJD are combined.}              
 \tablenotetext{a}{$N_{H}$ is determined either directly from fitting the X-ray emitting spectrum or using the column density derived from the Leiden/Argentine/Bonn (LAB) Survey of Galactic H$\textsc{i}$ (Kalberla et al. 2005)}              
 \tablenotetext{b}{Powerlaw index was derived from either fitting X-ray spectra from either \textit{ROSAT}, \textit{XMM} or \textit{Chandra} with a powerlaw model or was assumed to be equal to 4.5. If the following $\Gamma$ has an uncertainty, the powerlaw index was derived from fitting the X-ray spectrum extracted from \textit{ROSAT}.}              
 \tablenotetext{c}{Absorbed X-ray flux and X-ray luminosity is calculated for an energy range of 0.3-2.0 keV.}           
 \tablenotetext{d}{Value frozen to that obtained from modelling the \textit{Chandra} X-ray spectrum.}
 \tablenotetext{e}{Value set to the average of that obtained from modelling multiple \textit{Chandra} spectra.}

\end{deluxetable*}
\clearpage

%% file: xmm_table_fits.tex
\clearpage
\LongTables
\begin{deluxetable*}{lcccccccccccccc}
\tabletypesize{\footnotesize}
	\tablecolumns{7} 
	\tablewidth{0pt}
	\tablecaption{The best fit, or assumed parameters for our absorbed powerlaw model that was used to derive the 0.3-2.0 keV X-ray flux and luminosity of each TDE candidate from the \emph{XMM-Newton pointed} observations overlapping the position of the source. \label{xmmfits}}
	\tablehead{ \colhead{Name} &  \colhead{ObsID} &   \colhead{$N_{H}^a$} &  \colhead{$\Gamma$$^{b}$} &  \colhead{$\chi^{2}$ (dof) }&  \colhead{X-ray flux$^{c}$}&  \colhead{X-ray Luminosity$^{c}$} \\
		&  \colhead{label}  &  \colhead{($10^{22}$ cm$^{-2}$)} &   &  &  \colhead{ (erg/s/cm$^{2}$) }&   \colhead{(erg/s) }
	}           
	\startdata  

2MASXJ0203	&	x1	&	$	0.02		$	&	$	1.62	_{-0.10	}	^{+	0.11	}	$	&	0.99 (53)	&	$	(	2.52	_{-	2.4	}	^{+	2.6	}	)	\times 10^{	-13	}	$	&	$	(	2.29	_{-	2.2	}	^{+	2.4	}	)	\times 10^{	42	}	$	\\
2MASX J0249	&	x2	&	$	0.44	_{-0.14	}	^{+	0.15	}	$	&	$	7.70	_{-0.92	}	^{+	1.06	}	$	&	0.96 (53)	&	$	(	4.37	_{-	4.0	}	^{+	4.7	}	)	\times 10^{	-13	}	$	&	$	(	3.35	_{-	3.0	}	^{+	3.6	}	)	\times 10^{	41	}	$	\\
3XMM J152130.7+074916	&	x3	&	$	0.03		$	&	$	5.18	_{-0.30	}	^{+	0.31	}	$	&	1.36(72)	&	$	(	3.32	_{-	3.1	}	^{+	3.7	}	)	\times 10^{	-13	}	$	&	$	(	3.15	_{-	2.9	}	^{+	3.5	}	)	\times 10^{	43	}	$	\\
3XMM J152130.7+074916	&	x4	&	$	0.03		$	&	$	5.18		$	&		&	$	(	5.76		)	\times 10^{	-15	}	$	&	$	(	5.47		)	\times 10^{	41	}	$	\\
ASASSN-14li	&	x5	&	$	0.06		$	&	$	10.00	_{-0.50	}	^{f}	$	&	1.52 (100)	&	$	(	5.43	_{-	5.4	}	^{+	5.7	}	)	\times 10^{	-11	}	$	&	$	(	5.12	_{-	5.1	}	^{+	5.4	}	)	\times 10^{	43	}	$	\\
ASASSN-15lh	&	x6	&	$	0.03		$	&	$	4.50		$	&		&	$	(	6.42		)	\times 10^{	-14	}	$	&	$	(	1.12		)	\times 10^{	43	}	$	\\
D1-9	&	x7	&	$	0.02		$	&	$	4.50		$	&		&	$	(	1.70		)	\times 10^{	-14	}	$	&	$	(	6.76		)	\times 10^{	42	}	$	\\
D23H-1	&	x8	&	$	0.04		$	&	$	4.50		$	&		&	$	(	7.51		)	\times 10^{	-15	}	$	&	$	(	7.74		)	\times 10^{	41	}	$	\\
\enddata              
	\tablecomments{All uncertainties correspond to the 90\% confidence level and observations that have similar MJD are combined.}              
	\tablenotetext{a}{$N_{H}$ is determined either directly from fitting the X-ray emitting spectrum or using the column density derived from the Leiden/Argentine/Bonn (LAB) Survey of Galactic H$\textsc{i}$ (Kalberla et al. 2005)}              
	\tablenotetext{b}{Powerlaw index was derived from either fitting X-ray spectra from either \textit{ROSAT}, \textit{XMM} or \textit{Chandra} with a powerlaw model or was assumed to be equal to 4.5. If the following $\Gamma$ has an uncertainty, the powerlaw index was derived from fitting the X-ray spectrum extracted from \textit{ROSAT}.}              
	\tablenotetext{c}{Absorbed X-ray flux and X-ray luminosity is calculated for an energy range of 0.3-2.0 keV.}           
	\tablenotetext{d}{Value frozen to that obtained from modelling the \textit{XMM-Newton} X-ray spectrum.}
	\tablenotetext{e}{Value set to the average of that obtained from modelling multiple \textit{XMM-Newton} spectra.}
	\tablenotetext{f}{When modelling, we find that this parameter hits an upperlimit.}
	
\end{deluxetable*}
\clearpage

%% file: xmm_slew_table_fits.tex
\clearpage
\LongTables
\begin{deluxetable*}{lcccccccccccccc}
\tabletypesize{\footnotesize}
\tablecolumns{7} 
	\tablewidth{0pt}
	\tablecaption{The best fit, or assumed parameters for our absorbed powerlaw model that was used to derive the 0.3-2.0 keV X-ray flux and luminosity of each TDE candidates from the \emph{XMM-Newton slew} observations overlapping the position of the source. \label{xmmslewfits}}
	\tablehead{ \colhead{Name} &  \colhead{ObsID} &   \colhead{$N_{H}^a$} &  \colhead{$\Gamma$$^{b}$} &  \colhead{$\chi^{2}$ (dof) }&  \colhead{X-ray flux$^{c}$}&  \colhead{X-ray Luminosity$^{c}$} \\
		&  \colhead{label}  &  \colhead{($10^{22}$ cm$^{-2}$)} &   &  &  \colhead{ (erg/s/cm$^{2}$) }&   \colhead{(erg/s) }
	}           
	\startdata

2MASXJ0203	&	xs1	&	$	0.02		$	&	$	4.50		$	&		&	$	(	1.84	_{-	1.4	}	^{+	2.3	}	)	\times 10^{	-11	}	$	&	$	(	1.67	_{-	1.3	}	^{+	2.0	}	)	\times 10^{	44	}	$	\\
3XMM J152130.7+074916	&	xs2	&	$	0.03		$	&	$	5.18		$	&		&	$	(	1.70		)	\times 10^{	-11	}	$	&	$	(	1.61		)	\times 10^{	45	}	$	\\
3XMM J152130.7+074916	&	xs3	&	$	0.03		$	&	$	5.18		$	&		&	$	(	7.59		)	\times 10^{	-12	}	$	&	$	(	7.21		)	\times 10^{	44	}	$	\\
ASASSN-14ae	&	xs4	&	$	0.02		$	&	$	4.50		$	&		&	$	(	3.49		)	\times 10^{	-12	}	$	&	$	(	1.54		)	\times 10^{	43	}	$	\\
ASASSN-14ae	&	xs5	&	$	0.02		$	&	$	4.50		$	&		&	$	(	2.39		)	\times 10^{	-12	}	$	&	$	(	1.06		)	\times 10^{	43	}	$	\\
ASASSN-14ae	&	xs6	&	$	0.02		$	&	$	4.50		$	&		&	$	(	2.19		)	\times 10^{	-12	}	$	&	$	(	9.67		)	\times 10^{	42	}	$	\\
ASASSN-14ae	&	xs7	&	$	0.02		$	&	$	4.50		$	&		&	$	(	1.88		)	\times 10^{	-12	}	$	&	$	(	8.30		)	\times 10^{	42	}	$	\\
ASASSN-14li	&	xs8	&	$	0.02		$	&	$	4.50		$	&		&	$	(	2.53		)	\times 10^{	-12	}	$	&	$	(	2.39		)	\times 10^{	42	}	$	\\
\enddata              
	\tablecomments{All uncertainties correspond to the 90\% confidence level and observations that have similar MJD are combined.}              
	\tablenotetext{a}{$N_{H}$ is determined either directly from fitting the X-ray emitting spectrum or using the column density derived from the Leiden/Argentine/Bonn (LAB) Survey of Galactic H$\textsc{i}$ (Kalberla et al. 2005)}              
	\tablenotetext{b}{Powerlaw index was derived from either fitting X-ray spectra from either \textit{ROSAT}, \textit{XMM} or \textit{Chandra} with a powerlaw model or was assumed to be equal to 4.5. If the following $\Gamma$ has an uncertainty, the powerlaw index was derived from fitting the X-ray spectrum extracted from \textit{ROSAT}.}              
	\tablenotetext{c}{Absorbed X-ray flux and X-ray luminosity is calculated for an energy range of 0.3-2.0 keV.}           
	\tablenotetext{d}{Value frozen to that obtained from modelling the \textit{XMM-Newton} X-ray spectrum.}
	\tablenotetext{e}{Value set to the average of that obtained from modelling multiple \textit{XMM-Newton} spectra.}
	
\end{deluxetable*}
\clearpage

%% file: swift_table_fits.tex
\clearpage
\LongTables
\begin{deluxetable*}{lcccccccccccccc}
\tabletypesize{\footnotesize}
\tablecolumns{7} 
	\tablewidth{0pt}
	\tablecaption{The best fit, or assumed parameters for our absorbed powerlaw model that was used to derive the 0.3-2.0 keV X-ray flux and luminosity of each TDE candidate from the \emph{Swift XRT} observations overlapping the position of the source. \label{swiftfits}}
	\tablehead{ \colhead{Name} &  \colhead{ObsID} &   \colhead{$N_{H}^a$} &  \colhead{$\Gamma$$^{b}$} &  \colhead{$\chi^{2}$ (dof) }&  \colhead{X-ray flux$^{c}$}&  \colhead{X-ray Luminosity$^{c}$} \\
		&  \colhead{label}  &  \colhead{($10^{22}$ cm$^{-2}$)} &   &  &  \colhead{ (erg/s/cm$^{2}$) }&   \colhead{(erg/s) }
	}           
	\startdata  
2MASXJ0203	&	sw1	&	$	0.02		$	&	$	1.40	_{-0.33	}	^{+	0.35	}	$	&	1.30 (4)	&	$	(	4.01	_{-	3.5	}	^{+	4.6	}	)	\times 10^{	-13	}	$	&	$	(	3.65	_{-	3.2	}	^{+	4.2	}	)	\times 10^{	42	}	$	\\
2MASXJ0203	&	sw2	&	$	0.02		$	&	$	1.28	_{-0.46	}	^{+	0.46	}	$	&	1.30 (2)	&	$	(	4.61	_{-	4.0	}	^{+	5.3	}	)	\times 10^{	-13	}	$	&	$	(	4.19	_{-	3.6	}	^{+	4.8	}	)	\times 10^{	42	}	$	\\
2MASX J0249	&	sw3	&	$	0.44		$	&	$	4.50		$	&		&	$	(	1.30		)	\times 10^{	-13	}	$	&	$	(	9.95		)	\times 10^{	40	}	$	\\
2MASX J0249	&	sw4	&	$	0.44		$	&	$	4.50		$	&		&	$	(	1.21	_{-	0.4	}	^{+	2.0	}	)	\times 10^{	-13	}	$	&	$	(	9.26	_{-	3.3	}	^{+	15.2	}	)	\times 10^{	40	}	$	\\
2MASX J0249	&	sw5	&	$	0.44		$	&	$	4.50		$	&		&	$	(	7.52	_{-	3.1	}	^{+	11.9	}	)	\times 10^{	-14	}	$	&	$	(	5.76	_{-	2.4	}	^{+	9.1	}	)	\times 10^{	40	}	$	\\
2MASX J0249	&	sw6	&	$	0.44		$	&	$	4.50		$	&		&	$	(	1.15		)	\times 10^{	-13	}	$	&	$	(	8.81		)	\times 10^{	40	}	$	\\
2MASX J0249	&	sw7	&	$	0.44		$	&	$	4.50		$	&		&	$	(	1.07		)	\times 10^{	-13	}	$	&	$	(	8.19		)	\times 10^{	40	}	$	\\
2MASX J0249	&	sw8	&	$	0.44		$	&	$	4.50		$	&		&	$	(	8.24		)	\times 10^{	-14	}	$	&	$	(	6.31		)	\times 10^{	40	}	$	\\
\enddata              
	\tablecomments{All uncertainties correspond to the 90\% confidence level and observations that have similar MJD are combined.}              
	\tablenotetext{a}{$N_{H}$ is determined either directly from fitting the X-ray emitting spectrum or using the column density derived from the Leiden/Argentine/Bonn (LAB) Survey of Galactic H$\textsc{i}$ (Kalberla et al. 2005)}              
	\tablenotetext{b}{Powerlaw index was derived from either fitting X-ray spectra from either \textit{ROSAT}, \textit{XMM} or \textit{Chandra} with a powerlaw model or was assumed to be equal to 4.5. If the following $\Gamma$ has an uncertainty, the powerlaw index was derived from fitting the X-ray spectrum extracted from \textit{ROSAT}.}              
	\tablenotetext{c}{Absorbed X-ray flux and X-ray luminosity is calculated for an energy range of 0.3-2.0 keV.}           
	\tablenotetext{d}{Value frozen to that obtained from modelling the \textit{Swift} X-ray spectrum.}
	\tablenotetext{e}{Value set to the average of that obtained from modelling multiple \textit{Swift} spectra.}
	\tablenotetext{f}{When modeling, we find that this parameter hits an upperlimit.}
\end{deluxetable*}
\clearpage

%% file: t90l90.tex
\begin{table}[t]
	\begin{center}
		\caption{The $T_{90}$ and $L_{90}$ values derived for our \textit{X-ray TDE} and \textit{likely X-ray TDE} candidates. \label{t90l90values}}
	\begin{tabular}{lllll}
		\hline
Name &$ T_{90} (10^{7}$ seconds) & $L_{90}$ (erg/s)\\
\hline\hline
ASASSN-14li	&$	3.888	\pm	0.022	$&	$	(	3.45	\pm	1.25	)	\times	10^{	42	}	$	\\
Swift J1644+57	&$	0.678	\pm	0.263	$&	$	(	5.83	\pm	23.85	)	\times	10^{	45	}	$	\\
Swift J2058+05	&$	2.129	\pm	0.162	$&	$	(	4.63	\pm	2.43	)	\times	10^{	45	}	$	\\
XMMSL1 J0740-85	&$	2.121	\pm	0.073	$&	$	(	4.41	\pm	4.75	)	\times	10^{	41	}	$	\\
\hline
2MASX J0249	&$	2.718	\pm	0.005	$&	$	(	8.54	\pm	1.42	)	\times	10^{	40	}	$	\\
3XMM J152130.7+074916	&$	4.175	\pm	0.004	$&	$	(	6.38	\pm	0.23	)	\times	10^{	42	}	$	\\
IGR J17361-4441	&$	1.517	\pm	0.259	$&	$	(	4.66	\pm	4.52	)	\times	10^{	42	}	$	\\
NGC247	&$	1.497	\pm	0.003	$&	$	(	5.31	\pm	4.54	)	\times	10^{	37	}	$	\\
OGLE16aaa	&$	1.428	\pm	0.002	$&	$	(	4.70	\pm	1.86	)	\times	10^{	42	}	$	\\
PTF-10iya	&$	0.721	\pm	0.011	$&	$	(	2.26	\pm	3.61	)	\times	10^{	42	}	$	\\
SDSS J1201	&$	1.242	\pm	0.143	$&	$	(	3.43	\pm	16.61	)	\times	10^{	42	}	$	\\
SDSS J1311	&$	4.250	\pm	0.010	$&	$	(	1.60	\pm	0.17	)	\times	10^{	42	}	$	\\
SDSS J1323	&$	1.987	\pm	0.689	$&	$	(	5.37	\pm	10.71	)	\times	10^{	42	}	$	\\
	\hline
\end{tabular}
\end{center}
\end{table}

%% file: nh_nhgalc_values.tex
\begin{table}[t]
	\begin{center}
		\caption{The ratio of the measured $N_{\rm H}$ derived from modelling the X-ray spectra from each event and the corresponding Galactic $N_{\rm H}$ as derived from the Leiden/Argentine/Bonn (LAB) Survey of Galactic H$\textsc{i}$ \citep{2005A&A...440..775K}. These values were used to produce Figure~\ref{nhvshr}. \label{nhmnhgalcvalue}}
	\begin{tabular}{lcccc}
		\hline
Name &MJD& Instrument & $N_{\rm H}(\rm measured)/N_{\rm H}( \rm Galactic)$\\
\hline\hline
ASASSN-14li	&	56994.7	&	swift	&$	7.64	\pm	1.77	$	\\
ASASSN-14li	&	57046.7	&	swift	&$	8.75	\pm	0.63	$	\\
ASASSN-14li	&	57145.8	&	swift	&$	8.48	\pm	1.11	$	\\
ASASSN-14li	&	57228.1	&	swift	&$	8.56	\pm	2.25	$	\\
ASASSN-14li	&	57364.5	&	swift	&$	8.31	\pm	0.88	$	\\
ASASSN-14li	&	57425.5	&	swift	&$	4.92	\pm	3.64	$	\\
ASASSN-14li	&	57538.5	&	swift	&$	2.86	\pm	4.51	$	\\
\hline
\end{tabular}
\end{center}
\end{table}

%% file: hrvalues.tex
\begin{table}[t]
	\begin{center}
		\caption{The derived hardness ratio (HR) used to produce Figures~\ref{nhvari}, ~\ref{hr}and ~\ref{nhvshr}. Here HR 	is defined as (H-S)/(H+S), where H is the counts in the equivalent 2.0-10.0 keV energy band and S is the counts in the equivalent 0.3-2.0 keV energy band. There values were taken from Tables  \ref{rosathms}--\ref{swifthms}. Here we have not listed any observations for which we derive upperlimits for the HR ratio. Here uncertainties are one sigma. \label{hrvalues}}
	\begin{tabular}{lcccc}
		\hline
Name &MJD& Instrument & Hardness Ratio \\
\hline\hline
ASASSN-14li	&	56994.7	&	swift	&$	-0.995	\pm	0.225	$ \\
ASASSN-14li	&	56999.5	&	xmm	&$	-0.907	\pm	0.006	$ \\
ASASSN-14li	&	57046.7	&	swift	&$	-0.996	\pm	0.094	$ \\
ASASSN-14li	&	57145.8	&	swift	&$	-0.993	\pm	0.227	$ \\
ASASSN-14li	&	57228.1	&	swift	&$	-0.984	\pm	0.027	$ \\
ASASSN-14li	&	57364.5	&	swift	&$	-0.981	\pm	0.026	$ \\
ASASSN-14li	&	57425.5	&	swift	&$	-0.965	\pm	0.044	$ \\
\hline
\end{tabular}
\end{center}
\end{table}

%% file: opticaluvxrayratio.tex
\begin{table}[t]
	\begin{center}
		\caption{The integrated optical/UV and X-ray luminosities used to produce Figure~\ref{opticaltoxray}. \label{xtooratio}}
	\begin{tabular}{lcc}
		\hline
Name &X-ray luminosity [erg/s]  & Optical/UV  luminosity [erg/s]\\
 & (0.3-10.0 keV) & (0.002-0.1 keV) \\
\hline\hline
ASASSN-14li	&$(	3.37	\pm	0.01	)	\times	10^{	42	}	$	&$(	1.78	\pm	0.13	)	\times	10^{	42	}	$	\\
Swift J1644+57	&$(	3.51	\pm	1.69	)	\times	10^{	47	}	$	&$(	4.72	\pm	0.41	)	\times	10^{	40	}	$	\\
Swift J2058+05	&$(	4.97	\pm	4.03	)	\times	10^{	48	}	$	&$(	1.73	\pm	0.04	)	\times	10^{	43	}	$	\\
XMMSL1 J0740-85	&$(	0.94	\pm	1.90	)	\times	10^{	43	}	$	&$(	9.12	\pm	0.56	)	\times	10^{	40	}	$	\\
\hline

NGC247	&$(	0.75	\pm	0.27	)	\times	10^{	40	}	$	&$(	2.61	\pm	0.32	)	\times	10^{	38	}	$	\\
OGLE16aaa	&$(	3.61	\pm	1.56	)	\times	10^{	43	}	$	&$(	1.05	\pm	0.09	)	\times	10^{	43	}	$	\\
PTF-10iya	&$(	1.13	\pm	1.93	)	\times	10^{	43	}	$	&$(	3.03	\pm	0.11	)	\times	10^{	43	}	$	\\
SDSS J1201	&$(	23.38	\pm	19.71	)	\times	10^{	44	}	$	&$(	2.64	\pm	0.31	)	\times	10^{	42	}	$	\\
	\hline
\end{tabular}
\end{center}
\end{table}